\documentclass{aa}
\usepackage[varg]{txfonts}
\pdfoutput=1
\usepackage{graphicx}
\usepackage{multirow}
\usepackage[colorlinks=true,linkcolor=blue,citecolor=blue,urlcolor=blue]{hyperref}

\newcommand{\orcidicon}[1]{\unskip\protect\href{https://orcid.org/#1}{\protect\includegraphics[width=8pt,clip]{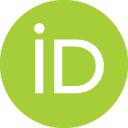}}}

\defcitealias{springel2010pur}{AREPO}
\defcitealias{Springel2005Gadget}{GADGET-3}
\defcitealias{hopkins15}{GIZMO}
\defcitealias{planck2015_xiii}{Planck 2015}
\defcitealias{Planck2014}{Planck 2013}

\begin{document} 

\title{And Then There WEre Baryons (ATWEB): The Baryon Budget of the Universe in Halos and the IGM over 13 Billion Years of Cosmic Time}

\titlerunning{ATWEB: The Open Baryon Cycle and the Baryon Budget of the Universe}

\author{M. Ayromlou
      \inst{1}\thanks{\email{ayromlou@uni-bonn.de}}\orcidicon{0000-0003-3783-2321}
      \and G. M. Voit\inst{2}\orcidicon{0000-0002-3514-0383}
      \and T. H. Reiprich\inst{1,3}\orcidicon{0000-0003-2047-2884}
      \and C. Porciani\inst{1,3}\orcidicon{0000-0002-7797-2508} \and D. Eckert\inst{4}\orcidicon{0000-0001-7917-3892}
      }

\institute{Argelander-Institut f\"ur Astronomie, Auf dem H\"ugel 71, D-53121 Bonn, Germany \label{1}
\and Michigan State University, Department of Physics and Astronomy, East Lansing, MI 48824, USA \label{2}
\and Cluster of Excellence “Our Dynamic Universe” (DYNAVERSE) \label{3}
\and Department of Astronomy, University of Geneva, Ch. d’Ecogia 16,
1290 Versoix, Switzerland \label{4}}

\date{}
 
\abstract
{
The distribution of baryonic matter in the Universe is shaped by more complex physics than the gravity-only dark matter. Observations of the local Universe show that, except in the most massive galaxy clusters, the baryon fraction within halos falls below the cosmic mean. This deficit can be characterized using the previously established closure radius, the halocentric distance within which the enclosed baryon fraction reaches the cosmic value. Using the TNG, EAGLE and SIMBA cosmological hydrodynamical simulations, we examine how, when, and on what scales the baryonic and dark matter distributions diverge. At $z=0$ less than $30\%$ of cosmic baryons reside within galaxy-hosting halos, compared with $\sim$45\% of the dark matter, leaving over $70\%$ of all baryons in the intergalactic medium (IGM).
Building on the closure radius, we introduce three further characteristic scales: the compensation radius, closure envelope, and compensation envelope. The compensation radius marks where the enclosed total mass recovers its gravity-only value, and today it typically lies inside the closure radius in all three models. The closure and compensation envelopes are the largest median radii across halo mass. Galaxy groups provide the largest contribution per dex of halo mass to IGM enrichment with halo-associated baryons and redistribute baryons over the largest distances. They set the closure envelope, which reaches $\sim2$--$6.5\,\mathrm{Mpc}$ at $z=0$, depending on the model. The closure radius usually increases with cosmic time at fixed halo mass. Tracking individual halos, we find open baryon cycles with substantial redistribution beyond halo boundaries across all halo masses, while even the progenitors of today's most massive clusters fall below the cosmic baryon fraction. The compensation envelope grows toward low redshift, reaching $\sim1.5$--$5.5\,\mathrm{Mpc}$ at $z=0$. Across cosmic time, it coincides with the scale on which the matter power spectrum recovers its gravity-only form, establishing a direct link between the spatial extent of baryon redistribution and the impact of feedback on large-scale structure.
}

\keywords{Galaxies: halos -- Galaxies: groups -- Galaxies: evolution -- intergalactic medium -- large-scale structure of Universe}

\maketitle

\section{Introduction}
\label{sec: Introduction}
In the $\Lambda$CDM cosmology, structures form through the gravitational collapse of overdense regions in the primordial matter density field \citep{Peebles1980large,Blumenthal1984Formation}. The anisotropic collapse gives rise to the large-scale cosmic web, in which sheets and filaments channel matter into the nodes where more massive halos form \citep{Zeldovich1970Gravitational,Shandarin1989Large,Bond1996How}. The spatial distribution of baryons and their subsequent redistribution by non-gravitational processes are thus shaped by the growth of this web-like structure.

Baryons and dark matter originated from the same primordial density fluctuations \citep[e.g. see][for a review]{mo2010galaxy}. In the hierarchical structure formation scenario, dark matter structures form first, and baryons subsequently fall into their gravitational potential wells, where they cool, form stars, and build galaxies \citep{white1978core,white1991galaxy}. Over time, baryons become subject to a variety of non-gravitational processes, such as radiative cooling, photoionization heating, star formation, and feedback from supernovae and active galactic nuclei (AGN), which cause their distribution to depart from that of the dark matter, particularly on small scales \citep{Silk1997Feedback,Springel2003,Voit2005Tracing,schaye10}.

In the context of galaxy formation and evolution, baryons in galaxies and their halos are often considered to be part of a baryon cycle, in which gas accretes from the intergalactic medium (IGM) through the circumgalactic medium (CGM) into the interstellar medium (ISM). Feedback-driven outflows return gas to the CGM, from which some of it may later be reaccreted by the galaxy \citep{Reiprich2013Outskirts,Tumlinson2017Circumgalactic,Donahue2022Baryon,Faucher-Giguere2023Key,Ramesh2026Magnetogenesis}. This cycle, however, is not necessarily closed: feedback can transport gas beyond the halo boundary, and some of this material may never return to its original halo. Simulations show that gas ejected in winds can be accreted by other galaxies \citep{Angles-Alcazar2017cosmic}, and that baryons can be transferred between the Lagrangian regions associated with different halos \citep{Borrow2020Cosmological}. We refer to this picture as an ``open baryon cycle'' (see Fig.~\ref{Fig: open_baryon_cycle}) and discuss its implications for halo baryon deficits and large-scale baryon redistribution throughout this paper.

Cosmological parameters inferred from the cosmic microwave background (CMB) imply a universal baryon fraction of $f_{\rm b,cosmic}=\Omega_{\rm b}/\Omega_{\rm m}\simeq0.16$ \citep{Planck2020cosmic}. Multiwavelength inventories of galaxies and groups, as well as many clusters, recover baryon fractions below this cosmic mean within standard apertures ($R_{\rm 500c}$ or $R_{\rm 200c}$) with the shortfall most pronounced at low halo masses, $\log_{10} (M_{\rm 200c}/\mathrm{M_{\rm \odot}})<14$ \citep{Persic1992baryon,Fukugita2004Cosmic,pratt2009galaxy,sun2009chandra,McGaugh2010cosmological,Eckert2021Feedback}. Cosmological hydrodynamical simulations show that stellar and AGN feedback can lower the baryon fraction, $f_{\rm b,200c}$, both by ejecting gas beyond the halo boundary and by preventing gas from accreting \citep{Ayromlou2021Comparing,Sorini2022How,Mitchell2022Baryonic,Davies2020Quenching,Davies2026COLIBRE}.

These halo baryon deficits are one aspect of the broader \emph{missing-baryons problem}, which is commonly discussed in two related senses. \emph{(i) Cosmic missing baryons}: the historical shortfall between the baryon density recovered from low-redshift inventories of stars and gas and that inferred from the CMB, which motivated searches for diffuse warm/hot gas in the CGM and IGM \citep{Shull2012Baryon,Nicastro2018Observations,Tanimura2019Detection,Zhang2024SRGeROSITA}. Measurements using the dispersion of fast radio bursts now recover a cosmic baryon density consistent with the CMB constraints \citep{Macquart2020Census,Connor2025Gas}. \emph{(ii) Halo baryon deficit}: within individual halos, the retained baryon fraction inside a specified aperture, such as $R_{\rm 500c}$ or $R_{\rm 200c}$, is typically below the cosmic mean $f_{\rm b,cosmic}$. In this sense, halos are ``missing'' the baryons they would contain if baryons followed the dark matter distribution with the cosmic abundance ratio. This paper connects the two aspects: we quantify how the baryons missing from halos relate to the cosmic baryon budget and to the baryons residing in the IGM.

Not all dark matter resides within galaxy-hosting halos; a substantial fraction lies in filamentary structures, sheets, and voids \citep{Aragon-Calvo2010Multiscale,Cautun2014Evolution}\footnote{In $\Lambda$CDM, the fraction of total dark matter in collapsed halos is higher than in galaxy-hosting halos, since halos span roughly 20 orders of magnitude in mass. Integrating the halo mass function over this range places $69$--$73\%$ in halos within $R_{\rm 200c}$ \citep[e.g. see][]{Zheng2026Dark}. We return to this point in Sect.~\ref{subsec: halo_contribution}.}. If baryons perfectly traced the dark matter distribution, a significant fraction of baryons would still inherently reside outside halos, in the IGM \citep[e.g.][]{Fukugita1998Cosmic, Hahn2007Properties, Tinker2008Toward, Ludlow2011Peaks, Libeskind2018Tracing}. However, a net baryon deficit within halos corresponds to an even larger fraction of cosmic baryons outside them relative to this expectation. Recent observational estimates using fast radio bursts (FRBs) suggest that $\sim 76\%$ of cosmic baryons reside in the IGM at $z<1$ \citep{Connor2025Gas}. Complementarily, recent X-ray observations with eROSITA have directly detected anisotropic distributions of warm/hot gas extending beyond $R_{\rm 200c}$, and even beyond $R_{\rm 100c}$, providing density, temperature, and metallicity measurements at these large radii \citep{Reiprich2021Abell3391,Dietl2024Discovery,Veronica2024eROSITA,Reiprich2025Fornax}.

The baryons missing from halos can be found by extending the search radius beyond the halo boundary. This characteristic scale, known as the \emph{closure radius} ($R_{\rm c}$), is defined as the halocentric distance within which the baryon fraction equals the cosmic mean \citep{Ayromlou2023Feedback}. These baryons have either been ejected or prevented from accretion or, most likely, a combination of the two. With current facilities, direct measurements of $R_{\rm c}$ are challenging. Nevertheless, indirect estimates have been attempted, for instance by extrapolating the X-ray gas density profiles of halos to larger radii \citep{Zhang2024Hot} or by combining Sunyaev-Zel'dovich (SZ) effect measurements with CMB lensing data \citep{Hadzhiyska2025Missing}. However, these measurements remain highly uncertain due to assumptions about both the gas and the dark matter distribution at large radii.

On the theoretical side, \cite{Ayromlou2023Feedback} introduced a universal scaling relation connecting the amplitude of the closure radius to the halo baryon fraction. Its functional form is robust across all modern cosmological hydrodynamical simulations, but its slope and redshift dependence vary with the implementation of feedback processes. Here, we extend this work by examining the evolution of the closure radius over cosmic time.

As a consequence of baryonic processes like feedback redistributing matter beyond individual halos, the total matter power spectrum is suppressed relative to gravity-only simulations on scales from sub-megaparsec to several megaparsecs \citep[e.g.][]{Schneider2015New,Chisari2018Impact,vanDaalen2020Exploring,Reischke2026First}. This suppression is a key uncertainty in precision cosmology, particularly for weak lensing surveys \citep{Huang2019Modelling,Martinelli2021Euclid,Secco2024Dark}. The same processes also affect galaxy formation \citep{Naab2017Theoretical,To2024Deciphering,Fu2025Role,Lucie-Smith2025Cosmological}. The scale and amplitude of power spectrum suppression depend on the details of baryonic physics. In particular, the mean baryon fraction in galaxy clusters and groups has been shown to predict this suppression in simulations \citep{vanDaalen2020Exploring,Miller2026Baryonic}. Therefore, interpreting large-scale structure observations and constraining galaxy formation models require understanding how the distributions of baryons and dark matter are displaced from one another across scales, and how this displacement evolves over time.

This work is part of the \textbf{ATWEB} (And Then There WEre Baryons) project, which aims to understand the distribution and dynamics of baryons and their departures from gravity-only dark matter across cosmic time. In this paper, we focus on identifying the missing baryons, both those inherently in the IGM and those displaced from halos. We study how, when, and on what scales baryons and dark matter are displaced from one another.

Building upon the closure radius, we present three new characteristic scales: the compensation radius ($R_{\rm comp}$), the innermost radius at which the enclosed total mass matches the gravity-only counterpart, and the closure and compensation envelopes ($R_{\rm c}^{\rm env}$ and $R_{\rm comp}^{\rm env}$), each defined as the maximum of the mass-binned median of the corresponding radius at a given redshift. $R_{\rm c}^{\rm env}$ marks the largest distance to which halo baryon deficits extend, and $R_{\rm comp}^{\rm env}$ the largest distance to which baryonic processes have a net effect on the enclosed total mass.

This paper is structured as follows: We first describe the methodology, including the simulations used, halo identification, and relevant measurements in Sect.~\ref{sec: Methodology}. We then present our main results: Sect.~\ref{sec: Baryon_Budget} quantifies the cosmic baryon budget and the contributions from different halo masses; Sect.~\ref{sec: Characteristic_Radii} examines the characteristic scales of baryon--dark matter redistribution; and Sect.~\ref{sec: LSS_Implications} examines the implications for large-scale structure, including the suppression of the matter power spectrum and the impact on halo masses. Sect.~\ref{sec: Summary} summarizes our findings and concludes.

\section{Methodology}
\label{sec: Methodology}
\subsection{Simulations}
\label{subsec: Simulations}
We use three publicly available cosmological hydrodynamical simulations: IllustrisTNG, EAGLE, and SIMBA, as well as their dark matter-only (DMO) counterparts. We choose them because (i) they have been widely studied and tested, and (ii) they differ significantly in their subgrid physics, particularly feedback mechanisms that are crucial for baryon redistribution. Maps of the baryon fraction and matter distribution in the three simulations at $z=0$ are shown in Appendix~\ref{app: matter_distribution}.

The simulations employ different codes for hydrodynamics and gravity, and distinct subgrid models. Nevertheless, they all contain key processes to model galaxy formation and evolution in a cosmological context: gas cooling, star formation, stellar feedback, stellar evolution and chemical enrichment, UV background radiation, and supermassive black hole formation (seeding), evolution (growth and merger), and feedback. Moreover, all three have been calibrated to reproduce key low-redshift galaxy observables, including the galaxy stellar mass function.

The three simulations identify halos with the Friends-of-Friends (FoF) algorithm \citep{Davis1985TheEvolution} and take the position of the most bound particle as the halo center, although they vary in detail\footnote{They also use slightly different algorithms to identify galaxies. Since galaxies are not used in this work, we do not discuss them further.}. To assess the impact of halo identification on our results, we compare the dark matter budgets of the three models in Sect.~\ref{subsec: baryon_budget} and their gravity-only distributions in Appendix~\ref{app: DMO_dist}.

In most cosmological simulations, including those used here, the dominant drivers of large-scale baryon redistribution are UV background radiation in low-mass halos ($8\lesssim \log_{10}(M_{\rm 200c}/\mathrm{M_\odot}) \lesssim 10$), stellar feedback in dwarf hosting halos ($10 \lesssim \log_{10}(M_{\rm 200c}/\mathrm{M_\odot}) \lesssim 12.5$), and AGN feedback in Milky Way-like halos and galaxy groups ($12 \lesssim \log_{10}(M_{\rm 200c}/\mathrm{M_\odot}) \lesssim 14$), while galaxy clusters ($\log_{10}(M_{\rm 200c}/\mathrm{M_\odot}) \gtrsim 14$) remain broadly unaffected on large scales \citep[see][]{Ayromlou2023Feedback}. Below, we provide a brief overview of each simulation. For more details on each model, we refer to the model papers.

\subsubsection{IllustrisTNG}
\label{subsubsec: IllustrisTNG}
\begin{figure*}
    \centering
    \includegraphics[width=0.95\textwidth]{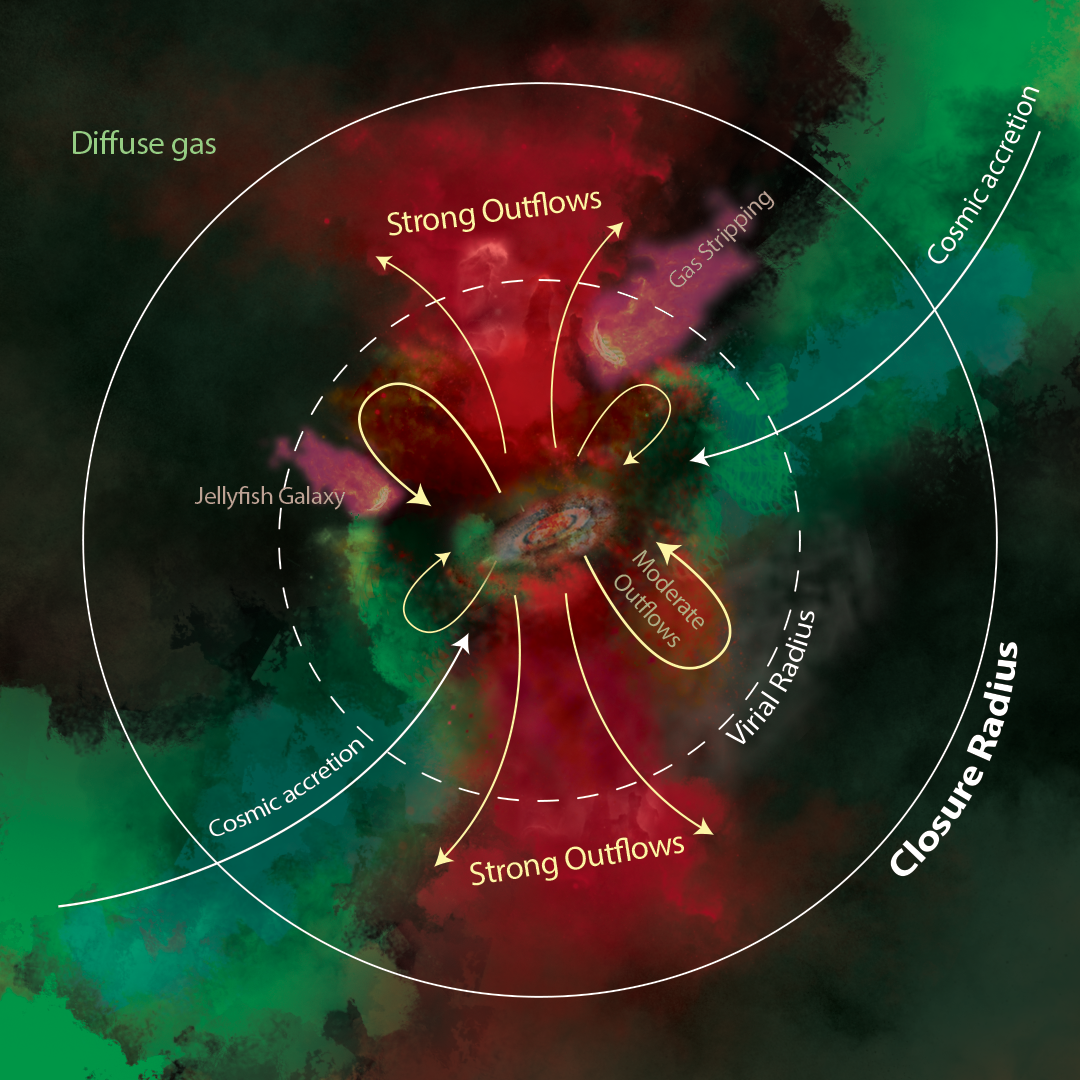}
    \caption{Open Baryon Cycle: The schematic figure illustrates the ``open'' cycle of baryonic matter within and beyond the halo boundary ($R_{\rm 200c}$). The emphasis on an ``open'' cycle is to account for the ejection of baryons beyond the halo boundary, some of which may never return to the halo. The figure highlights the processes of gas accretion, moderate and strong feedback-driven outflows, the halo virial radius, the \textbf{closure radius}, and the possible interaction between the halo environment, including its outskirts, with the galaxies moving through it. The green regions illustrate the inflow of the matter while the red represent the outflows. The dark regions show the diffuse gas with no strong net inflow or outflow, on average. The figure is schematic, represents the case of a galaxy group of $\log_{10}(M_{\rm 200c}/\mathrm{M_{\odot}}) \sim 13.5$, and is not to scale.}
\label{Fig: open_baryon_cycle}
\end{figure*}
The IllustrisTNG \citep[hereafter TNG;][]{marinacci2018first,naiman2018first,nelson2018first,pillepich2018First} project consists of a suite of cosmological magneto-hydrodynamical simulations based on the moving-mesh code \textsc{AREPO} \citep{springel2010pur}. The TNG calibration and model description are presented in \cite{weinberger17} and \cite{pillepich2018Simulating}.

The suite spans three box sizes: 50 Mpc (TNG50), 100 Mpc (TNG100), and 300 Mpc (TNG300). The model was calibrated at TNG100 resolution; we adopt this box (110 cMpc on a side, $m_{\rm DM}=7.5\times10^{6}\,\mathrm{M_\odot}$, $m_{\rm baryon}=1.4\times10^{6}\,\mathrm{M_\odot}$; \citetalias{planck2015_xiii} cosmology) in our analysis. We trace halo assembly with the \textsc{Sublink} merger trees \citep{rodriguez2015merger}.

\subsubsection{EAGLE}
\label{subsubsec: EAGLE}
EAGLE \citep{Crain2015TheEagle,Schaye2015eagle} employs a modified version of the smoothed particle hydrodynamics (SPH) code \textsc{GADGET-3} \citep{Springel2005Gadget}. The EAGLE model has been widely used, including for baryon deficits in halos \citep{Davies2020Quenching} and zooms of galaxy clusters \citep{Bahe2017Hydrangea,Barnes2017Cluster-EAGLE}. We use the main 100 Mpc run ($m_{\rm DM}=9.7\times10^{6}\,\mathrm{M_\odot}$, $m_{\rm baryon}=1.8\times10^{6}\,\mathrm{M_\odot}$; \citetalias{Planck2014} cosmology).

\subsubsection{SIMBA}
\label{subsubsec: SIMBA}
SIMBA is a cosmological hydrodynamical simulation suite \citep{Dave2019SIMBA} based on the \textsc{GIZMO} code \citep{hopkins15}, which uses a meshless finite mass (MFM) method for hydrodynamics. It is built upon its predecessor, the MUFASA simulation \citep{Dave2016MUFASA}. SIMBA includes multiple boxes, resolutions, and physics variations. We use the flagship 147 Mpc run ($m_{\rm DM}=9.6\times10^{7}\,\mathrm{M_\odot}$, $m_{\rm baryon}=1.8\times10^{7}\,\mathrm{M_\odot}$; \citetalias{planck2015_xiii} cosmology). As SIMBA halo masses below $10^{11}\,\mathrm{M_\odot}$ are not reliable, we focus our interpretation of SIMBA above this mass.

\subsection{Halo properties and notation}
\label{subsec: Halo properties}

Throughout this paper we distinguish two conceptually different classes of quantities, and reserve a different symbol for each. Lower-case $f$ denotes a \emph{baryon fraction}, i.e. a baryonic mass divided by the total mass of the same region; upper-case $F$ denotes a \emph{share of the cosmic budget}, i.e. the mass of a component in a given class of region divided by the total cosmic mass of that component. Here we list them explicitly to avoid confusion:
\begin{itemize}
      \item $f_{\rm b,cosmic} \equiv \Omega_{\rm b}/\Omega_{\rm m} \simeq 0.157$: the cosmic baryon fraction.
      \item $f_{\rm b}(<R) \equiv M_{\rm b}(<R)/M_{\rm tot}(<R)$: the baryon fraction enclosed within a halocentric radius $R$. We use $f_{\rm b,200c} \equiv f_{\rm b}(<R_{\rm 200c})$ for the halo baryon fraction, and analogously $f_{\rm b,500c}$ and $f_{\rm gas,500c}$.
      \item $F_{\rm b}^{\rm FoF}$ and $F_{\rm b}^{\rm IGM} = 1 - F_{\rm b}^{\rm FoF}$: the shares of all cosmic baryons that reside inside and outside halos, respectively, with halos defined as FoF groups (see below). We use $F_{\rm DM}^{\rm FoF}$ for the corresponding share of the cosmic dark matter, and $F_{\rm b,gas}^{\rm FoF}$ and $F_{\rm b,\star}^{\rm FoF}$ for the shares of the cosmic \emph{baryon} budget held in halos in each phase, so that $F_{\rm b,gas}^{\rm FoF} + F_{\rm b,\star}^{\rm FoF} = F_{\rm b}^{\rm FoF}$.
\end{itemize}

Here $R_{\rm 200c}$ ($R_{\rm 500c}$) is the radius within which the mean enclosed density equals 200 (500) times the critical density of the Universe at that redshift, and $M_{\rm 200c}$ ($M_{\rm 500c}$) is its enclosed mass.

We use two definitions of the halo boundary, each matched to its purpose. For the \emph{global} census of Sect.~\ref{subsec: baryon_budget} we take a halo to be the full FoF group: the sample is selected by a threshold on the spherical-overdensity mass of the central, $M_{\rm 200c} \ge 10^{10}\,\mathrm{M_\odot}$, but all material in the FoF group is counted, including beyond $R_{\rm 200c}$, and matter in smaller halos ($M_{\rm 200c} < 10^{10}\,\mathrm{M_\odot}$) is counted as IGM. This avoids assigning genuinely collapsed structures to the IGM: in FoF catalogs only central subhalos carry a spherical-overdensity radius, so a satellite group beyond the $R_{\rm 200c}$ of a more massive system would otherwise be missed. The FoF group is nevertheless only a proxy for the gravitationally bound region: it retains bound substructure that an $R_{\rm 200c}$ aperture would discard, and includes only negligible diffuse, unbound material \citep[see][]{springel2001populating}; the bound-domain formalism \citep{Voit2025Bound} separates these contributions explicitly. For all \emph{per-halo} quantities (the mass functions of Sect.~\ref{subsec: halo_contribution}, the baryon fractions and characteristic radii of Sects.~\ref{sec: Characteristic_Radii} and~\ref{sec: LSS_Implications}) we use spherical apertures at $R_{\rm 200c}$ or $R_{\rm 500c}$, matched to observational apertures and to the closure-radius definition in \cite{Ayromlou2023Feedback}.

Each hydrodynamical simulation has a corresponding DMO run. We match halos between the two by position and mass: for each halo we take the closest system in the other run with $0.5 \le M_1/M_2 \le 2$ within $1\,\mathrm{cMpc}$, requiring the match to be mutual. The matched fraction approaches $100$ per cent for massive systems and remains above $90$ per cent at the lowest masses.

\subsection{Observations}
\label{subsec: Observations}
We compare the simulation predictions with various observational data, described below together with the modifications we apply to translate them into the quantities we need.
\subsubsection{IGM baryon share}
\label{subsubsec: IGM baryon share}

We define the IGM baryon share as the fraction of all cosmic baryons that reside outside halos,
\begin{equation}
F_{\rm b}^{\rm IGM} \equiv \frac{\Omega_{\rm b,IGM}}{\Omega_{\rm b}} = 1 - \frac{\Omega_{\rm b,FoF}}{\Omega_{\rm b}} = 1 - F_{\rm b}^{\rm FoF},
\end{equation}
where $\Omega_{\rm b,IGM}$ and $\Omega_{\rm b,FoF}$ are the cosmic baryon density parameters of baryons outside and inside halos, respectively. We stress that $F_{\rm b}^{\rm IGM}$ is a share of the cosmic baryon budget and not a baryon fraction in the sense of $f_{\rm b,200c}$: it is normalized by $\Omega_{\rm b}$ rather than by the mass of the region itself. We employ several observational estimates of $F_{\rm b}^{\rm IGM}$, primarily derived from FRB dispersion measures. \cite{Connor2025Gas} used DSA-110 localized FRBs and forward modeling to partition baryons between halos and the diffuse intergalactic medium, finding $F_{\rm b}^{\rm IGM} = 0.76^{+0.10}_{-0.11}$ at $z<1$. \cite{Wang2023BaryonFraction} modeled both the intergalactic and host-galaxy contributions to the dispersion measure using simulation-based probability distributions calibrated on IllustrisTNG, obtaining $0.927 \pm 0.075$ from 17 localized FRBs.

Allowing for redshift evolution, \cite{Lin2023BaryonFraction} performed a Bayesian analysis of 18 localized FRBs and found $F_{\rm b}^{\rm IGM} = F_{\rm b,0}^{\rm IGM}[1+\alpha z/(1+z)]$, where $F_{\rm b,0}^{\rm IGM} = 0.92^{+0.06}_{-0.12}$ and $\alpha=0.49^{+0.59}_{-0.47}$, supporting a predominantly diffuse low-redshift baryon reservoir. The fit is loosely constrained, however, reaching $F_{\rm b}^{\rm IGM}=1$ at $z \simeq 0.2$; they regard this normalization as higher than expected and adopt $F_{\rm b,0}^{\rm IGM} = 0.84$ in their reduced fits, for which $\alpha$ is consistent with zero.
In contrast, the FLIMFLAM analysis of \cite{Khrykin2024FlimflamDR1} reconstructed the foreground structure along 8 FRB sightlines with spectroscopic galaxy surveys, separating IGM and CGM contributions sightline by sightline. They inferred a lower value, $F_{\rm b}^{\rm IGM} = 0.59^{+0.11}_{-0.10}$, highlighting systematic uncertainties in CGM modeling and baryon partitioning.

These analyses do not all isolate the same component, and the distinction matters for the comparison. \cite{Connor2025Gas} and \cite{Khrykin2024FlimflamDR1} separate gas in intervening halos from the diffuse IGM explicitly. \cite{Wang2023BaryonFraction} and \cite{Lin2023BaryonFraction} likewise report IGM baryon fractions, but their dispersion-measure distributions are derived from theoretical treatments of the IGM and galaxy halos together, without independently separating the two, so their comparison with our outside-FoF share depends on the adopted treatment of halo gas. A further, smaller offset arises because the partitioned analyses use halo apertures of order $R_{\rm 200c}$ rather than FoF groups. 

We also use the classic estimates of \cite{Shull2012Baryon}, who combined Ly$\alpha$ forest and O\,VI surveys with cosmological simulations to partition low-redshift baryons between the photoionized Ly$\alpha$ forest, the warm-hot intergalactic medium (WHIM), and a collapsed phase (galaxies, groups, clusters, and the CGM), leaving an apparent shortfall of $\sim 30\%$. For comparison with our simulation measurements we adopt the complement of their collapsed phase, $18 \pm 4\%$, giving $F_{\rm b}^{\rm IGM} = 0.82 \pm 0.04$, as the closest observational analogue of the halo-associated budget defined in Sect.~\ref{subsec: Halo properties}. This attributes the unaccounted baryons to material outside halos, the natural reading given that the candidate reservoirs are hotter WHIM and weak absorbers, but it should be understood as an assumption rather than a measurement.

\subsubsection{Halo baryon fraction}
\label{subsubsec: Halo baryon fraction}
The halo baryon fraction $f_{\rm b}(<R)$ (Sect.~\ref{subsec: Halo properties}) counts all baryonic mass (gas, stars, and black holes) within a specified radius, typically $R_{\rm 200c}$ or $R_{\rm 500c}$. The halo gas fraction, $f_{\rm gas}(<R) = M_{\rm gas}(<R)/M_{\rm tot}(<R)$, is often used instead, especially in X-ray and SZ studies of clusters and groups. We employ various observational estimates, primarily from X-ray and SZ measurements:
\begin{itemize}
      \item eROSITA eFEDS X-ray data with stacking of optically selected GAMA groups, complemented by CHEX-MATE SZ-selected clusters, where measurements of the hot gas fraction are provided within $R_{\rm 500c}$ and $R_{\rm 200c}$ \citep{Popesso2024Hot}.
      \item X-COP sample of 13 massive, local ($z<0.1$) Planck SZ-selected clusters with joint XMM-Newton and Planck (SZ) analysis, where hydrostatic gas fractions alone are provided, within both $R_{\rm 500c}$ and $R_{\rm 200c}$ \citep{Eckert2019Non}.
      \item LoCuSS sample of 41 RASS-selected X-ray luminous clusters \citep{Ebeling1998ROSAT,Ebeling2000ROSAT,Bohringer2004REFLEX} with Subaru weak-lensing $M_{\rm 500c}$ \citep{Okabe2016LoCuSS} as the reference mass, reports X-ray gas fractions within $R_{\rm 500c}$ from Chandra and XMM-Newton \citep{Mulroy2019LoCuSS}.
      \item 91 SPT SZ-selected clusters ($0.2 < z < 1.25$) with weak-lensing calibrated masses, where total baryon fractions within $R_{\rm 500c}$ are provided, combining hot gas from X-ray with stellar mass from optical and infrared photometry, including an ICL correction \citep{chiu2018baryon}.
      \item Local ($0.01 \le z \le 0.035$) RASS-selected galaxy group sample with XMM-Newton follow-up, with hot gas fractions within $R_{\rm 500c}$ \citep{lovisari2015scaling}.
      \item X-ray selected groups and clusters (XMM-Newton) with X-ray gas masses and optical stellar mass measurements including BCG and intracluster light (ICL), where total baryon fractions are provided within $R_{\rm 500c}$ \citep{gonzalez2013galaxy}.
      \item The Canadian Cluster Comparison Project (CCCP) sample of 50 rich clusters ($0.15<z<0.55$) with Canada-France-Hawaii Telescope (CFHT) weak-lensing masses and Chandra/XMM-Newton X-ray data, where hot gas fractions within $R_{\rm 500c}$ are provided \citep{Mahdavi2013Joint}.
      \item 5 optically selected poor clusters with XMM-Newton X-ray data plus optical stellar mass (BCG and ICL) from \cite{Gonzalez2007Census}, where total baryon fractions within $R_{\rm 500c}$ are provided \citep{sanderson2013baryon}.
      \item 91 X-ray selected COSMOS groups plus 27 local clusters, with multi-band photometry stellar masses and gas fractions adopted from local hydrostatic X-ray scaling relations, providing baryon fractions within $R_{\rm 500c}$ \citep{giodini2009stellar}.
\end{itemize}

For halo-scale gas measurements, we use the aforementioned X-ray and SZ constraints for the hot phase, and estimate the cooler-gas contribution from SIMBA's simulated temperature-mass distribution. The cool-gas contribution is small in clusters but grows toward lower masses: $\sim$$90$--$95\%$ of the gas in groups and clusters is hot, versus less than half in Milky Way-like halos \citep[see][]{Rohr2025Cooler,Staffehl2025Abundance}. The choice of SIMBA is motivated by its lower halo gas fractions compared to other models, which are slightly more consistent with the observed data (see Sect.~\ref{subsec: halo_contribution}). However, since this correction is derived from one of the models under comparison, at low halo masses, where the cool-gas contribution is largest, the corrected observational points are not fully independent of SIMBA. Where the stellar mass is not available, we adopt the stellar-mass--halo-mass relation of the L-Galaxies semi-analytical model \citep{Ayromlou2021Galaxy}, calibrated to the observed stellar mass function at $z\leq 2$ and broadly consistent with fits like \cite{Moster2018EMERGE} and \cite{Behroozi2019UNIVERSEMACHINE}. This yields an observational dataset of halo baryon fractions spanning a wide range of halo masses.

\subsection{Matter power spectrum}
\label{subsec: Matter Power Spectrum}
We analyze the matter power spectrum, \(P(k)\), which quantifies the statistical distribution of matter density fluctuations in the Universe as a function of spatial scale, parametrized by the wavenumber \(k\). The matter power spectrum is a fundamental statistic of large-scale structure and encodes information about both the primordial density field and the subsequent nonlinear evolution driven by gravity and baryonic physics.

The matter overdensity field is defined as $\delta(\mathbf{x}) = (\rho(\mathbf{x}) - \bar{\rho})/\bar{\rho}$, where \(\rho(\mathbf{x})\) is the total matter density and \(\bar{\rho}\) its cosmic mean. The power spectrum is the variance of the Fourier modes of \(\delta\),
\begin{equation}
P(k) \equiv \langle |\delta(\mathbf{k})|^2 \rangle ,
\end{equation}
where the angle brackets denote an ensemble average over all Fourier modes with the same magnitude \(k = |\mathbf{k}|\), assuming statistical homogeneity and isotropy.

We compute $P(k)$ for both the DMO and hydrodynamical simulations using \textsc{Pylians} \citep{Pylians}, and quantify the impact of baryonic physics through the ratio $P_{\rm hydro}(k)/P_{\rm DMO}(k)$, isolating deviations from gravity-only physics. Density field is assigned to a $512^3$ grid, giving access to wavenumbers up to $k \simeq 11$--$16\,\mathrm{Mpc^{-1}}$, depending on box size.

\section{The baryon budget of the Universe}
\label{sec: Baryon_Budget}
\begin{figure*}
      \centering
      \includegraphics[width=1\columnwidth]{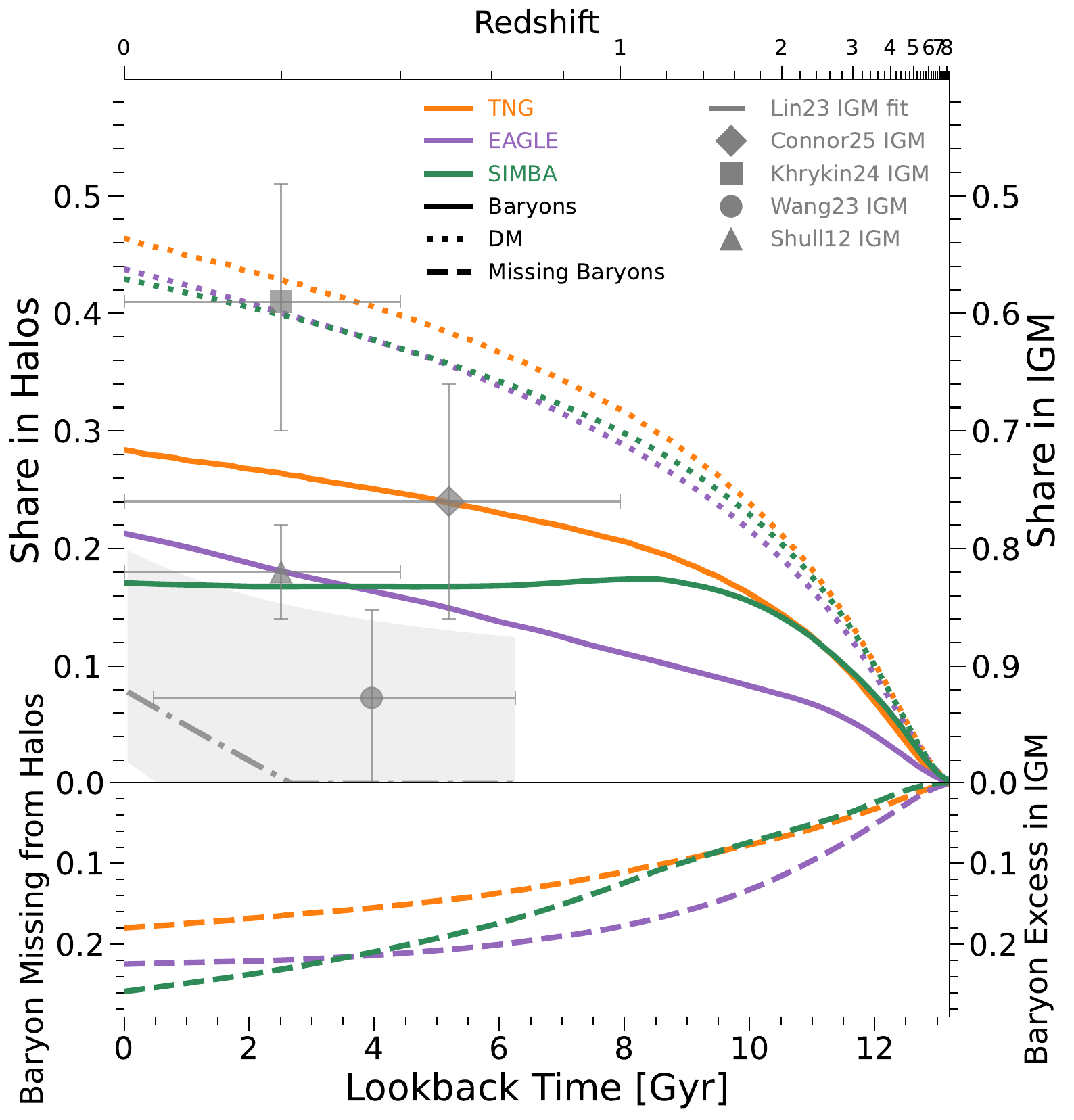}\hspace{0.02\textwidth}
      \includegraphics[width=0.98\columnwidth]{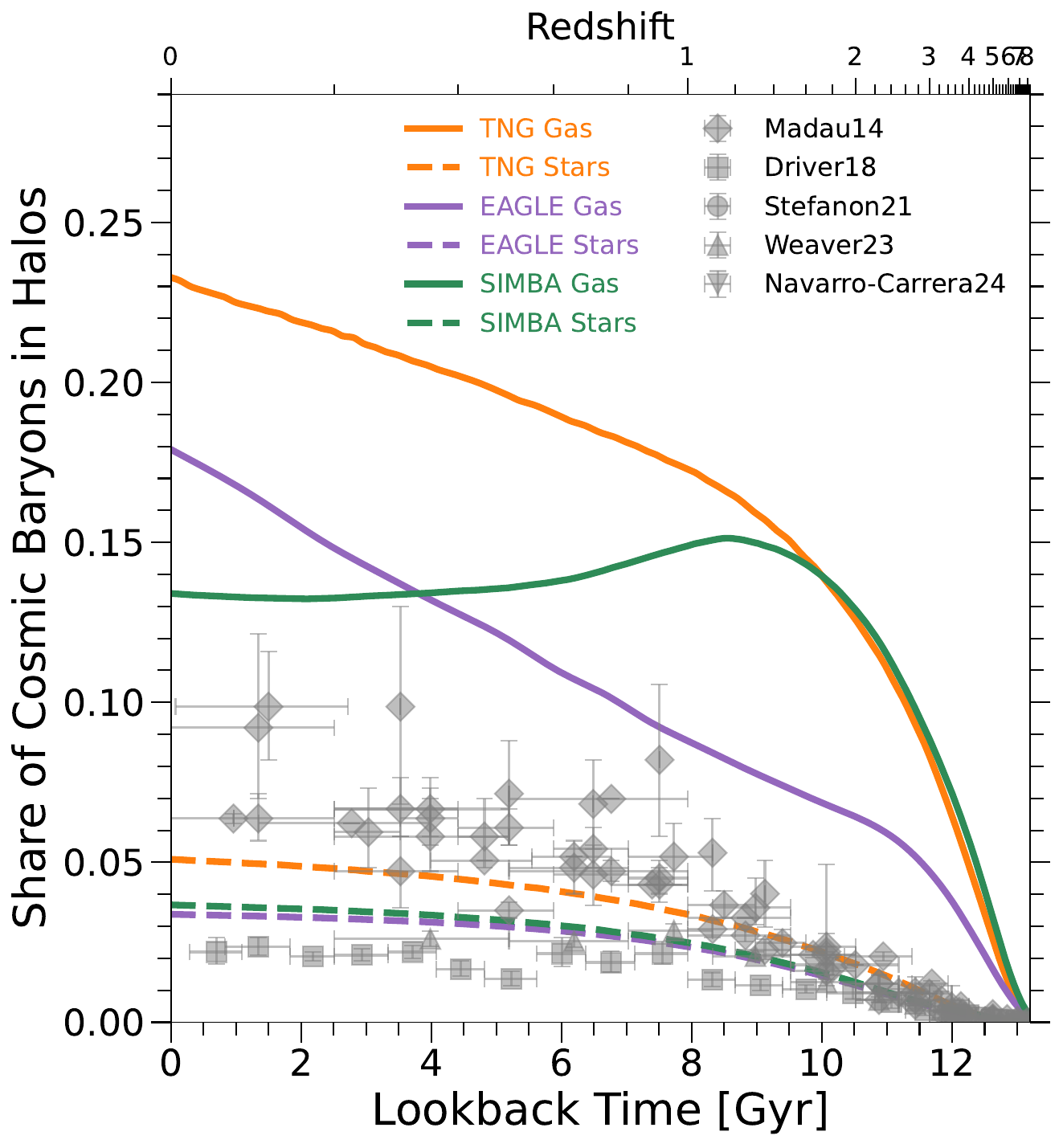}
      \caption{\textbf{Left:} Share of each cosmic matter component residing inside FoF halos, $F^{\rm FoF}$ (left axis), as a function of lookback time (lower x-axis) and redshift (upper x-axis), for TNG (orange), EAGLE (purple), and SIMBA (green). Solid lines show baryons, $F_{\rm b}^{\rm FoF}$, and dotted lines dark matter, $F_{\rm DM}^{\rm FoF}$. The reversed right-hand axis gives the corresponding IGM shares, $F^{\rm IGM}=1-F^{\rm FoF}$, so that every curve may be read as either quantity. Gray markers show observational estimates of the IGM baryon share from FRB and pre-FRB measurements, plotted against the right-hand axis. The gray dashed-dotted line is the redshift-dependent fit of \cite{Lin2023BaryonFraction}, with the shaded band its uncertainty. The lower sub-panel shows the baryon deficit relative to the expectation for baryons tracing dark matter, $F_{\rm DM}^{\rm FoF}-F_{\rm b}^{\rm FoF}=F_{\rm b}^{\rm IGM}-F_{\rm DM}^{\rm IGM}$ (dashed lines); both axes increase downwards. \textbf{Right:} Shares of the cosmic baryon budget residing in halos as gas, $F_{\rm b,gas}^{\rm FoF}$ (solid lines), and stars, $F_{\rm b,\star}^{\rm FoF}$ (dashed lines). Black holes contribute less than $0.1\%$ and are omitted. Gray markers show observational estimates of the cosmic stellar mass density, normalized to the cosmic baryon density.}
\label{Fig: matter_fraction}
\end{figure*}
While the standard baryon cycle describes the inflow of baryons into galaxies, their conversion into stars, and subsequent feedback-driven outflows, it overlooks that baryons can be ejected beyond the halo boundary, leading to a net loss of baryons from halos. The open baryon cycle accounts for baryon redistribution on scales larger than individual halos, out to the closure radius (see Fig.~\ref{Fig: open_baryon_cycle}). The baryons missing from halos consist of both ejected baryons and those prevented from accreting in the first place; beyond $R_{\rm 200c}$ (dashed circle), out to the closure radius (solid circle), the environment can additionally strip gas from galaxies moving through it. A central theme is how baryonic processes, particularly feedback, drive the baryon distribution away from that of the gravity-only dark matter.
\subsection{Baryon budget of the Universe}
\label{subsec: baryon_budget}
Baryons are not only missing from halos; a significant fraction inherently resides outside them, in the IGM, because not all dark matter is contained within halos. Fig.~\ref{Fig: matter_fraction} shows the distribution of dark matter and baryons in the Universe as a function of lookback time (lower axis) and redshift (upper axis) in three simulations and several observations (only baryons). All halos with $M_{\rm 200c} \geq 10^{10} \, \rm M_{\odot}$ are included; this threshold applies to the global census of this section, whereas the mass-resolved distributions of Sect.~\ref{subsec: halo_contribution} onwards extend down to $10^{9} \, \rm M_{\odot}$. The redistribution of baryons from halos into the IGM is directly visible in the projected baryon-fraction maps of Fig.~\ref{Fig: matter_maps} (Appendix~\ref{app: matter_distribution}), where halos appear baryon-poor against a baryon-enriched IGM.

The left panel of Fig.~\ref{Fig: matter_fraction} shows baryons (solid lines) and dark matter (dotted lines) split into two components: within halos (the full FoF groups) and within the IGM (everything outside them; see Sect.~\ref{subsec: Halo properties} for why the global census uses FoF groups rather than $R_{\rm 200c}$ apertures). By construction the two shares are complementary, $F^{\rm IGM} = 1 - F^{\rm FoF}$, so each curve can be read as either share on the two axes. For the baryonic component we also show observational constraints on the IGM baryon share, from the classic cosmic baryon budget \citep{Shull2012Baryon} and recent FRB measurements \citep{Lin2023BaryonFraction,Wang2023BaryonFraction,Khrykin2024FlimflamDR1,Connor2025Gas}, shown in gray. Of these, \cite{Connor2025Gas} and \cite{Khrykin2024FlimflamDR1} separate halo gas from the diffuse IGM, \cite{Wang2023BaryonFraction} and \cite{Lin2023BaryonFraction} constrain the combined diffuse gas component, and the \cite{Shull2012Baryon} point is the complement of their collapsed phase rather than a direct measurement (Sect.~\ref{subsubsec: IGM baryon share}).

At the present time, slightly less than half of the dark matter resides within halos, between $43\%$ (SIMBA) and $46\%$ (TNG), with the remainder in the IGM. The three models differ by only a few per cent here, a spread that could arise both from baryonic effects and from differences in halo identification and mass resolution. If baryons perfectly followed the dark matter, we would expect a similar fraction of baryons to be in halos. However, less than $30\%$ of baryons are found within halos at $z=0$, with over $70\%$ residing in the IGM, according to the simulations and most of the observations. The simulations differ considerably in the exact partitioning, with TNG retaining the largest baryon share in halos ($28\%$), EAGLE an intermediate value ($21\%$), and SIMBA the smallest ($17\%$); the corresponding IGM shares, read from the reversed right-hand axis, are $72\%$, $79\%$, and $83\%$. The variation reflects the different feedback implementations. The simulated IGM shares are qualitatively consistent with \cite{Connor2025Gas} and the \cite{Shull2012Baryon} complement within their uncertainties; \cite{Wang2023BaryonFraction} and \cite{Lin2023BaryonFraction} lie above all three models, partly reflecting their joint treatment of halo and IGM gas (Sect.~\ref{subsubsec: IGM baryon share}), while the FLIMFLAM result of \cite{Khrykin2024FlimflamDR1} lies below all three.

The lower left panel of Fig.~\ref{Fig: matter_fraction} isolates the baryons that have been displaced from halos, i.e. the difference between the dark matter and baryon shares in halos, $F_{\rm DM}^{\rm FoF} - F_{\rm b}^{\rm FoF}$. This is identically the excess of baryons in the IGM over the expectation for baryons tracing the dark matter, $F_{\rm b}^{\rm IGM} - F_{\rm DM}^{\rm IGM}$, which is the reading given by the right-hand axis of the lower left panel. In all three models this quantity grows monotonically toward the present day, reaching $18\%$ of all cosmic baryons in TNG, $22\%$ in EAGLE, and $26\%$ in SIMBA at $z=0$. These displaced shares are not directly comparable to the missing-baryon budget of Sect.~\ref{subsec: halo_contribution} as they differ in the aperture and reference expectation, baryons tracing the dark matter versus the cosmic baryon fraction of the hydrodynamical $M_{\rm 200c}$. EAGLE displaces the most baryons at early times, but SIMBA overtakes it below a lookback time of $4\,\mathrm{Gyr}$ ($z \simeq 0.35$), reflecting the late onset of its strongly ejective AGN feedback.

The dark matter share in halos increases monotonically toward the present, from below $2\%$ at $z>7$ to $\sim 32\%$ at $z=1$, and to $43$--$46\%$ at $z=0$. Baryons follow a similar trend in all simulations, but with a consistently lower share in halos than the dark matter. Importantly, the share of cosmic baryons residing outside halos exceeds the corresponding share of the total matter at all redshifts, so the material outside halos is baryon-rich relative to the cosmic mean. Its baryon-to-total-matter ratio grows from parity at $z\gtrsim6$, when little material has yet collapsed, to between $\simeq 1.3$ and $\simeq 1.4$ times $f_{\rm b,cosmic}$ at $z=0$, lowest in TNG and highest in SIMBA (see Appendix~\ref{app: IGM_fb}). We note that the three models agree on the dark matter share to within $<4$ percentage points while differing by 11 percentage points in the baryon share, which is consistent with the model-to-model spread in $F_{\rm b}^{\rm IGM}$ being driven primarily by baryonic physics, although it does not by itself exclude a contribution from differences in halo identification or mass resolution. The observed IGM fractions are too uncertain to constrain the redshift evolution of these quantities.

The right panel of Fig.~\ref{Fig: matter_fraction} breaks down the baryonic content within halos into gas (solid lines) and stars (dashed lines). At each redshift the two components sum to the total baryon share within halos shown in the left panel. Black holes are not shown, as they contribute less than $0.1\%$ of the cosmic baryon budget at all redshifts in every model. At the present time, stars account for $5.1\%$ of all cosmic baryons in TNG, $3.4\%$ in EAGLE, and $3.7\%$ in SIMBA. The gas within halos shows much larger absolute differences among the simulations: $23\%$ in TNG, $18\%$ in EAGLE, and only $13\%$ in SIMBA. The stellar shares therefore span less than two percentage points, whereas the gas shares span nearly ten, so the model-to-model variation in the cosmic baryon budget is dominated by the gas rather than the stars. The comparatively small stellar spread is expected, since all three models were calibrated to reproduce the observed galaxy stellar mass function at low redshift. Gray markers show observational determinations of the cosmic stellar mass density, normalized to the cosmic baryon density \citep{Madau2014Cosmic,Driver2018GAMA,Stefanon2021Galaxy,Weaver2023COSMOS2020,Navarro-Carrera2024Constraints}. These determinations span a factor of several among themselves, from $\simeq 2\%$ to $\simeq 10\%$ of the cosmic baryon budget at $z=0$, a range much wider than the differences between the models, all of which fall within it. The stellar component is therefore not currently a discriminating measurement, whereas the gas is.

While the stellar component within halos is always increasing with decreasing redshift across all simulations, the gas shows a more complex behavior. In TNG and EAGLE, the gas share within halos increases with decreasing redshift, but in SIMBA it peaks at a lookback time of $\simeq 8.7\,\mathrm{Gyr}$ ($z \simeq 1.2$): it first increases from high redshift up to that epoch, and then declines toward $z=0$. This behavior reflects how feedback processes operate in these simulations: in SIMBA, AGN feedback becomes particularly efficient at ejecting gas from halos at lower redshifts, leading to a reduction in the gas fraction within halos over time. In contrast, TNG and EAGLE have feedback implementations that allow for more gas retention within halos at late times, consistent with previous findings of SIMBA having stronger AGN feedback compared to TNG and EAGLE \citep{Ayromlou2023Feedback}.

Given the differences in the IGM baryon share among the simulations, it can serve as a powerful observational constraint on theoretical models. Therefore, more precise measurements of the IGM baryon share across cosmic time can directly constrain the physics of feedback and large-scale baryon redistribution.

\subsection{Contribution of different halo masses to the baryon and dark matter budget of the Universe}
\label{subsec: halo_contribution}
\begin{figure*}
      \centering
      \includegraphics[width=0.49\textwidth]{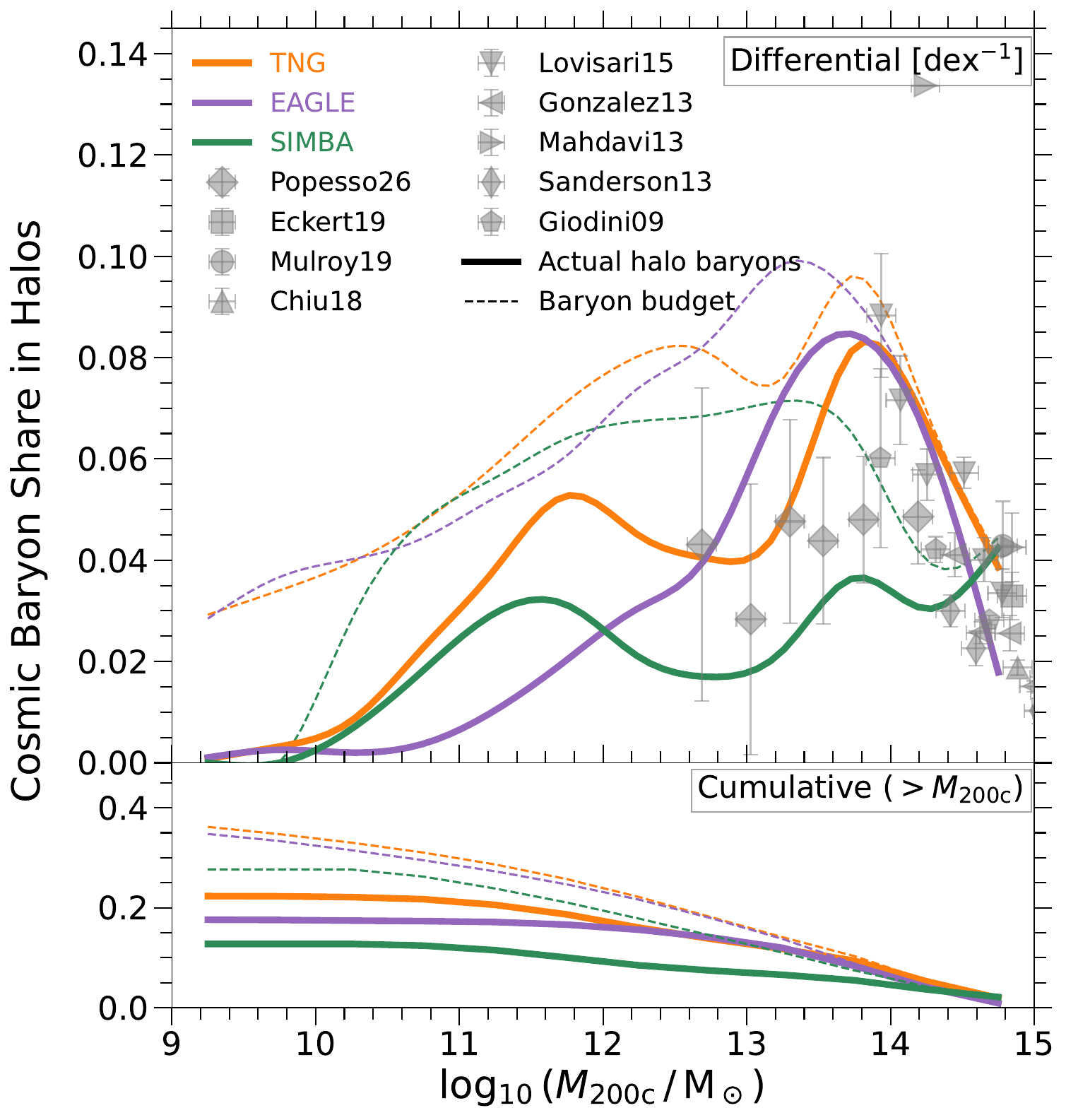}\hspace{0.01\textwidth}
      \includegraphics[width=0.49\textwidth]{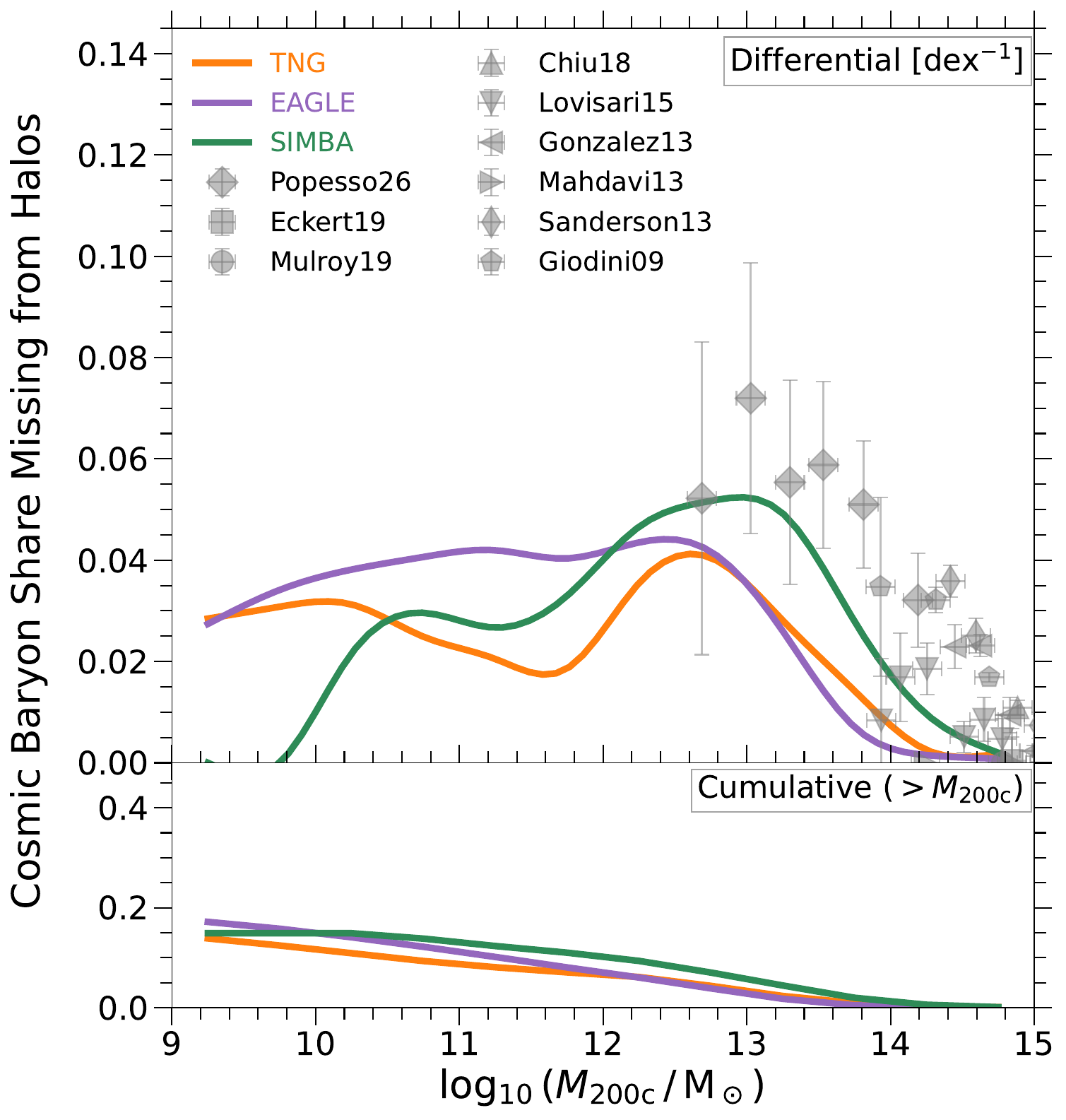}
      \caption{Contribution of halos of different mass to the cosmic baryon budget at $z=0$, for TNG (orange), EAGLE (purple), and SIMBA (green). \textbf{Left:} baryons retained within $R_{\rm 200c}$ (solid lines). \textbf{Right:} baryons missing from halos relative to the expected mass $f_{\rm b,cosmic}M_{\rm 200c}$. In each column, the upper sub-panel shows the differential distribution $p$ per dex of halo mass; its integral over a mass interval gives the share of cosmic baryons retained within (left) or missing from (right) halos in that interval. The lower sub-panel shows the corresponding cumulative share $\mathcal{F}(>M_{\rm 200c})$ contributed by halos above the indicated mass. Thin dashed lines in the left panel show the expected budget; the retained and missing distributions sum to this budget (Eqs.~\ref{eq: cumulative_b}--\ref{eq: pdf_explicit}). Gray markers show our estimates from X-ray observations (Sect.~\ref{subsubsec: Halo baryon fraction}) and are intended for a qualitative comparison.}
\label{Fig: matter_dist_halo_contribution_z0}
\end{figure*}
\begin{figure*}
      \centering

      \includegraphics[width=0.33\textwidth]{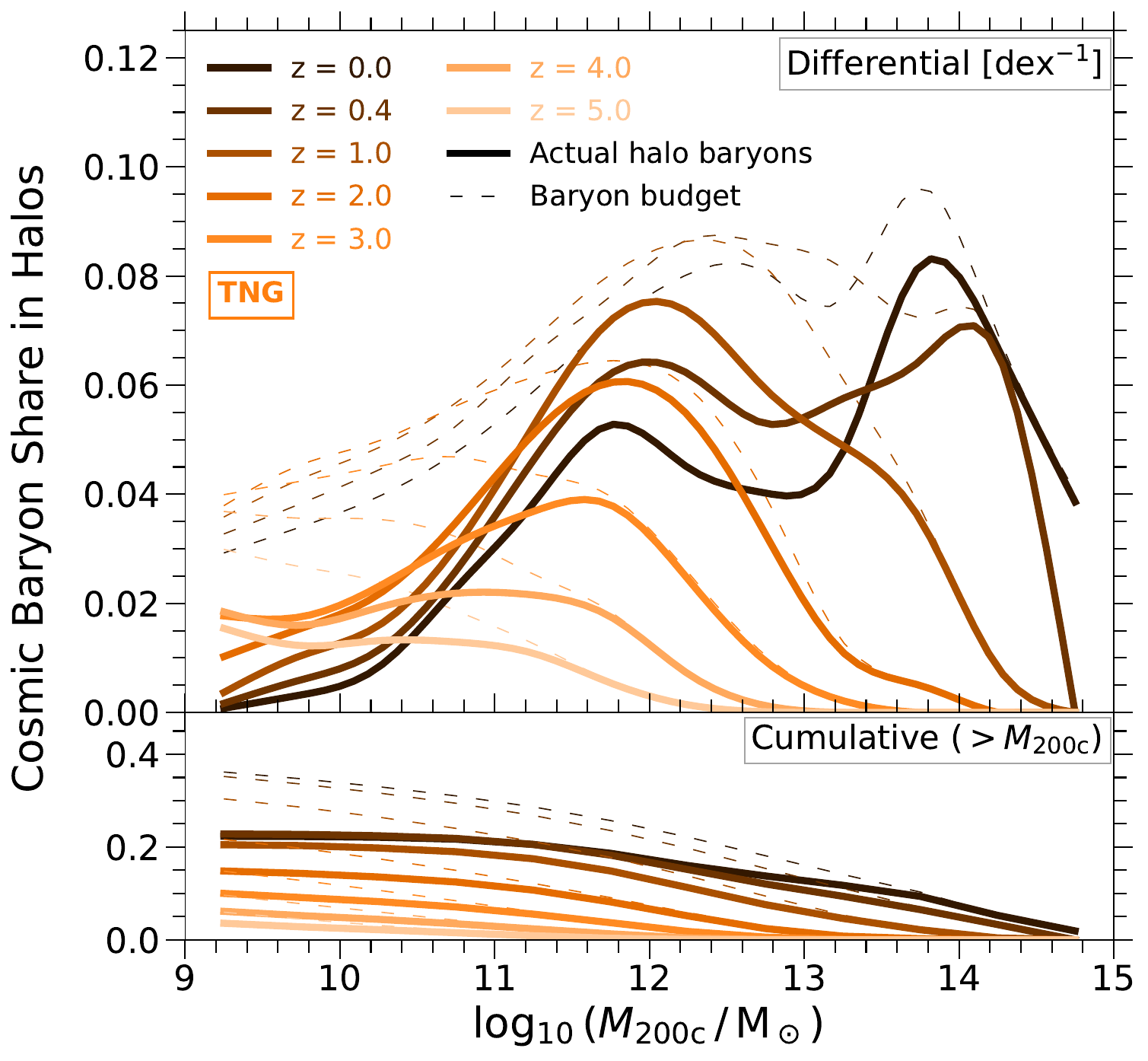}
      \includegraphics[width=0.33\textwidth]{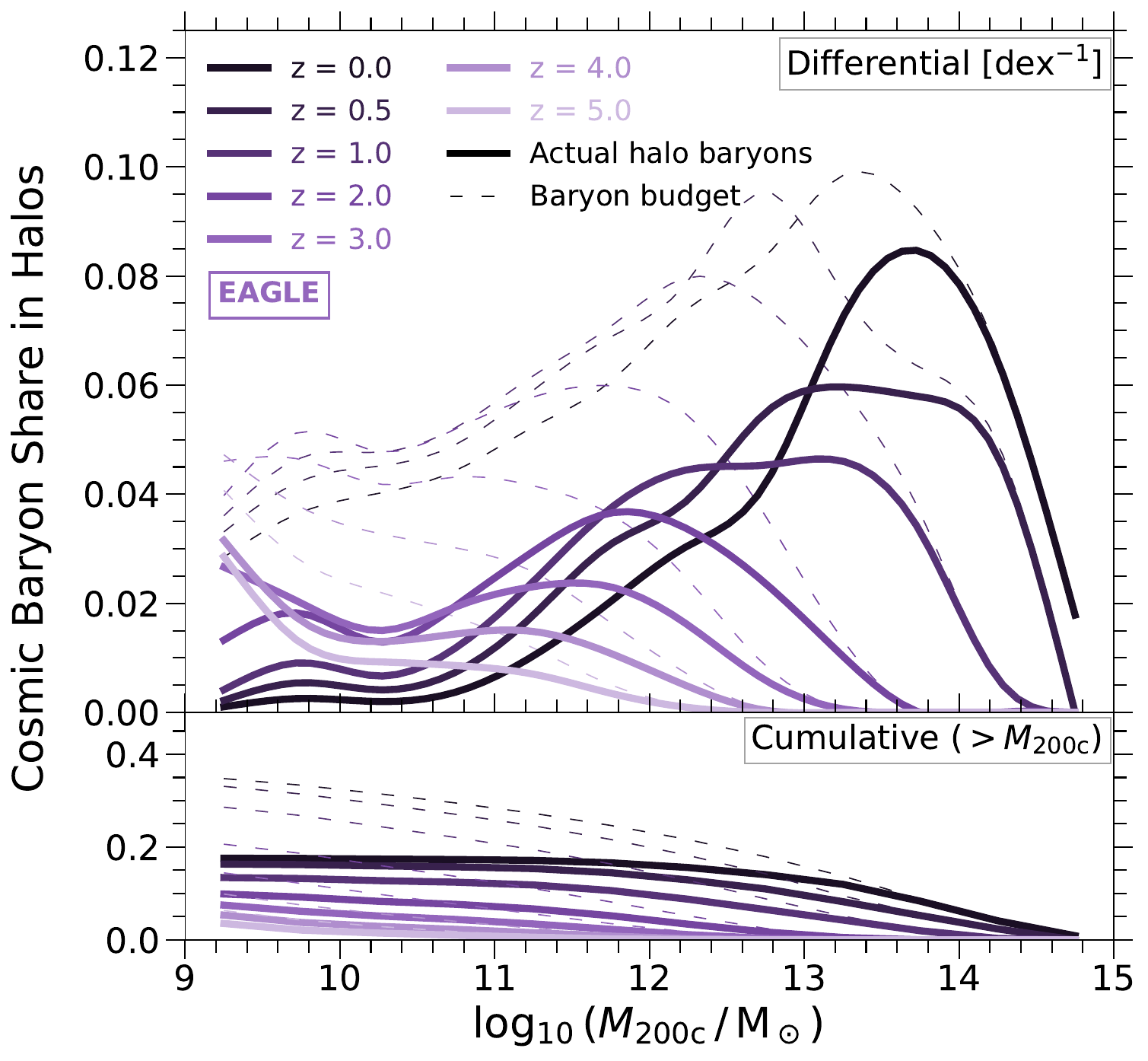}
      \includegraphics[width=0.33\textwidth]{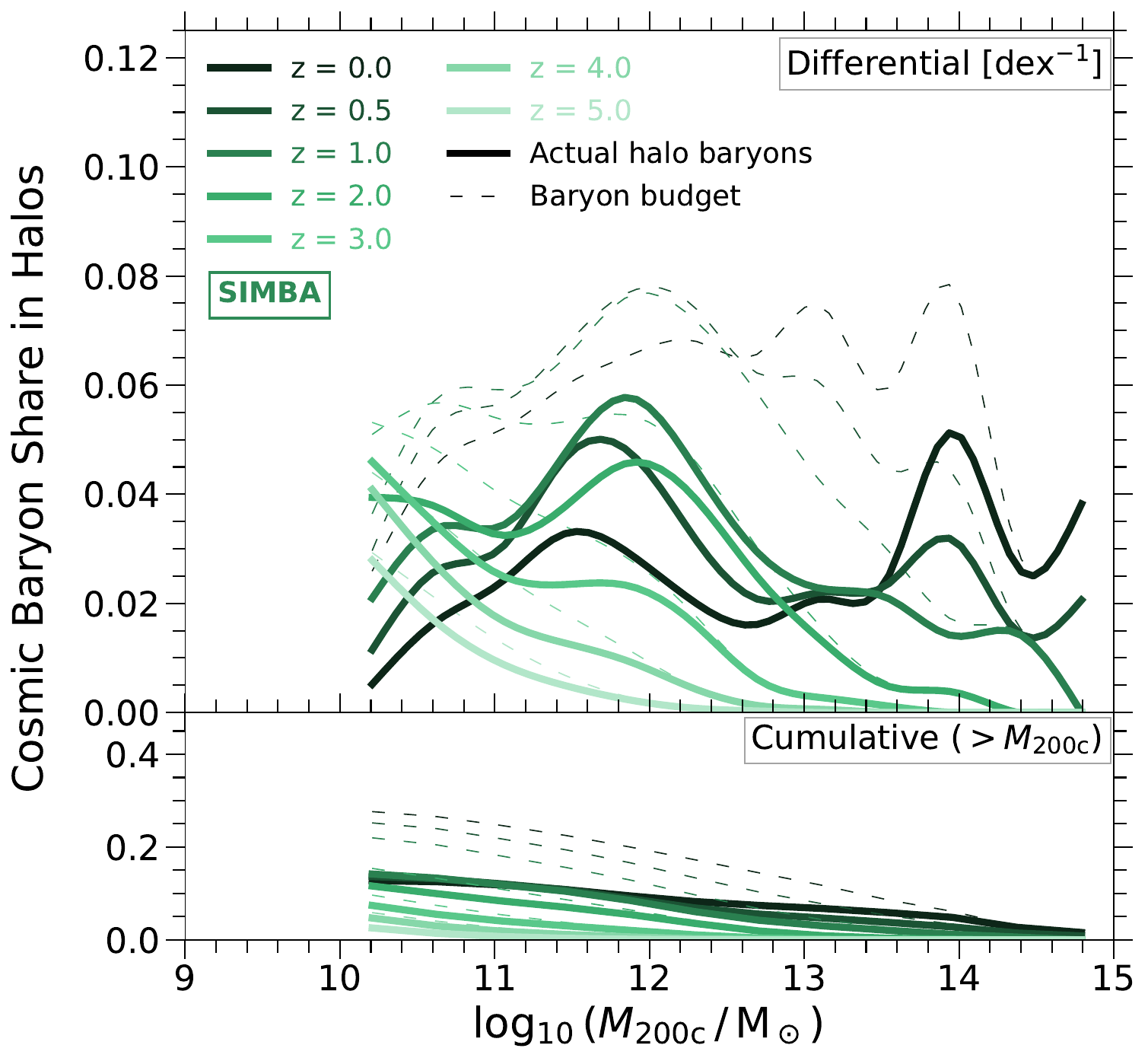}

      \includegraphics[width=0.33\textwidth]{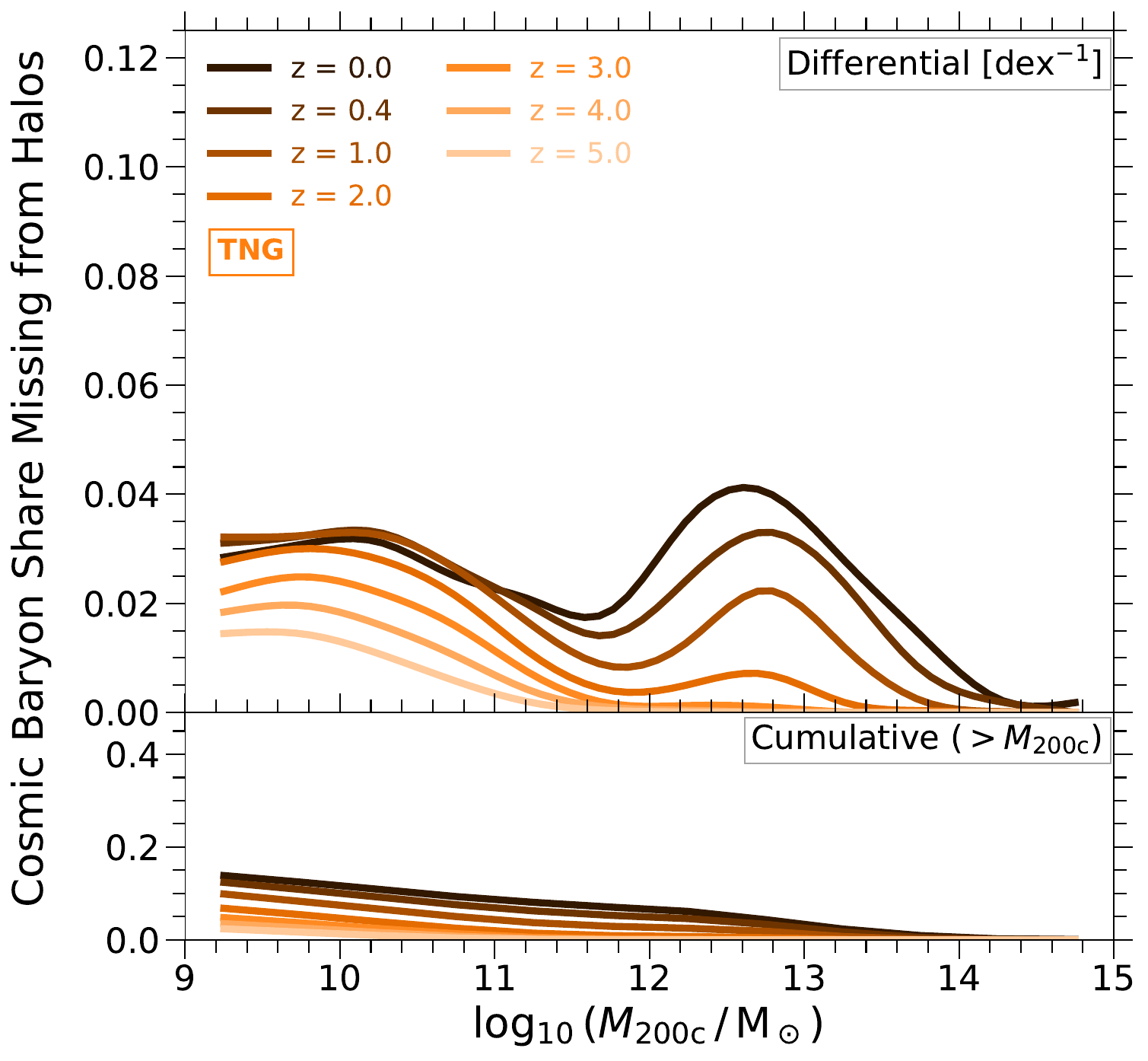}
      \includegraphics[width=0.33\textwidth]{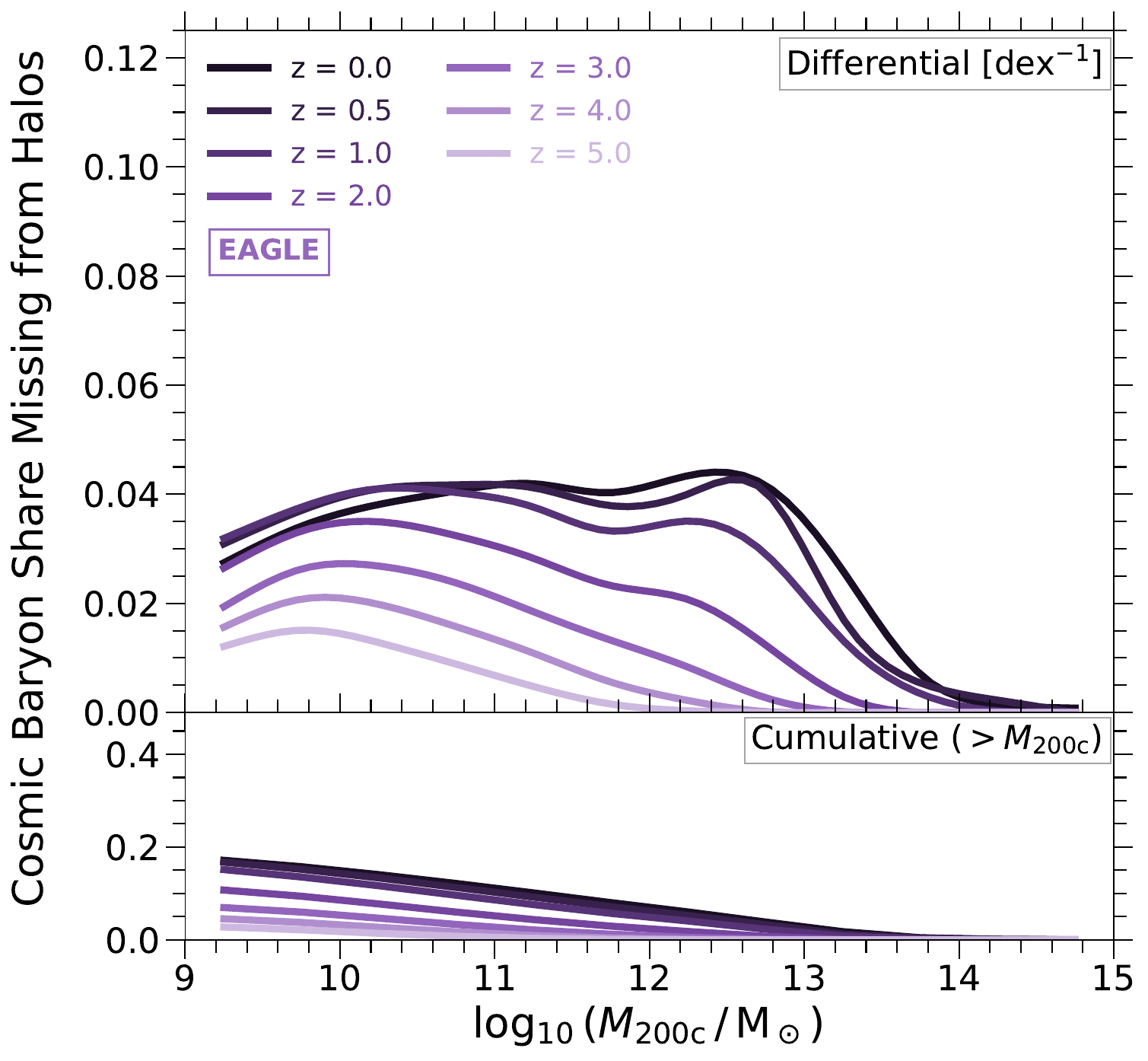}
      \includegraphics[width=0.33\textwidth]{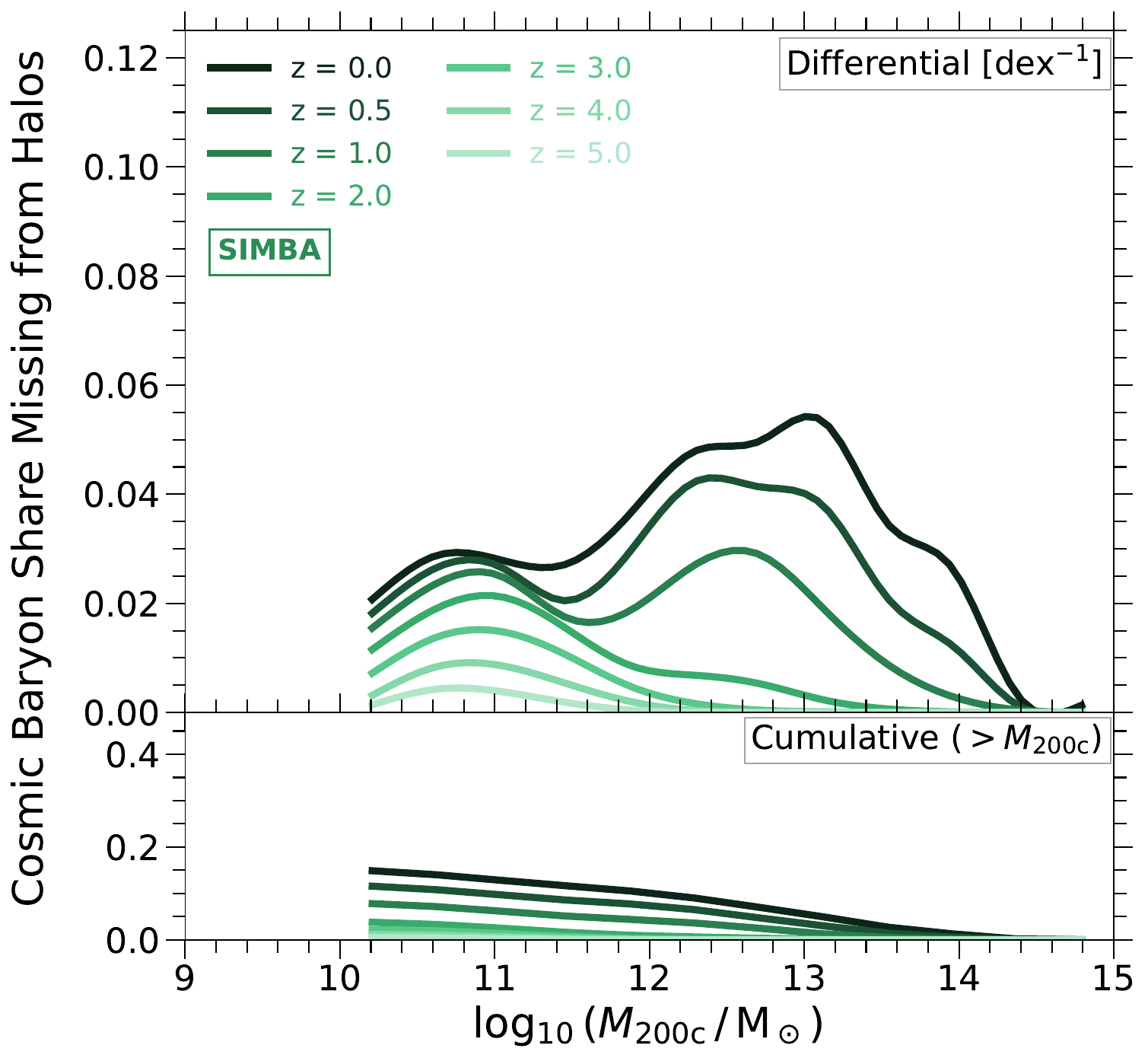}

      \caption{Redshift evolution of the halo contributions to the baryon budget of the Universe (shown at $z=0$ in Fig.~\ref{Fig: matter_dist_halo_contribution_z0}), for TNG (left column), EAGLE (middle column), and SIMBA (right column). \textbf{Top row:} baryons retained within $R_{\rm 200c}$ (solid lines). \textbf{Bottom row:} baryons missing from halos relative to the expected halo baryon budget $f_{\rm b,cosmic}M_{\rm 200c}$ (solid lines). As in Fig.~\ref{Fig: matter_dist_halo_contribution_z0}, each panel carries a differential upper sub-panel giving $p$ per dex of halo mass and a cumulative lower sub-panel giving $\mathcal{F}(>M_{\rm 200c})$. Thin dashed lines in the top row show the expected budget; the retained and missing distributions sum to this budget at each mass and redshift. Line shade encodes redshift within each simulation's color, from $z=0$ (darkest) to $z=5$ (lightest). See Eqs.~(\ref{eq: cumulative_b})--(\ref{eq: pdf_explicit}) for the definitions. The corresponding distributions for the DMO runs are shown in Appendix~\ref{app: DMO_dist}.}
\label{Fig: matter_dist_halo_contribution}
\end{figure*}
We quantify how halos of different mass contribute to the cosmic baryon and dark matter budgets across redshift in the three simulations (columns of Figs.~\ref{Fig: matter_dist_halo_contribution_z0} and~\ref{Fig: matter_dist_halo_contribution}), and compare with observational constraints. To address how much of each budget is held by halos of a given mass, we begin from the \emph{cumulative} share held by all halos more massive than $M$. For the baryons this is
\begin{equation}
    \mathcal{F}_{\rm b,halo}(>M) \equiv
    \frac{1}{f_{\rm b,cosmic}\,\bar{\rho}_{\rm m}}
    \int_{M}^{\infty} \left\langle M_{\rm b,halo}(M')\right\rangle\,
    \frac{dn}{dM'}\, {\rm d}M' ,
    \label{eq: cumulative_b}
\end{equation}
where $\left\langle M_{\rm b,halo}(M')\right\rangle$ is the \emph{mean} baryonic mass within $R_{\rm 200c}$ of halos of mass $M'$, $dn/dM'$ is the halo mass function, and $\bar{\rho}_{\rm m} \equiv \Omega_{\rm m}\,\rho_{\rm crit}$ is the cosmic mean matter density. The prefactor normalizes by the cosmic baryon density, so $\mathcal{F}_{\rm b,halo}$ is dimensionless; its value at the lower halo-mass limit gives the total share of cosmic baryons within $R_{\rm 200c}$ of halos. The corresponding quantities for the dark matter, for the dark-matter-only run, and for the baryons missing from halos follow by substituting $\left\langle M_{\rm DM,halo}\right\rangle$, $\left\langle M_{\rm DMO,halo}\right\rangle$, or $\left\langle M_{\rm b,halo}^{\rm expected}\right\rangle - \left\langle M_{\rm b,halo}\right\rangle$ for the numerator, and $(1-f_{\rm b,cosmic})\bar{\rho}_{\rm m}$, $\bar{\rho}_{\rm m}$, or $f_{\rm b,cosmic}\bar{\rho}_{\rm m}$ for the normalization, where $M_{\rm b,halo}^{\rm expected} \equiv f_{\rm b,cosmic}\,M_{\rm 200c}$ is the baryon mass a halo would contain if its baryons were present at the cosmic fraction.

The differential distributions plotted in Figs.~\ref{Fig: matter_dist_halo_contribution_z0} and~\ref{Fig: matter_dist_halo_contribution} are the negative derivatives of these cumulative shares with respect to logarithmic halo mass,
\begin{equation}
    p_{\rm b,halo}(M_{\rm 200c}) \equiv
    -\frac{{\rm d}\mathcal{F}_{\rm b,halo}(>M_{\rm 200c})}{{\rm d}\log_{10} M_{\rm 200c}} ,
    \label{eq: pdf_def}
\end{equation}
and analogously for $p_{\rm DM,halo}$, $p_{\rm DMO,halo}$ and $p_{\rm b,missing}$. Each $p$ gives a share of the corresponding cosmic budget per dex of halo mass. Its integral over a mass interval gives the share residing within, or missing from, halos in that interval, and $\int p\,{\rm d}\log_{10} M$ over the sampled mass range recovers the cumulative share at its lower mass limit. Equivalently,
\begin{equation}
    p_{\rm b,halo}(M_{\rm 200c}) =
    \frac{dn}{d\log_{10} M_{\rm 200c}} \times
    \frac{\left\langle M_{\rm b,halo}(M_{\rm 200c})\right\rangle}
         {f_{\rm b,cosmic}\,\bar{\rho}_{\rm m}} .
    \label{eq: pdf_explicit}
\end{equation}
In the following, we refer to $p_{\rm b,halo}$ and $p_{\rm b,missing}$ as the contributions of baryons within and missing from halos, respectively, and analogously for the dark matter and DMO cases.

In the simulations, all required quantities are measured directly from the halo catalogs and particle data (see Sect.~\ref{subsec: Simulations}). For observations, we require the gas mass, stellar mass, and halo mass of the relevant systems, extracted from literature datasets as described in Sect.~\ref{subsubsec: Halo baryon fraction}. The cosmic density of each component follows from CMB cosmological parameters; for consistency with the simulations we adopt \cite{planck2015_xiii} for the observations as well.

For the simulations, the halo mass function $dn/dM$ is measured directly from each run, so that each set of curves is internally self-consistent and no cross-simulation calibration enters. All baryonic quantities, including the baryons missing from halos, are binned by the hydrodynamical $M_{\rm 200c}$ and weighted by the same run's hydrodynamical mass function; the DMO mass function enters only the DMO reference curves.

For the observational estimates no measured mass function is available, and we instead adopt the Millennium mass function \citep{springel2005simulations}, rescaled to the Planck cosmology \citep{planck2015_xiii} following \cite{angulo2010one,Angulo2012Scaling}, as provided by L-Galaxies \citep{Ayromlou2021Galaxy}. This is a deliberate choice: a gravity-only mass function keeps the observational estimates independent of the three models they are compared against, though the true mass function of the observed Universe is not the gravity-only one. Its magnitude can be gauged from the hydrodynamical-to-DMO halo mass ratio in Fig.~\ref{Fig: Halo_Mass_Ratio}; the observational points should be read with this caveat in mind. Given current observational limitations, particularly for gas, we show observational results only at $z\sim 0$ (Sect.~\ref{subsubsec: Halo baryon fraction}); the simulation curves span $0 \le z \le 5$.

Fig.~\ref{Fig: matter_dist_halo_contribution_z0} shows the contribution of halos of different mass to the cosmic baryon budget at $z=0$ for the simulations (lines) and observational estimates (markers); baryons within halos on the left, baryons missing from halos on the right. The inferred halo contribution varies by up to a factor of 2--3 between observational data sets, but all show a clear trend for baryons within halos: the contribution is small for massive clusters because they are rare, rises toward lower-mass clusters and groups, peaks around $\log_{10}(M_{\rm 200c}/\mathrm{M_{\odot}})\sim 13.8$, and then declines again toward lower masses, with one data point from \cite{Popesso2024Hot} suggesting another upturn at $\log_{10}(M_{\rm 200c}/\mathrm{M_{\odot}})\sim 12.8$. This behavior is broadly consistent with TNG and EAGLE, peaking at $\log_{10}(M_{\rm 200c}/\mathrm{M_{\odot}})\sim 13.5$--$13.8$. SIMBA peaks at a similar mass but at less than half the amplitude, and rises again.

The right panel of Fig.~\ref{Fig: matter_dist_halo_contribution_z0} shows the contribution of baryons missing from halos, i.e. baryons redistributed to the IGM that would have remained inside halos if baryons traced dark matter exactly. This component does not include baryons that are intrinsically outside halos (Sect.~\ref{subsec: baryon_budget}), so the two panels together account only for the baryon budget associated with halos. In all three simulations, the largest contribution per dex to the missing-baryon budget comes from galaxy groups, peaking near $\log_{10}(M_{\rm 200c}/\mathrm{M_{\odot}})\sim 12.5$ in TNG and EAGLE and near $\sim 13.0$ in SIMBA, because these systems combine large baryon deficits and large closure radii with a high abundance. The observational estimates are broadly consistent with the simulations, although they generally imply more baryons displaced into the IGM. SIMBA agrees best with these data, consistent with its stronger AGN-driven gas ejection, lower halo baryon fractions, and larger closure radii for groups and clusters (see Sect.~\ref{subsec: closure_radius}), bearing in mind the SIMBA-derived cool-gas correction of Sect.~\ref{subsubsec: Halo baryon fraction}.

Fig.~\ref{Fig: matter_dist_halo_contribution} extends this comparison to different redshifts. The top and bottom rows show the baryons retained within and missing from halos, for each simulation in turn; darker lines correspond to lower redshift, from $z=5$ (lightest) to $z=0$ (darkest). The corresponding DMO distributions are given in Appendix~\ref{app: DMO_dist}.

We first establish how the \emph{total} matter budget is distributed without baryonic effects: as shown in Appendix~\ref{app: DMO_dist}, the three DMO runs agree closely above $\log_{10}(M_{\rm 200c}/\mathrm{M_\odot})\simeq 11$, all placing $\simeq 31\%$ of the cosmic matter budget within $R_{\rm 200c}$ of halos above that mass. This is expected, since in the absence of baryonic physics the halo mass function depends only on cosmology, and the three suites adopt closely similar cosmological parameters. Where the gravity-only runs agree, model-to-model differences in the baryonic distributions of Figs.~\ref{Fig: matter_dist_halo_contribution_z0} and~\ref{Fig: matter_dist_halo_contribution} are therefore unlikely to reflect residual differences in cosmology, mass function, or halo identification. All three show the same basic trend: at $z=0$, galaxy groups with $13 \lesssim \log_{10}(M_{\rm 200c}/\mathrm{M_\odot})\lesssim 14$ contribute most to the total matter budget, because low-mass halos, while abundant, contain relatively little mass individually, whereas high-mass halos, although massive, are rare. The peak shifts toward lower halo masses with increasing redshift, as expected from hierarchical structure formation, and the overall contribution of halos increases with decreasing redshift because halos are progressively less developed at earlier times.

The back-reaction of baryonic physics on the total matter distribution can be read directly from the same figures. Because the factor $f_{\rm b,cosmic}$ cancels against the normalization, the dashed budget curves of Figs.~\ref{Fig: matter_dist_halo_contribution_z0} and~\ref{Fig: matter_dist_halo_contribution} are the total matter distribution of the hydrodynamical runs, on the same normalization as the DMO distributions of Appendix~\ref{app: DMO_dist}, so comparing the two measures like-for-like how baryonic processes rearrange the total matter. The overall behavior is similar, with groups dominating at $z=0$ and the peak moving to lower masses at higher redshift, but the amplitudes are systematically lower in the hydrodynamical runs: at $z=0$ the total matter share within $R_{\rm 200c}$ of resolved halos falls from $41\%$ to $37\%$ in TNG, from $42\%$ to $35\%$ in EAGLE, and from $31\%$ to $28\%$ in SIMBA (cumulative sub-panels of Fig.~\ref{Fig: matter_dist_halo_contribution_z0} and Fig.~\ref{Fig: DMO_dist_z0}). This is consistent with feedback-driven baryon removal weakening halo potential wells, suppressing subsequent accretion of matter into halos (Sect.~\ref{subsec: halo_mass_ratio}).

\begin{figure*}
      \centering
      \setlength{\tabcolsep}{0pt}
      \begin{tabular}{@{}lll@{}}
      \includegraphics[width=0.33\textwidth]{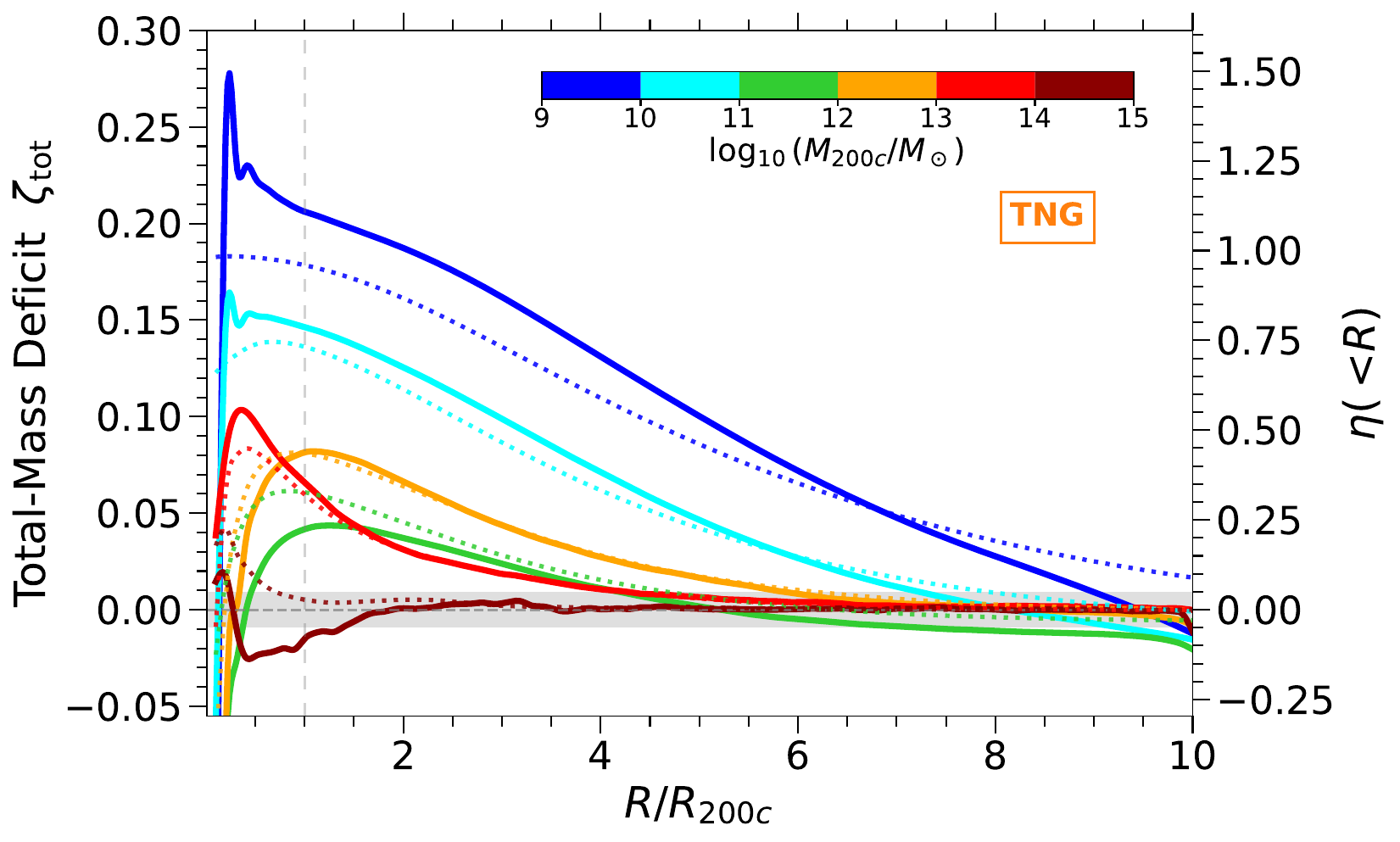} &
      \includegraphics[width=0.33\textwidth]{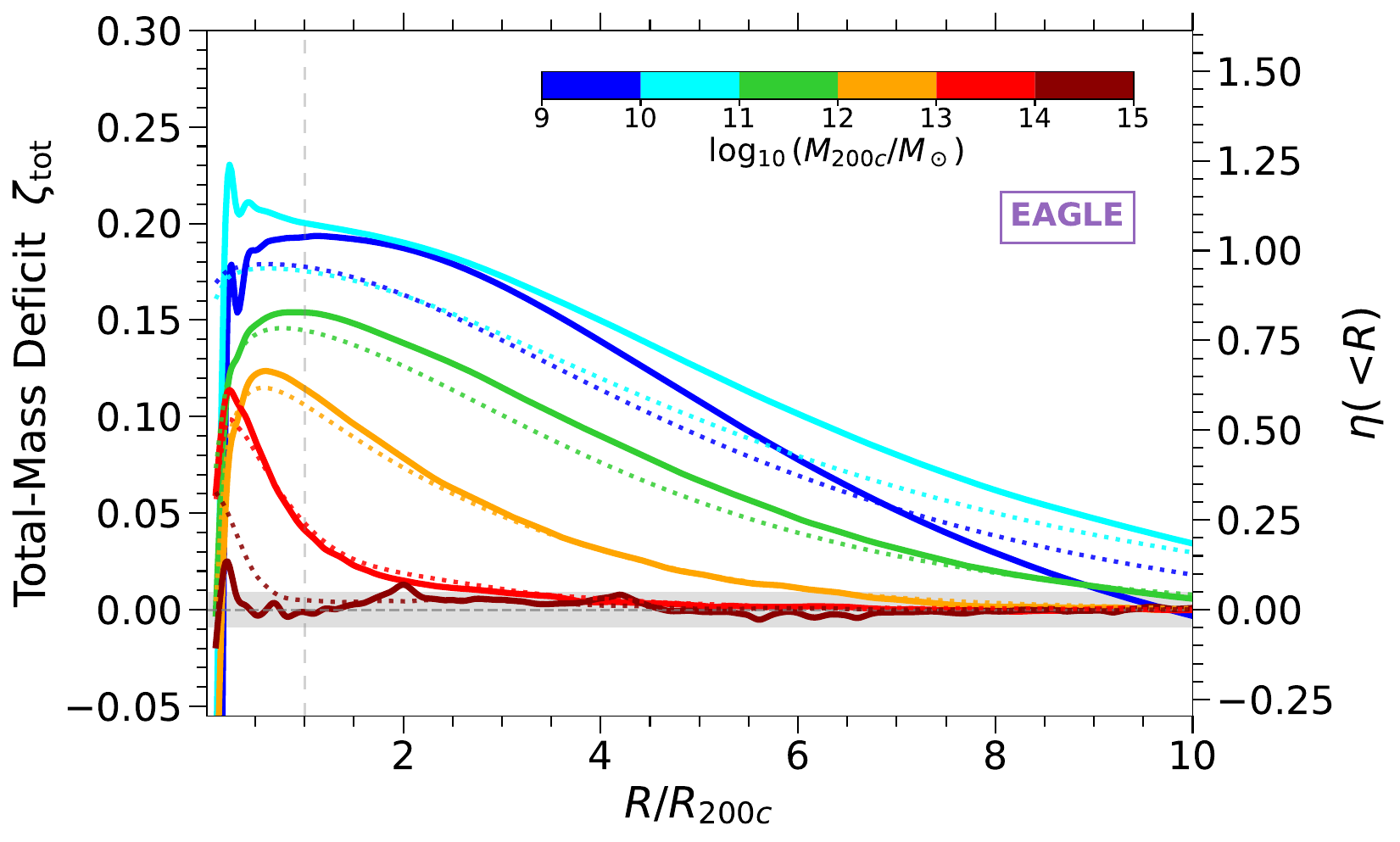} &
      \includegraphics[width=0.33\textwidth]{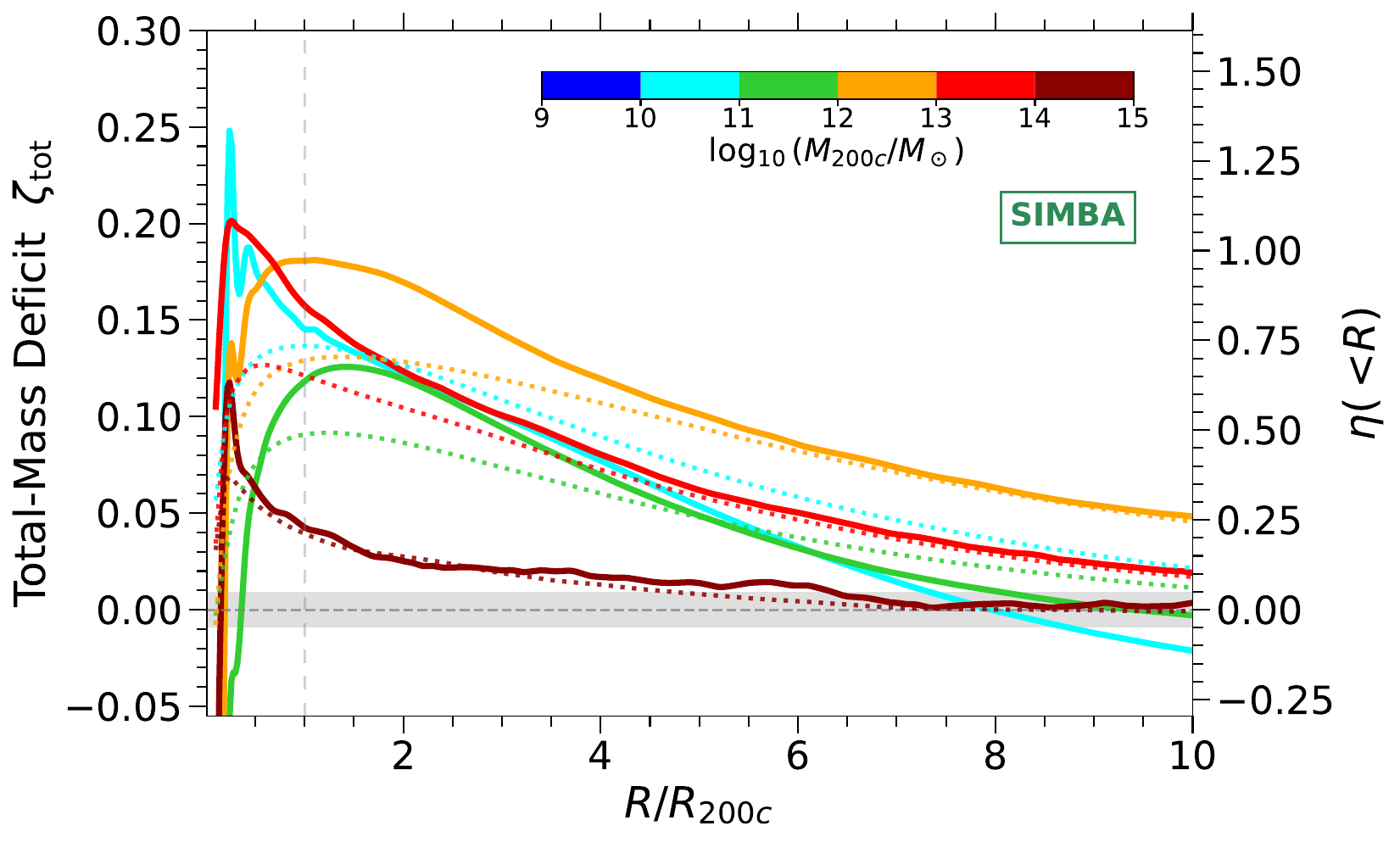} \\[0pt]
      \hspace*{-0.4mm}\includegraphics[width=0.295\textwidth]{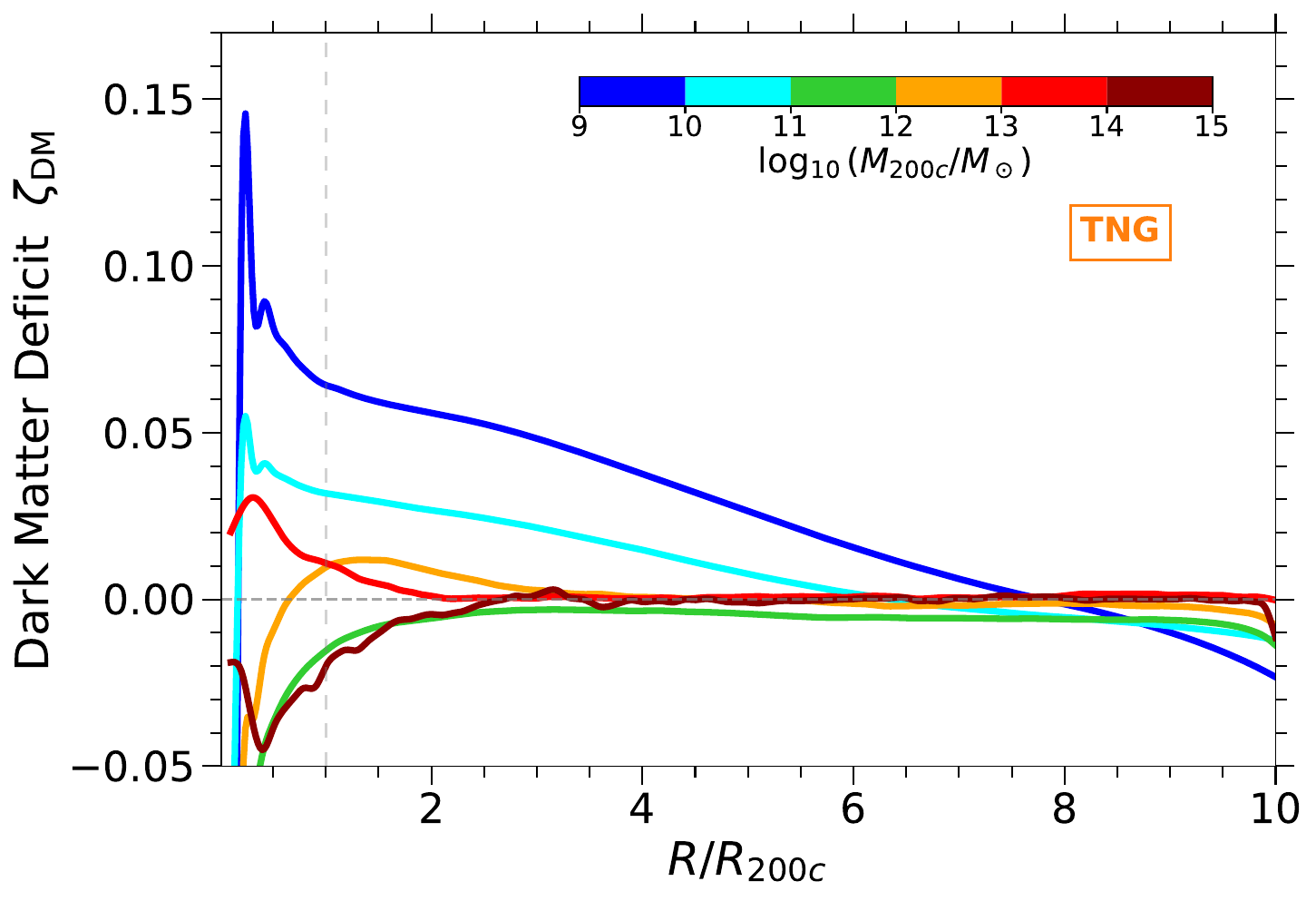} &
      \hspace*{-0.4mm}\includegraphics[width=0.295\textwidth]{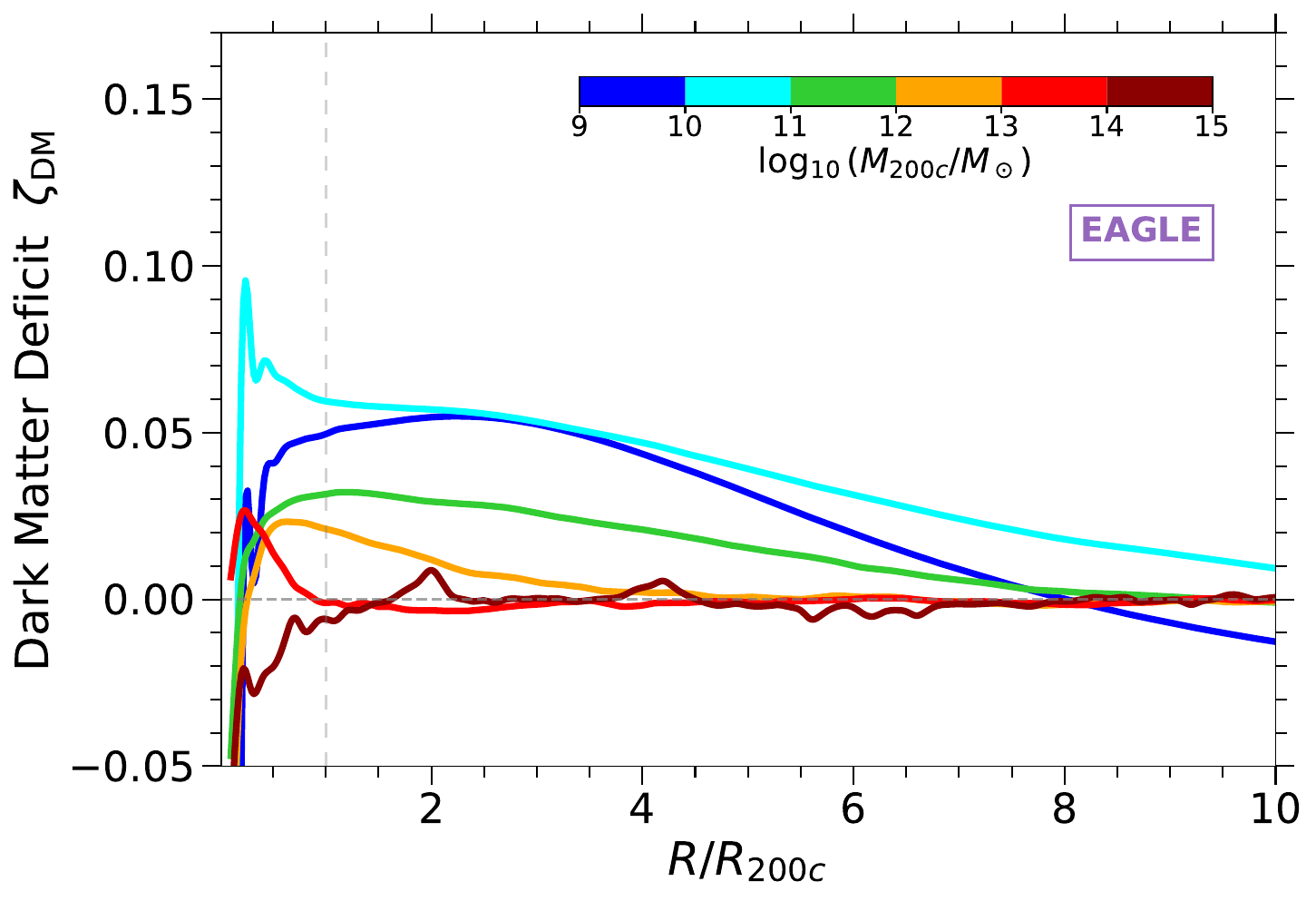} &
      \hspace*{-0.4mm}\includegraphics[width=0.295\textwidth]{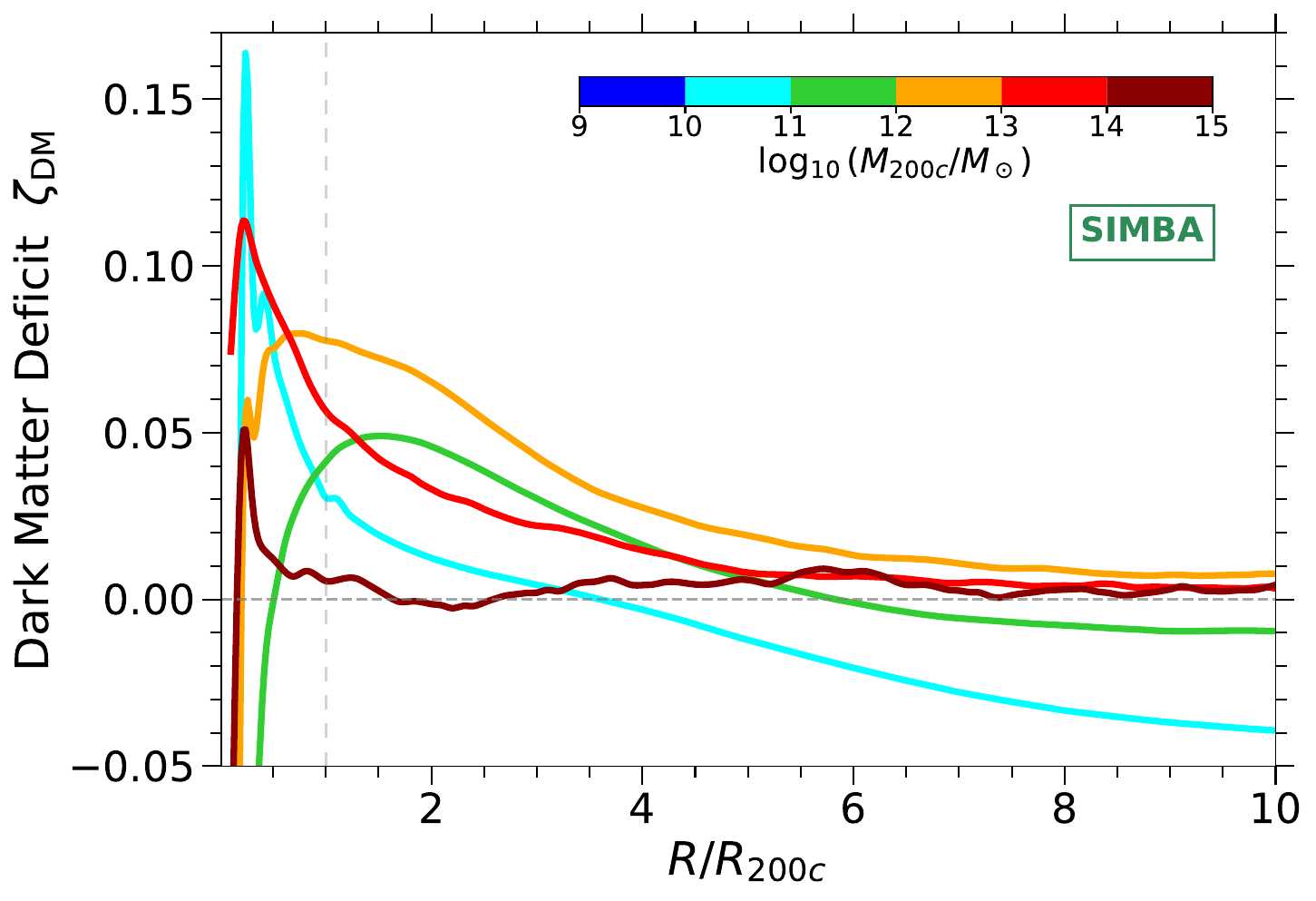}
      \end{tabular}

      \caption{Deficit profiles ($\zeta$ and $\eta$) at $z=0$, for TNG (left), EAGLE (middle), and SIMBA (right). \textbf{Top row:} total-mass deficit $\zeta_{\rm tot}$ (solid lines, left axis) and baryon-fraction deficit $\eta$ (dotted lines, right axis). \textbf{Bottom row:} dark matter deficit $\zeta_{\rm DM}$. Curves show medians of individual-halo profiles as functions of halocentric radius $R/R_{\rm 200c}$; colors denote halo mass bins as indicated by the color bars. Gray bands show the adopted tolerances for $\zeta_{\rm tot}$ and $\eta$, with the two axes scaled so these bands approximately coincide. Vertical dashed lines mark $R_{\rm 200c}$ and horizontal dashed lines mark zero deficit. Positive and negative values indicate deficits and excesses relative to the respective reference values (Appendix~\ref{app: tolerance_definitions}).}
\label{Fig: Deficit_Profiles_zeta}
\end{figure*}

The cosmic baryon shares in and missing from halos show a similar hierarchy but with stronger model dependence. At $z=0$, baryons retained within halos peak in groups and low-mass clusters, around $\log_{10}(M_{\rm 200c}/\mathrm{M_{\odot}})\sim 13.5$--$13.8$, at slightly higher mass than the dark matter peak. This shift indicates that baryons are removed more efficiently from lower-mass halos. TNG and SIMBA also show a secondary peak near $\log_{10}(M_{\rm 200c}/\mathrm{M_{\odot}})\sim 11.8$, whereas EAGLE remains single-peaked. The double-peak structure persists up to $z\sim 1$ in TNG and SIMBA and likely marks the onset of their more ejective AGN feedback above $\log_{10}(M_{\rm 200c}/\mathrm{M_{\odot}})\sim 12$; by $z>1$, even these models revert to a single peak because AGN feedback is less dominant and massive halos are rarer. Apparent upturns at the highest masses or near the resolution limit should not be over-interpreted, as they are driven by cosmic variance or resolution effects. The same caveat applies to the bumpiness of the curves at $\log_{10}(M_{\rm 200c}/\mathrm{M_{\odot}}) \gtrsim 13$, which reflects the small number of such halos in the simulation volumes rather than genuine structure in the mass distribution.

The redshift evolution of the retained baryons is also more sensitive to the feedback model than the dark matter evolution. In TNG and EAGLE, the overall contribution of baryons within halos grows steadily toward lower redshift, whereas in SIMBA it peaks at $z\sim 1$ and then declines modestly toward the present. This decline is consistent with the particularly strong AGN feedback in SIMBA, which ejects gas from halos and reduces their baryon content, and with its rapid rise in the quenched fraction of massive galaxies at $z<1.5$ \citep{Doven2026Environmental}.

The strongest signature of baryon redistribution appears in the bottom row (baryons missing from halos). At $z=0$, groups with $12 \lesssim \log_{10}(M_{\rm 200c}/\mathrm{M_\odot})\lesssim 13.5$ give the largest contribution per dex to this budget in all simulations, with the peak near $\log_{10}(M_{\rm 200c}/\mathrm{M_{\odot}})\sim 12.5$ in TNG and EAGLE and near $\log_{10}(M_{\rm 200c}/\mathrm{M_{\odot}})\sim 13$ in SIMBA. This directly points to AGN feedback as the main mechanism enriching the IGM with baryons. At the same time, lower-mass halos remain important contributors: in EAGLE the distribution is nearly flat below $\log_{10}(M_{\rm 200c}/\mathrm{M_{\odot}})\sim 13$, while TNG and SIMBA show a dip at $\log_{10}(M_{\rm 200c}/\mathrm{M_{\odot}})\sim 11.5$--$12$ where AGN-driven ejection becomes efficient, with stellar feedback dominating below that mass. Because low-mass halos are so abundant, halos around $\log_{10}(M_{\rm 200c}/\mathrm{M_{\odot}})\sim 10$ still contribute nearly as much as the group peak in TNG and EAGLE. We do not draw this comparison for SIMBA, whose halos below $10^{11}\,\mathrm{M_\odot}$ are unreliable (Sect.~\ref{subsubsec: SIMBA}). Up to $z\sim 1$, the overall shape changes only modestly, but toward higher redshift the dominant contribution shifts to lower masses; by $z>4$, groups no longer contribute significantly in any simulation and low-mass halos dominate.

The cumulative sub-panels give the totals directly. Summed over all resolved halos at $z=0$, the share of the cosmic baryon budget retained within $R_{\rm 200c}$ is $23\%$ in TNG, $18\%$ in EAGLE and $13\%$ in SIMBA, while the share missing from halos is $14\%$, $17\%$ and $15\%$ respectively. Together these give the total baryon budget associated with halos, $37\%$, $35\%$ and $28\%$, the value approached by the dashed curves at the low-mass end. The ordering is expected: SIMBA, with the strongest AGN feedback, retains the least and has displaced more than half of its halo-associated baryons beyond $R_{\rm 200c}$, whereas TNG retains the most.

We stress that the quantities in this section are defined with respect to $R_{\rm 200c}$ apertures, so that the ``baryons missing from halos'' are those displaced beyond $R_{\rm 200c}$. This differs from the global census of Sect.~\ref{subsec: baryon_budget}, where halos are FoF groups whose effective radii are larger than $R_{\rm 200c}$ by a factor of roughly $1.5$--$2$; a fraction of the baryons counted as missing here therefore still lies within the FoF group of their host and is not yet part of the IGM as defined there. The expected baryon mass is also anchored differently in the two cases, on the enclosed total mass here and on the enclosed dark matter in Sect.~\ref{subsec: baryon_budget}; both definitions are applied consistently to all three models.

\begin{figure*}
      \centering

      \includegraphics[width=0.33\textwidth]{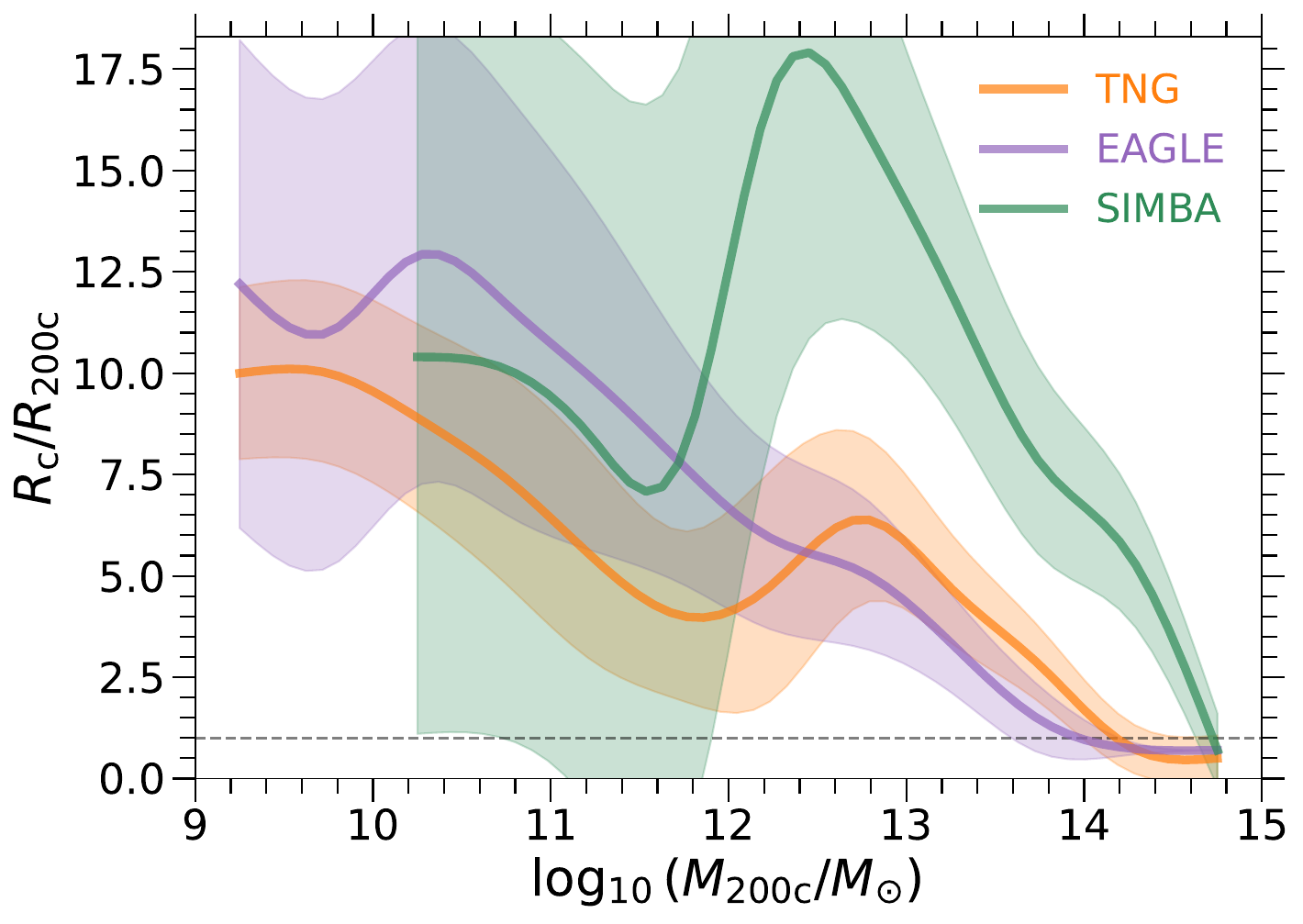}
      \includegraphics[width=0.31\textwidth]{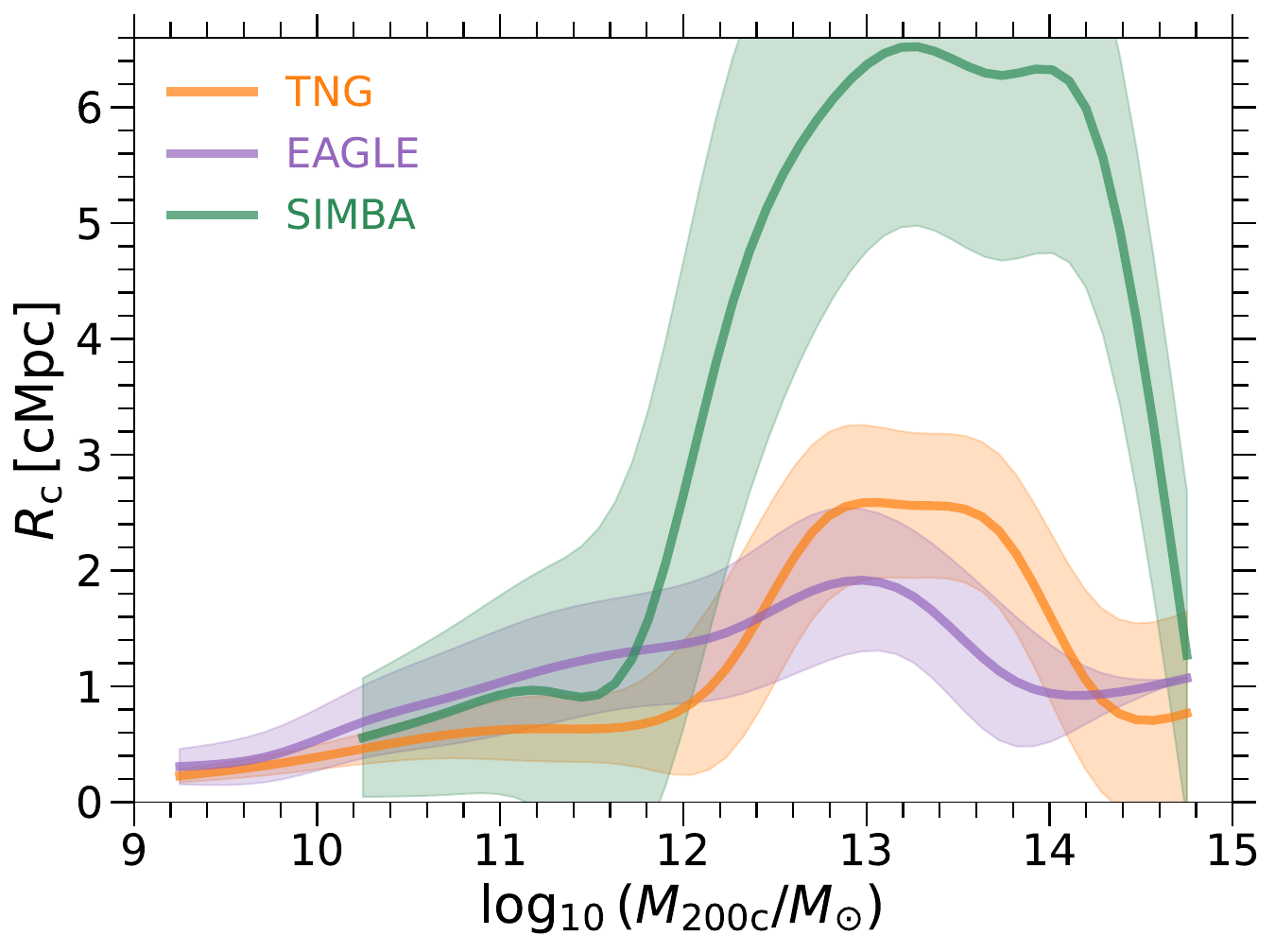}
      \includegraphics[width=0.32\textwidth]{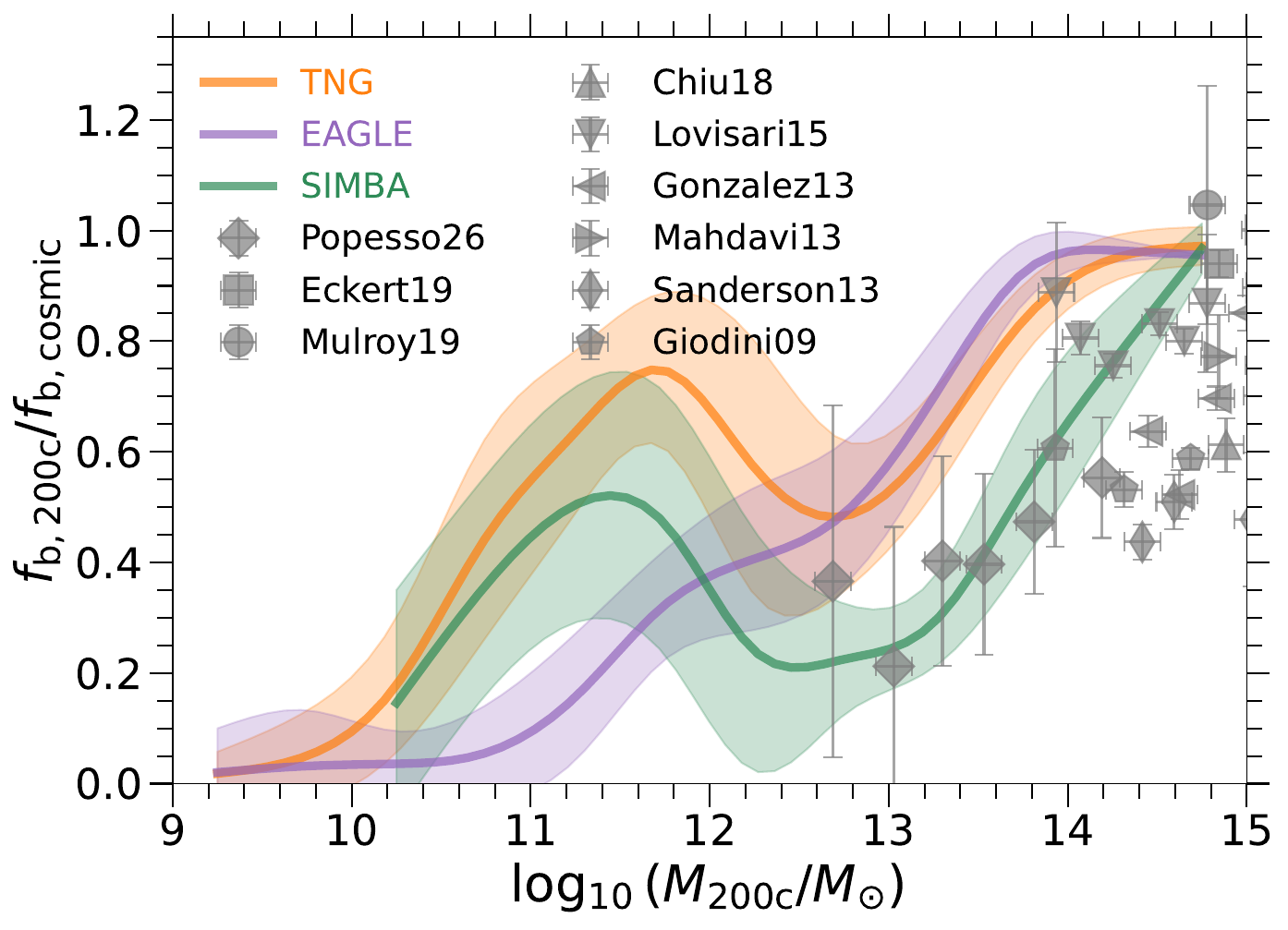}

      \includegraphics[width=0.33\textwidth]{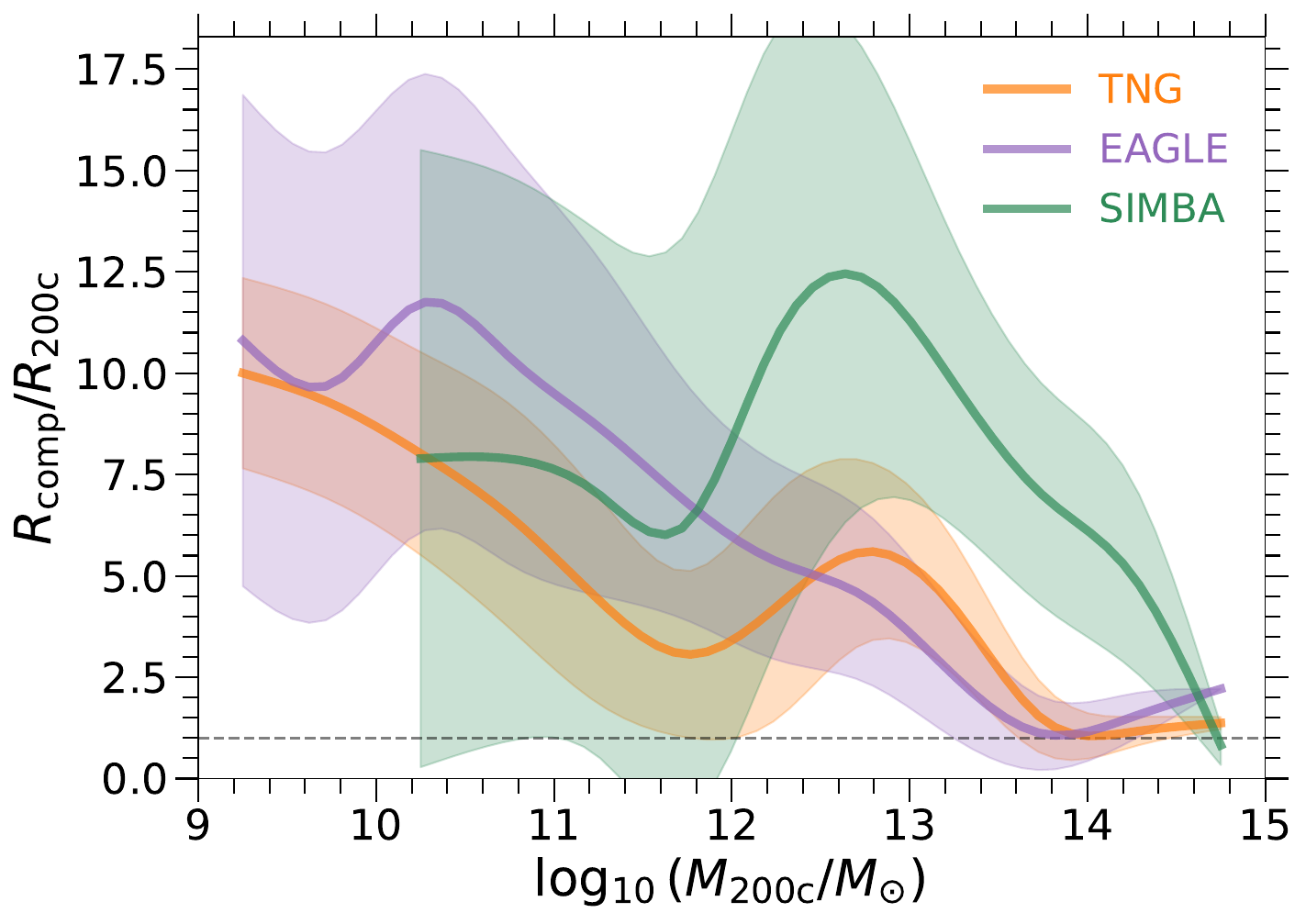}
      \includegraphics[width=0.31\textwidth]{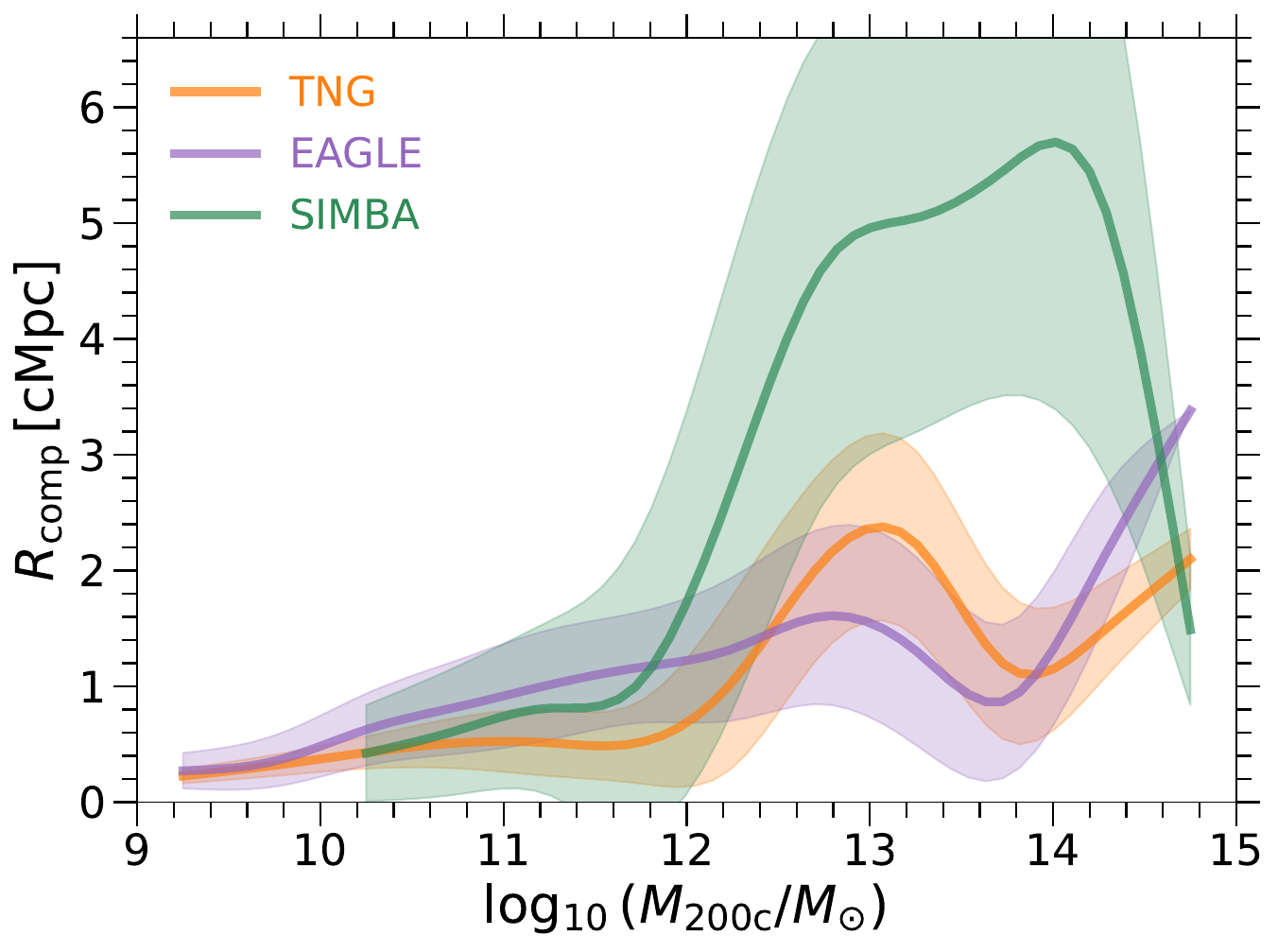}
      \includegraphics[width=0.32\textwidth]{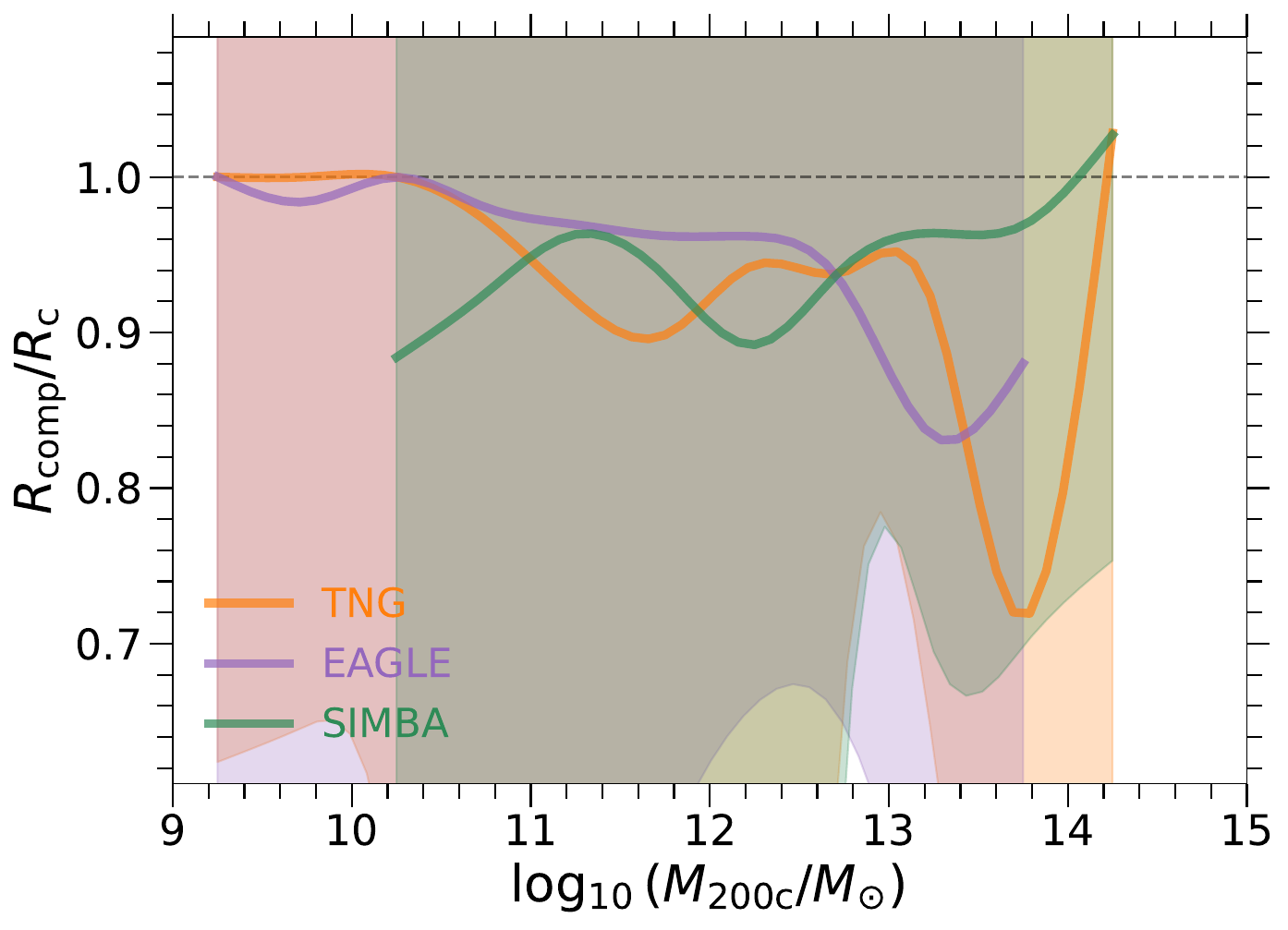}

      \caption{Characterizing the impact of baryon redistribution as a function of halo mass $M_{\rm 200c}$ at $z=0$, for TNG (orange), EAGLE (purple), and SIMBA (green). \textbf{Top row, left to right:} closure radius normalized by the halo radius, $R_{\rm c}/R_{\rm 200c}$; closure radius in comoving units, $R_{\rm c}\,[\rm cMpc]$; and baryon fraction within $R_{\rm 200c}$ normalized by the cosmic mean, $f_{\rm b,200c}/f_{\rm b,cosmic}$. \textbf{Bottom row, left to right:} normalized compensation radius, $R_{\rm comp}/R_{\rm 200c}$; compensation radius in comoving units, $R_{\rm comp}\,[\rm cMpc]$; and the ratio of compensation radius to the closure radius, $R_{\rm comp}/R_{\rm c}$. Solid curves show medians and shaded bands the 16th--84th percentiles within halo mass bins. Horizontal dashed lines mark radii equal to $R_{\rm 200c}$ in the left column and equal compensation and closure radii in the bottom-right panel. Gray markers in the top-right panel show observational measurements of the halo baryon fraction.}
\label{Fig: Rc_Rcomp_z0}
\end{figure*}

These totals are set by the mass range the simulations resolve, and that limit affects the two components very differently. Integrating the halo mass function over all halo masses expected in $\Lambda$CDM places $69$--$73\%$ of the dark matter within halos \citep{Zheng2026Dark}, far above the $43$--$46\%$ measured in Sect.~\ref{subsec: baryon_budget} for halos above $10^{10}\,\mathrm{M_\odot}$. For the baryons the correction is much smaller: by $10^{10}\,\mathrm{M_\odot}$ the halo baryon fraction has fallen to a few per cent of the cosmic value (Fig.~\ref{Fig: Rc_Rcomp_z0}) and the contribution of such halos to the retained budget approaches zero (Fig.~\ref{Fig: matter_dist_halo_contribution_z0}), as expected when UV-background photoheating prevents gas accretion below that mass. We therefore expect that extending the census beyond the resolved halo population would add little: over $70\%$ of cosmic baryons would lie outside collapsed structures of any mass, while most of the dark matter is bound within them.

\section{Baryon--dark matter redistribution}
\label{sec: Characteristic_Radii}
To quantify the spatial scales over which baryonic processes displace baryons relative to the dark matter, we employ four characteristic radii, organized into two complementary tracks: closure radius and closure envelope; compensation radius and compensation envelope. Below, we describe them in detail.

The closure radius ($R_{\rm c}$) was introduced in \cite{Ayromlou2023Feedback} and identifies the halocentric distance at which the enclosed baryon fraction equals the cosmic mean. Here we extend its analysis to higher redshifts and to individual halo evolutionary tracks. Taking the maximum of the smoothed, mass-binned median relation $R_{\rm c}(M_{\rm 200c})$ at a given redshift defines the \emph{closure envelope}, $R_{\rm c}^{\rm env}(z)$, the largest scale to which the baryon deficit of the halo population extends (Sect.~\ref{subsec: characteristic radii method}).

We introduce the compensation radius ($R_{\rm comp}$) as the halocentric distance at which the total enclosed masses in the hydrodynamical and DMO simulations converge. While $R_{\rm c}$ measures the recovery of the \emph{relative} baryon-to-total-mass proportion, $R_{\rm comp}$ measures the recovery of the \emph{absolute} total mass, including any response of the dark matter to baryonic processes. The two per-halo scales are complementary: $R_{\rm c}$ can be smaller or larger than $R_{\rm comp}$ depending on whether baryons and dark matter are proportionally or differentially displaced. Finally, the \emph{compensation envelope}, $R_{\rm comp}^{\rm env}(z)$, is the corresponding maximum of the smoothed, mass-binned median relation.

\subsection{Characteristic radii and their envelopes}
\label{subsec: characteristic radii method}
\begin{figure*}
      \centering
      \setlength{\tabcolsep}{0pt}
      \begin{tabular}{@{}rrr@{}}
      \includegraphics[width=0.33\textwidth]{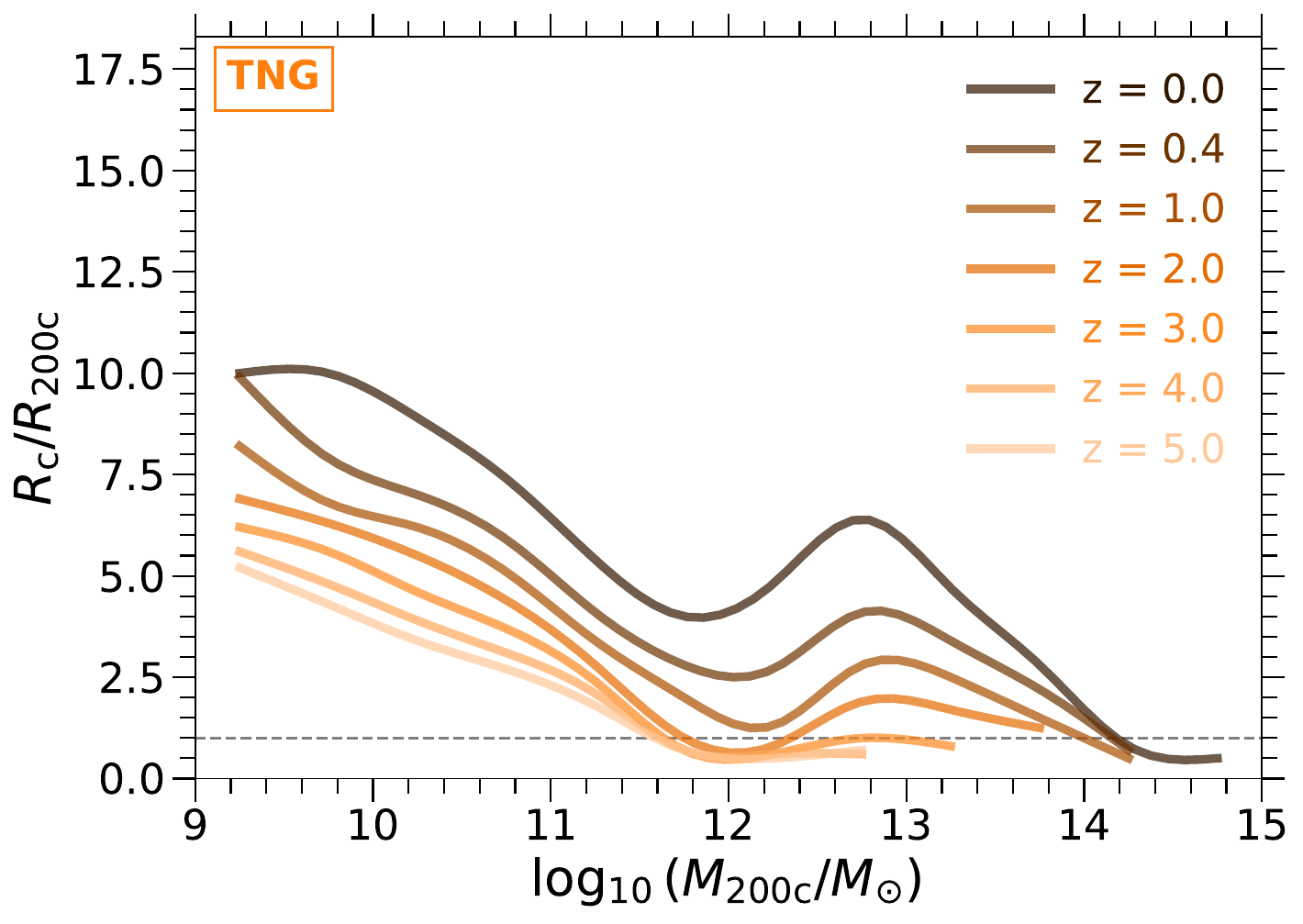} &
      \includegraphics[width=0.33\textwidth]{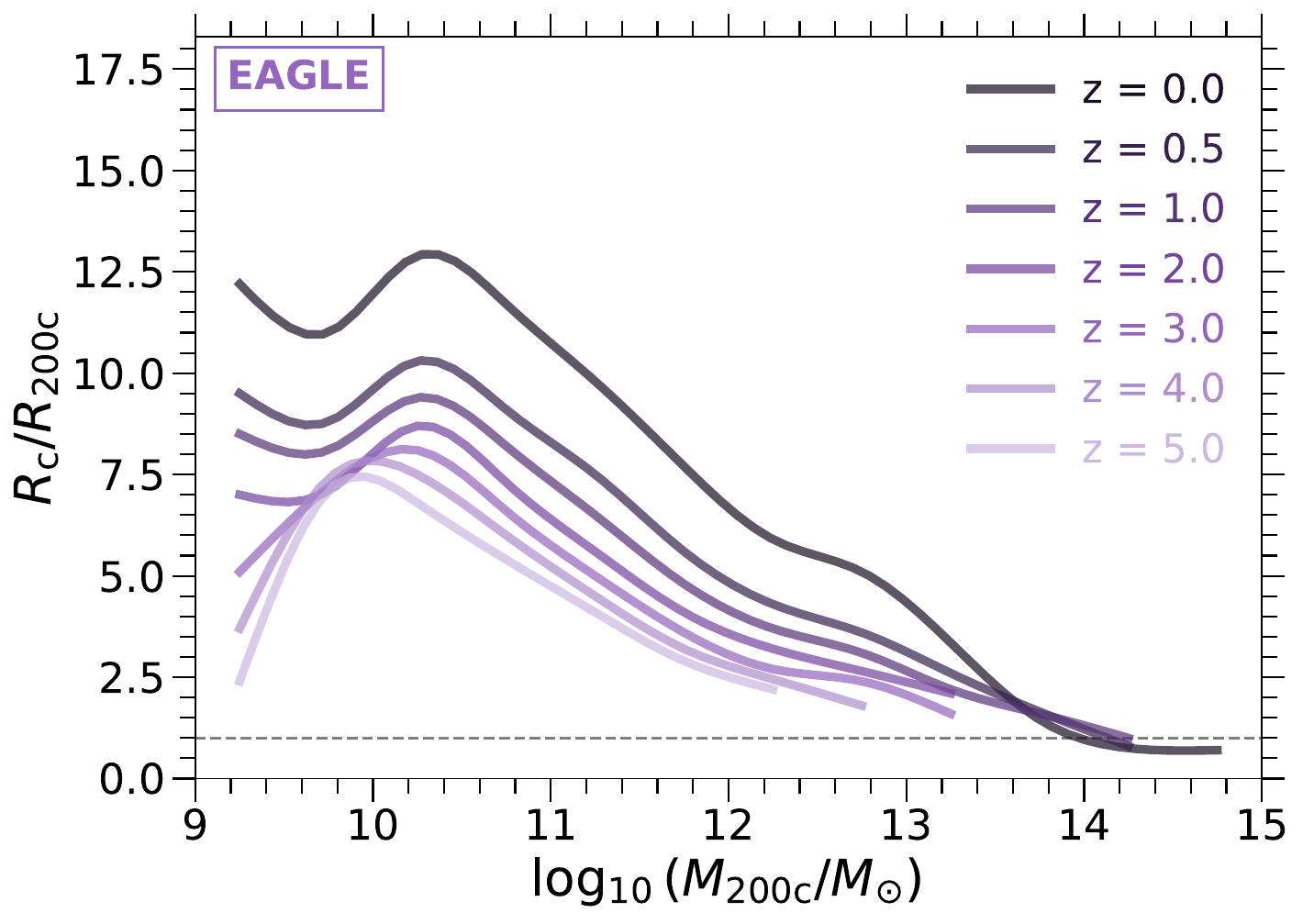} &
      \includegraphics[width=0.33\textwidth]{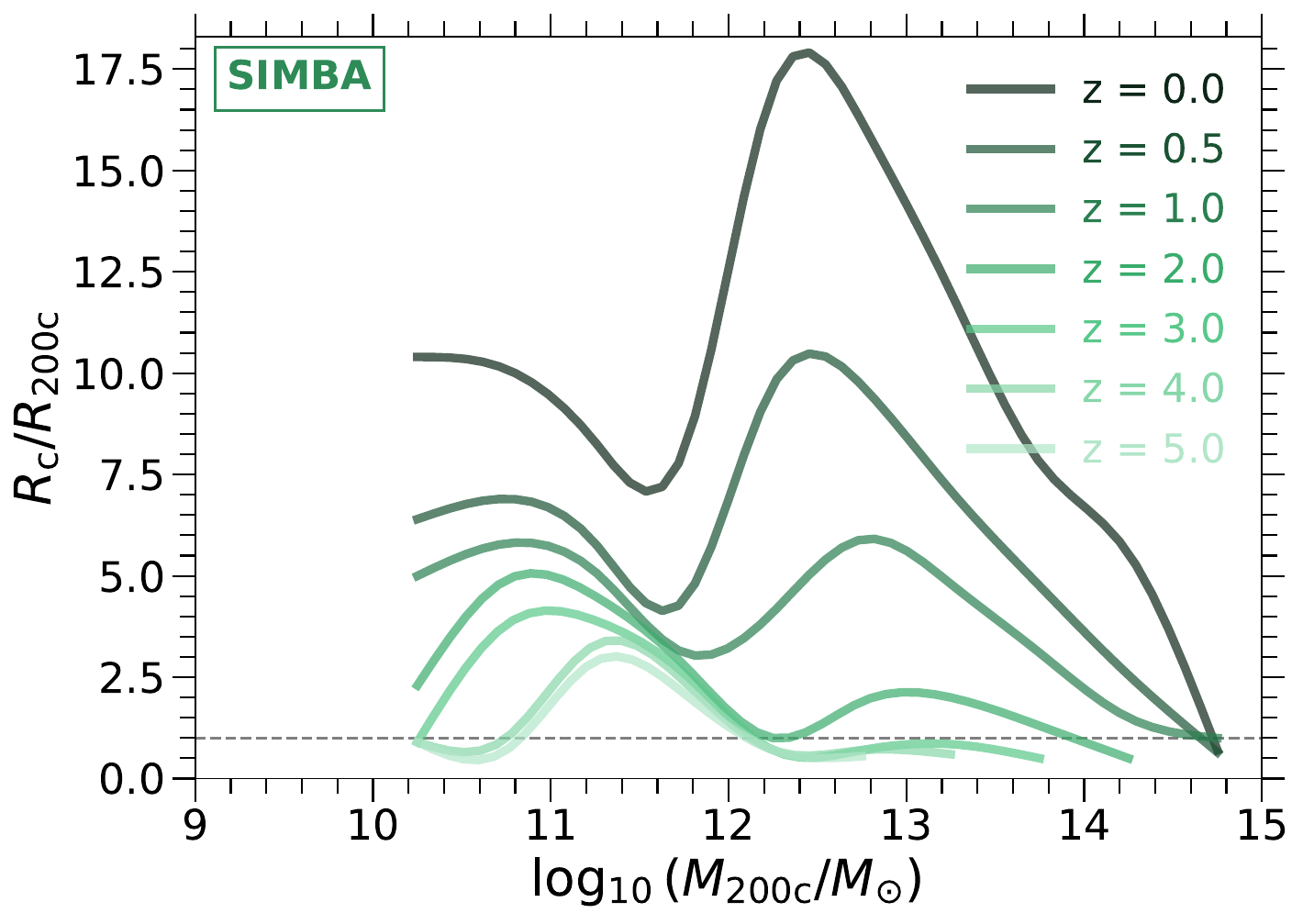} \\[0pt]
      \includegraphics[width=0.3165\textwidth]{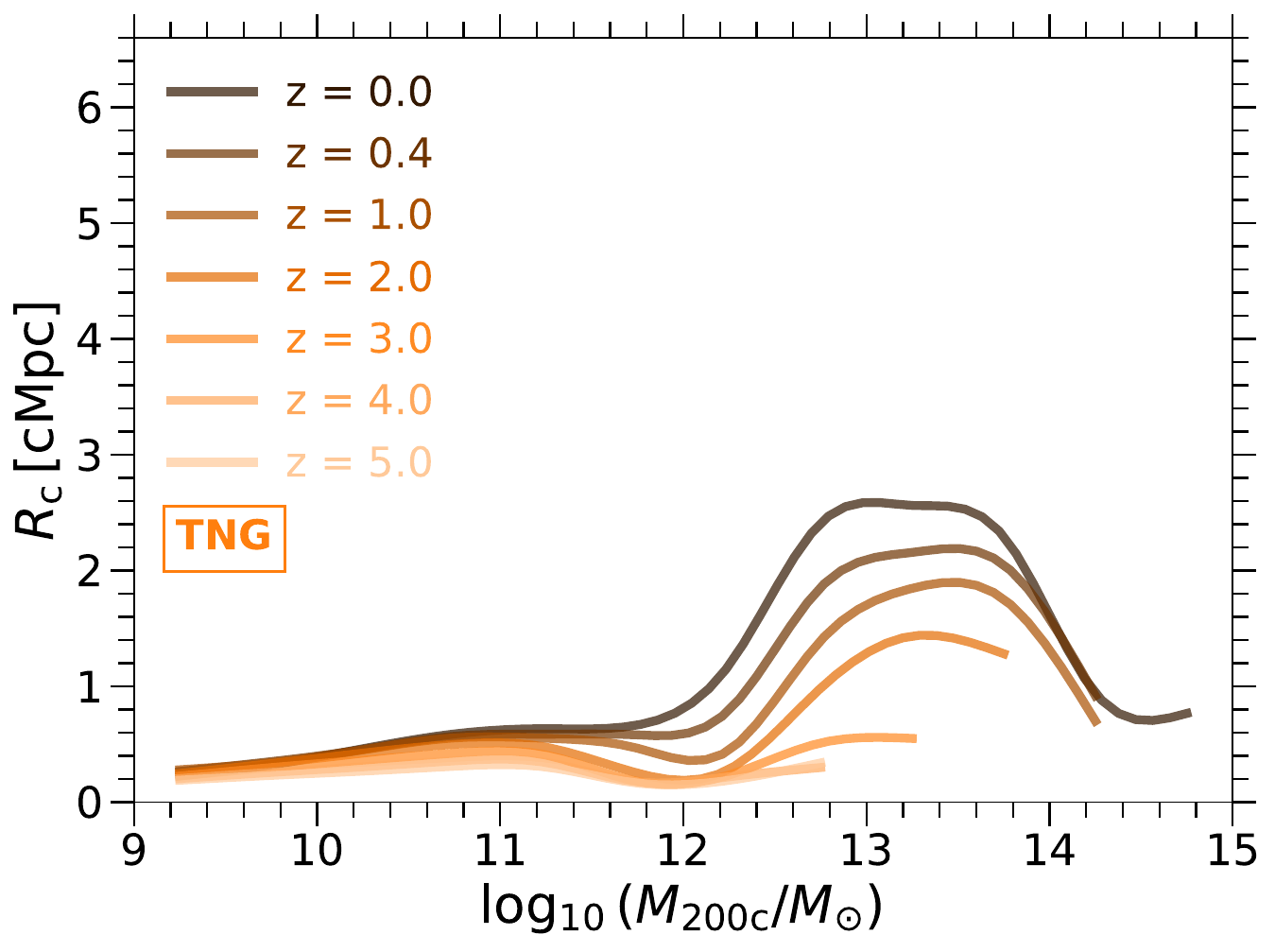} &
      \includegraphics[width=0.3165\textwidth]{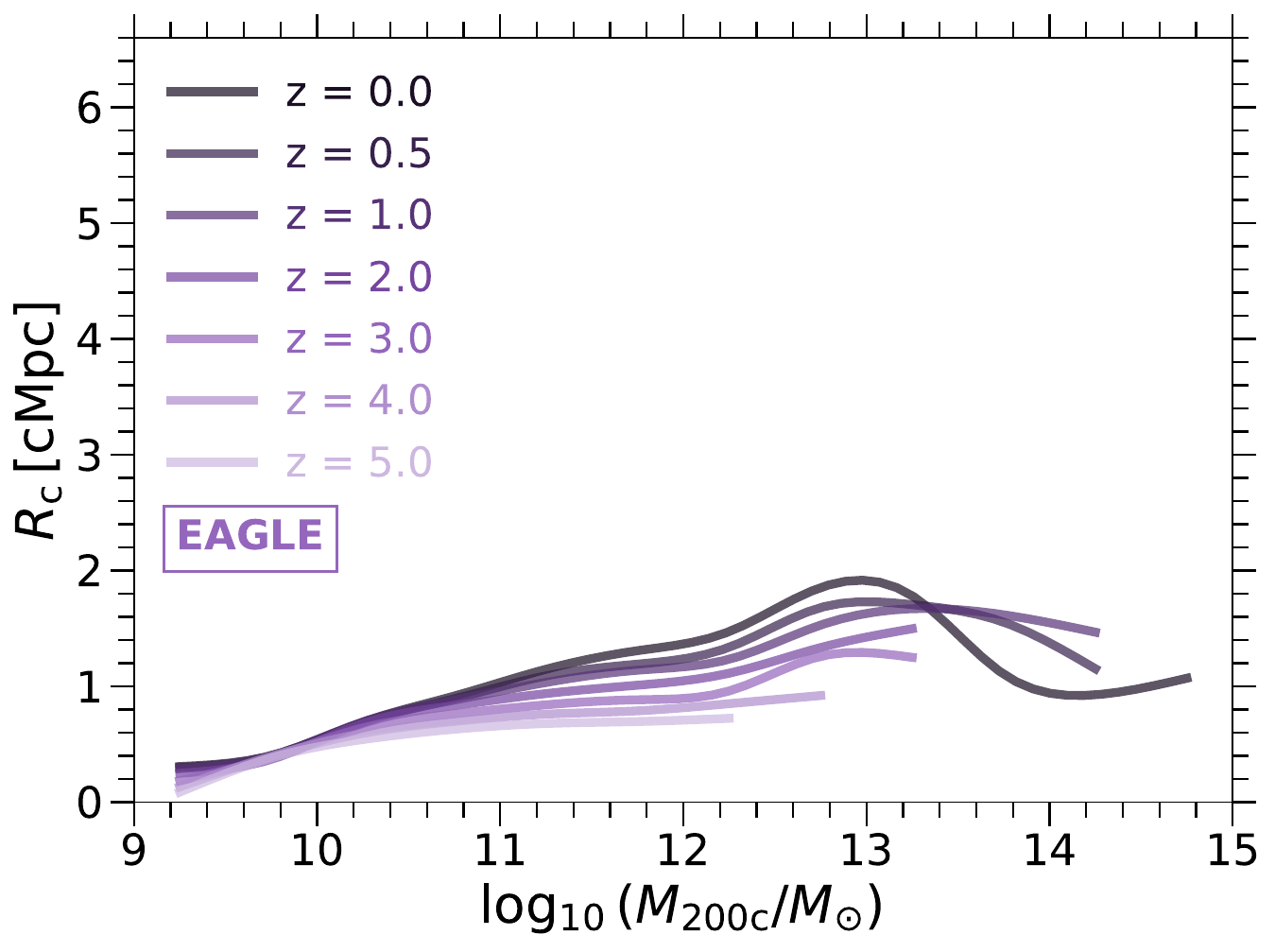} &
      \includegraphics[width=0.3165\textwidth]{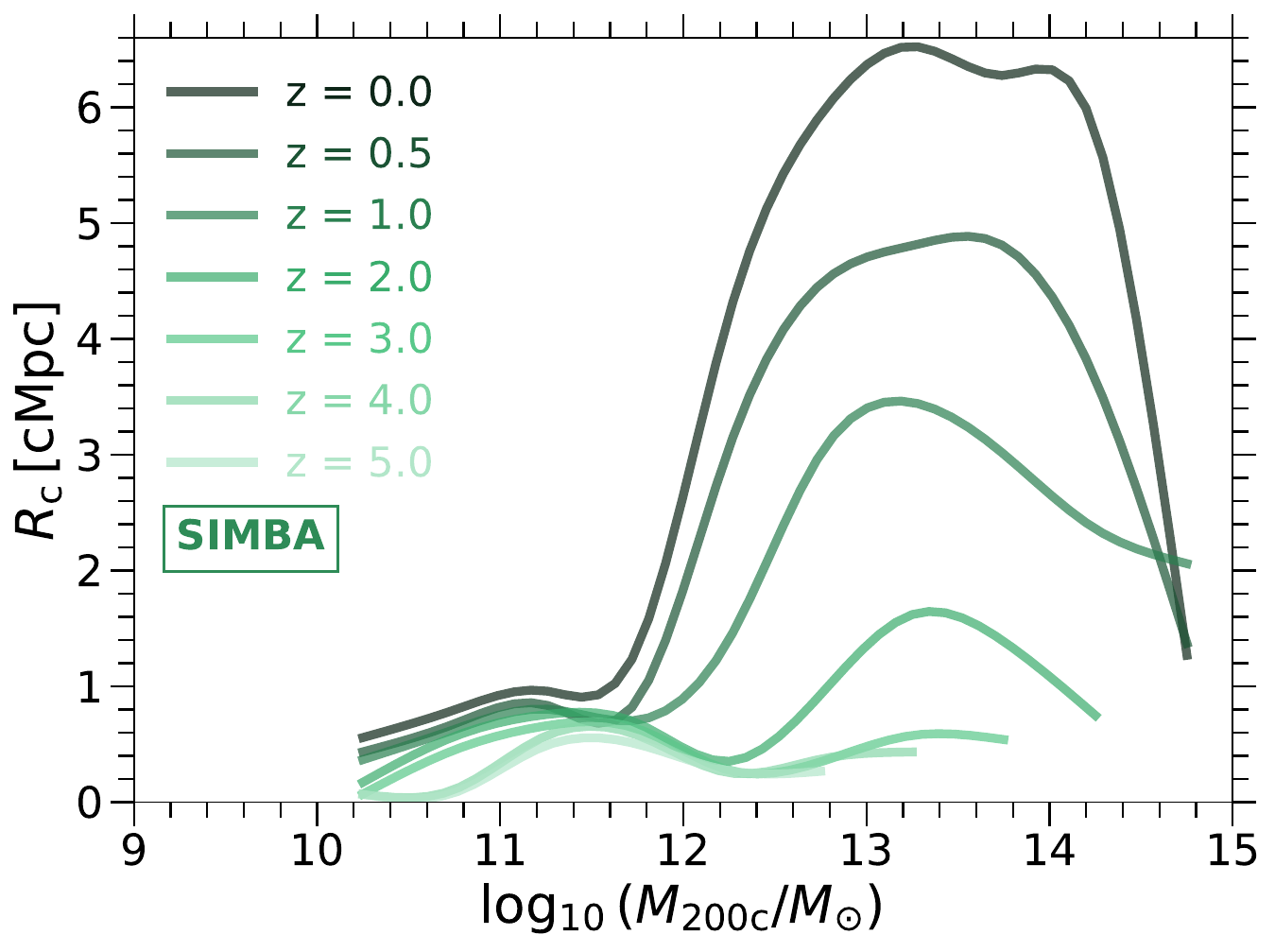} \\[0pt]
      \includegraphics[width=0.33\textwidth]{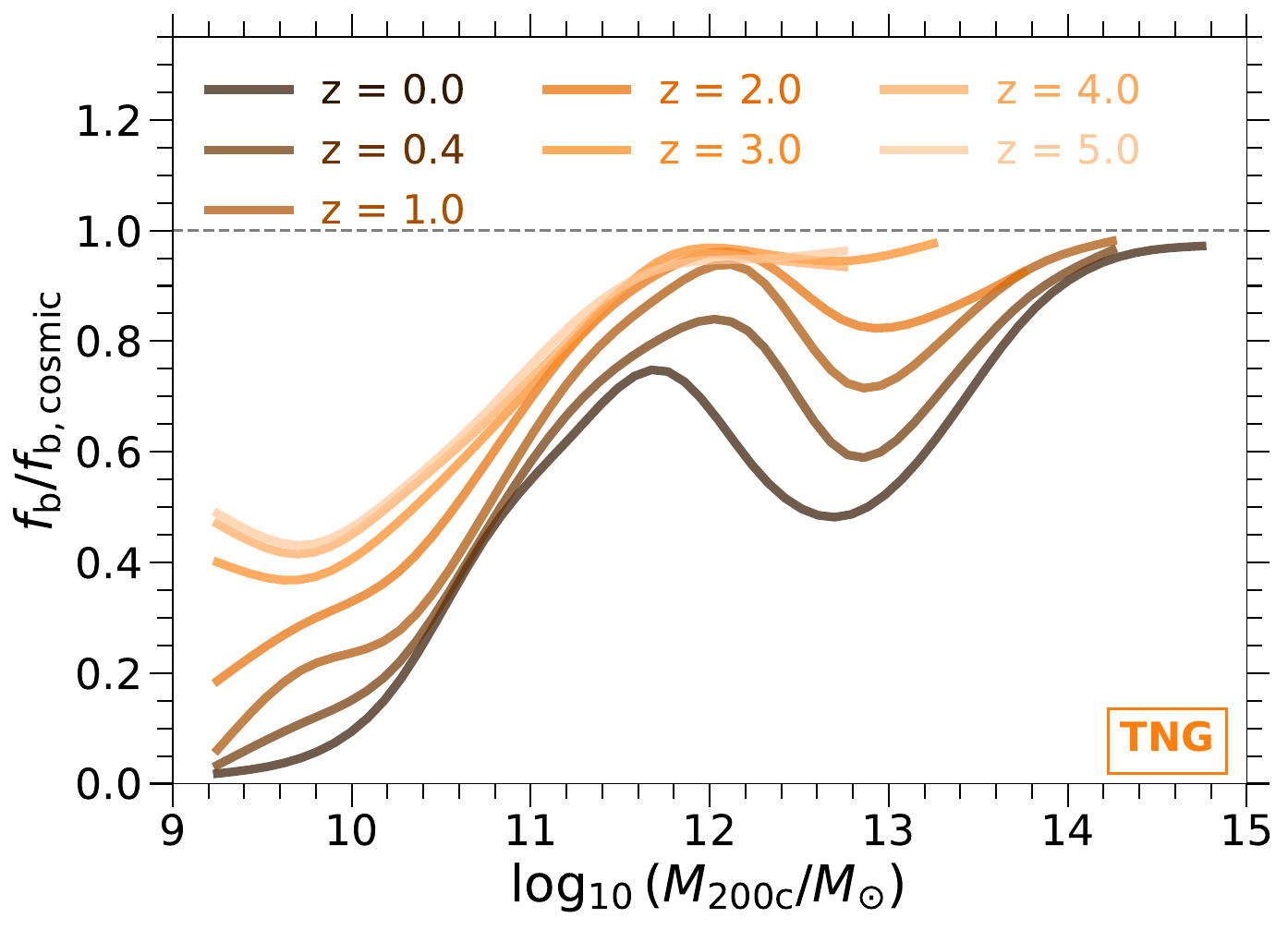} &
      \includegraphics[width=0.33\textwidth]{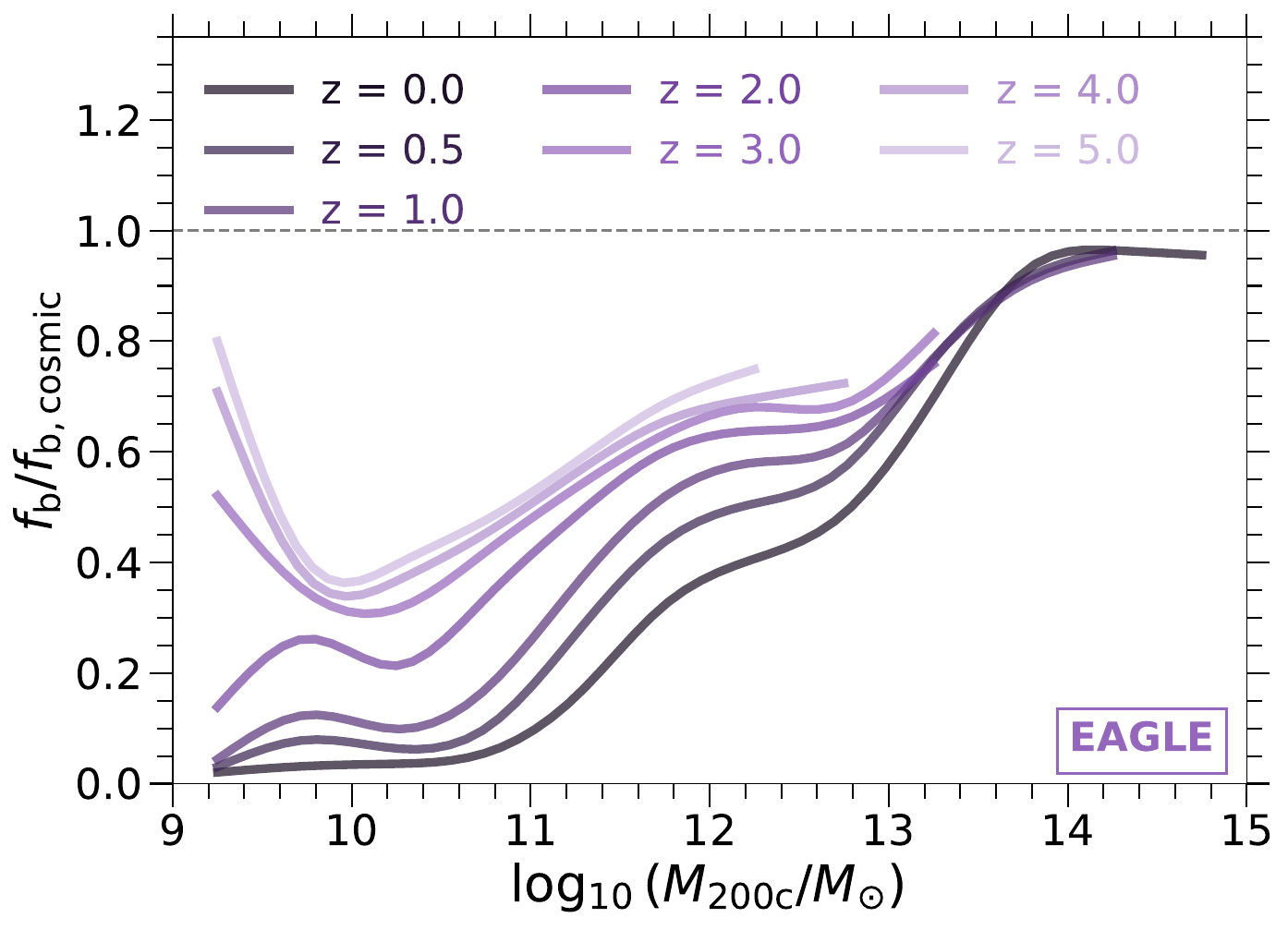} &
      \includegraphics[width=0.33\textwidth]{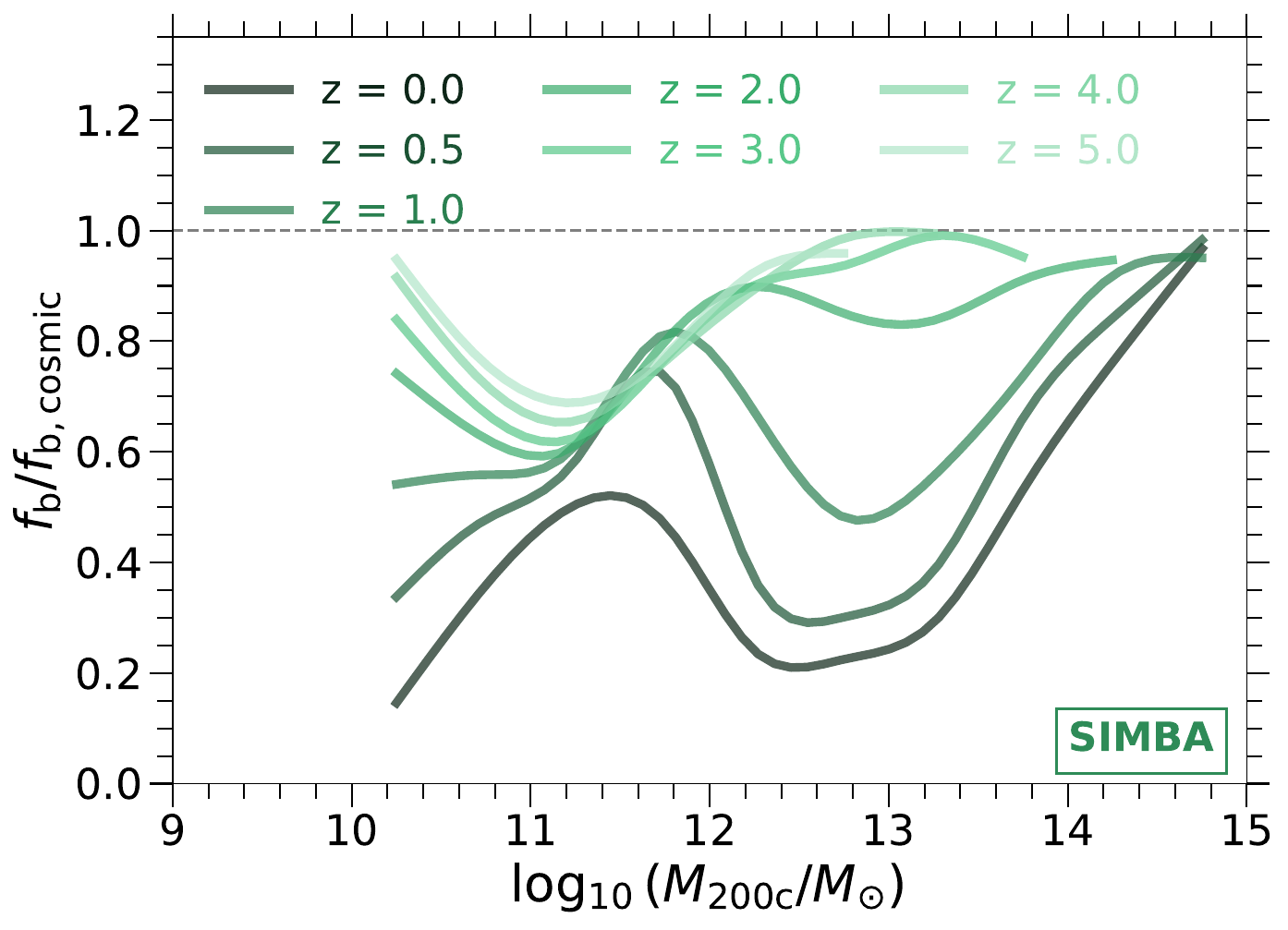}
      \end{tabular}

      \caption{Redshift evolution of the normalized closure radius ($R_{\rm c}/R_{\rm 200c}$, top row), the closure radius in comoving units ($R_{\rm c}\,[\rm cMpc]$, middle row), and the halo baryon fraction ($f_{\rm b,200c}/f_{\rm b,cosmic}$, bottom row), for TNG (left column), EAGLE (middle column), and SIMBA (right column). Curves show medians; darker lines correspond to lower redshift, from $z=5$ (lightest) to $z=0$ (darkest). Dashed lines mark unity in the top and bottom rows.}
\label{Fig: Rc_z}
\end{figure*}
For each halo at each redshift we measure two characteristic radii from its spherically averaged radial mass profile. A characteristic radius may also lie inside $R_{\rm 200c}$, when the corresponding deficit does not leave its tolerance band within the halo.

The closure radius, $R_{\rm c}$, is the innermost radius at which the enclosed baryon fraction reaches the cosmic mean within a relative tolerance of $5\%$, that is, at which the baryon fraction deficit $\eta$ satisfies
\begin{equation}
    \left| \eta(R_{\rm c}) \right| \equiv
    \left| 1 - \frac{f_{\rm b}(<R_{\rm c})}{f_{\rm b,cosmic}} \right| \le 0.05 ,
    \label{eq: Rc_operational}
\end{equation}
following \cite{Ayromlou2023Feedback}. The closure radius is enforced to be at least $0.5\,R_{\rm 200c}$ to avoid spurious results from the innermost radial bins; this condition matters only for very massive clusters and occurs very rarely.

The compensation radius, $R_{\rm comp}$, is measured relative to the halo's matched counterpart in the corresponding DMO run (Sect.~\ref{subsec: Halo properties}). $R_{\rm comp}$ is then the innermost radius at which the enclosed-mass ratio lies within $0.93\%$ of unity (see Appendix~\ref{app: tolerance_derivation}), that is, at which the total-mass deficit $\zeta_{\rm tot}$ satisfies
\begin{equation}
    \left| \zeta_{\rm tot}(R_{\rm comp}) \right| \equiv
    \left| 1 - \frac{M_{\rm hydro}(<R_{\rm comp})}{M_{\rm DMO}(<R_{\rm comp})} \right| \le 0.0093 .
    \label{eq: Rcomp_operational}
\end{equation}
We additionally define the dark matter deficit, $\zeta_{\rm DM}(R) \equiv 1 - M_{\rm DM}^{\rm hydro}(<R)/M_{\rm DM}^{\rm DMO}(<R)$, which isolates the back-reaction of baryonic processes on the dark matter distribution (Appendix~\ref{app: tolerance}).
This tolerance is not independent of the closure tolerance: it is the $5\%$ closure band transported onto the enclosed total mass through the exact relation between the two, making the two radii coincide, to within about $1\%$ in $\eta$, when baryonic processes displace only baryons with no dark matter response (Appendix~\ref{app: tolerance}).

To characterize how these scales evolve with cosmic time we construct the envelopes $R_{\rm c}^{\rm env}(z)$ and $R_{\rm comp}^{\rm env}(z)$. At each snapshot the per-halo radii are binned in $\log_{10}M_{\rm 200c}$ (15 equal-width bins over $9 \le \log_{10}(M_{\rm 200c}/{\rm M_\odot}) \le 15$) and the median is taken in each bin; bins containing fewer than 10 halos are discarded to suppress shot noise at the high-mass end. The median profile is then smoothed with a Gaussian kernel of width $\sigma = 1$ bin (${\approx}\,0.4$~dex), and the global maximum of the smoothed curve defines the envelope at that redshift. Shaded bands around the envelopes give the 16th--84th percentiles of halo radii in the peak bin, i.e., the halo-to-halo scatter at the characteristic mass.

The profiles from which both radii are read are shown in Fig.~\ref{Fig: Deficit_Profiles_zeta} (top row), together with the tolerance bands of Eqs.~(\ref{eq: Rc_operational}) and~(\ref{eq: Rcomp_operational}). The two axes are scaled so that the bands coincide, making the comparison between the two criteria direct: at a given halo mass, whichever of $\eta$ and $\zeta_{\rm tot}$ enters the shaded region first sets the smaller of the two radii. Within and just outside $R_{\rm 200c}$ the dark matter deficit $\zeta_{\rm DM}$ is predominantly \emph{positive} in all three models, reaching $0.05$--$0.16$ depending on model and halo mass, so that the enclosed dark matter falls short of its gravity-only expectation there. Its mass dependence differs substantially between the models, however: the largest values occur in the lowest-mass bins in TNG and EAGLE, but at group and cluster masses in SIMBA. Toward the radii at which the two criteria are satisfied, $\zeta_{\rm DM}$ (bottom row) falls to $\sim 10^{-3}$ and changes sign in several mass bins. The ordering of $R_{\rm comp}$ and $R_{\rm c}$ is therefore governed by a quantity of the same order as the adopted tolerance, and individual mass bins should not be over-interpreted. %

\subsection{The closure radius across cosmic time}
\label{subsec: closure_radius}
\begin{figure*}
      \centering
      \includegraphics[width=0.33\textwidth]{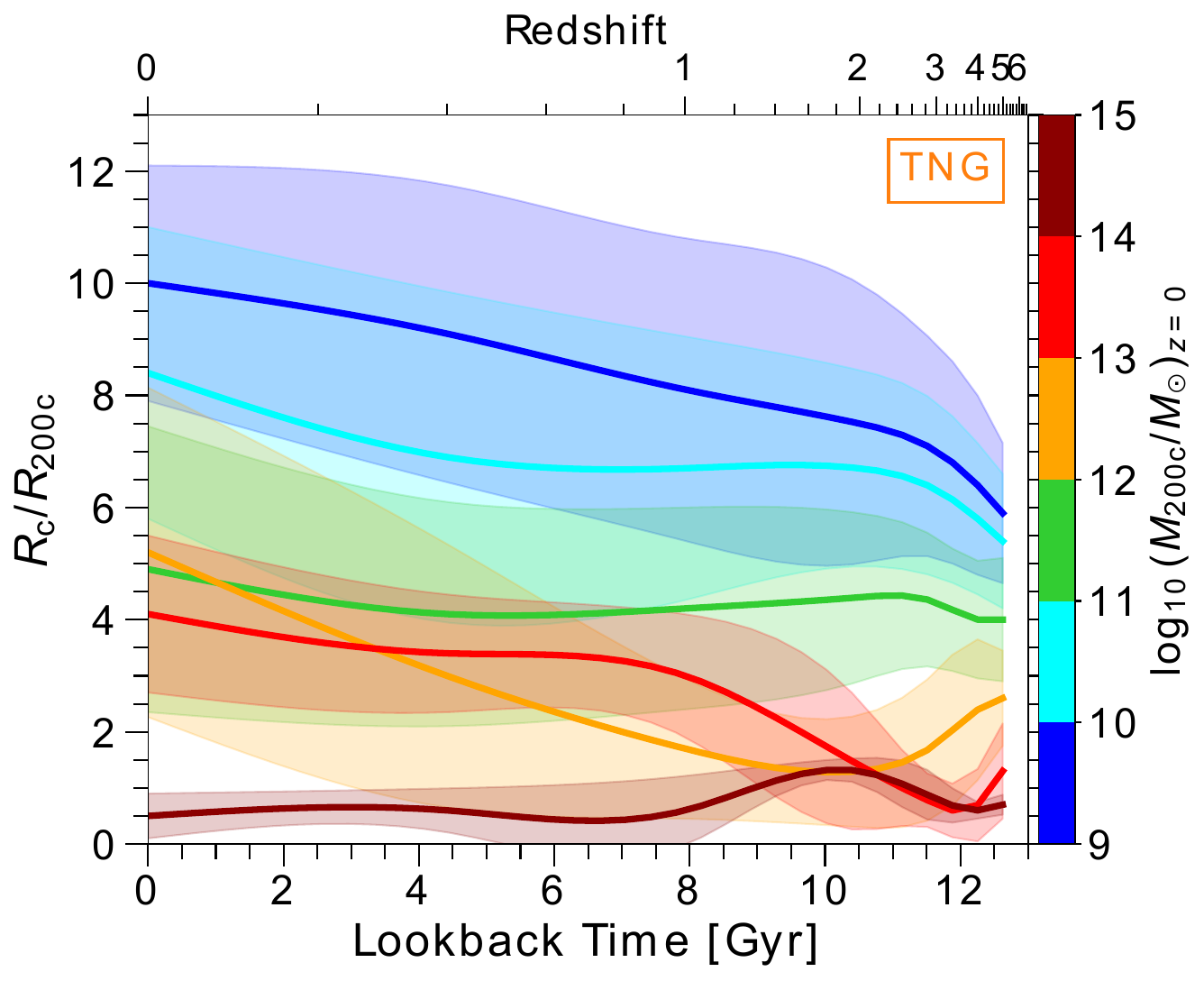}
      \includegraphics[width=0.33\textwidth]{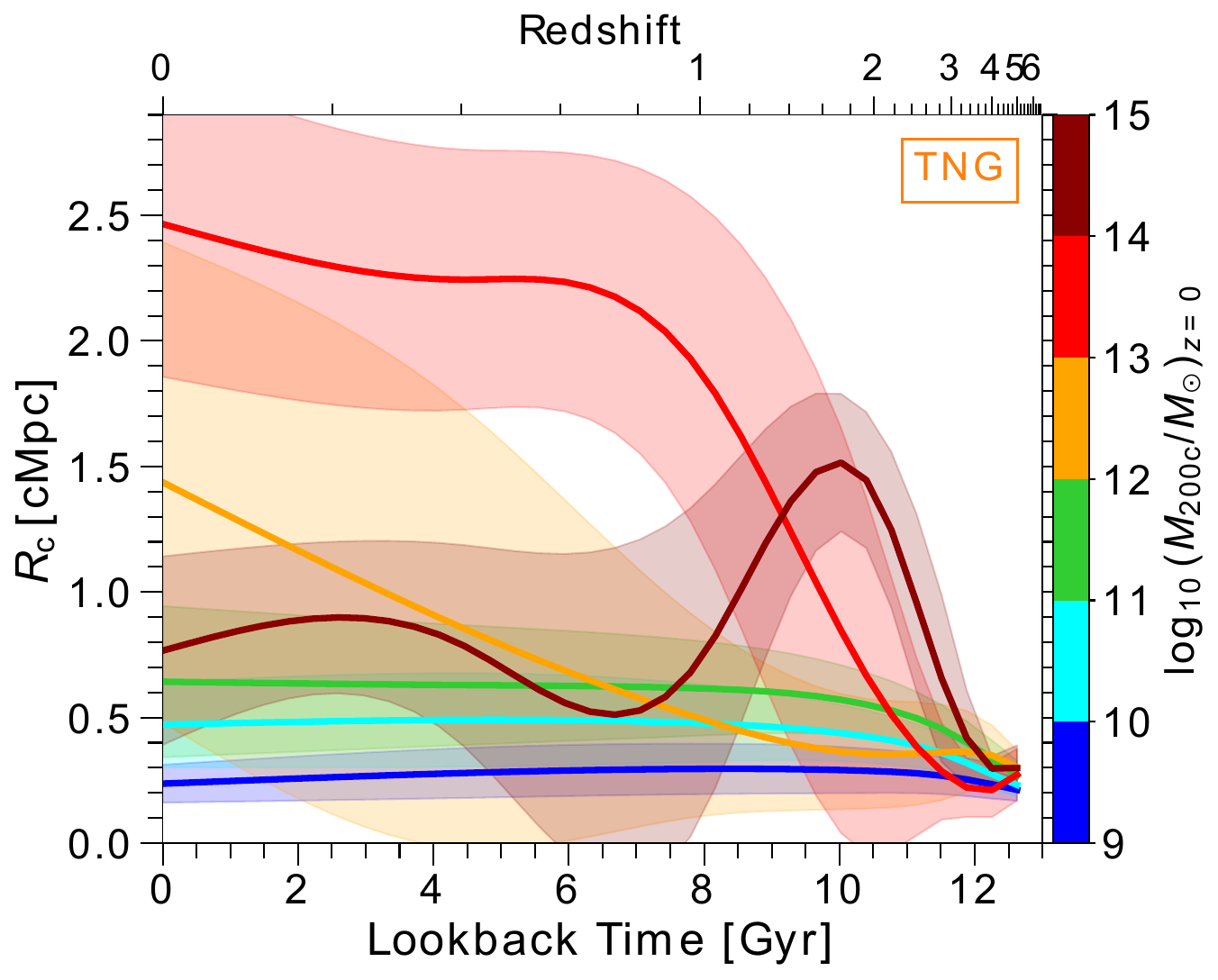}
      \includegraphics[width=0.33\textwidth]{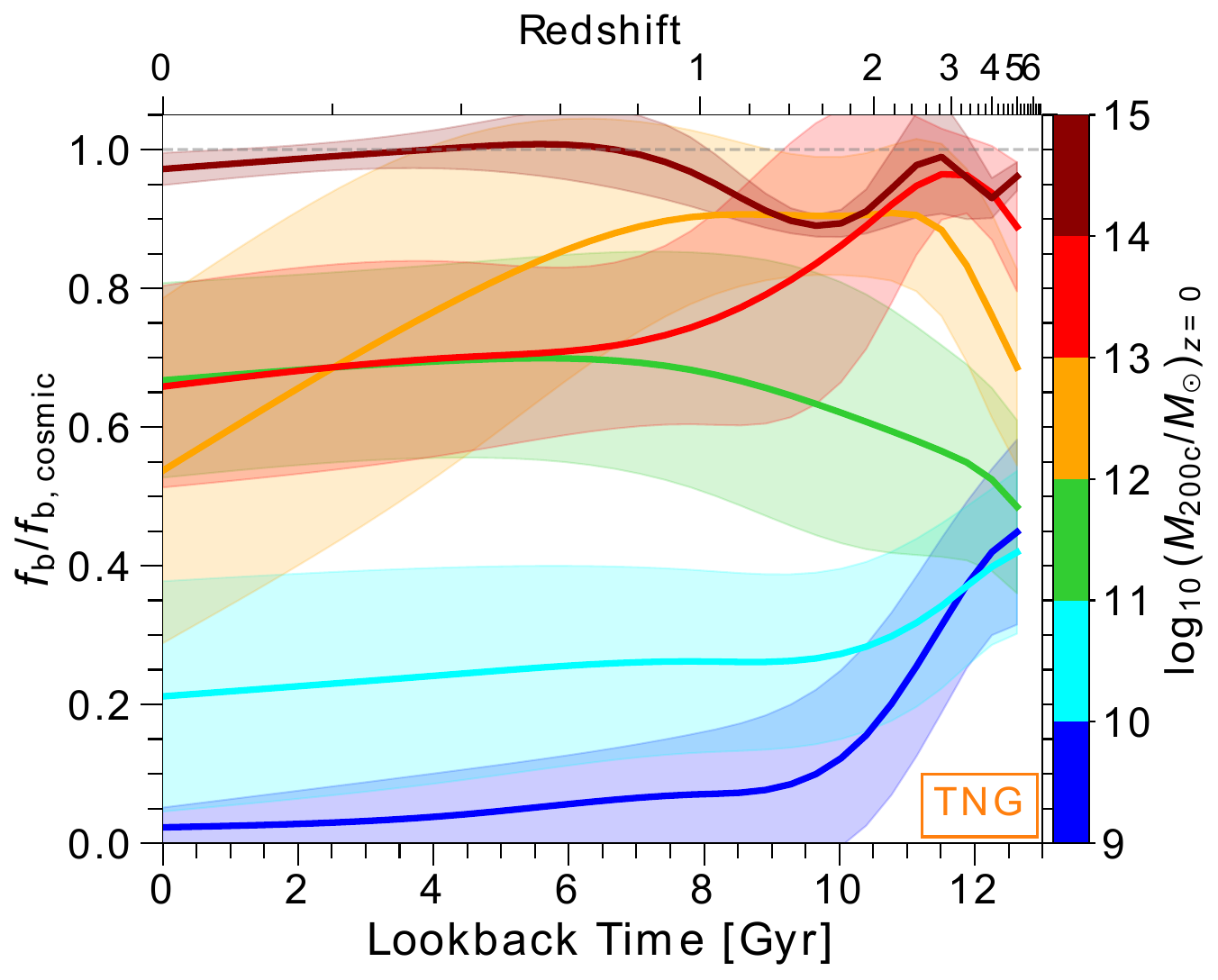}
      \includegraphics[width=0.33\textwidth]{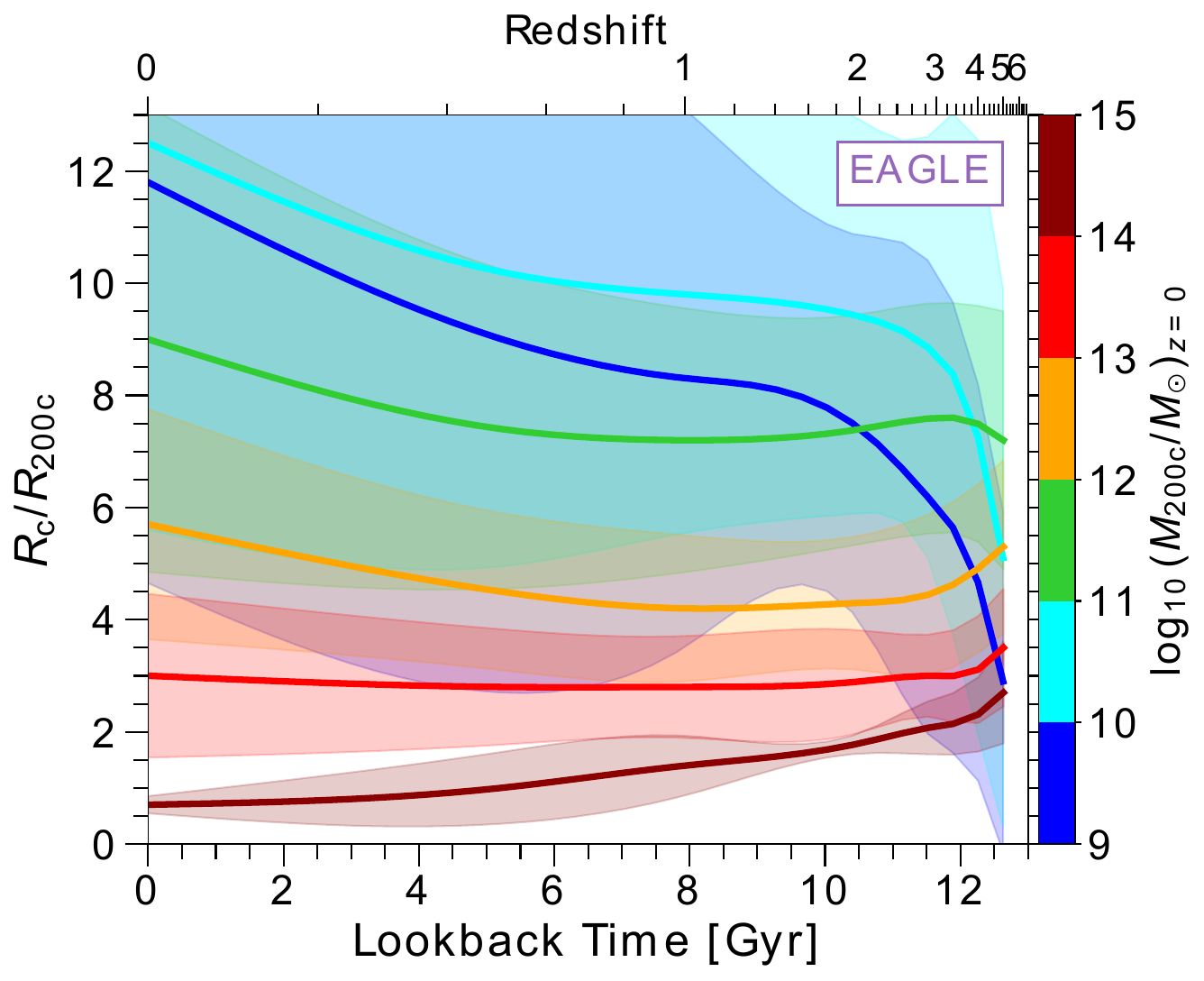}
      \includegraphics[width=0.33\textwidth]{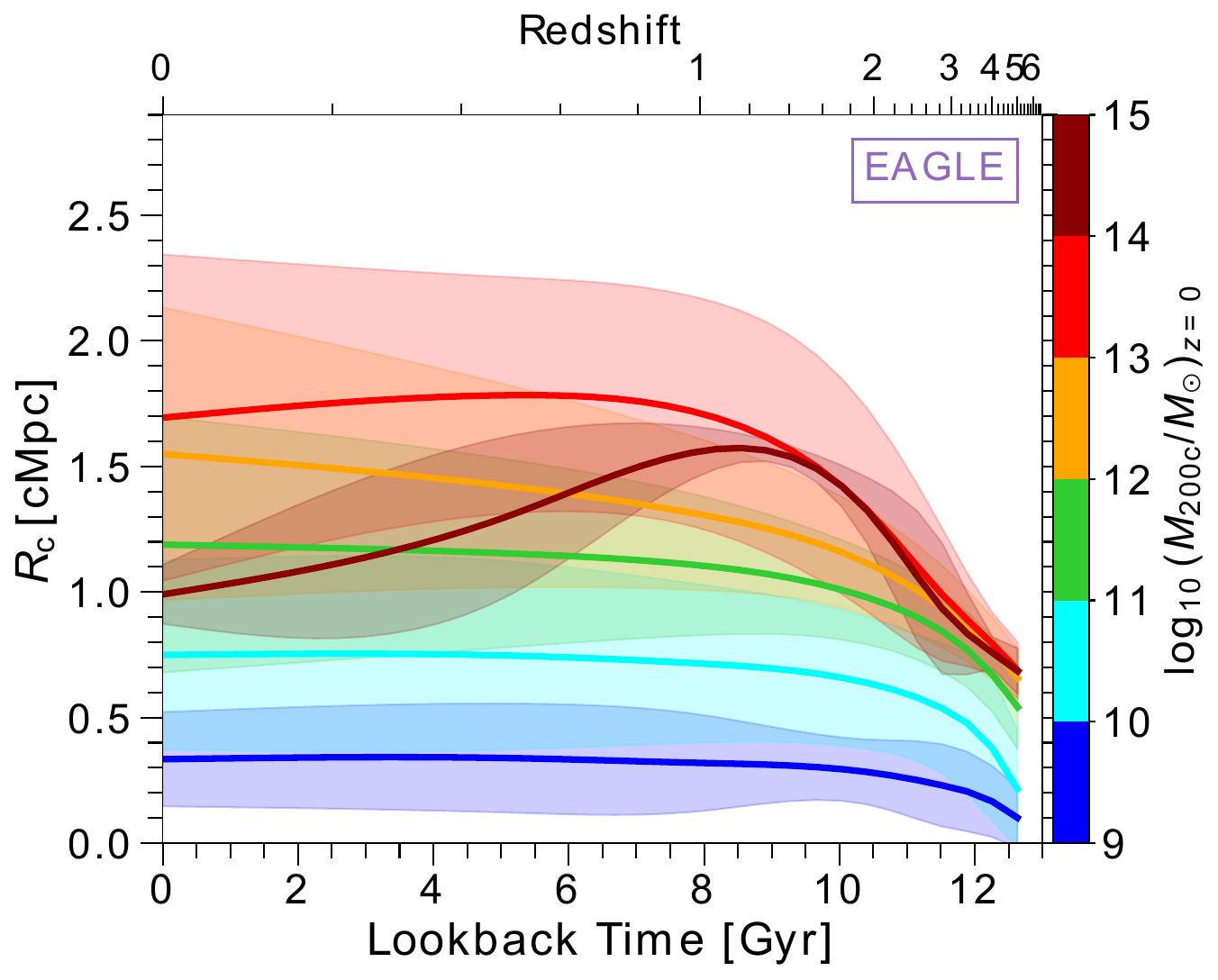}
      \includegraphics[width=0.33\textwidth]{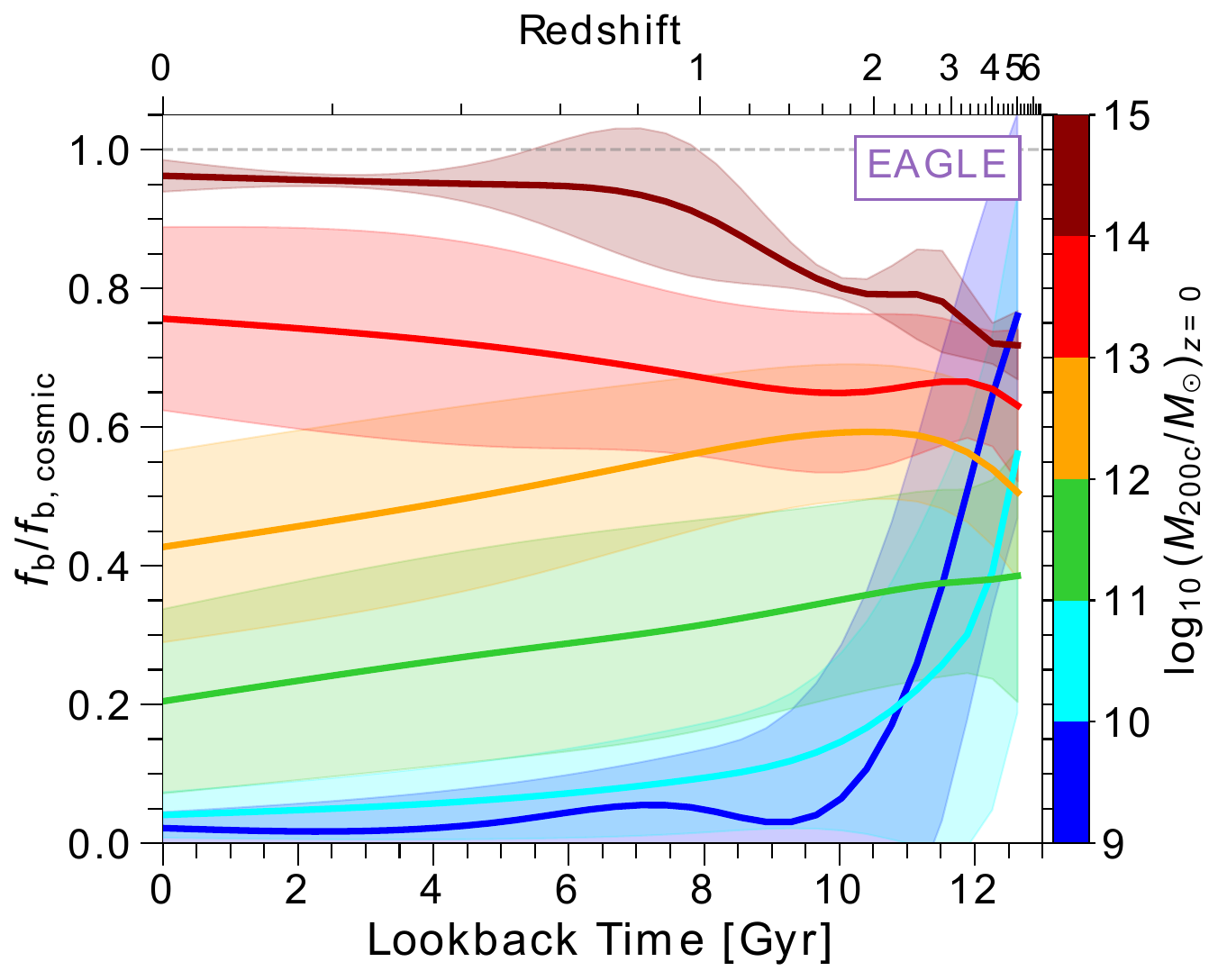}
      \caption{Evolution of baryon redistribution along halo assembly histories in TNG (top row) and EAGLE (bottom row), as a function of lookback time (lower x-axis) and redshift (upper x-axis). Columns show the normalized closure radius ($R_{\rm c}/R_{\rm 200c}$, left), the closure radius in comoving units ($R_{\rm c}\,[\rm cMpc]$, middle), and the normalized halo baryon fraction ($f_{\rm b,200c}/f_{\rm b,cosmic}$, right). Halos are grouped by their mass $M_{\rm 200c}$ at $z=0$ (colors), and their main progenitors are followed back in time. Curves show medians and shaded bands the 16th--84th percentile halo-to-halo scatter. Dashed lines in the right column mark the cosmic baryon fraction.}
\label{Fig: Rc_tree}
\end{figure*}
Fig.~\ref{Fig: Rc_Rcomp_z0} collects both characteristic radii at $z=0$ for the three simulations. Its top row shows the closure radius, normalized and in comoving units, together with the halo baryon fraction that drives it; the bottom row shows the corresponding compensation radius and the ratio of the two, and is discussed in Sect.~\ref{subsec: compensation_radius}. Here we consider the top row, and Fig.~\ref{Fig: Rc_z}, which gives its redshift evolution for each simulation separately.

At $z=0$ the observed halo baryon fraction within $R_{\rm 200c}$, normalized to the cosmic value (top right), shows a clear trend with halo mass: for $\log_{10}(M_{\rm 200c}/\mathrm{M_{\odot}})\gtrsim 13$, the halo baryon fraction increases with halo mass, and reaches complete closure for massive clusters with $\log_{10}(M_{\rm 200c}/\mathrm{M_{\odot}})\gtrsim 15$. At $\log_{10}(M_{\rm 200c}/\mathrm{M_{\odot}}) \lesssim 13$, the observed halo baryon fraction appears to increase with decreasing halo mass, but the data are too uncertain to report a reliable trend. Nevertheless, if that is the case, it would suggest a strong change in the efficiency of AGN feedback at around $\log_{10}(M_{\rm 200c}/\mathrm{M_{\odot}})\sim 12$--$13$, consistent with the predictions from TNG and SIMBA. If instead no such upturn exists, it would indicate a smoother mass dependence of the feedback efficiency, closer to EAGLE. More precise measurements of the halo baryon fraction at $\log_{10}(M_{\rm 200c}/\mathrm{M_{\odot}})\lesssim 13$ are therefore needed to constrain theoretical models of AGN feedback.

The halo baryon fraction and normalized closure radius at $z=0$ (top right and top left) were discussed in detail in \cite{Ayromlou2023Feedback} and are shown here for completeness. Briefly, SIMBA agrees best with most of the observations, favoring stronger AGN feedback that ejects gas from halos more efficiently. At low masses this comparison is not fully independent of SIMBA, whose cool-gas correction is applied to the observations (Sect.~\ref{subsubsec: Halo baryon fraction}). TNG and EAGLE predict closely similar $f_{\rm b,200c}$ at $z=0$, despite implementing AGN feedback very differently, with TNG using an ejective kinetic mode and EAGLE a thermal mode. This degeneracy disappears at lower halo masses, where TNG predicts significantly higher $f_{\rm b,200c}$ than EAGLE.

At $z=0$, the normalized closure radius, $R_{\rm c}/R_{\rm 200c}$, shows a trend completely the inverse of the halo baryon fraction: in low-mass halos ($9 \lesssim \log_{10}(M_{\rm 200c}/\mathrm{M_{\odot}})\lesssim 12$), it decreases with mass. This trend continues in EAGLE, but in TNG and SIMBA it breaks: a dip at $\log_{10}(M_{\rm 200c}/\mathrm{M_{\odot}})\sim 12$, a rise to a peak at $\sim 12.8$, then a decline again. The dip marks the mass at which AGN feedback becomes excessively ejective compared to other halo mass ranges, and the peak the mass at which gravity begins to dominate over feedback, allowing halos to retain their gas and possibly recapture some of the ejected gas.

The closure radius in comoving units ($R_{\rm c}$) shows a markedly different trend. At $z=0$ (top middle), it increases with halo mass up to $\log_{10}(M_{\rm 200c}/\mathrm{M_{\odot}})\simeq 13$, and decreases toward more massive halos thereafter. Regardless of the simulation, $R_{\rm c}$ peaks at galaxy-group masses, $\log_{10}(M_{\rm 200c}/\mathrm{M_{\odot}})\sim 12.8$--$13.2$, so groups set the largest scale of baryon redistribution in the Universe. The same groups also give the largest contribution per dex to the missing baryons in the IGM (see Sect.~\ref{subsec: halo_contribution}). This places galaxy groups at the center of the baryon redistribution physics.

At higher redshifts, the shape of all of these quantities, including the halo baryon fraction, normalized closure radius, and closure radius in comoving units, remains relatively intact, with overall higher baryon fractions and smaller closure radii at higher redshifts across all simulations. This is consistent with halos being less evolved at earlier times, and therefore having had less time to eject and redistribute baryons to larger scales.

\begin{figure*}
      \centering
      \setlength{\tabcolsep}{0pt}
      \begin{tabular}{@{}rrr@{}}
      \includegraphics[width=0.33\textwidth]{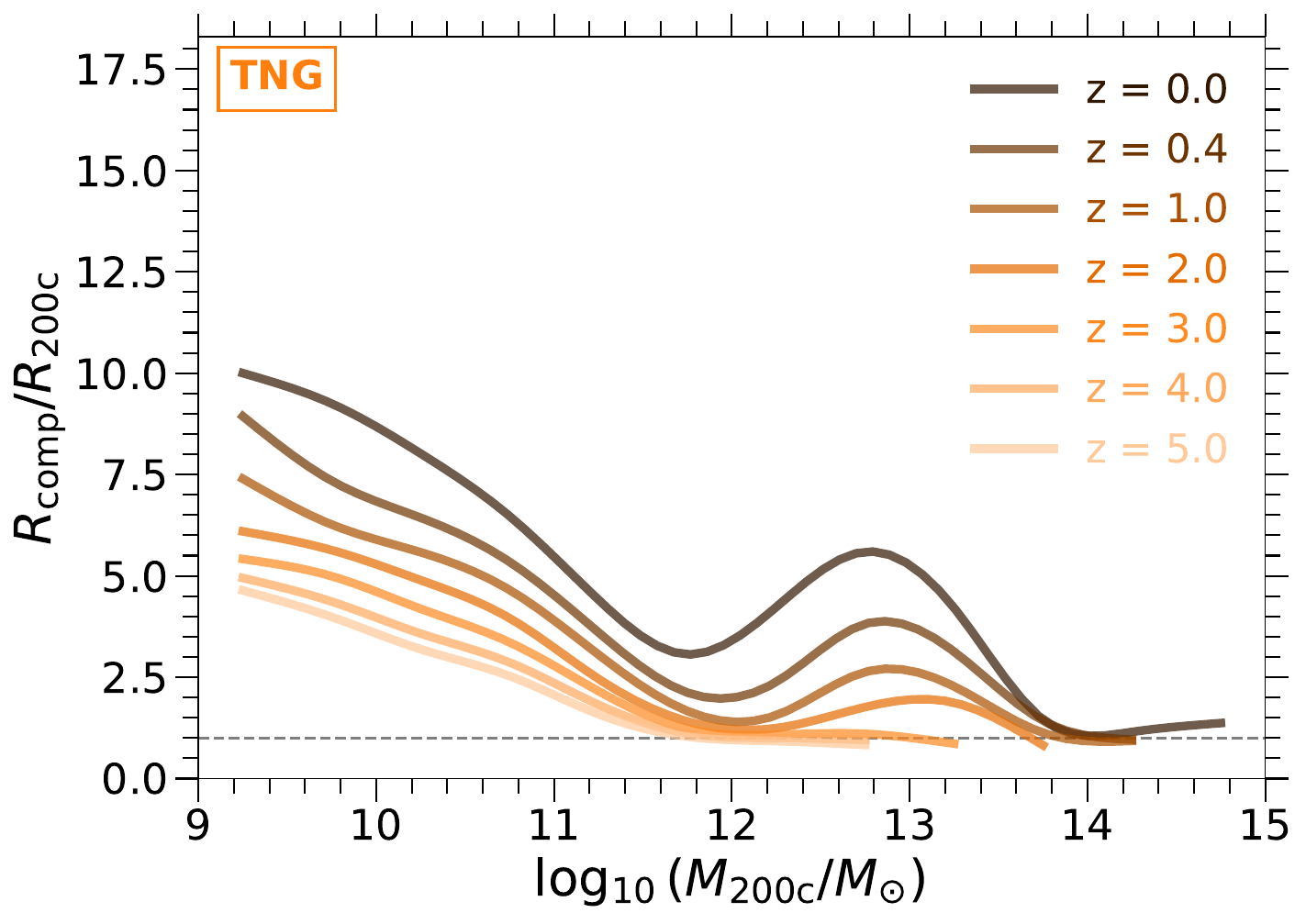} &
      \includegraphics[width=0.33\textwidth]{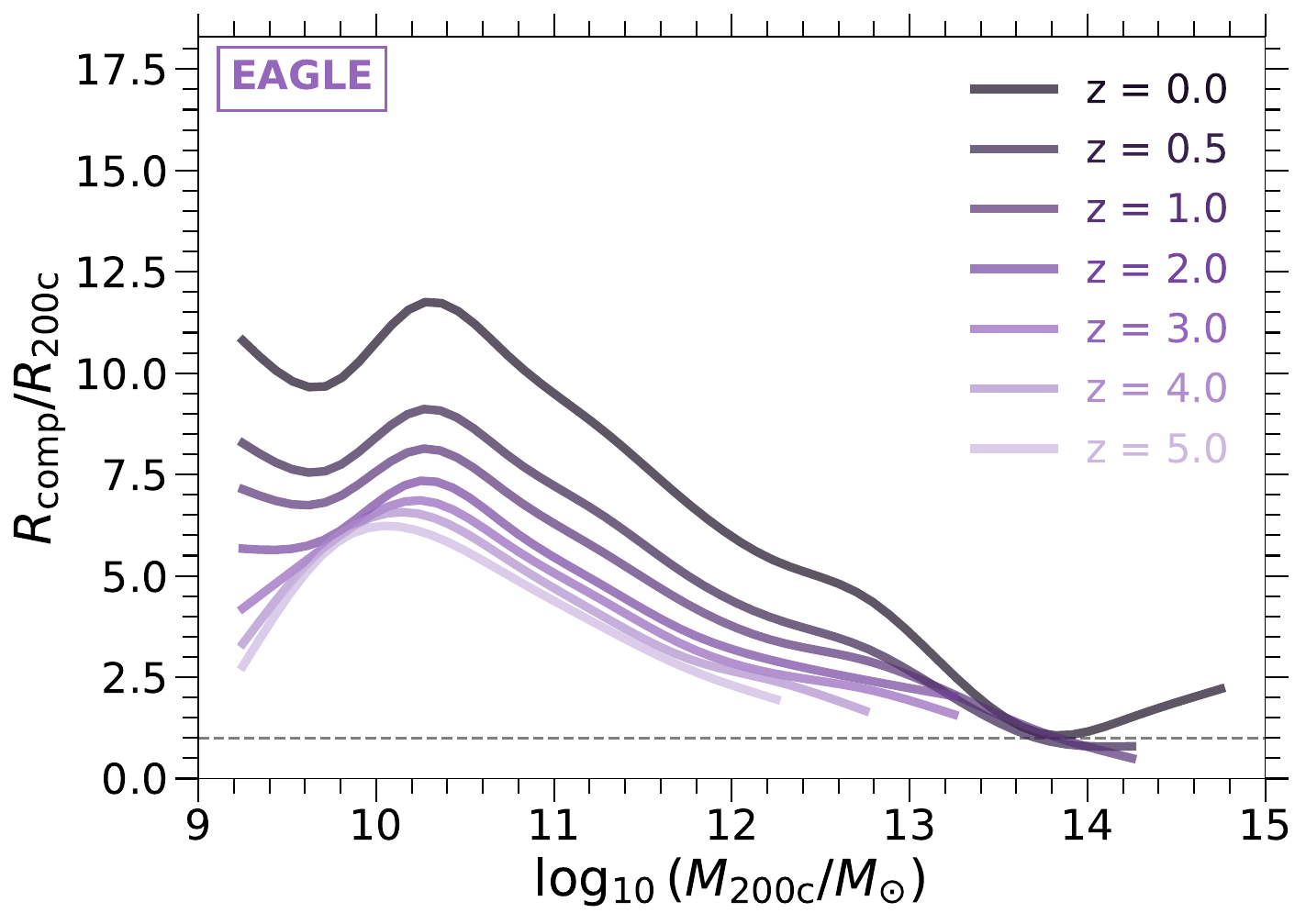} &
      \includegraphics[width=0.33\textwidth]{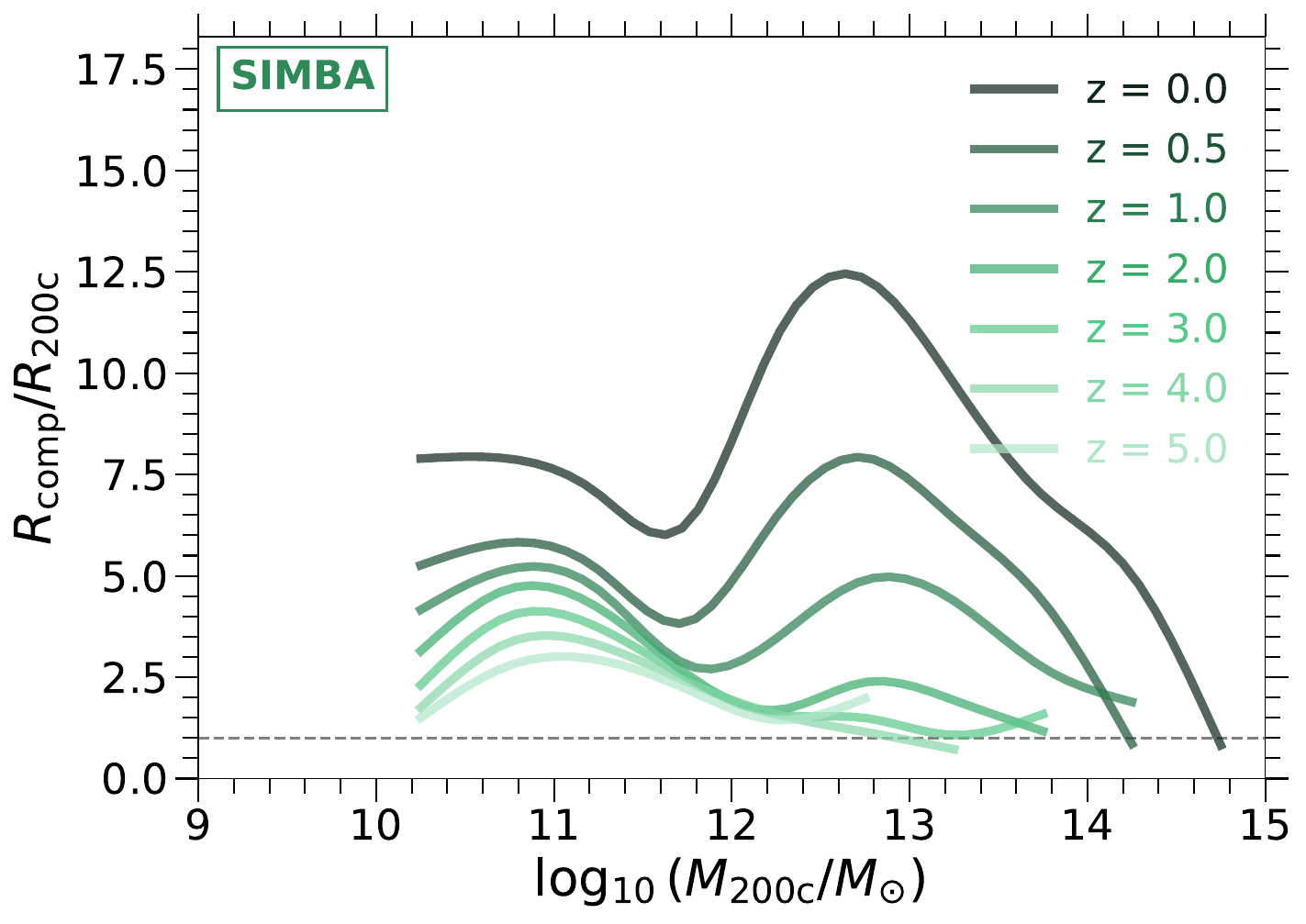} \\[0pt]
      \includegraphics[width=0.3165\textwidth]{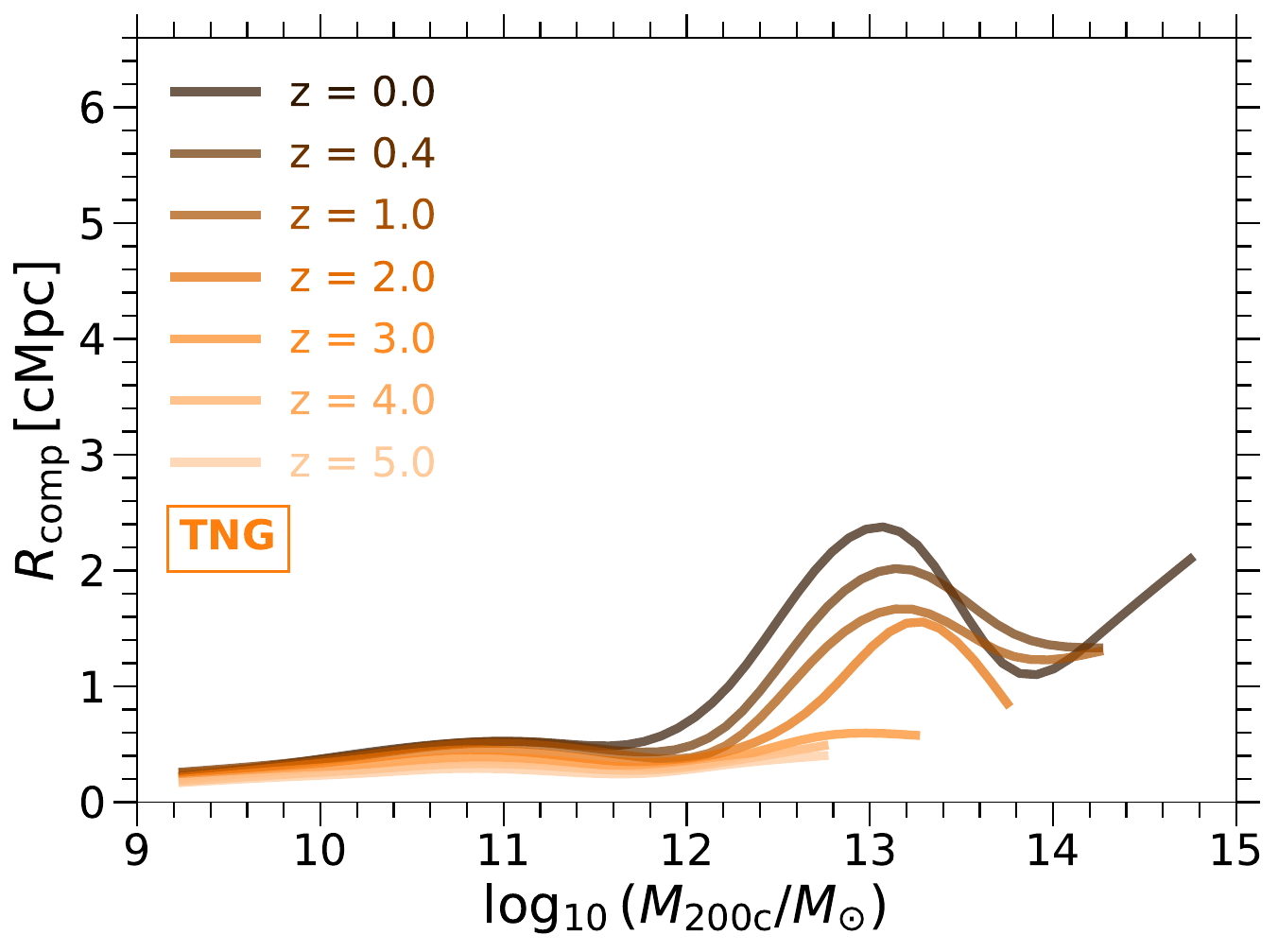} &
      \includegraphics[width=0.3165\textwidth]{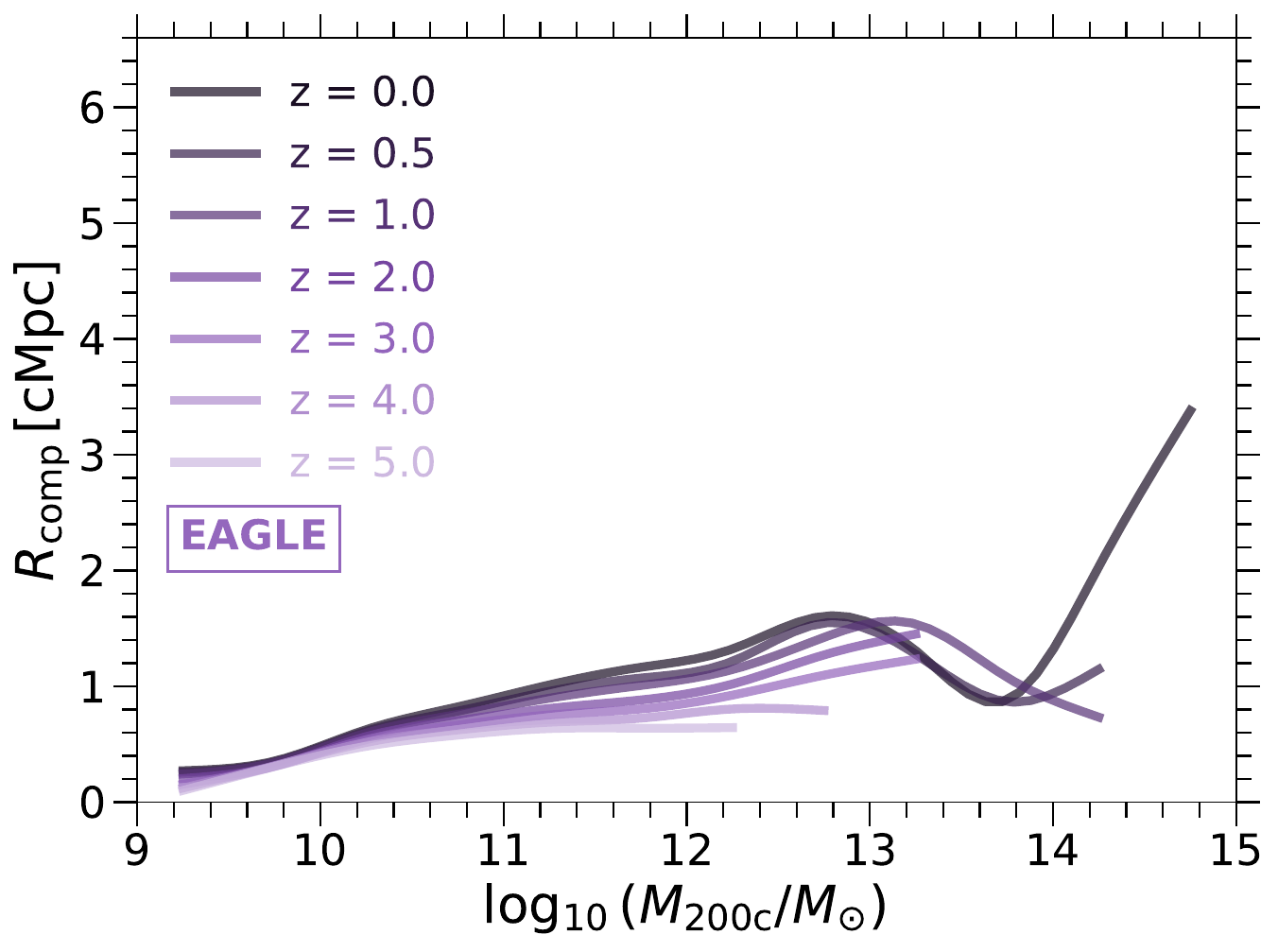} &
      \includegraphics[width=0.3165\textwidth]{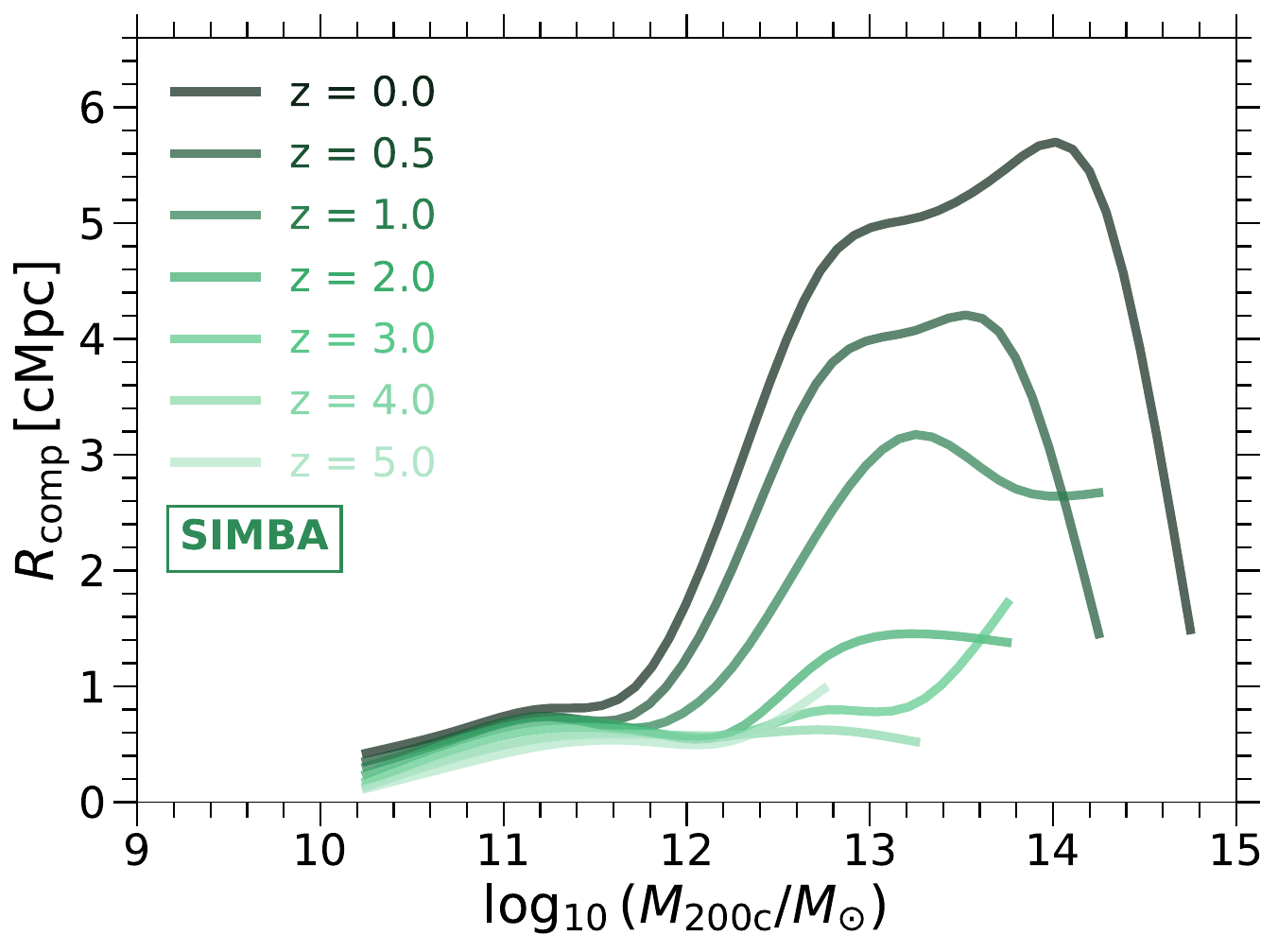}
      \end{tabular}

      \caption{Redshift evolution of the normalized compensation radius ($R_{\rm comp}/R_{\rm 200c}$, top row) and the compensation radius in comoving units ($R_{\rm comp}\,[\rm cMpc]$, bottom row), for TNG (left column), EAGLE (middle column), and SIMBA (right column). Curves show medians; darker lines correspond to lower redshift, from $z=5$ (lightest) to $z=0$ (darkest). Dashed lines mark $R_{\rm comp}=R_{\rm 200c}$ in the top row.}
\label{Fig: Rcomp_z}
\end{figure*}

Across all redshifts and simulations, halos with $\log_{10}(M_{\rm 200c}/\mathrm{M_{\odot}})\lesssim 12$ show little baryon redistribution beyond $\sim 1\,\mathrm{cMpc}$, although the normalized closure radius can be large because the halo itself is small. The same holds for very massive clusters ($\log_{10}M_{\rm 200c}/\mathrm{M_{\odot}}\gtrsim 14.5$), where almost no baryon redistribution occurs beyond their boundary or $1\,\mathrm{cMpc}$ scales. This again identifies galaxy groups as the key baryon redistributors in the Universe, at least out to $z \sim 3$. At $z>3$, no significant baryon redistribution occurs beyond scales of order $1\,\mathrm{cMpc}$ in any halo mass range in any simulation. In TNG and SIMBA, halos with $\log_{10}(M_{\rm 200c}/\mathrm{M_{\odot}})\gtrsim 12$ already have near-cosmic baryon fractions at $z>3$, and the normalized closure radii are close to unity, indicating no significant net baryon redistribution beyond the halo boundary. Nonetheless, this does not exclude continued exchange across the closure radius. EAGLE shows the largest baryon redistribution at $z>3$, owing to its stronger stellar feedback, with no halos at that epoch reaching the cosmic baryon fraction within their $R_{\rm 200c}$.

In all simulations, the peak of the comoving closure radius shifts slightly toward higher halo masses with increasing redshift, but it remains around galaxy groups overall, at least up to $z=3$, where groups are still present in these simulation boxes. The peak amplitude also decreases with redshift, again consistent with less baryon redistribution at higher redshifts. In all simulations the comoving closure radii at $z>3$ are of order $1\,\mathrm{cMpc}$ or smaller, and the corresponding physical scale is even smaller: at $z=3$, $1\,\mathrm{cMpc}$ corresponds to only $250\,\mathrm{kpc}$ in physical units, so effectively no large-scale baryon redistribution occurs at $z>3$ in any simulation. Whether this holds in the real Universe remains an open question. Since models of galaxy formation significantly underpredict the fraction of massive quenched galaxies at $z>3$ \citep{Lagos2025Diverse,Vani2025Probing,Vani2026Galaxies}, they may also underpredict the strength of feedback and therefore the extent of baryon redistribution. Future observations of halo and IGM baryon shares together with galaxy star formation activity at $z>3$ can shed light on this.

We moreover investigate the time evolution of the closure radius along halo assembly histories, using the merger trees of TNG and EAGLE to track the main progenitors of $z=0$ halos back in time. Fig.~\ref{Fig: Rc_tree} shows the normalized and comoving closure radius and the halo baryon fraction against lookback time and redshift, for six $z=0$ mass bins ($9\leq\log_{10}(M_{\rm 200c}/{\rm M_\odot})\leq 15$). Individual halo tracks are shown in Appendix~\ref{app: App1}.

Both simulations show the same qualitative behavior. In units of the halo radius (left column), $R_{\rm c}/R_{\rm 200c}$ grows toward the present in every mass bin below $10^{14}\,\mathrm{M_\odot}$, reaching $10$--$12$ for the least massive systems at $z=0$, whereas for the most massive bin it instead declines slightly. In comoving units (middle column) the largest radii belong throughout to halos that end up with $10^{13} \lesssim M_{\rm 200c}/\mathrm{M_\odot} \lesssim 10^{14}$, reaching $\simeq 2.5\,\mathrm{cMpc}$ in TNG and $\simeq 1.9\,\mathrm{cMpc}$ in EAGLE at the present day.

The baryon fraction tracks (right column) show that this growth is accompanied by a declining enclosed baryon fraction, and not by halo growth alone. The decline is clearest in the two least massive bins of both models: in TNG the least massive halos fall from $f_{\rm b,200c}/f_{\rm b,cosmic} \simeq 0.45$ at a lookback time of $12.6\,\mathrm{Gyr}$ to $\simeq 0.02$ today. The intermediate bins are not monotonic. In TNG the $10^{11}$--$10^{12}\,\mathrm{M_\odot}$ track rises to $\simeq 0.70$ around a lookback time of $7\,\mathrm{Gyr}$ and ends at $\simeq 0.66$, above its value at the earliest times sampled, while the $10^{12}$--$10^{13}\,\mathrm{M_\odot}$ track peaks near $9\,\mathrm{Gyr}$ before falling to $\simeq 0.54$. A declining baryon fraction need not correspond to a loss of baryonic mass, since $M_{\rm 200c}$ and $R_{\rm 200c}$ also grow along each track. The most massive bin stays closest to the cosmic value, but in neither model does it remain within the $5\%$ closure tolerance at all times: in TNG it leaves the band at intermediate lookback times, reaching $\simeq 0.89$ near $10\,\mathrm{Gyr}$ before recovering, whereas in EAGLE it drops below the band beyond a lookback time of $\simeq 6\,\mathrm{Gyr}$ and declines steadily to $\simeq 0.72$ at the earliest times sampled. Similar behavior is seen in COLIBRE, where gas fractions of cluster progenitors decline before recovering toward $z=0$ \citep{Davies2026COLIBRE}.

Notably, TNG and EAGLE yield similar median cluster baryon fractions at $z=0$ (dark red lines, right column of Fig.~\ref{Fig: Rc_tree}), but differ substantially at earlier times. In EAGLE, the median baryon fraction of cluster progenitors declines steadily with increasing lookback time, whereas in TNG it evolves non-monotonically, with a dip at $z\sim 1.5$. Similar present-day baryon fractions can therefore arise from different evolutionary histories. These differences likely reflect the different implementations of physics rather than variations in halo formation history, which contribute to halo-to-halo scatter in quantities such as gas kinematics \citep{Ayromlou2024Atlas}. The closure radius tracks likewise differ between the models (Fig.~\ref{Fig: Rc_tree}, middle column).

We stress that these are histories selected by present-day mass. At early times each track follows progenitors substantially less massive than the bin label, and the trends should therefore not be read as the evolution of halos at fixed mass. Only progenitors above the resolved mass limit of $10^{9}\,\mathrm{M_\odot}$ are followed, so the number of tracks contributing to a median decreases toward higher redshift. This affects the lowest-mass bins, whose progenitors drop below the limit earliest and whose early medians are therefore biased toward the more massive members of the bin; the intermediate and massive bins remain well sampled.

\subsection{The compensation radius across cosmic time}
\label{subsec: compensation_radius}
\begin{figure*}
      \centering
      \includegraphics[width=0.35\textwidth]{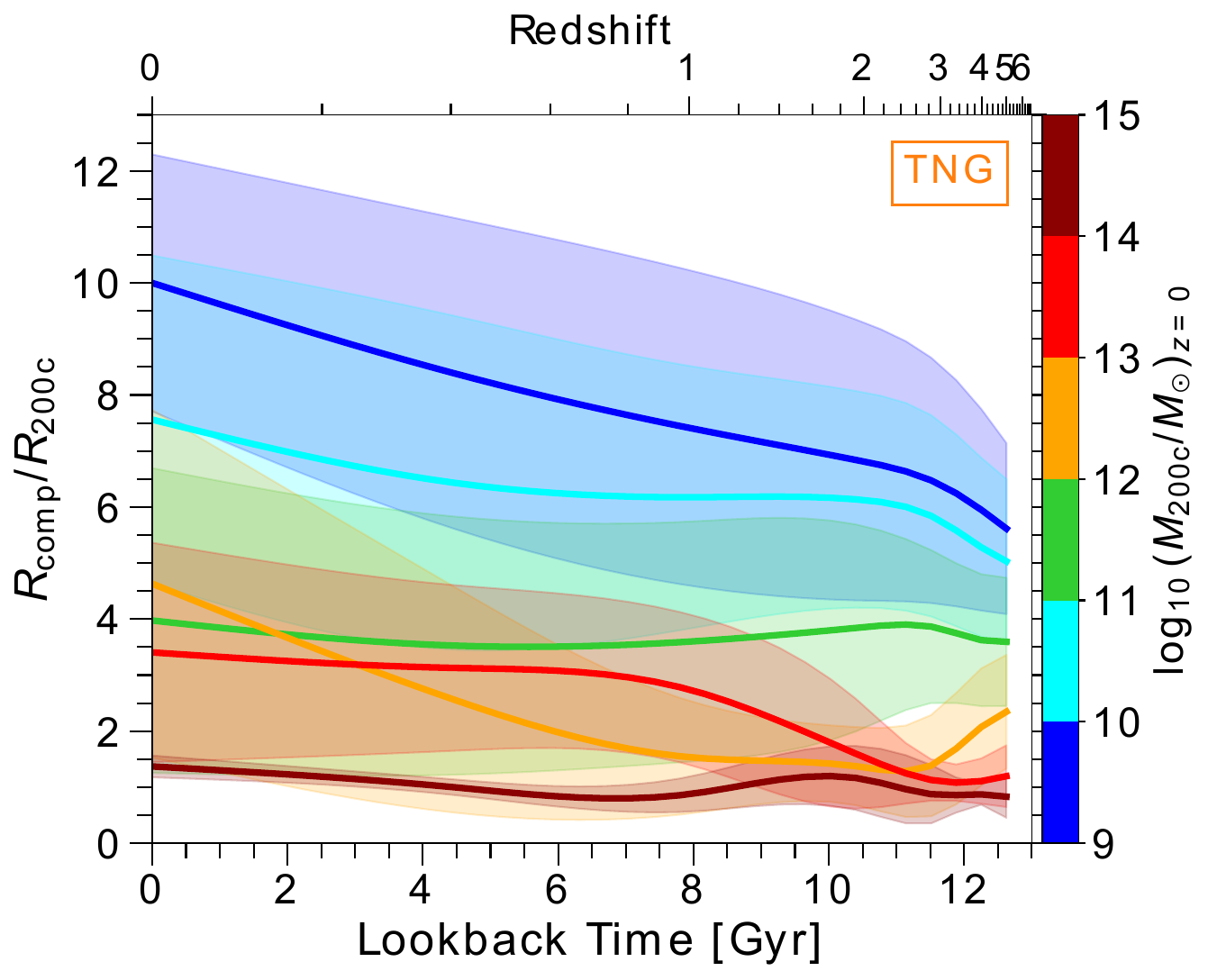} \hspace{0.5cm}
      \includegraphics[width=0.35\textwidth]{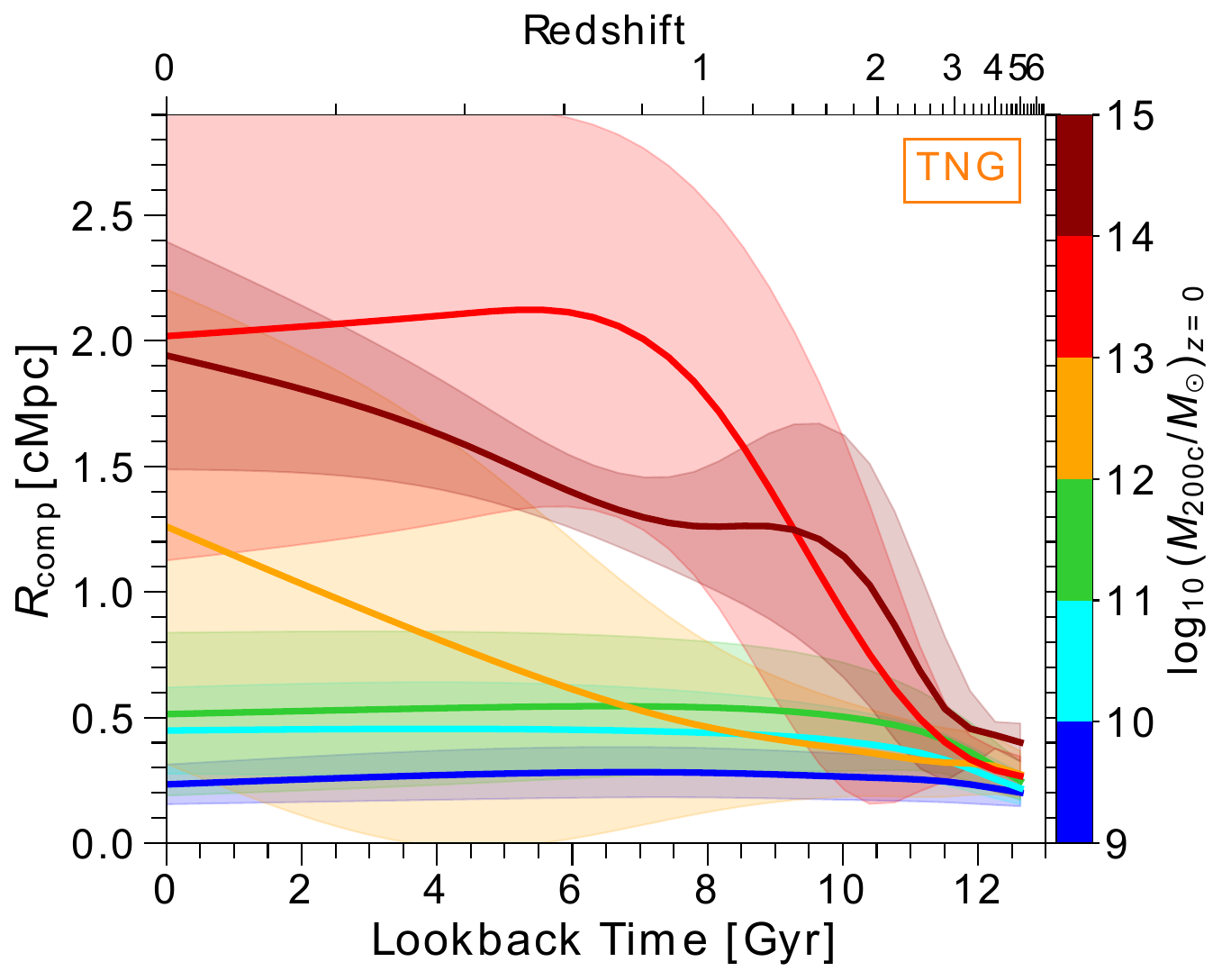}
      \includegraphics[width=0.35\textwidth]{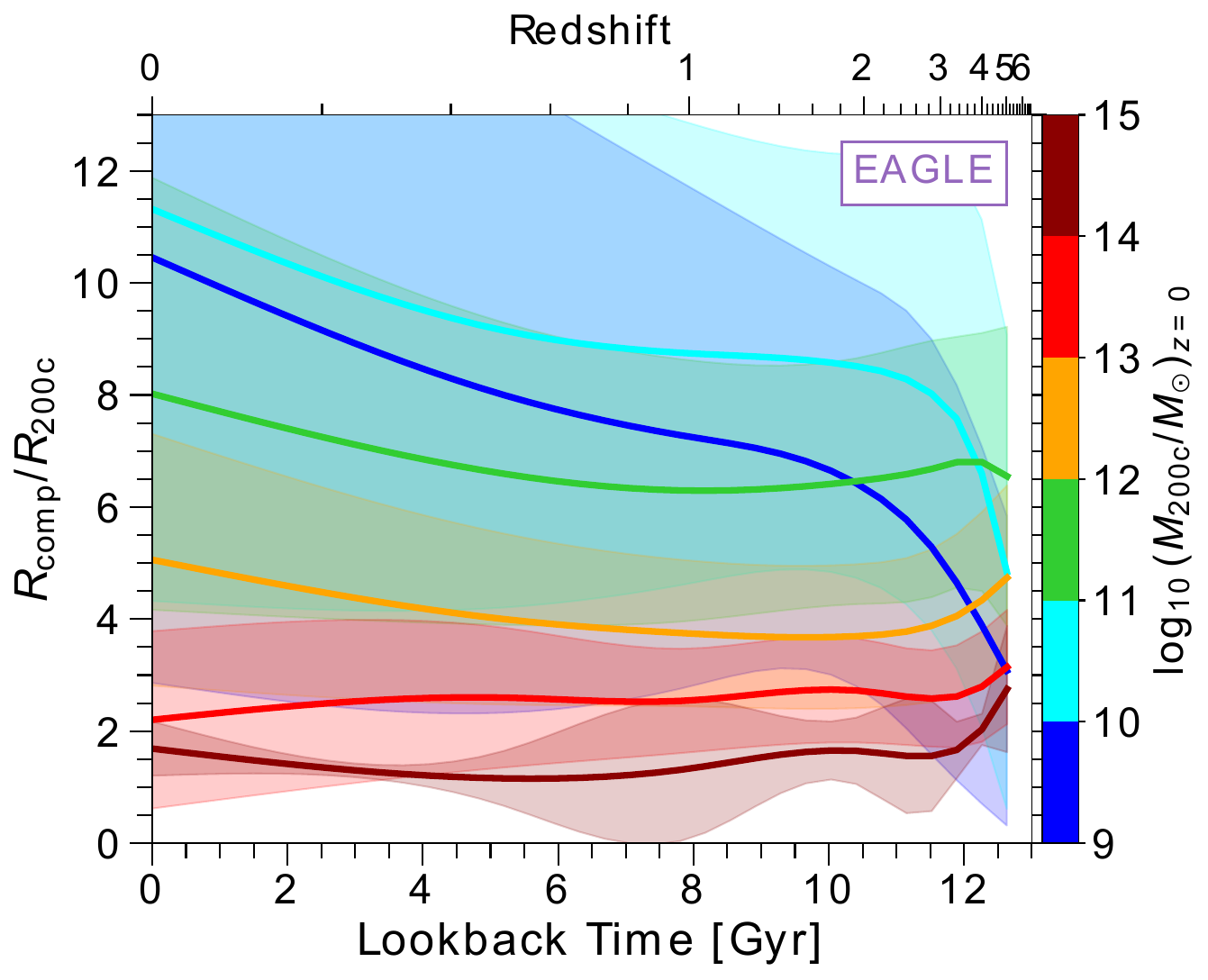} \hspace{0.5cm}
      \includegraphics[width=0.35\textwidth]{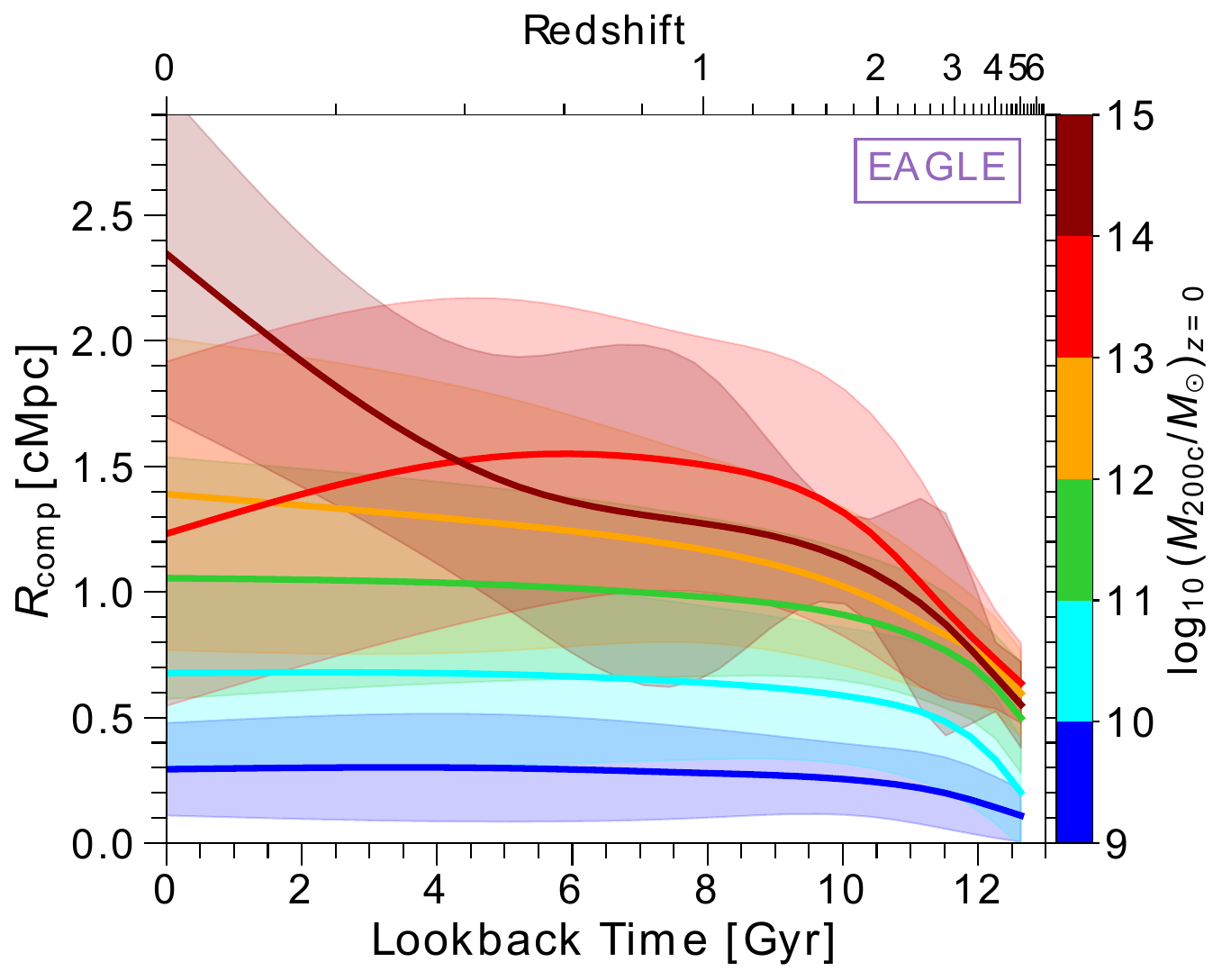}
      \caption{Evolution of the compensation radius (left: normalized, right: comoving) along halo assembly histories in TNG (top row) and EAGLE (bottom row), as a function of lookback time (lower x-axis) and redshift (upper x-axis). Halos are grouped by their mass $M_{\rm 200c}$ at $z=0$ (colors), and their main progenitors are followed back in time. Curves show medians and shaded bands the 16th--84th percentile halo-to-halo scatter.}
\label{Fig: Rcomp_tree}
\end{figure*}

The compensation radius (Eq.~\ref{eq: Rcomp_operational}) probes the recovery of the total mass, including any dark matter response, so the total enclosed mass can differ from its gravity-only value even where the baryon fraction has already closed. $R_{\rm c}$ is then smaller than $R_{\rm comp}$ where baryons and dark matter are proportionally depleted, leaving the baryon fraction unchanged but the total mass suppressed, and larger where the dark matter is instead in excess of its gravity-only expectation; which of the two applies is set by the sign of $\zeta_{\rm DM}(R_{\rm c})$, as detailed in Appendix~\ref{app: tolerance}.

The bottom row of Fig.~\ref{Fig: Rc_Rcomp_z0} shows the compensation radius at $z=0$. Its overall shape resembles that of the closure radius: $R_{\rm comp}$ stays below $1\,\mathrm{cMpc}$ for halos less massive than $\log_{10}(M_{\rm 200c}/\mathrm{M_\odot})\simeq 12$, rises steeply above that mass, and reaches its largest values at group scales, $\simeq 2.4\,\mathrm{cMpc}$ in TNG and $\simeq 1.6\,\mathrm{cMpc}$ in EAGLE near $\log_{10}(M_{\rm 200c}/\mathrm{M_\odot})\simeq 13$, and $\simeq 6\,\mathrm{cMpc}$ in SIMBA, whose maximum is broad and lies closer to the group-cluster boundary. In units of the halo radius, $R_{\rm comp}/R_{\rm 200c}$ is largest for the least massive halos, reaching $8$--$12$, and falls to of order unity for the most massive clusters.

The ratio of the two radii is shown directly in the bottom-right panel. Over the mass range where all three simulations are well sampled, $11 \lesssim \log_{10}(M_{\rm 200c}/\mathrm{M_\odot}) \lesssim 13.5$, the median ratio lies below unity in every model, between about $0.8$ and $1$: the enclosed total mass recovers its gravity-only value at a smaller radius than the enclosed baryon fraction recovers the cosmic value. The ratio approaches unity at the lowest masses in TNG and EAGLE. Above $\log_{10}(M_{\rm 200c}/\mathrm{M_\odot}) \simeq 13.5$ the two radii diverge and the ratio becomes erratic, dipping lowest in TNG and exceeding unity in both TNG and SIMBA at the highest masses; the number of such systems in these volumes is small, so we do not interpret this behavior. We stress that this is a statement about the population and not about individual halos: the halo-to-halo scatter on $R_{\rm comp}/R_{\rm c}$ is comparable to or larger than the offset of the median from unity at every mass, so that individual halos scatter across unity throughout. What the measurement establishes is that the median halo compensates before it closes.

Fig.~\ref{Fig: Rcomp_z} shows how these trends evolve. As for the closure radius, $R_{\rm comp}$ is smaller at earlier times at fixed halo mass in all three models, and the group-scale peak in comoving units weakens progressively toward higher redshift. Fig.~\ref{Fig: Rcomp_tree} follows the compensation radius along the same assembly histories as Fig.~\ref{Fig: Rc_tree}. The tracks parallel the closure radius: the largest comoving radii belong to the most massive systems and grow toward the present, reaching $\simeq 2\,\mathrm{cMpc}$ at $z=0$ in TNG for halos that end up above $10^{13}\,\mathrm{M_\odot}$, while the lowest-mass bins stay below $0.5\,\mathrm{cMpc}$ throughout. Furthermore, $R_{\rm comp}/R_{\rm 200c}$ grows toward the present in most halo mass bins, but flattens or decreases mildly in EAGLE groups and clusters. The two tracks therefore differ at the highest masses: halos ending in the $10^{14}$--$10^{15}\,\mathrm{M_\odot}$ bin have a small closure radius but a compensation radius comparable to that of the $10^{13}$--$10^{14}$ bin. These bins contain the fewest systems and the widest halo-to-halo scatter, so we record the difference without interpreting.

\subsection{The closure and compensation envelopes}
\label{subsec: envelopes}
\begin{figure}
      \centering
      \includegraphics[width=0.49\textwidth]{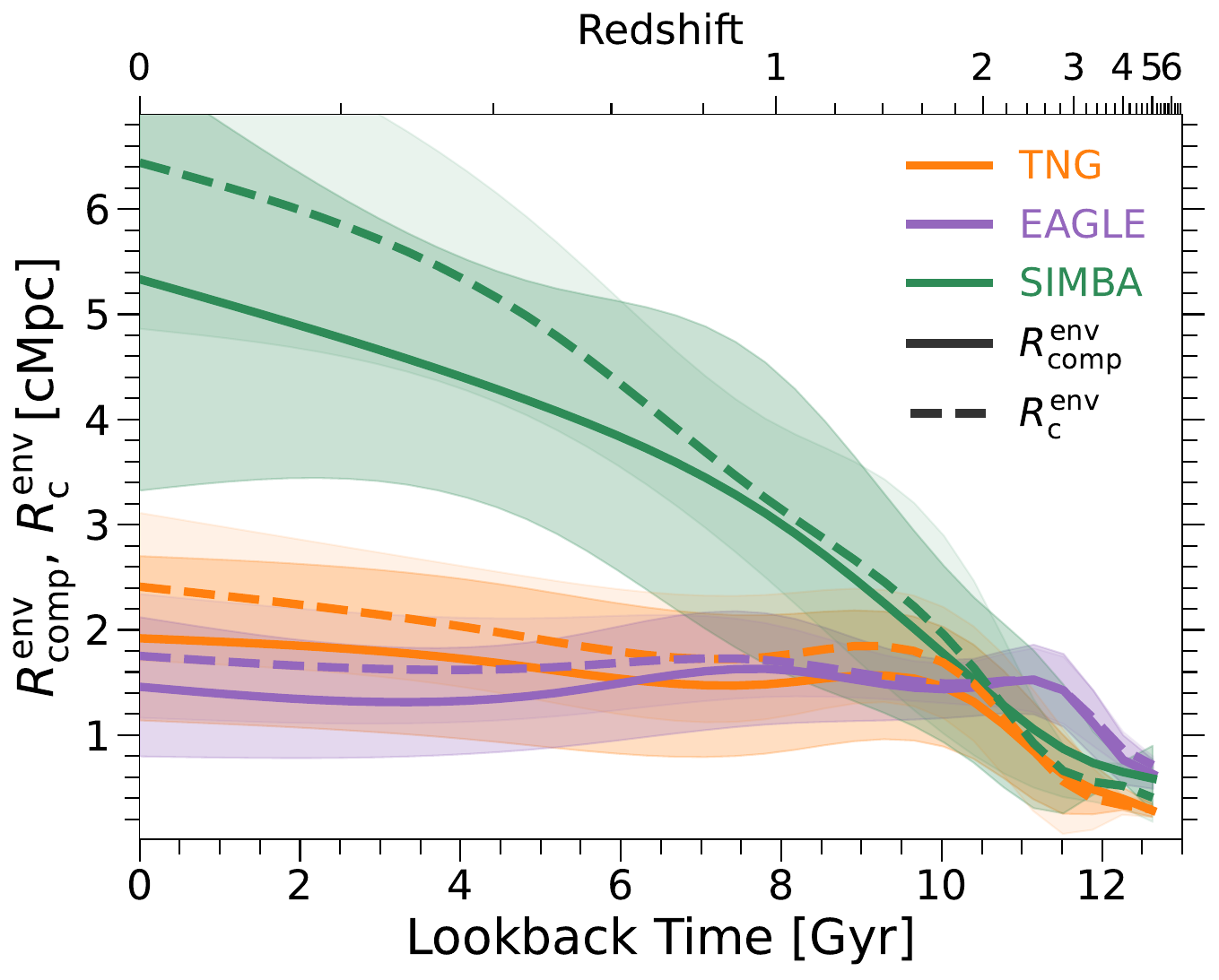}

      \caption{Closure (dashed lines) and compensation (solid lines) envelopes in comoving units as functions of lookback time (lower x-axis) and redshift (upper x-axis), for the TNG (orange), EAGLE (purple), and SIMBA (green) simulations. The envelopes are the maxima of the smoothed median mass-binned profiles (Sect.~\ref{subsec: characteristic radii method}); shaded bands show the halo-to-halo scatter (16th--84th percentile) of the radii within the peak mass bin, not the statistical uncertainty on the envelope.}
\label{Fig: Rc_max}
\end{figure}
\begin{figure}
      \centering
      \includegraphics[width=0.47\textwidth]{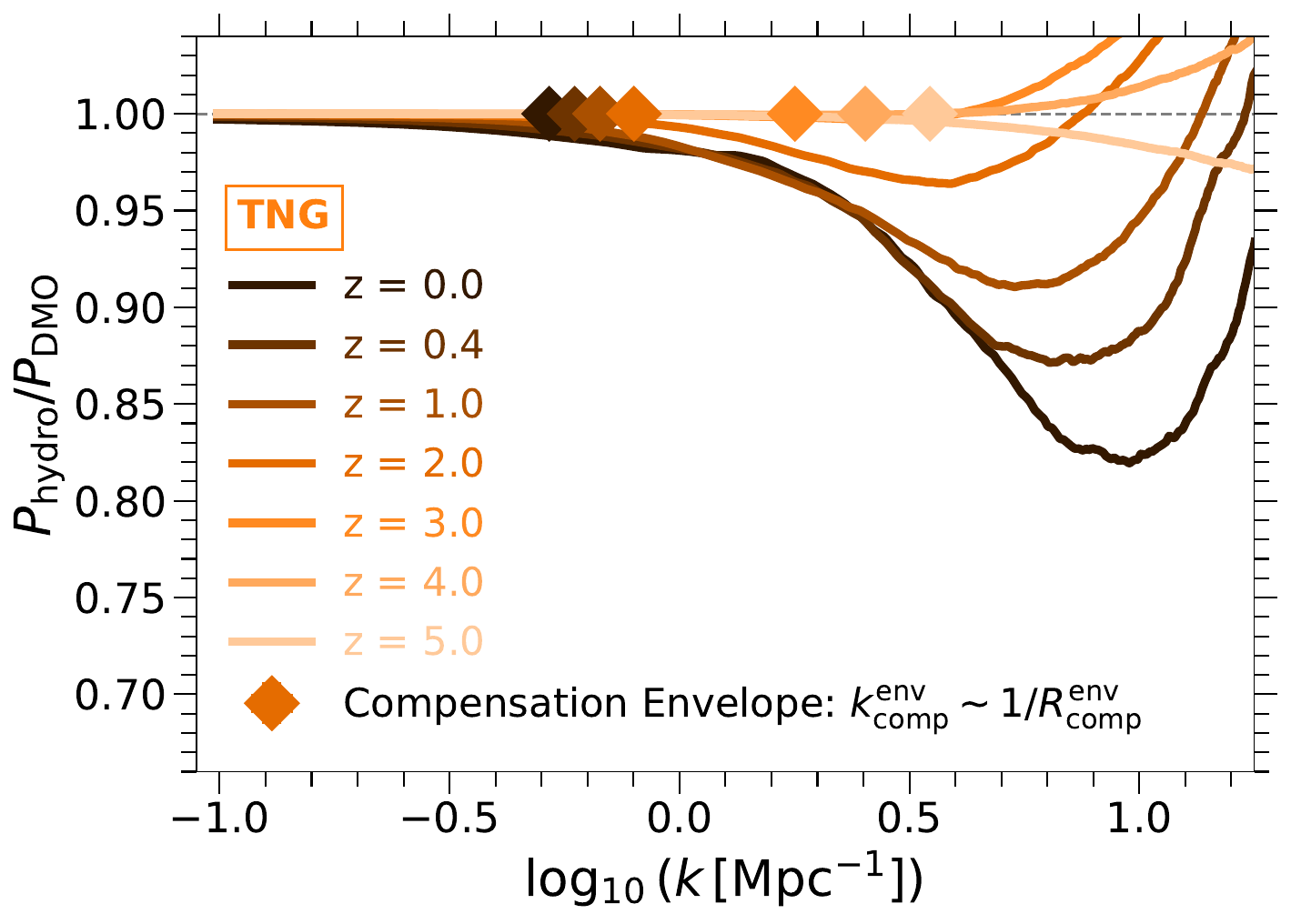}
      \includegraphics[width=0.47\textwidth]{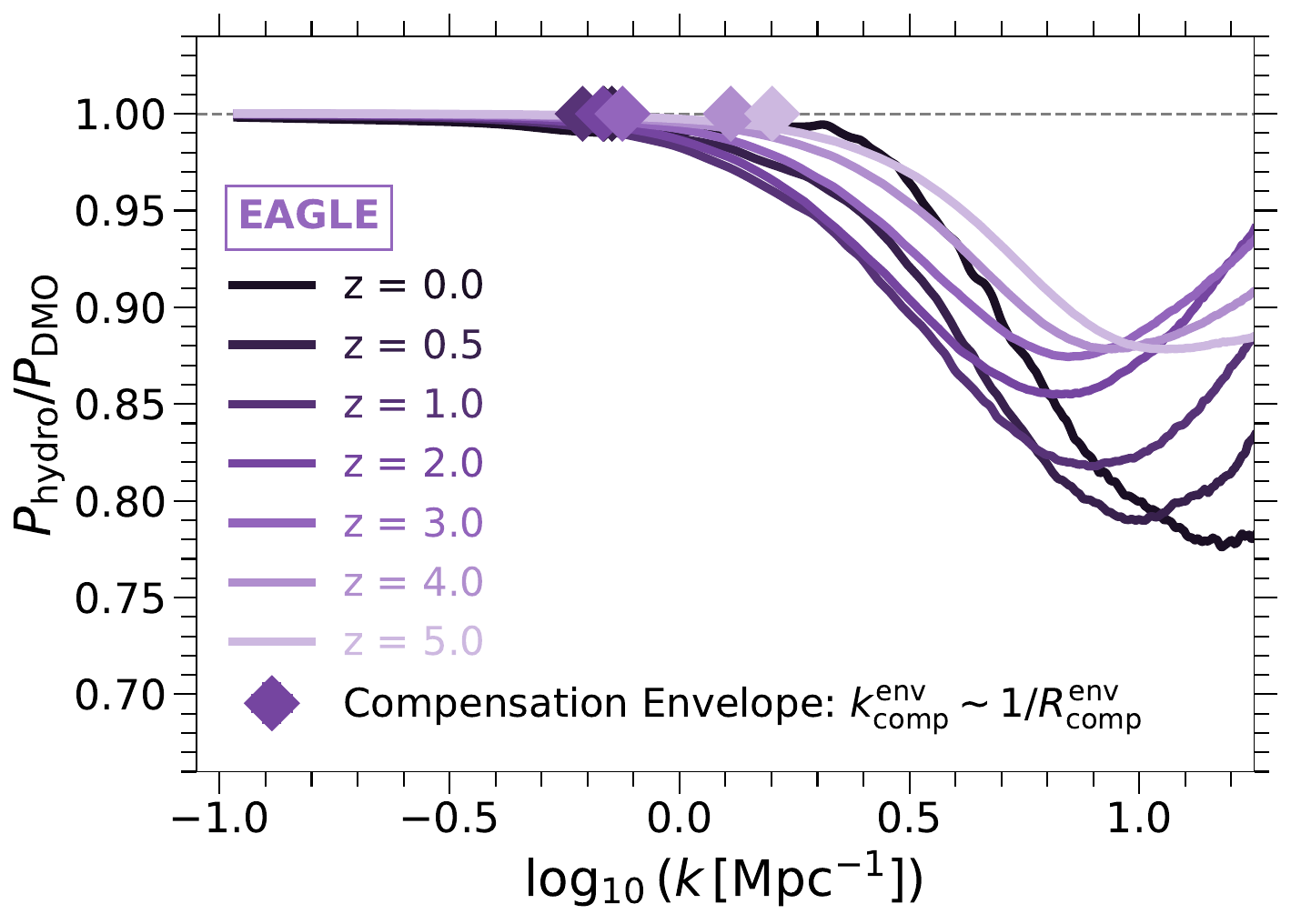}
      \includegraphics[width=0.47\textwidth]{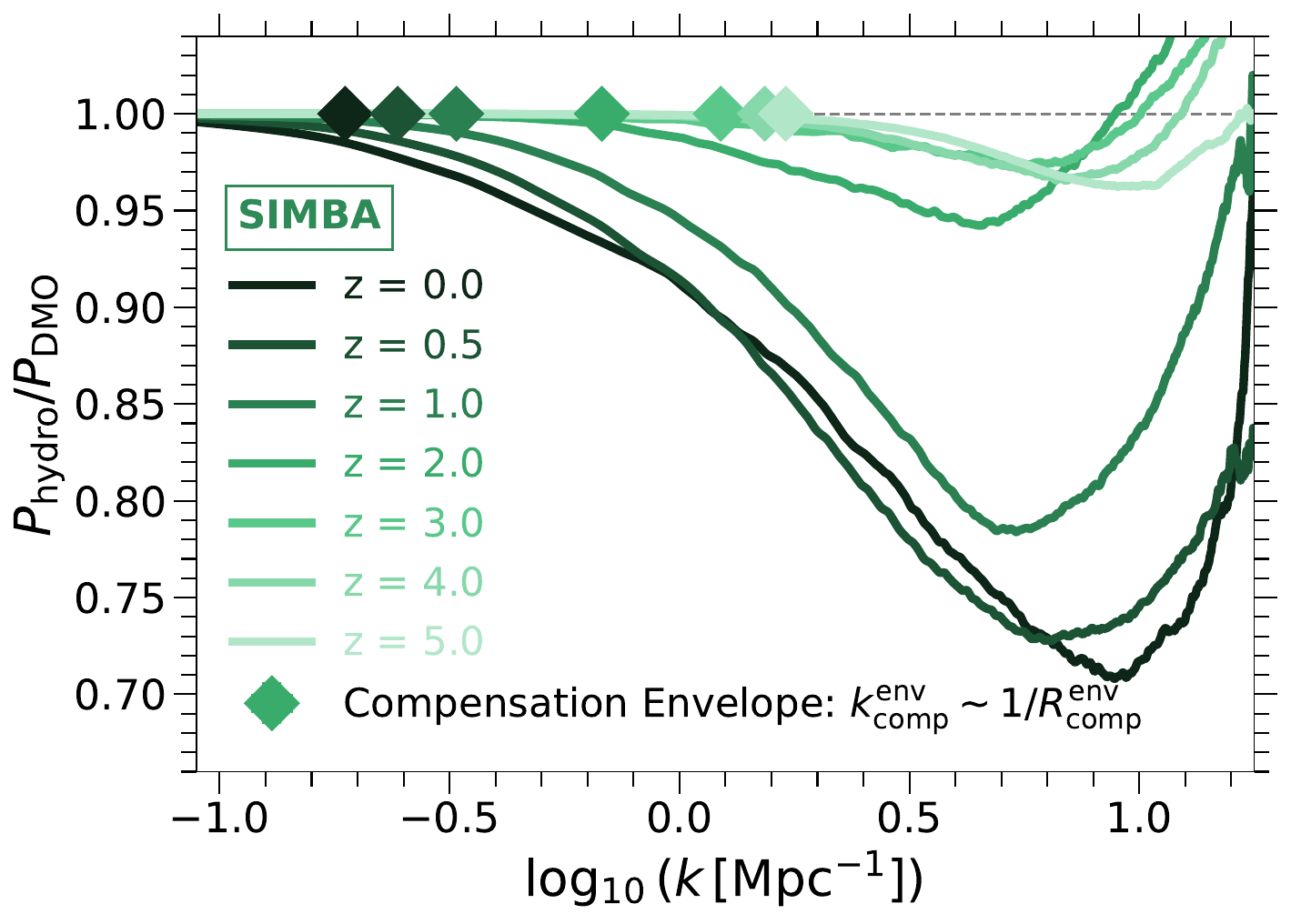}
      \caption{Impact of baryon redistribution on matter clustering. Curves show the ratio of the total matter power spectrum in each hydrodynamical simulation to that of its dark-matter-only counterpart, $P_{\rm hydro}(k)/P_{\rm DMO}(k)$, for TNG (top), EAGLE (middle), and SIMBA (bottom). Darker colors indicate lower redshift, from $z=5$ (lightest) to $z=0$ (darkest). Diamonds mark the inverse compensation envelope, $k_{\rm comp}^{\rm env}=1/R_{\rm comp}^{\rm env}$, at each redshift. The compensation envelope closely coincides with the scale where $P_{\rm hydro}(k)/P_{\rm DMO}(k)$ approaches unity.}
\label{Fig: Pk}
\end{figure}
While the closure and compensation radii describe individual halos, the global impact of baryon redistribution is captured by taking their maxima over the halo population. The closure envelope, $R_{\rm c}^{\rm env}(z)$, and the compensation envelope, $R_{\rm comp}^{\rm env}(z)$, represent the largest distances to which baryonic effects on the baryon fraction and on the total mass, respectively, extend at a given redshift. As discussed in Sects.~\ref{subsec: halo_contribution} and~\ref{subsec: closure_radius}, galaxy groups are the key baryon redistributors and therefore largely set these envelopes.

Fig.~\ref{Fig: Rc_max} shows both envelopes as functions of lookback time and redshift. At $z\gtrsim 3$, both envelopes are of order $1\,\mathrm{cMpc}$ or below in all models, and both grow toward the present. The growth is not strictly monotonic: EAGLE's compensation envelope has a shallow minimum near a lookback time of $3\,\mathrm{Gyr}$ and a local maximum near $7.5\,\mathrm{Gyr}$. At $z\gtrsim 1.5$ the models are very similar, and they diverge at $z\lesssim 1.5$, with SIMBA reaching much larger scales than TNG and EAGLE, reflecting its stronger AGN feedback and more efficient baryon redistribution.

At $z=0$ the compensation envelope reaches $1.9\,\mathrm{cMpc}$ in TNG, $1.5\,\mathrm{cMpc}$ in EAGLE, and $5.4\,\mathrm{cMpc}$ in SIMBA, compared with closure envelopes of $2.4$, $1.8$, and $6.4\,\mathrm{cMpc}$ respectively. The compensation envelope is therefore the \emph{smaller} of the two at $z=0$ in all three models, by a similar factor of $\simeq 0.8$ despite the threefold spread in their absolute values. The unsmoothed maxima of the $z=0$ median relations (Fig.~\ref{Fig: Rc_Rcomp_z0}) give only a slightly larger ratio, $\simeq 0.9$: smoothing lowers narrow peaks more than broad ones, and the $R_{\rm comp}$ peak is the narrower. The ordering is unaffected, but the ratio's precise value should not be over-interpreted. In the idealized limit of Eq.~(\ref{eq: sign_test}), a compensation radius interior to the closure radius corresponds to $\zeta_{\rm DM}(R_{\rm c}) < 0$, the enclosed dark matter slightly exceeding its gravity-only expectation there, so that the total mass recovers before the baryon fraction does. This is mass accounting; the dynamical origin of the dark matter response is not established here. The ordering is also not universal in time: in EAGLE the two envelopes cross at lookback times between about $6$ and $10\,\mathrm{Gyr}$, where the compensation envelope is formally the larger of the two, and at $z \gtrsim 2$ all three models converge to within their halo-to-halo scatter.

We emphasize that both envelopes depend on the minimum-count cut described in Sect.~\ref{subsec: characteristic radii method}. In Fig.~\ref{Fig: Rc_Rcomp_z0}, $R_{\rm comp}$ rises again at the highest halo masses in TNG and EAGLE, reaching values larger than the quoted envelopes, but the mass bins concerned contain too few halos to enter the estimate. They should therefore be read as the largest scales attained by well-sampled parts of the halo population, not as absolute maxima over all halos.

\section{Implications for large-scale structure}
\label{sec: LSS_Implications}

\begin{figure}
      \centering
      \includegraphics[width=0.46\textwidth]{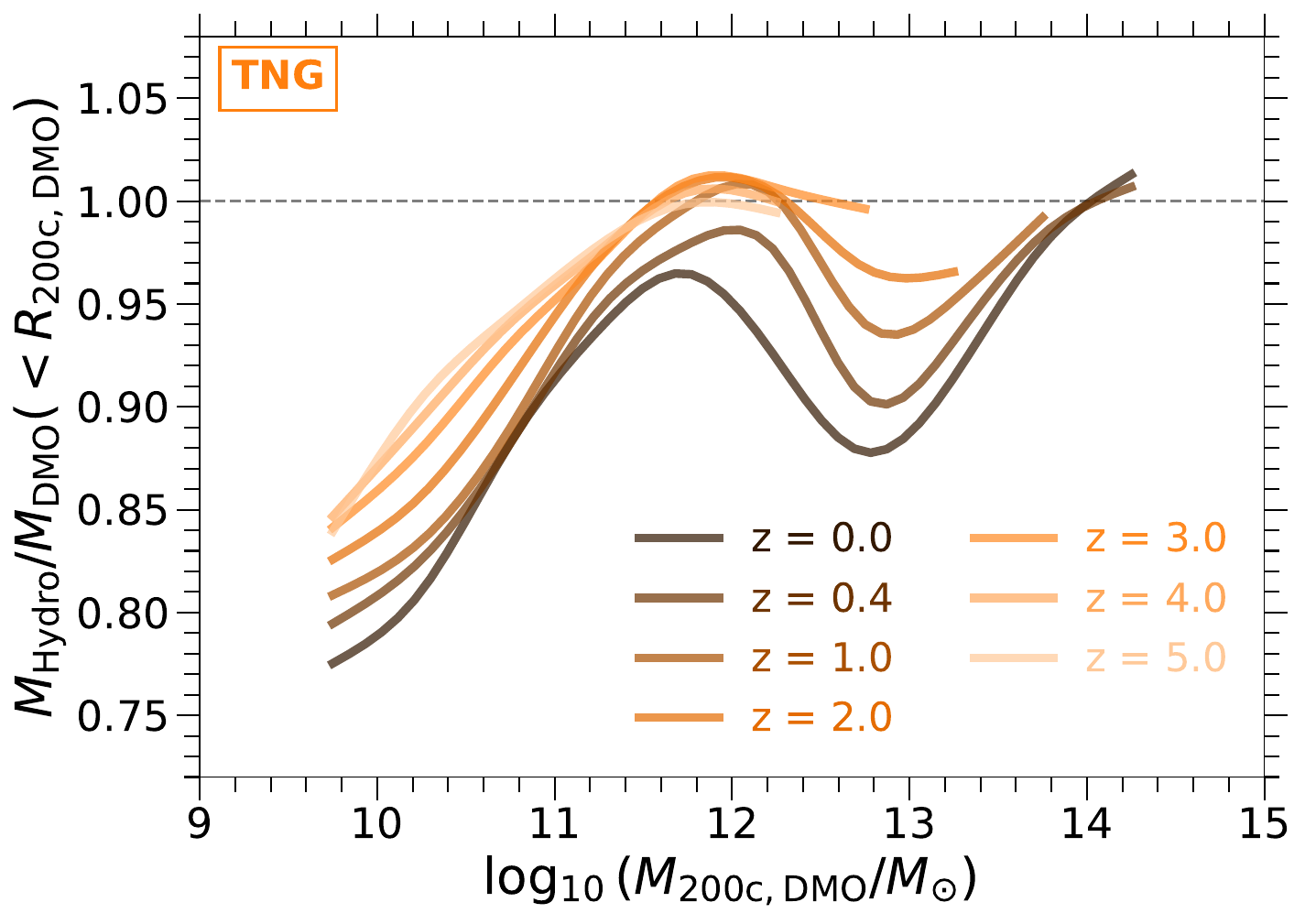}
      \includegraphics[width=0.46\textwidth]{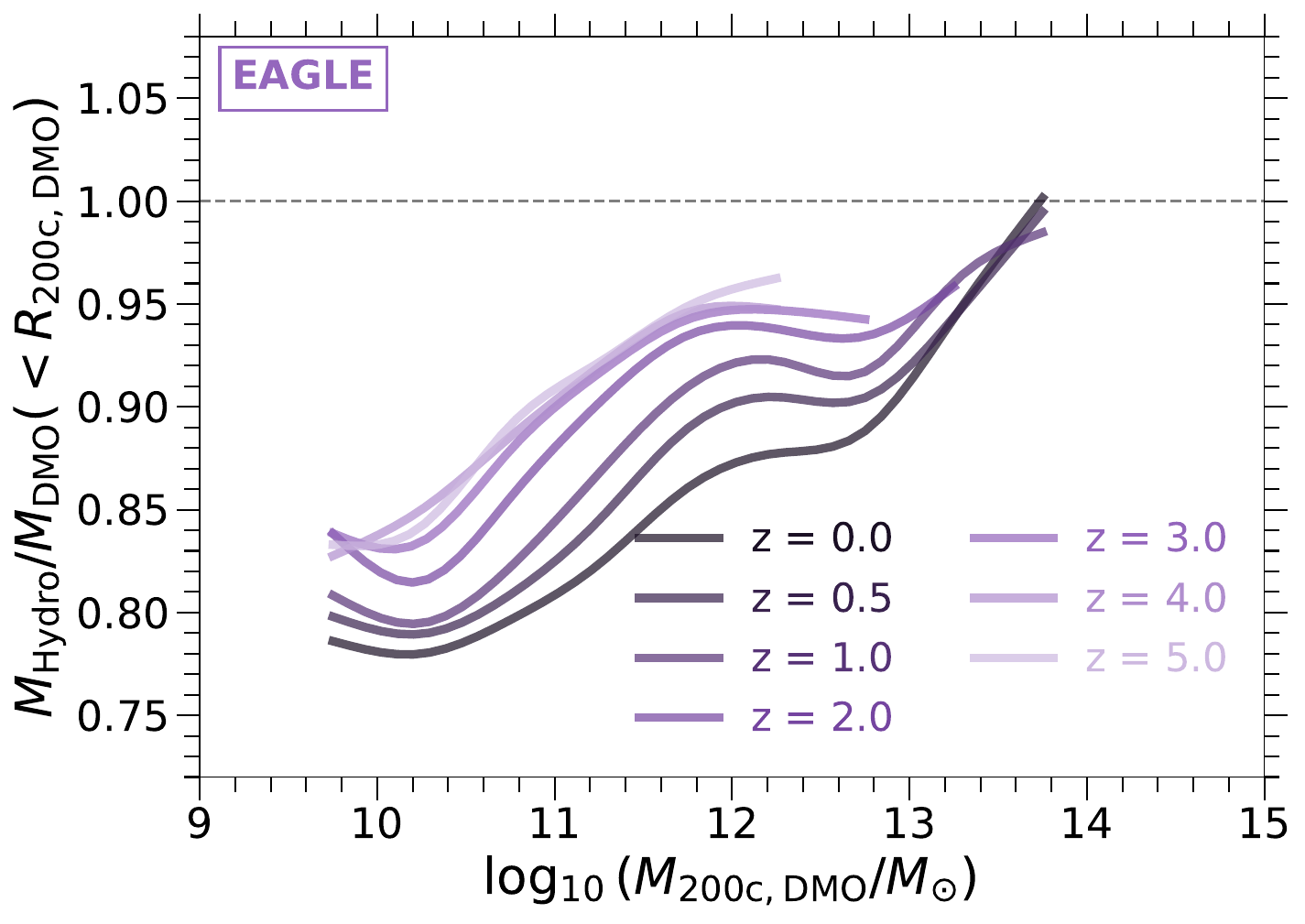}
      \includegraphics[width=0.46\textwidth]{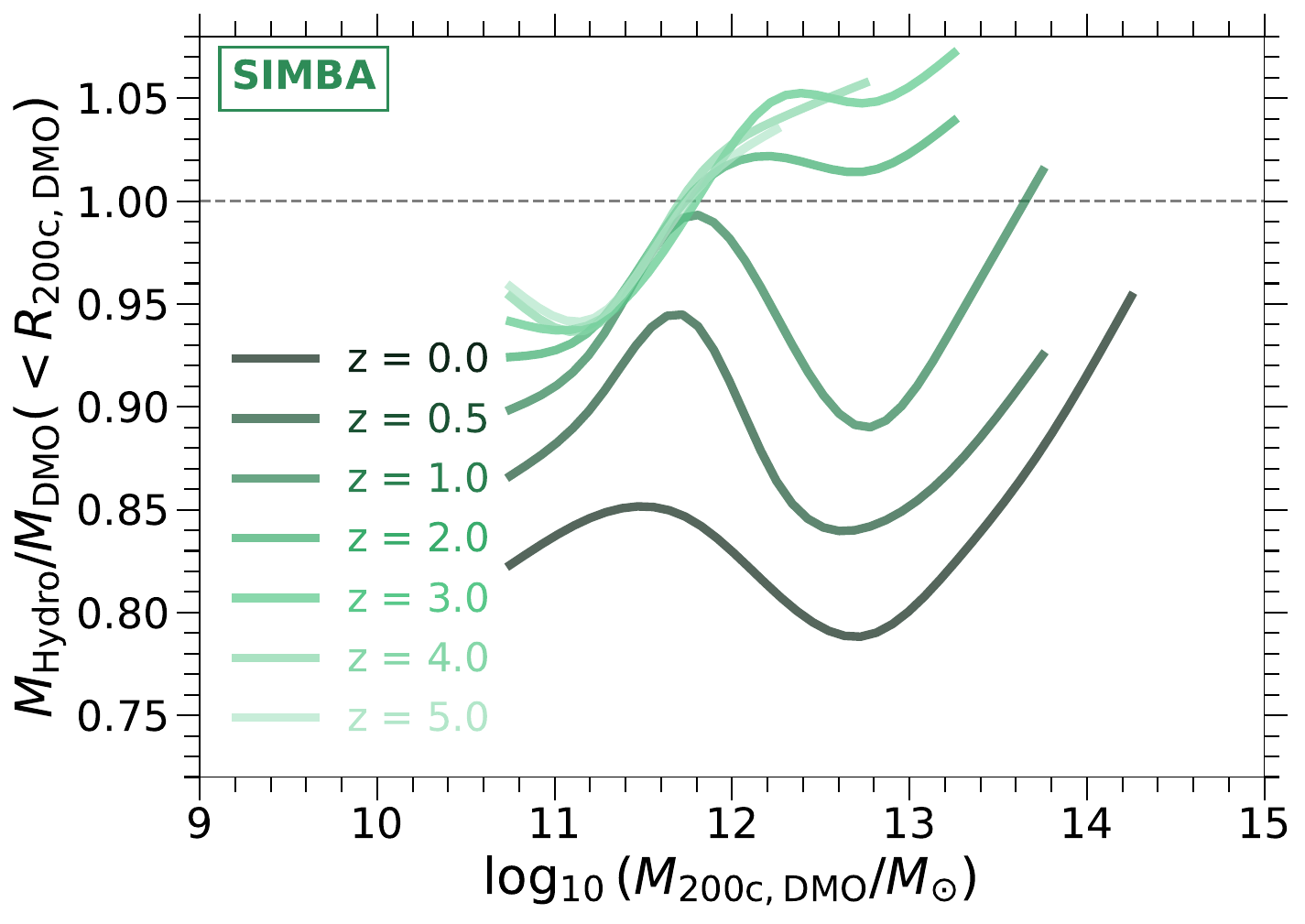}
      \caption{Impact of baryon redistribution on enclosed halo mass. Curves show the median ratio of total enclosed masses in matched hydrodynamical and dark-matter-only halos, both measured within $R_{\rm 200c,DMO}$, as a function of $M_{\rm 200c,DMO}$ for TNG (top), EAGLE (middle), and SIMBA (bottom). Darker lines correspond to lower redshift, from $z=5$ (lightest) to $z=0$ (darkest). Dashed lines mark unity.}
\label{Fig: Halo_Mass_Ratio}
\end{figure}
The characteristic radii introduced in Sect.~\ref{sec: Characteristic_Radii} have consequences on scales beyond the individual halo. Here we examine two key quantities: (i) the suppression of the matter power spectrum, which is directly measurable and is a leading systematic uncertainty for weak lensing surveys, and (ii) the change in halo mass relative to a gravity-only universe, which is not itself an observable but sets how theoretical halo masses relate to those inferred from data. Both are governed by the total matter distribution, so the relevant scale throughout is the compensation envelope.

\subsection{Matter power spectrum suppression}
\label{subsec: Pk}
Fig.~\ref{Fig: Pk} shows the ratio of the power spectrum in the hydrodynamical simulations to that in their DMO counterparts at different redshifts. The diamonds mark the wavenumbers associated with the compensation envelope, $k_{\rm comp}^{\rm env}=1/R_{\rm comp}^{\rm env}$, at each redshift\footnote{Other scale associations include $R\sim\pi/k$ \citep[e.g.][]{Eggemeier2017Cosmology}; the Fourier wavelength is $\lambda=2\pi/k$.}. From the figure, the compensation envelope closely coincides with the scale where the power spectrum ratio approaches unity: at $k_{\rm comp}^{\rm env}$ the ratio is within $2\%$ of unity at all redshifts and in all simulations.

Overall, the power spectrum ratio is close to unity at small $k$ (large scales), where baryon redistribution is negligible. At $k\gtrsim k_{\rm comp}^{\rm env}$, baryonic processes, and baryon redistribution in particular, suppress the power spectrum by roughly $20$--$30\%$ at $z=0$, with the deepest suppression falling near $\log_{10} k \simeq 1$ ($k\simeq 10\,\mathrm{Mpc^{-1}}$) in all simulations. It is strongest in SIMBA, consistent with its more extensive baryon redistribution. Beyond the minimum the ratio rises again, reflecting the enhancement of clustering on these scales by gas cooling and galaxy formation.

The power spectrum suppression weakens at high redshifts, consistent with the smaller compensation envelopes and less baryon redistribution at those epochs. The one exception is EAGLE, where the $z=0$ suppression is weaker than at $z=1$--$3$ over some range of wavenumbers, likely reflecting its stronger stellar feedback at high redshifts and weaker AGN feedback at low redshifts compared to TNG and SIMBA. Importantly, while TNG and SIMBA show almost no power spectrum suppression at $z>3$, EAGLE shows a significant suppression even at $z=5$. This is intriguing because at these redshifts, most of the baryon redistribution in EAGLE happens through stellar feedback, whose impact is often neglected. Therefore, measurements of high-z matter power spectrum, for example from the Lyman-alpha forest, can provide powerful constraints on the strength of feedback and baryon redistribution at these early epochs.

\subsection{Impact of baryon redistribution on halo masses}
\label{subsec: halo_mass_ratio}

The impact of baryon redistribution and feedback processes on the halo mass is crucial for interpreting observations and for theoretical modeling. Fig.~\ref{Fig: Halo_Mass_Ratio} shows the ratio of halo mass in the hydrodynamical simulations to that in their DMO counterparts as a function of halo mass at different redshifts. In most papers in the literature, this ratio is measured by dividing the $M_{\rm 200c}$ of matched halos in the hydrodynamical and their DMO counterparts \citep{springel2018first, Davies2020Quenching}. However, this approach is affected by the fact that once $M_{\rm 200c}$ changes the corresponding $R_{\rm 200c}$ changes with it, so that the two masses of the same object are measured at two different scales. To avoid this issue, we measure the halo mass ratio by measuring the mass of the hydrodynamical halo at the same scale as the DMO halo, i.e. at the $R_{\rm 200c}$ of the DMO halo. This way, we can directly measure how much baryon redistribution and feedback processes have changed the mass of a halo at a fixed scale. 

At $z=0$, the ratio is close to unity at cluster masses, where baryon redistribution is negligible. At lower halo masses, the halo mass is suppressed and the halo mass ratios show a trend very similar to halo baryon fractions, reflecting that the underlying physical processes impacting both of these are similar. In TNG and SIMBA, the ratio increases with halo mass up to $\log_{10}(M_{\rm 200c}/\mathrm{M_{\odot}})\sim 11.5$, then drops sharply around $\log_{10}(M_{\rm 200c}/\mathrm{M_{\odot}})\sim 12.5$--$13$, where AGN feedback becomes efficient at ejecting gas from halos, and increases again toward higher halo masses. EAGLE instead increases monotonically with halo mass, without the intermediate dip, reaching unity by $\log_{10}(M_{\rm 200c}/\mathrm{M_{\odot}})\simeq 13.7$.

The redshift evolution of the ratio also shows a strong model dependence. Across all simulations, the ratio almost always increases with redshift. For TNG and EAGLE, this means less deviation of the hydrodynamical halo mass from DMO runs. In SIMBA, however, halos at $z\gtrsim 2$ show even higher masses in the hydrodynamical runs than in the DMO runs. This suggests that some baryonic processes accelerated the infall of matter into halos at higher redshifts, although the effect is small, at the $5\%$ level. A possible explanation is SIMBA's weak feedback at high redshifts, which allows rapid cooling of the gas and therefore more efficient accretion of matter, particularly gas into halos.

\section{Summary and conclusions}
\label{sec: Summary}
In this paper, we investigated the baryon budget of the Universe and the redistribution of baryons over the past 13 Gyr of cosmic evolution. We examined how baryons are partitioned between gravitationally bound structures (halos) and the IGM, and how baryonic processes drive their distribution away from that of gravity-only dark matter. To do so, we analyzed the TNG, EAGLE, and SIMBA cosmological hydrodynamical simulations together with observational constraints on the baryon content of halos and the IGM. Our main findings are as follows:
\begin{itemize}
      \item Baryons are systematically less concentrated in halos than dark matter at all cosmic epochs. By $z=0$, $43$--$46\%$ of the dark matter resides in halos with $\log_{10}(M_{\rm 200c}/\mathrm{M_\odot}) \geq 10$, whereas only $17$--$28\%$ of baryons do; the remaining $72$--$83\%$ of baryons lie in the IGM, in broad agreement with FRB and earlier baryon-census constraints (Fig.~\ref{Fig: matter_fraction}, left panel). %
      \item Most of the model-to-model variation in the cosmic baryon budget arises from gas rather than stars: the different feedback implementations redistribute halo baryons into the IGM to different degrees, with SIMBA's strong AGN feedback producing the highest IGM baryon share (Fig.~\ref{Fig: matter_fraction}).
      \item Galaxy groups dominate both sides of the halo baryon budget: at $z=0$ the baryons retained within halos peak at $\log_{10}(M_{\rm 200c}/\mathrm{M_\odot})\sim 13.5$--$13.8$, while the baryons redistributed into the IGM peak at somewhat lower masses, $12 \lesssim \log_{10}(M_{\rm 200c}/\mathrm{M_\odot}) \lesssim 13.5$. Both distributions evolve modestly to $z\sim 1$ and shift to lower masses at earlier times, so that by $z>4$ groups no longer contribute significantly (Figs.~\ref{Fig: matter_dist_halo_contribution_z0}, \ref{Fig: matter_dist_halo_contribution}).
\end{itemize}

To quantify the spatial extent of baryon redistribution, we introduce a framework of four characteristic radii: the closure radius, $R_{\rm c}$, introduced in \cite{Ayromlou2023Feedback}, and three new scales defined in this work. The closure radius is the radius around a halo within which the enclosed baryon fraction reaches the cosmic mean. The compensation radius, $R_{\rm comp}$, is the radius within which the total enclosed mass recovers the value it would have in a gravity-only universe. The closure envelope, $R_{\rm c}^{\rm env}(z)$, and the compensation envelope, $R_{\rm comp}^{\rm env}(z)$, are the maxima of these radii over the halo population at each redshift; they set the largest scales on which baryonic effects operate. Using this framework, we evaluate the extent of baryon redistribution across halo mass and cosmic time, and find the following:

\begin{itemize}
      \item The halo baryon fraction and the closure radius evolve with both halo mass and redshift: at fixed mass, the baryon deficiency grows with cosmic time, lowering baryon fractions and enlarging closure radii, demonstrating the cumulative imprint of feedback (Figs.~\ref{Fig: Rc_Rcomp_z0} and~\ref{Fig: Rc_z}).
      \item The largest closure radii, $2$--$7\,\mathrm{cMpc}$ depending on the simulation, occur in galaxy groups ($\log_{10}M_{\rm 200c}/\mathrm{M_\odot}\sim 13$) at $z\lesssim 2$; at $z>3$, no simulation shows baryon redistribution beyond $\sim 1\,\mathrm{cMpc}$ at any halo mass (Figs.~\ref{Fig: Rc_Rcomp_z0} and~\ref{Fig: Rc_z}).
      \item Tracking present-day halos back in time, the normalized closure radius ($R_{\rm c}/R_{\rm 200c}$) grows toward the present in every mass bin below $10^{14}\,\mathrm{M_\odot}$, accompanied by a falling enclosed baryon fraction, while in the most massive bin it declines slightly. Clusters are closest to the cosmic value at $z=0$ but do not hold it throughout: their progenitors were baryon-depleted at intermediate lookback times in TNG and beyond $\simeq 6\,\mathrm{Gyr}$ in EAGLE (Fig.~\ref{Fig: Rc_tree}).
      \item Similar present-day halo baryon fractions and closure radii across simulations can arise from different evolutionary histories (Fig.~\ref{Fig: Rc_tree}).
      \item Along the same histories, the normalized compensation radius ($R_{\rm comp}/R_{\rm 200c}$) grows toward the present in most halo mass bins, but flattens or decreases mildly in EAGLE groups and clusters (Figs.~\ref{Fig: Rcomp_z} and \ref{Fig: Rcomp_tree}).
      \item At $z=0$ the compensation radius is the smaller of the two per-halo scales over the mass range where all three simulations are well sampled, with a median $R_{\rm comp}/R_{\rm c}$ of $0.8$--$1$: the enclosed total mass recovers its gravity-only value before the enclosed baryon fraction recovers the cosmic value (Fig.~\ref{Fig: Rc_Rcomp_z0}). %
      \item Both envelopes grow toward low redshift, from $\lesssim 1\,\mathrm{cMpc}$ at $z\gtrsim 3$ to $\sim 1.5$--$5.5\,\mathrm{cMpc}$ at $z=0$ for the compensation envelope, compared with $\sim 2$--$6.5\,\mathrm{cMpc}$ for the closure envelope, depending on the simulation; the compensation envelope is consistently the smaller of the two by a rather similar factor in all three simulations (Fig.~\ref{Fig: Rc_max}).
      \item The same processes that drive baryon redistribution also suppress the matter power spectrum: at $z=0$, $P_{\rm hydro}(k)$ falls below $P_{\rm DMO}(k)$ by up to $20$--$30\%$ in the three simulations, but the suppression becomes significantly weaker toward higher redshift. The suppression scale is set by the compensation envelope: $k_{\rm comp}^{\rm env}=1/R_{\rm comp}^{\rm env}$ closely matches the scale at which $P_{\rm hydro}(k)/P_{\rm DMO}(k)$ returns to unity, within $2\%$ at every redshift in all three models (Fig.~\ref{Fig: Pk}). %
      \item Relative to matched DMO halos and within the same physical scale, hydrodynamical halos are suppressed in mass, due to feedback. In TNG and SIMBA the suppression deepens into a distinct dip around the scale at which AGN feedback is most efficient, while in EAGLE the ratio increases monotonically with halo mass. This effect is more pronounced at low redshift than at high redshift (Fig.~\ref{Fig: Halo_Mass_Ratio}).
\end{itemize}

Our results demonstrate that the classical baryon cycle is fundamentally open: baryons are transported from halos into the IGM, and the baryon and dark matter distributions diverge progressively over cosmic time. Galaxy groups dominate this transport, with the largest contribution per dex of halo mass, consistent with earlier work linking group-scale baryon content to the suppression of the matter power spectrum \citep[e.g.,][]{Miller2026Baryonic}. Groups combine large closure radii, strong sensitivity to feedback, and substantial baryon content, making them the most informative systems for studying baryon redistribution. Future measurements of their baryon content, such as those from the X-GAP project \citep{Eckert2024Galaxy}, together with improved constraints on the IGM baryon share, offer the clearest route to determining where the baryons missing from halos reside. The framework of four characteristic radii constructed here quantifies the spatial extent of baryon redistribution, on both per-halo and population scales; a key finding is that the compensation envelope coincides with the scale on which the matter power spectrum approaches its gravity-only form, which provides a diagnostic of feedback physics in galaxy formation models.

\section*{Data availability}
Radial profiles, closure and compensation radii, and their envelopes are measured using the parallel code PROFI, which we make publicly available as part of this work at \url{https://github.com/Ayromlou/PROFI}.
The IllustrisTNG, EAGLE, and SIMBA simulations are publicly available. The derived data products and analysis codes used here will be made available upon reasonable request to the corresponding author.

\begin{acknowledgements}
MA acknowledges support from the Argelander-Institut f\"ur Astronomie and the University of Bonn through the independent Argelander Fellowship. This research is partially funded by the Deutsche Forschungsgemeinschaft (DFG, German Research Foundation) under Germany’s Excellence Strategy – EXC 3037 – 533607693 – Our Dynamic Universe. We also acknowledge the Collaborative Research Center 1601 (SFB 1601 sub-project C5) funded by the Deutsche Forschungsgemeinschaft (DFG, German Research Foundation) – 500700252. MA also thanks Romeel Dav{\'e} for his assistance with the SIMBA simulation data and Fatemeh Deyhim for her help in designing the illustration. The computations and analyses presented in this work were performed on the Marvin supercomputer at the University of Bonn.
\end{acknowledgements}

\bibliographystyle{aa}
\bibliography{refbibtex}

\newpage
\begin{appendix}

\section{Illustration of the matter distribution in cosmological simulations}
\label{app: matter_distribution}
\begin{figure*}
        \centering
        \includegraphics[width=0.32\textwidth]{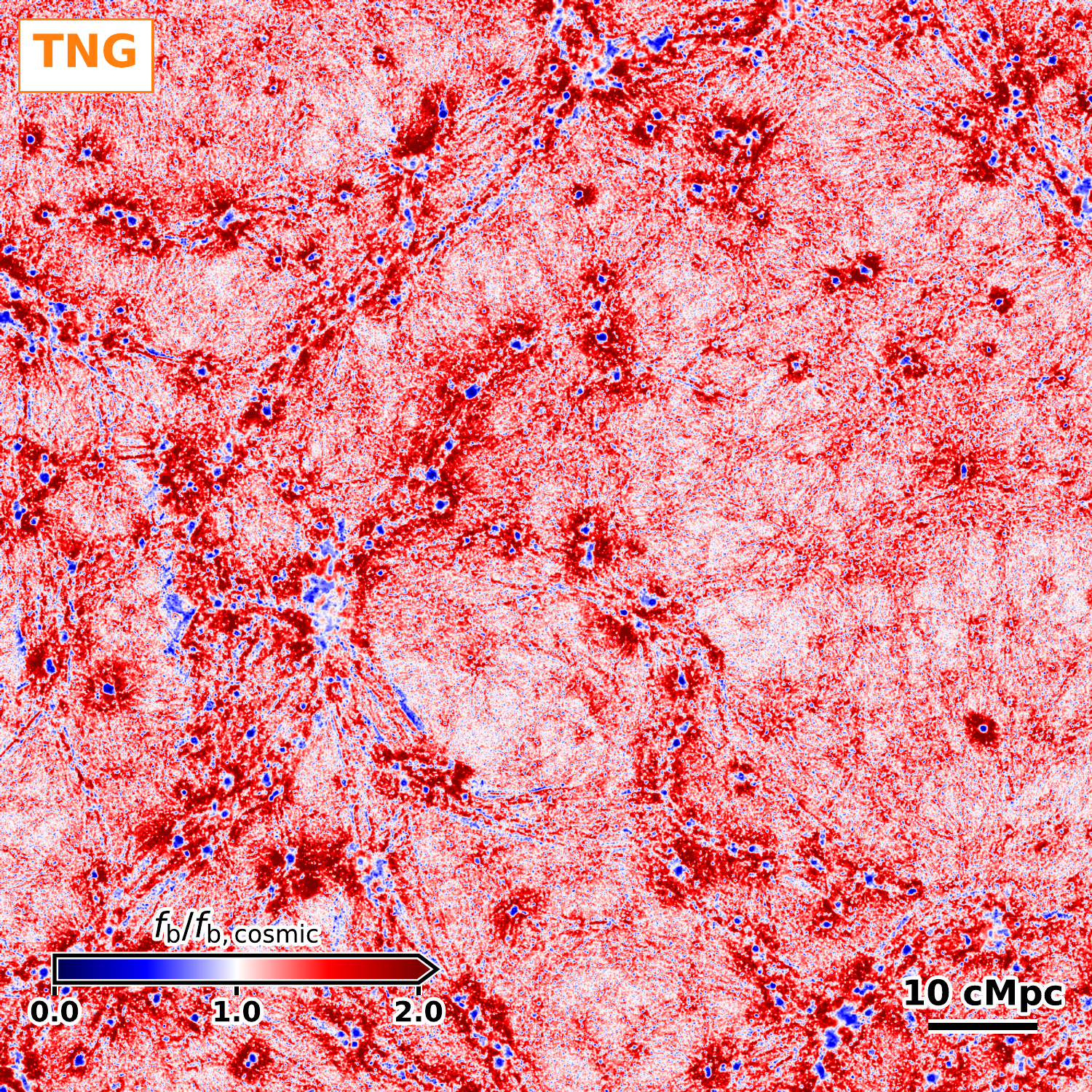}
        \includegraphics[width=0.32\textwidth]{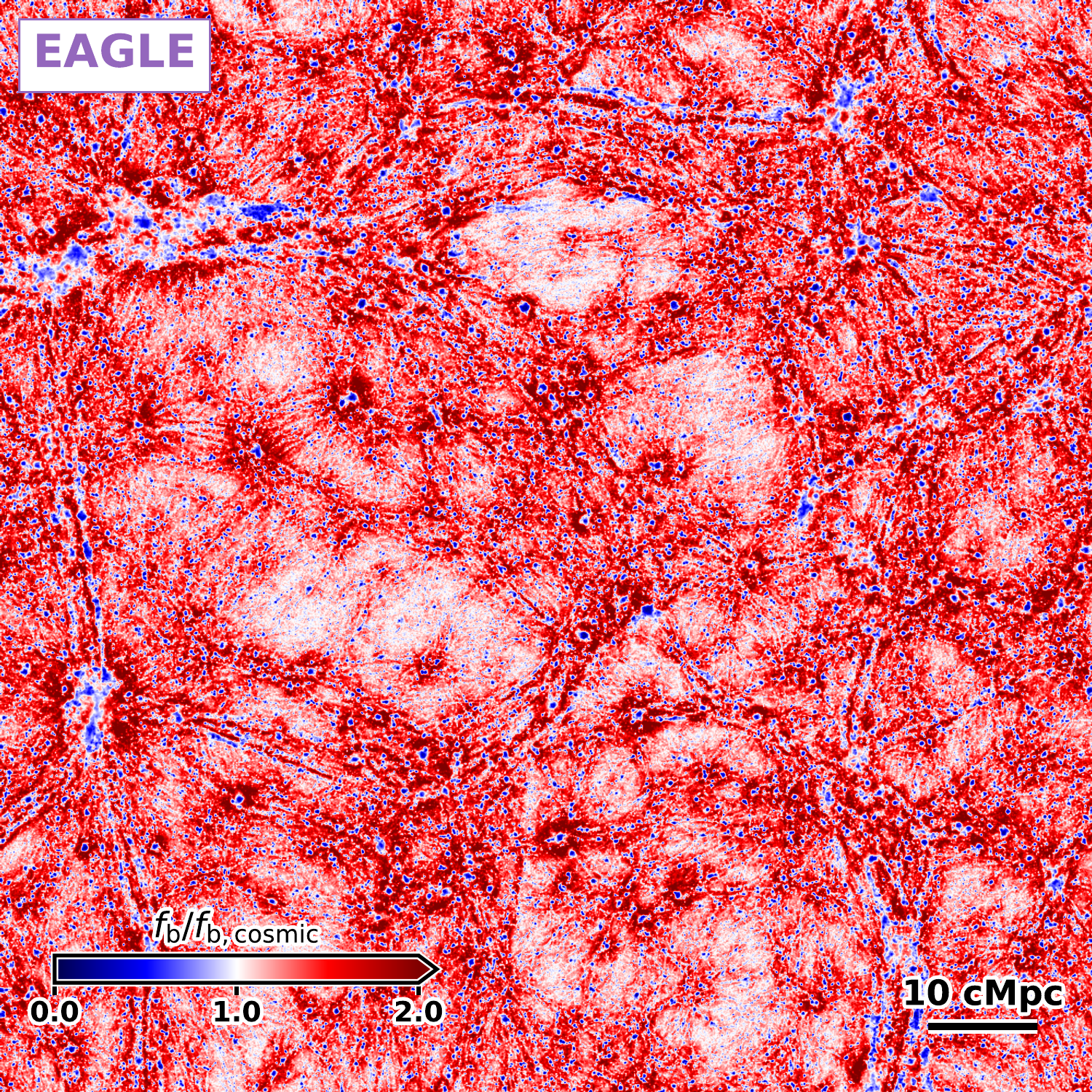}
        \includegraphics[width=0.32\textwidth]{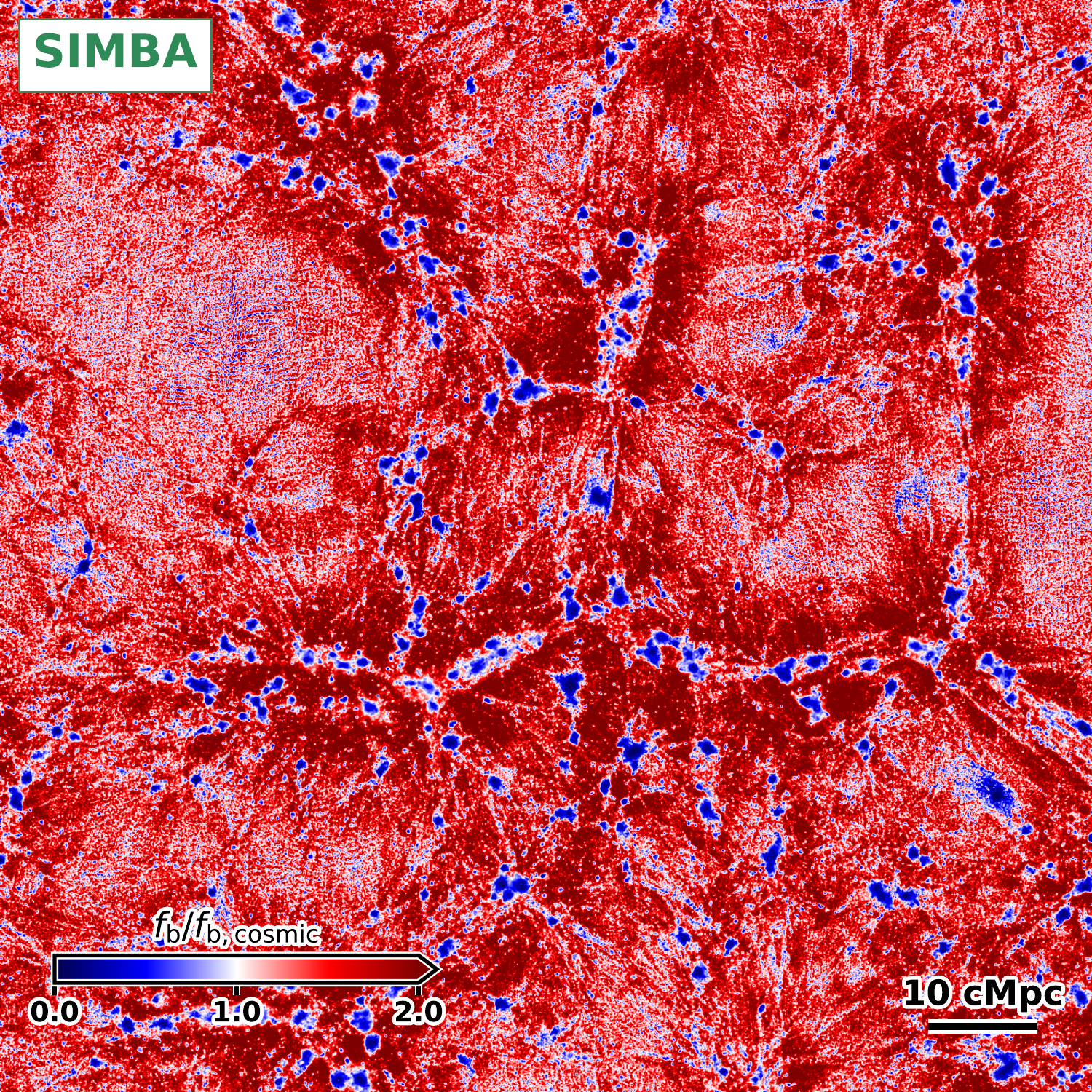}
        \includegraphics[width=0.32\textwidth]{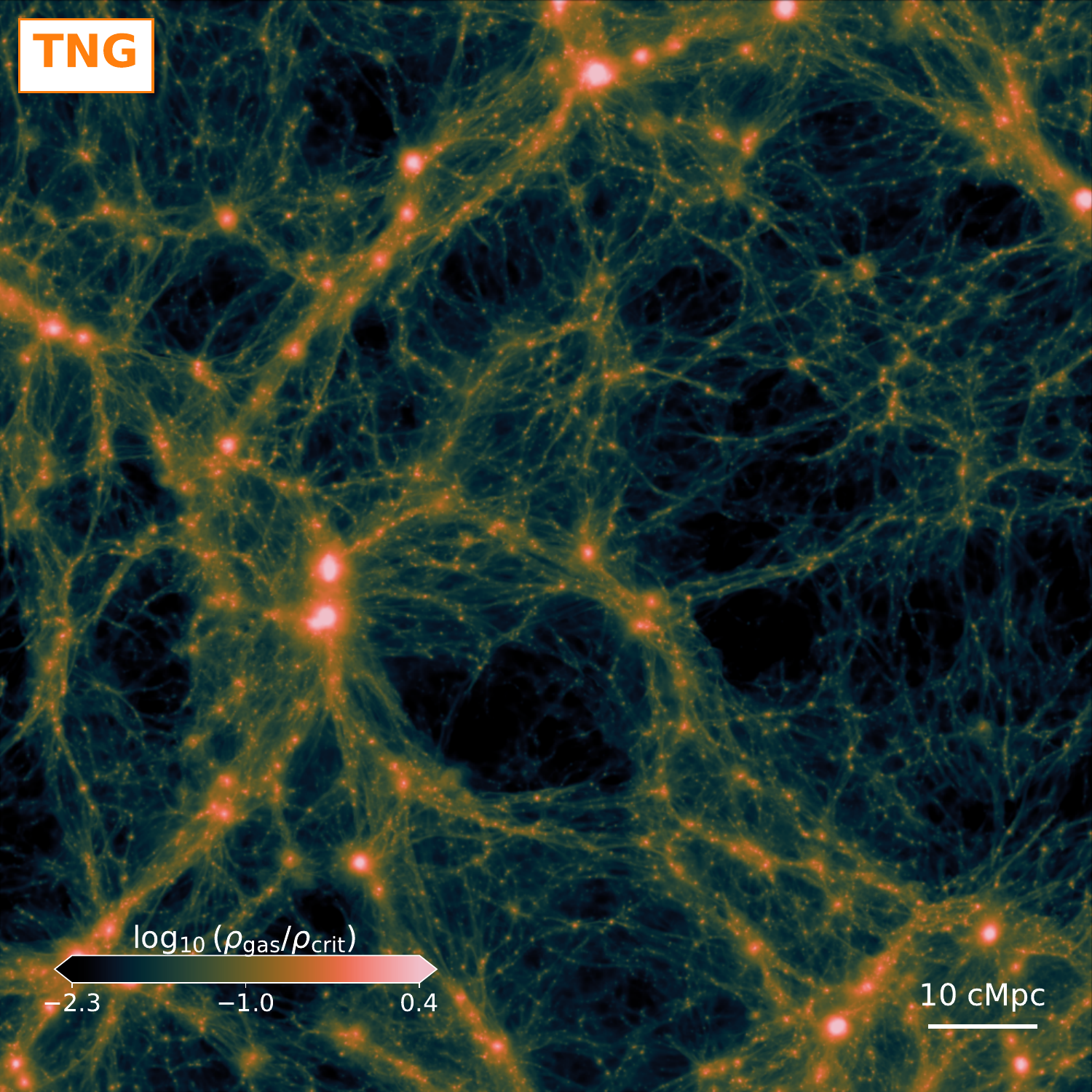}
        \includegraphics[width=0.32\textwidth]{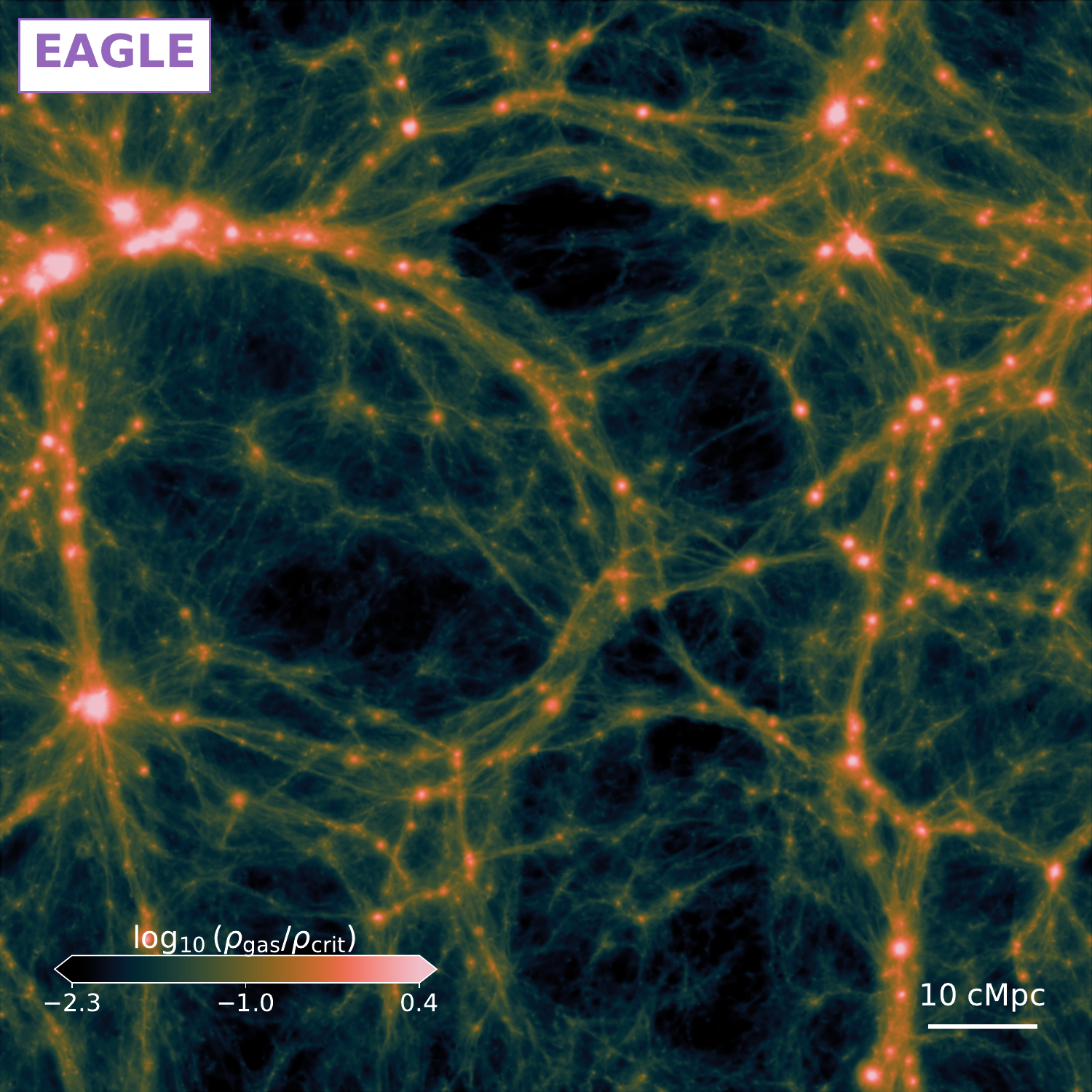}
        \includegraphics[width=0.32\textwidth]{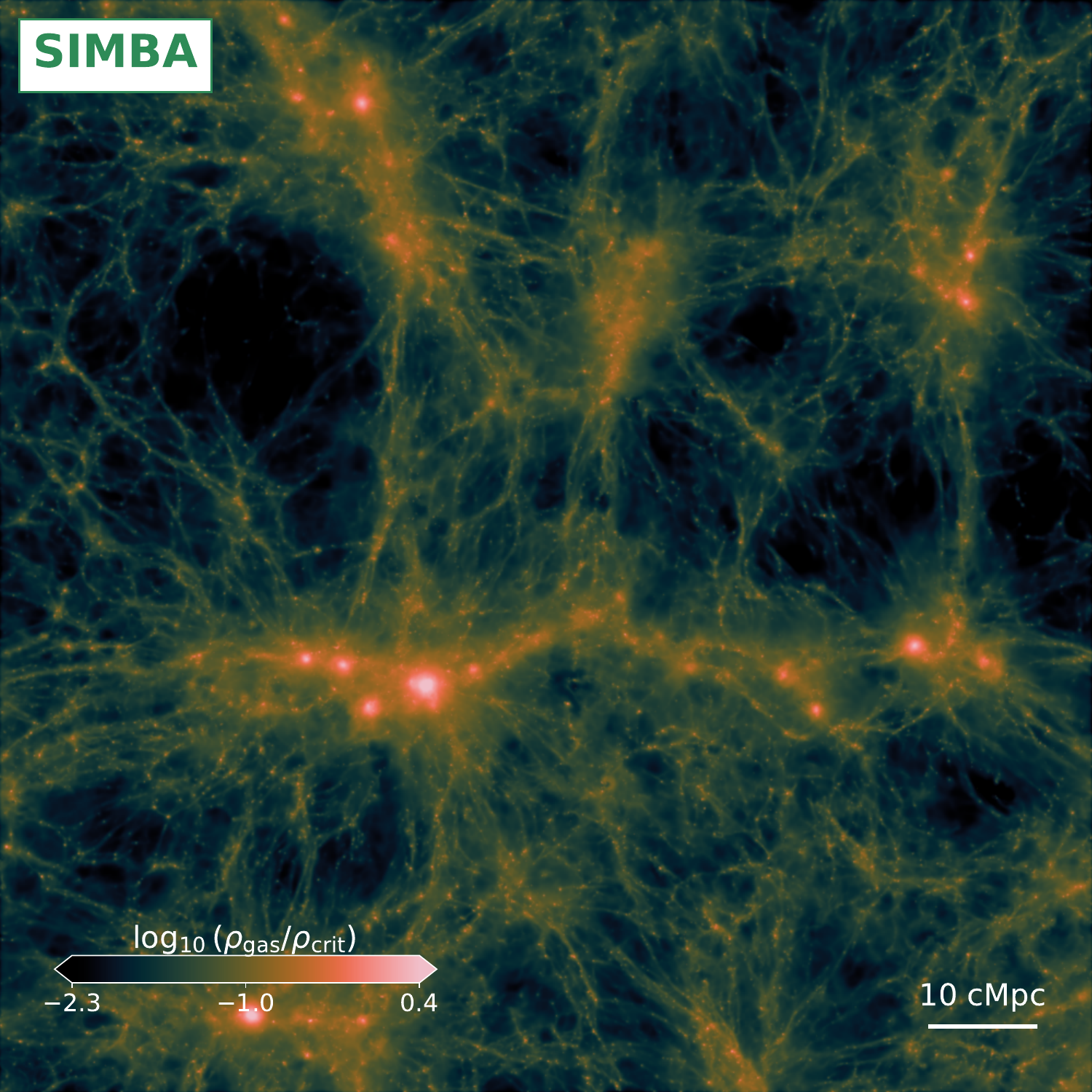}
        \includegraphics[width=0.32\textwidth]{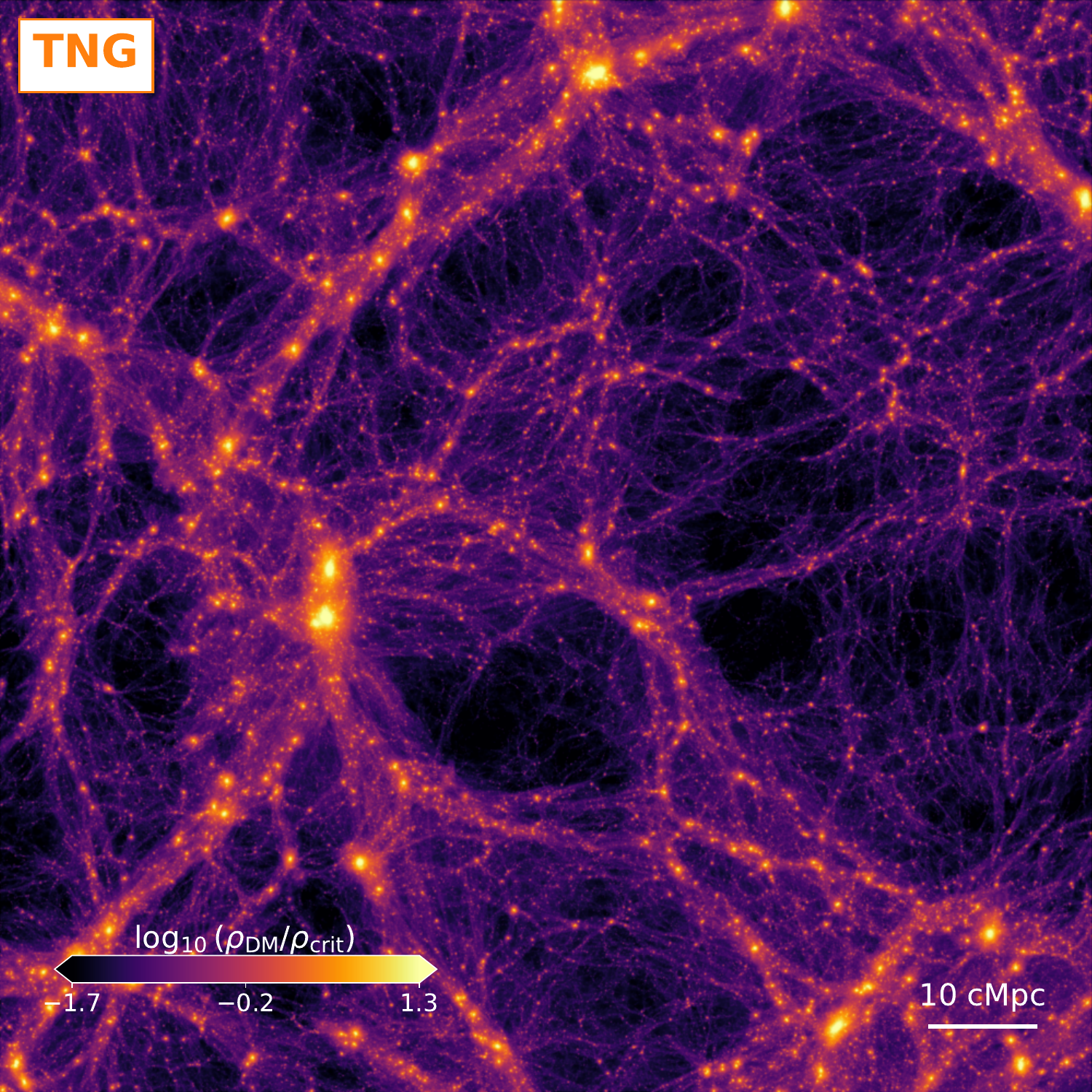}
        \includegraphics[width=0.32\textwidth]{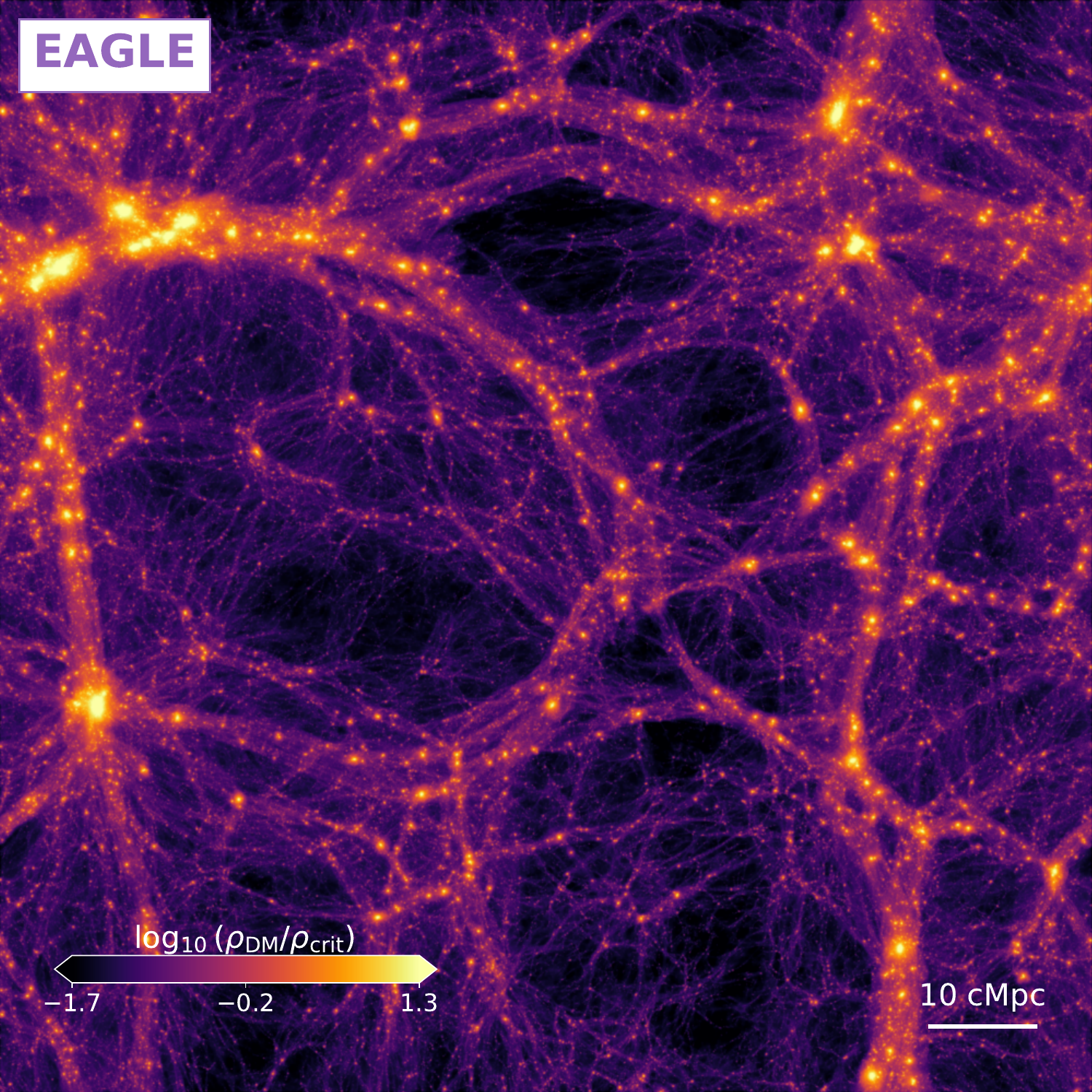}
        \includegraphics[width=0.32\textwidth]{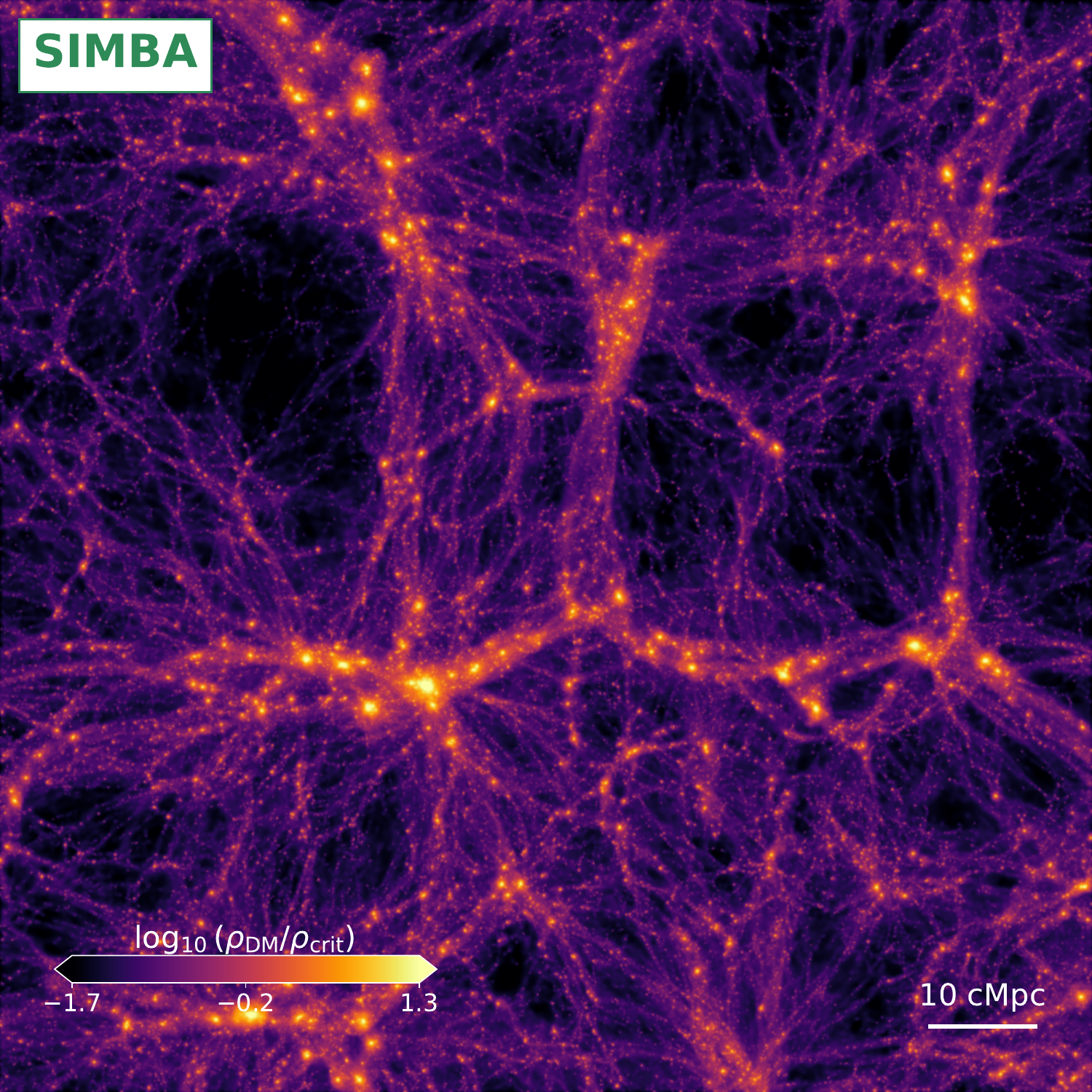}

        \caption{Illustration of the matter distribution in the TNG (left), EAGLE (center), and SIMBA (right) simulations at $z=0$. The top row shows the baryon fraction, $f_{\rm b}/f_{\rm b,cosmic}$, the middle row shows the gas density, and the bottom row shows the dark matter density, the latter two as $\log_{10}(\rho/\rho_{\rm crit})$. The maps are $100\,\mathrm{cMpc}$ on a side and $20\,\mathrm{cMpc}$ thick. The baryon fraction is systematically lower than the cosmic value in halos (bluer in the maps), while it is higher in the IGM (redder in the maps), reflecting the redistribution of baryons from halos into the IGM.}
\label{Fig: matter_maps}
\end{figure*}

Fig.~\ref{Fig: matter_maps} illustrates the central phenomenon of this paper directly. The maps show a slice of each simulation at $z=0$, $100\,\mathrm{cMpc}$ on a side and $20\,\mathrm{cMpc}$ in projection depth, with rows giving the projected baryon fraction relative to the cosmic mean, the gas density, and the dark matter density. Two features are immediately visible. First, the baryon fraction is anticorrelated with the matter density: halos and the densest filament spines are baryon-poor (blue), while the surrounding IGM is baryon-enriched (red), showing that baryons displaced from halos remain in the large-scale structure rather than disappearing from the budget. Second, the strength of this pattern differs considerably between the models, as the two environments discussed below illustrate. These maps are qualitative; the corresponding quantitative statements, in terms of cosmic baryon shares and characteristic radii, are given in Sects.~\ref{sec: Baryon_Budget} and~\ref{sec: Characteristic_Radii} of the main text.

The filamentary structure shows the redistribution most directly. In all three simulations the dense filament spines, which host the collapsed halos, are baryon-poor relative to the cosmic mean, appearing blue in the maps. The filament outskirts, by contrast, are baryon-enriched, appearing red throughout the web. The enrichment is not confined to the immediate outskirts of each filament but extends well beyond them, and its extent varies between the models: broadest in SIMBA, where AGN-driven outflows push baryons far into the surrounding structure, and narrowest in EAGLE and TNG, where baryons remain closer to the systems that ejected them. Filaments therefore act as a major channel for baryon redistribution from bound structures into the IGM, consistent with closure radii reaching several cMpc (Sect.~\ref{sec: Characteristic_Radii}).

Even the voids carry an imprint of the large-scale baryon redistribution. In none of the three models are voids baryon-poor: their baryon fraction typically remains above the cosmic mean, appearing light red in the maps, in contrast to the dark matter maps, where the same regions are strongly underdense. The enrichment is again strongest in SIMBA, consistent with its more extensive baryon redistribution and larger closure envelope of $\sim 6.5 \,\mathrm{cMpc}$ (Sect.~\ref{subsec: envelopes}), although the absolute baryon content of voids stays small because they contain little mass.

\section{Tolerance-matched definitions of the closure and compensation radii}
\label{app: tolerance}

Both the closure radius and the compensation radius are defined as the radius at which a profile crosses a reference value. In practice neither crossing is sharp, so each definition requires a tolerance. The tolerance for the closure radius was fixed in \cite{Ayromlou2023Feedback} and is retained unchanged here. This appendix derives the corresponding tolerance for the compensation radius, so that the two radii can be compared without introducing a threshold-dependent bias. The result requires no additional free parameter: the tolerance band of the closure radius is transported to the compensation radius through an exact relation between the two quantities.

\subsection{Definitions}
\label{app: tolerance_definitions}

All masses below are enclosed within a spherical aperture of radius $R$ centered on the halo. We write the enclosed baryon fraction as
\begin{equation}
    f_{\rm b}(R) \equiv \frac{M_{\rm b}^{\rm hydro}(R)}{M_{\rm tot}^{\rm hydro}(R)} ,
    \qquad
    M_{\rm tot}^{\rm hydro} = M_{\rm b}^{\rm hydro} + M_{\rm DM}^{\rm hydro} ,
\end{equation}
and define the \emph{baryon-fraction deficit}
\begin{equation}
    \eta(R) \equiv 1 - \frac{f_{\rm b}(R)}{f_{\rm b,cosmic}} ,
    \label{eq: eta_def}
\end{equation}
which measures the fractional shortfall of the enclosed baryon fraction relative to the cosmic value. Following \cite{Ayromlou2023Feedback}, the closure radius is the radius at which this deficit first falls within a tolerance $\epsilon$,
\begin{equation}
    \left| \eta(R_{\rm c}) \right| \le \epsilon , \qquad \epsilon = 0.05 ,
    \label{eq: Rc_tol}
\end{equation}
that is, $f_{\rm b}(R_{\rm c})$ lies within $\left(1 \pm \epsilon\right) f_{\rm b,cosmic}$. We emphasize that this criterion is two-sided: the enclosed baryon fraction may also exceed the cosmic value, for instance when the aperture encloses baryons ejected by neighboring systems, and such an aperture is equally regarded as closed.

Analogously, we define the \emph{total-mass deficit}
\begin{equation}
    \zeta_{\rm tot}(R) \equiv
    \frac{M^{\rm DMO}(R) - M_{\rm tot}^{\rm hydro}(R)}{M^{\rm DMO}(R)} = 1 - \frac{M_{\rm tot}^{\rm hydro}(R)}{M^{\rm DMO}(R)} ,
    \label{eq: zeta_tot}
\end{equation}
which measures the fractional shortfall of the enclosed total mass relative to its gravity-only counterpart, and which likewise may take either sign. The compensation radius is the radius at which this deficit first falls within a corresponding tolerance $\epsilon'$,
\begin{equation}
    \left| \zeta_{\rm tot}(R_{\rm comp}) \right| \le \epsilon' .
    \label{eq: Rcomp_tol}
\end{equation}
The remainder of this appendix determines $\epsilon'$ from $\epsilon$. The two profile quantities $\eta(R)$ and $\zeta_{\rm tot}(R)$ thus play strictly parallel roles, each defining one of the two radii, each vanishing when its reference value is recovered, and each positive when the corresponding quantity is deficient. They differ in what they require: $\eta$ is measured entirely within the hydrodynamical run, whereas $\zeta_{\rm tot}$ additionally requires the matched dark-matter-only counterpart.

Finally, it is convenient to express the mass deficit of each component separately, relative to the mass that the aperture would enclose if that component traced the dark-matter-only distribution exactly:
\begin{equation}
    \zeta_{\rm b}(R) \equiv
    \frac{f_{\rm b,cosmic} M^{\rm DMO} - M_{\rm b}^{\rm hydro}}
         {f_{\rm b,cosmic} M^{\rm DMO}} ,
    \label{eq: zeta_b}
\end{equation}
\begin{equation}
    \zeta_{\rm DM}(R) \equiv
    \frac{\left(1-f_{\rm b,cosmic}\right) M^{\rm DMO} - M_{\rm DM}^{\rm hydro}}
         {\left(1-f_{\rm b,cosmic}\right) M^{\rm DMO}} ,
    \label{eq: zeta_dm}
\end{equation}
so that each is normalized to its own component. Here $\zeta_{\rm b}$ measures the displacement of the baryons themselves, while $\zeta_{\rm DM}$ measures the back-reaction of baryonic processes on the dark matter: it vanishes if the dark matter distribution is unaffected, is positive if the dark matter is depleted relative to its gravity-only expectation, and is negative if it is enhanced, for example by adiabatic contraction. Since $M_{\rm tot}^{\rm hydro} = M_{\rm b}^{\rm hydro} + M_{\rm DM}^{\rm hydro}$, and since the two components carry the cosmic mass weights $f_{\rm b,cosmic}$ and $1-f_{\rm b,cosmic}$ in the dark-matter-only run, the three deficits satisfy the exact identity
\begin{equation}
    \zeta_{\rm tot}(R) = f_{\rm b,cosmic}\,\zeta_{\rm b}(R)
    + \left(1-f_{\rm b,cosmic}\right) \zeta_{\rm DM}(R) ,
    \label{eq: zeta_identity}
\end{equation}
that is, the total-mass deficit is the mass-weighted average of the deficits of the two components.

\subsection{The consistency requirement}
\label{app: tolerance_principle}

The tolerances applied to $\eta$ and to $\zeta_{\rm tot}$ cannot simply be set equal, because the two quantities have very different dynamic ranges. The quantity $\eta$ measures the displacement of baryons \emph{relative to} the dark matter and is of order unity in baryon-deficient halos, whereas $\zeta_{\rm tot}$ measures the deficit of the \emph{total} mass and is correspondingly smaller: a halo that loses all of its baryons but retains its dark matter has $\zeta_{\rm tot} = f_{\rm b,cosmic} \simeq 0.16$, and larger values require the dark matter to be depleted as well. A tolerance of $5\%$ applied directly to $\zeta_{\rm tot}$ would therefore be a substantial fraction of the signal.

We fix the tolerance of the compensation radius by imposing the following requirement:
\begin{quote}
    The two radii must coincide identically in the limiting case in which baryonic processes displace only the baryons and produce no back-reaction on the dark matter distribution.
\end{quote}
This is the natural choice, because it ensures that a measured difference between $R_{\rm comp}$ and $R_{\rm c}$ reflects the back-reaction of baryonic physics on the dark matter, rather than an arbitrary choice of thresholds. The requirement fixes the exact reference band derived in Sect.~\ref{app: tolerance_band}; the symmetric band adopted in practice approximates it.

\subsection{Exact relation between the two deficits}
\label{app: tolerance_derivation}

The enclosed baryon mass can be rewritten in terms of $\eta$ and $\zeta_{\rm tot}$ using Eqs.~(\ref{eq: eta_def}) and (\ref{eq: zeta_tot}),
\begin{equation}
    M_{\rm b}^{\rm hydro} = f_{\rm b}\, M_{\rm tot}^{\rm hydro}
    = f_{\rm b,cosmic}\left(1-\eta\right)\left(1-\zeta_{\rm tot}\right) M^{\rm DMO} ,
    \label{eq: Mb_step}
\end{equation}
while the enclosed dark matter mass follows directly from Eq.~(\ref{eq: zeta_dm}),
\begin{equation}
    M_{\rm DM}^{\rm hydro} = \left(1 - f_{\rm b,cosmic}\right) \left(1 - \zeta_{\rm DM}\right) M^{\rm DMO} .
    \label{eq: Mdm_step}
\end{equation}
Adding Eqs.~(\ref{eq: Mb_step}) and (\ref{eq: Mdm_step}) and using $M_{\rm tot}^{\rm hydro} = (1-\zeta_{\rm tot})M^{\rm DMO}$ gives a closed equation for $\zeta_{\rm tot}$,
\begin{equation}
    \left(1-\zeta_{\rm tot}\right)\left[1 - f_{\rm b,cosmic}\left(1-\eta\right)\right] = \left(1 - f_{\rm b,cosmic}\right)\left(1 - \zeta_{\rm DM}\right) ,
\end{equation}
which can be solved exactly to give
\begin{equation}
    \zeta_{\rm tot}(R) = \frac{f_{\rm b,cosmic}\,\eta(R) + \left(1-f_{\rm b,cosmic}\right)\zeta_{\rm DM}(R)}{1 - f_{\rm b,cosmic}\left[1-\eta(R)\right]}.
    \label{eq: master}
\end{equation}
Equation~(\ref{eq: master}) is exact, involves no approximation beyond the definitions above, and holds for either sign of $\eta$ and $\zeta_{\rm DM}$. Its numerator has the same mass-weighted form as the identity of Eq.~(\ref{eq: zeta_identity}), with the baryonic deficit $\zeta_{\rm b}$ replaced by $\eta$: the total-mass deficit is driven by the baryonic displacement and the dark matter back-reaction, weighted by their respective cosmic mass fractions. We write $\mathcal{B} \equiv f_{\rm b,cosmic}/(1-f_{\rm b,cosmic}) = 0.186$ for the cosmic baryon-to-dark-matter mass ratio, which sets the relative weight of the two terms. Dividing the numerator and denominator of Eq.~(\ref{eq: master}) by $1-f_{\rm b,cosmic}$ then gives the compact form
\begin{equation}
    \zeta_{\rm tot} = \frac{\mathcal{B}\,\eta + \zeta_{\rm DM}}{1 + \mathcal{B}\,\eta} ,
    \label{eq: master_rearranged}
\end{equation}
whose denominator is positive for any aperture with $f_{\rm b}(<R) < 1$, so that $\zeta_{\rm tot}$ takes the sign of $\mathcal{B}\,\eta + \zeta_{\rm DM}$.

\subsection{The tolerance-matched band}
\label{app: tolerance_band}

In the no-back-reaction limit, $\zeta_{\rm DM} = 0$, Eq.~(\ref{eq: master}) reduces to a one-to-one mapping between the two deficits,
\begin{equation}
    \zeta_{\rm tot}(\eta) = \frac{f_{\rm b,cosmic}\,\eta}
                       {1 - f_{\rm b,cosmic}\left(1-\eta\right)} .
    \label{eq: zeta_tot_of_eta}
\end{equation}
This mapping is monotonically increasing in $\eta$ and vanishes with it, so the two deficits always share the same sign, and the closure band $\left|\eta\right| \le \epsilon$ maps onto a single connected interval containing zero. Transporting the two edges of the closure band through Eq.~(\ref{eq: zeta_tot_of_eta}) therefore yields the tolerance-matched band of Eq.~(\ref{eq: Rcomp_tol}),
\begin{equation}
    \zeta_{\rm tot}^{+} \equiv \zeta_{\rm tot}(+\epsilon)
    = \frac{f_{\rm b,cosmic}\,\epsilon}{1 - f_{\rm b,cosmic}\left(1-\epsilon\right)} ,
    \label{eq: band_hi}
\end{equation}
\begin{equation}
    \zeta_{\rm tot}^{-} \equiv \zeta_{\rm tot}(-\epsilon)
    = \frac{-f_{\rm b,cosmic}\,\epsilon}{1 - f_{\rm b,cosmic}\left(1+\epsilon\right)} .
    \label{eq: band_lo}
\end{equation}
Because Eq.~(\ref{eq: zeta_tot_of_eta}) is monotonic, the two criteria $\left|\eta(R)\right| \le \epsilon$ and $\zeta_{\rm tot}(R) \in [\zeta_{\rm tot}^{-},\,\zeta_{\rm tot}^{+}]$ are then satisfied at exactly the same radii, so that $R_{\rm comp} = R_{\rm c}$ identically in this limit, as required in Sect.~\ref{app: tolerance_principle}. Note that this equality is exact and not merely valid to first order in $\epsilon$.

Adopting $f_{\rm b,cosmic} = 0.157$ and $\epsilon = 0.05$, Eqs.~(\ref{eq: band_hi}) and (\ref{eq: band_lo}) give
\begin{equation}
    -0.00940 \;\le\; \zeta_{\rm tot} \;\le\; +0.00923 ,
    \label{eq: band_numbers}
\end{equation}
i.e. a $5\%$ tolerance on the enclosed baryon fraction corresponds to a tolerance of just under $1\%$ on the enclosed total mass. The band is mildly asymmetric because Eq.~(\ref{eq: zeta_tot_of_eta}) is nonlinear, but the magnitudes of its two edges differ by less than $2\%$ of their mean. We therefore adopt in practice the symmetric linearization
\begin{equation}
    \left| \zeta_{\rm tot} \right| \le \mathcal{B}\,\epsilon = 0.0093 .
    \label{eq: band_linear}
\end{equation}
With the compact form of Eq.~(\ref{eq: master_rearranged}), the compensation criterion of Eq.~(\ref{eq: Rcomp_tol}), taken with the symmetric band $\left|\zeta_{\rm tot}\right| \le \mathcal{B}\epsilon$ of Eq.~(\ref{eq: band_linear}), is \emph{exactly} equivalent to
\begin{equation}
    \left| \eta(R) + \mathcal{B}^{-1}\,\zeta_{\rm DM}(R) \right|
    \le \epsilon \left[ 1 + \mathcal{B}\,\eta(R) \right] .
    \label{eq: interpretation_exact}
\end{equation}
Dropping the bracket gives the more transparent statement
\begin{equation}
    \left| \eta(R) + \mathcal{B}^{-1}\,\zeta_{\rm DM}(R) \right| \le \epsilon ,
    \qquad \epsilon = 0.05 .
    \label{eq: interpretation}
\end{equation}
The fractional correction thereby neglected is $\mathcal{B}\,\eta$, which depends on the value of $\eta$ at the radius under consideration and not on its distance from $\epsilon$: it is $0.93\%$ at $\eta = \epsilon$, but $1.9\%$ at $\eta = 0.1$ and $3.7\%$ at $\eta = 0.2$. Since compensation can be reached while $\eta$ still exceeds $\epsilon$, Eq.~(\ref{eq: interpretation}) should be read as an interpretation of the criterion rather than as the criterion itself.
In other words, the compensation radius applies the closure criterion of \cite{Ayromlou2023Feedback}, with the same tolerance, to the sum of the baryonic displacement and the dark matter back-reaction, the latter weighted by the ratio of the cosmic dark matter to baryon fractions. Equation~(\ref{eq: interpretation}) makes the leverage of the back-reaction explicit: since $\mathcal{B}^{-1} = 5.37$, a change of only $\left|\zeta_{\rm DM}\right| = 0.93\%$ in the dark matter contributes as much to the criterion as a $5\%$ deficit in the enclosed baryon fraction. We use Eq.~(\ref{eq: Rcomp_tol}) with the symmetric band of Eq.~(\ref{eq: band_linear}) as the operational definition, Eq.~(\ref{eq: band_numbers}) as its exactly matched counterpart, and Eq.~(\ref{eq: interpretation}) as its interpretation.

\subsection{Why the matching matters}
\label{app: tolerance_bias}

Because $\zeta_{\rm tot}$ is a small quantity, $R_{\rm comp}$ depends on the adopted band: tightening it pushes $R_{\rm comp}$ outward and would produce $R_{\rm comp} > R_{\rm c}$ as a definitional artifact rather than as a physical result, while loosening it does the reverse. Equations~(\ref{eq: band_hi}) and (\ref{eq: band_lo}) remove this freedom, since the band is fixed entirely by $\epsilon$ and $f_{\rm b,cosmic}$. The residual sensitivity to the band width is set by the logarithmic slope of the deficit profile at the crossing,
\begin{equation}
    \frac{\Delta \ln R_{\rm comp}}{\Delta \ln \left|\zeta_{\rm tot}\right|_{\rm band}}
    \simeq \left[ \frac{{\rm d}\ln\left|\zeta_{\rm tot}(R)\right|}{{\rm d}\ln R}
      \bigg|_{R_{\rm comp}} \right]^{-1} ,
    \label{eq: sensitivity}
\end{equation}
and analogously for $R_{\rm c}$ with $\eta$ in place of $\zeta_{\rm tot}$. This is a local relation, valid for small changes that leave the same crossing selected. The derivative is negative wherever the deficit decreases outwards, so a wider band moves the radius inwards and a narrower one outwards, as stated above. A shallow deficit profile implies that a modest change in the band displaces the radius substantially, so this slope sets the precision with which any measured difference between $R_{\rm comp}$ and $R_{\rm c}$ can be interpreted. Evaluated with the adopted numerical value $0.0093$ and $\zeta_{\rm DM} = 0$, the symmetric band of Eq.~(\ref{eq: band_linear}) reproduces the exactly matched band to within about $1\%$ in $\eta$, the offset being $0.8\%$ at the positive edge and $1.0\%$ at the negative one.

\subsection{Practical implementation}
\label{app: tolerance_practice}

Three aspects of the measurement require care. First, both criteria are two-sided, so neither $\eta(R)$ nor $\zeta_{\rm tot}(R)$ need approach zero monotonically from one side, and either may cross its band edges more than once. The convention adopted for $R_{\rm comp}$ must therefore mirror the convention adopted for $R_{\rm c}$ in \cite{Ayromlou2023Feedback}. Both are taken to be the innermost qualifying radius, found by searching outwards from the innermost radial bin, as described in Sect.~\ref{subsec: characteristic radii method}. A first crossing does not guarantee that the profile remains within its band at all larger radii.

Second, the sign of $\zeta_{\rm tot}$ is not by itself diagnostic. Because the enclosed baryon fraction may exceed the cosmic value, $\zeta_{\rm tot} < 0$ can arise either from baryon enrichment of the aperture, $\eta < 0$, or from a genuine enhancement of the dark matter, $\zeta_{\rm DM} < 0$. Equation~(\ref{eq: master_rearranged}) shows that these two contributions are degenerate in $\zeta_{\rm tot}$ alone, and Sect.~\ref{app: tolerance_decomposition} describes how we break that degeneracy.

Third, the band of Eq.~(\ref{eq: band_linear}) is narrow, so $R_{\rm comp}$ is more sensitive to numerical noise in $\zeta_{\rm tot}(R)$ than $R_{\rm c}$ is to noise in $\eta(R)$. Contributions include imperfect halo matching, residual offsets between the two halo centers, the different particle masses of the two runs, and Poisson noise in the enclosed DMO mass. The sensitivity is greater for the dark matter than for the baryons, since by Eq.~(\ref{eq: interpretation}) the dark matter enters the criterion with a weight of $5.37$. Individual compensation radii should therefore be read with these uncertainties in mind, which is one reason we work throughout with medians over halo populations rather than with single objects.

\subsection{Decomposing the total-mass deficit}
\label{app: tolerance_decomposition}

The identity of Eq.~(\ref{eq: zeta_identity}) separates the two contributions to the total-mass deficit. Since $\zeta_{\rm b}$ and $\zeta_{\rm DM}$ are obtained from the same profiles, the decomposition
\begin{equation}
    \underbrace{\zeta_{\rm tot}}_{\substack{\text{total-mass}\\ \text{deficit}}}
    =
    f_{\rm b,cosmic}
    \underbrace{\zeta_{\rm b}}_{\substack{\text{baryon}\\ \text{displacement}}}
    +
    \left(1-f_{\rm b,cosmic}\right)
    \underbrace{\zeta_{\rm DM}}_{\substack{\text{dark matter}\\ \text{back-reaction}}}
\end{equation}
accounts for the difference between the two radii in terms of the two component masses, and applies directly to the quantity that defines the compensation radius. It is an accounting identity, and does not by itself identify the dynamical process responsible. In particular, Eq.~(\ref{eq: master}) implies
\begin{equation}
    R_{\rm comp} > R_{\rm c}
    \quad \Longleftrightarrow \quad
    \zeta_{\rm DM}(R_{\rm c}) > 0 ,
    \label{eq: sign_test}
\end{equation}
so that, for a halo whose deficits are positive and decrease outwards, the compensation radius exceeds the closure radius when the dark matter is itself depleted relative to its gravity-only expectation, and falls short of it when the dark matter is instead enhanced. Equation~(\ref{eq: sign_test}) is an idealization, holding for the exactly matched band and for entry into that band from the positive-deficit side. Where the enclosed baryon fraction exceeds the cosmic value, so that the criteria are met on the negative branch, the ordering can reverse at fixed sign of $\zeta_{\rm DM}$; and under the symmetric band a sufficiently small positive $\zeta_{\rm DM}$ is also compatible with $R_{\rm comp} < R_{\rm c}$. We therefore read $\zeta_{\rm b}$ and $\zeta_{\rm DM}$ directly from the measured profiles, rather than inferring their signs from the ordering of the two radii. We note that $\zeta_{\rm b}$ is not identical to $\eta$, since the former is normalized by the expected baryon mass $f_{\rm b,cosmic}M^{\rm DMO}$ and the latter by the actual enclosed mass $M_{\rm tot}^{\rm hydro}$; the two are related by
\begin{equation}
    \zeta_{\rm b} = 1 - \left(1-\eta\right)\left(1-\zeta_{\rm tot}\right)
    \simeq \eta + \zeta_{\rm tot} .
    \label{eq: zeta_b_vs_eta}
\end{equation}
We therefore use $\eta$ and the band of Eq.~(\ref{eq: band_linear}) to \emph{define} $R_{\rm c}$ and $R_{\rm comp}$, preserving exact continuity with \cite{Ayromlou2023Feedback}, and use $\zeta_{\rm b}$ and $\zeta_{\rm DM}$ only as a diagnostic of which physical effect sets the difference between them.

\section{Matter distribution in the dark-matter-only runs}
\label{app: DMO_dist}

\begin{figure}
      \centering
      \includegraphics[width=\columnwidth]{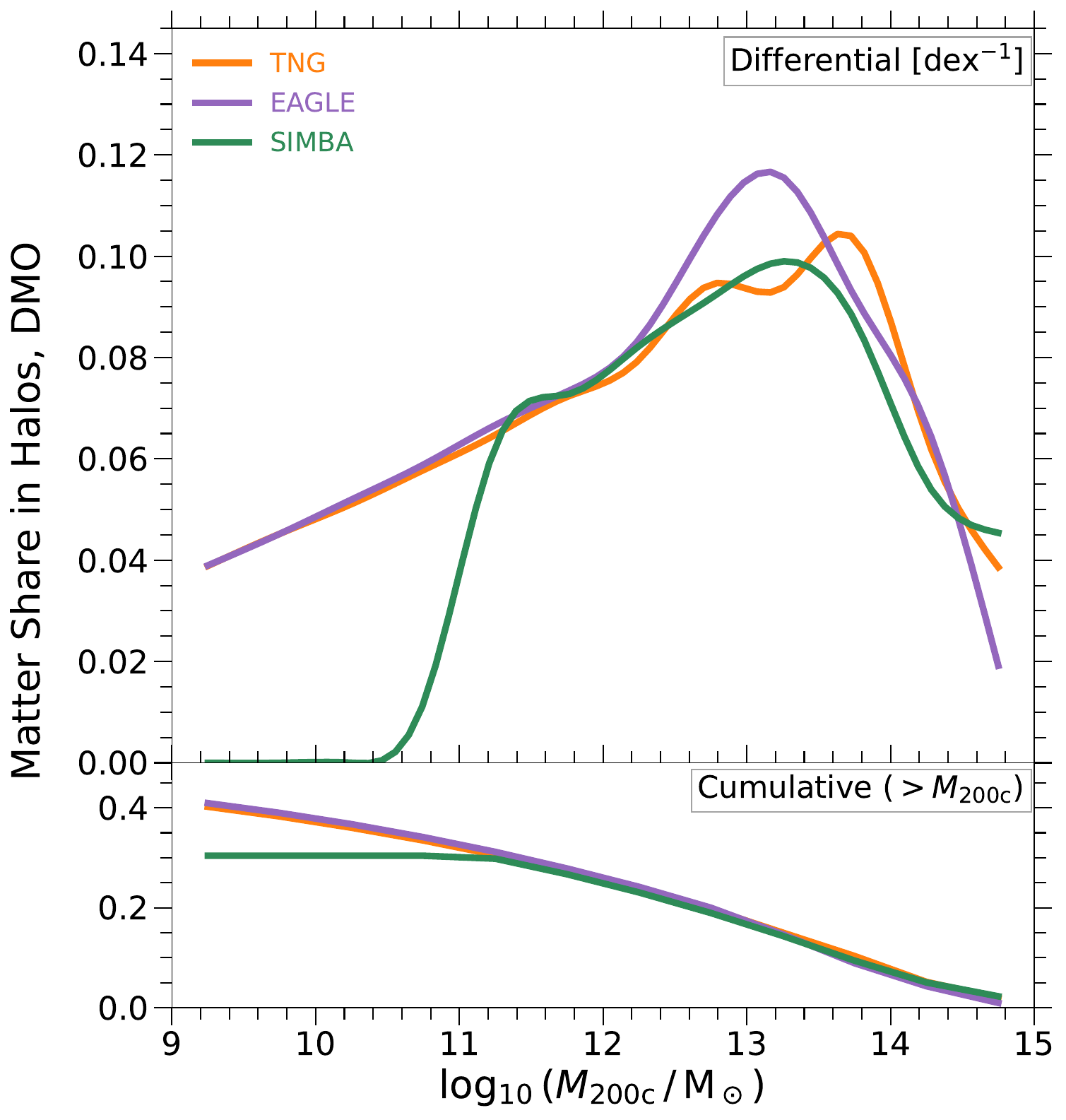}
      \caption{Contribution of halos of different mass to the cosmic \emph{total matter} budget at $z=0$ in the dark-matter-only runs of TNG (orange), EAGLE (purple), and SIMBA (green). The layout matches Fig.~\ref{Fig: matter_dist_halo_contribution_z0}: the upper sub-panel shows the differential distribution $p_{\rm DMO,halo}$ per dex of halo mass, the lower one the cumulative share held by all halos more massive than $M_{\rm 200c}$. This provides the baryon-free reference against which the hydrodynamical results of Sect.~\ref{subsec: halo_contribution} should be read.}
      \label{Fig: DMO_dist_z0}
\end{figure}

\begin{figure}
      \centering
      \includegraphics[width=0.44\textwidth]{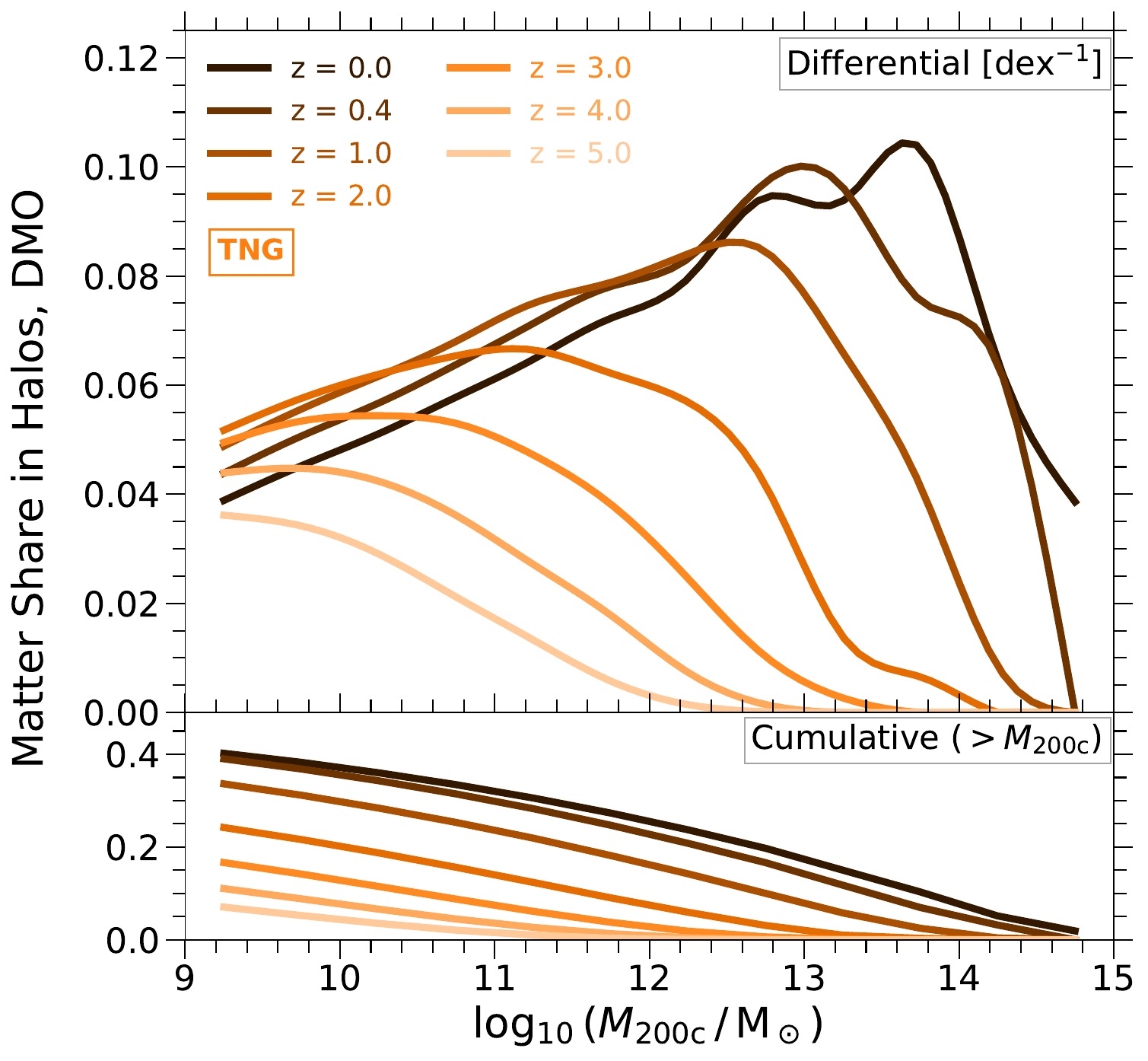}
      \includegraphics[width=0.44\textwidth]{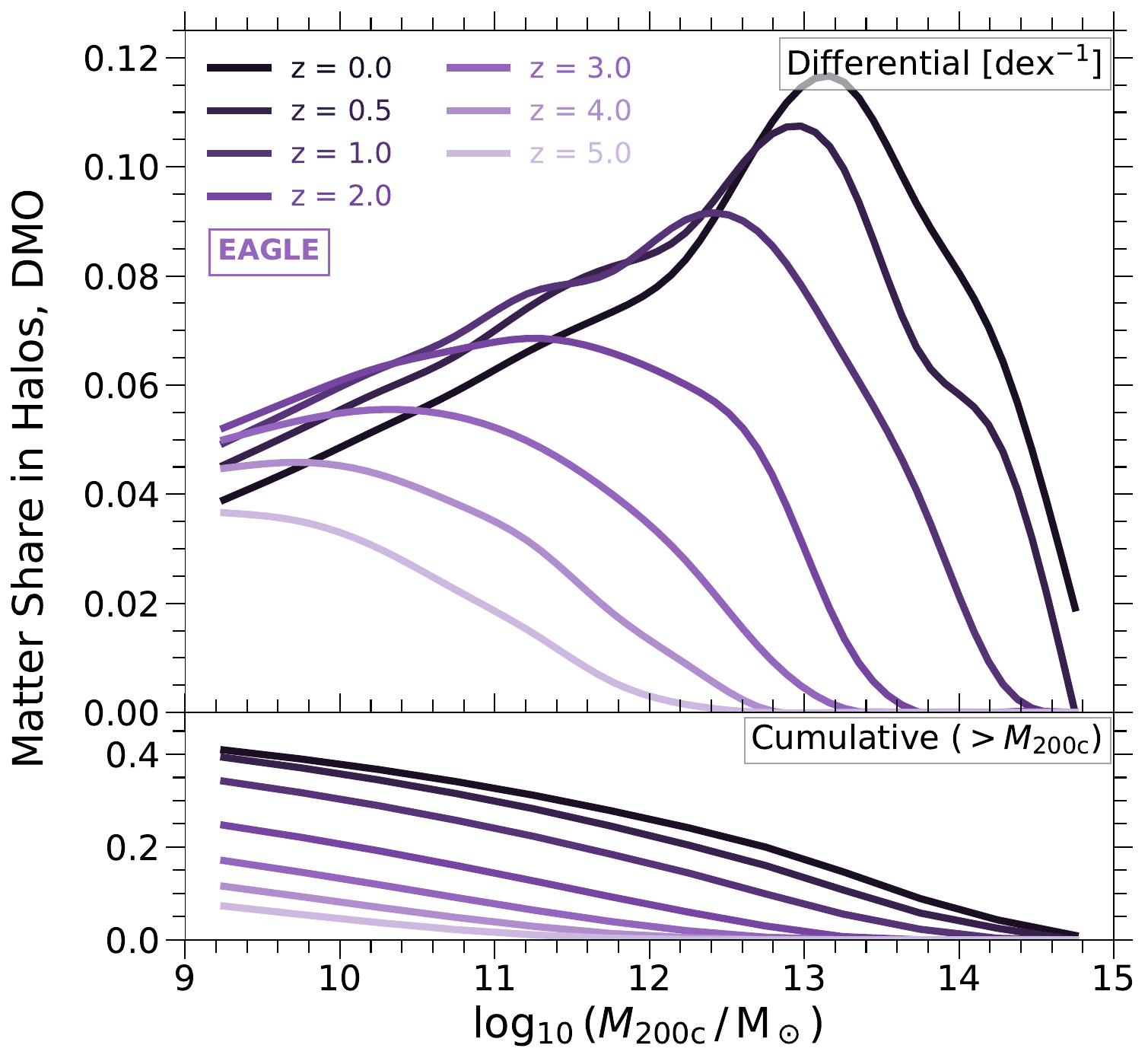}
      \includegraphics[width=0.44\textwidth]{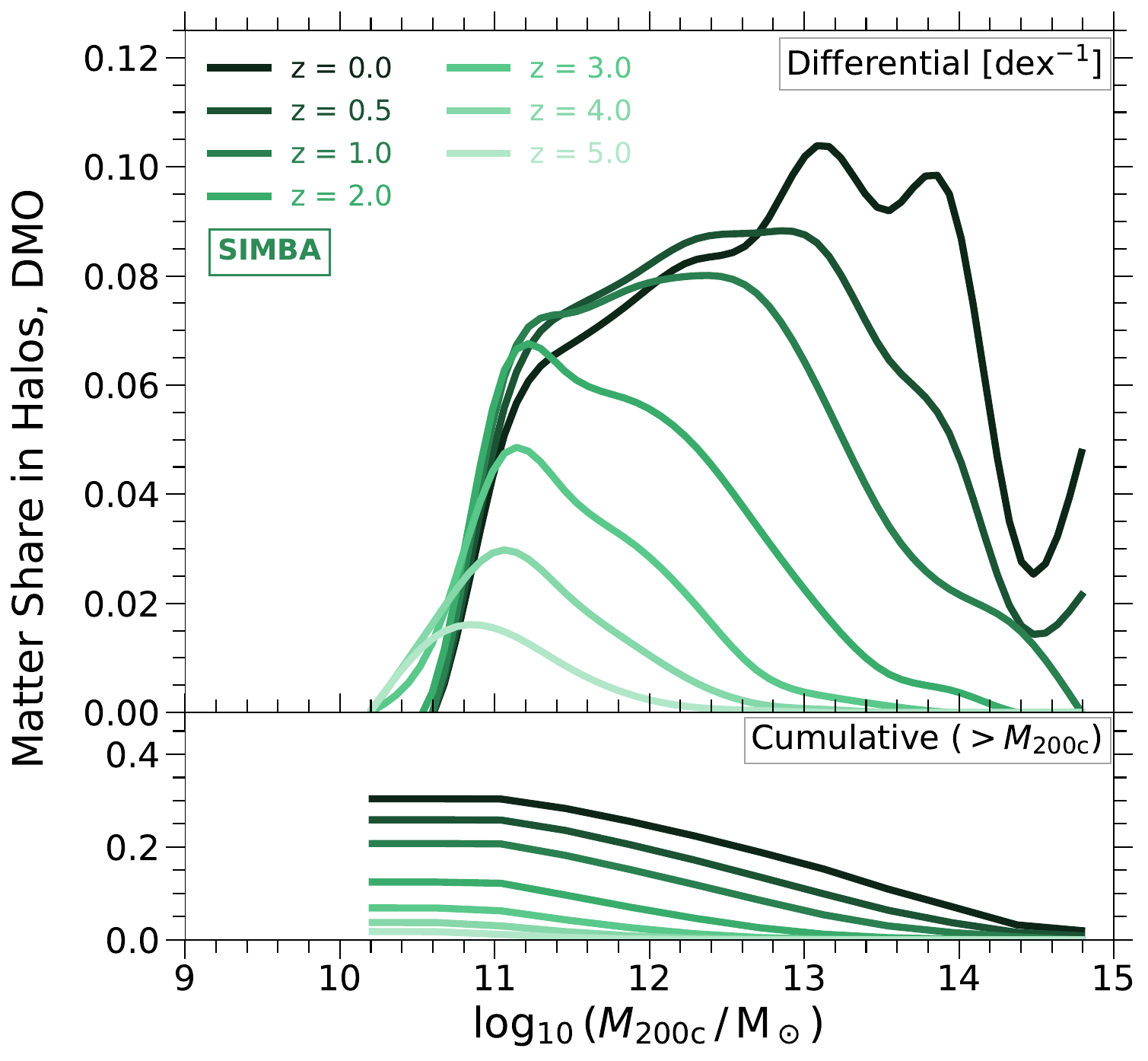}
      \caption{Redshift evolution of the quantity shown in Fig.~\ref{Fig: DMO_dist_z0}, for TNG (top), EAGLE (middle), and SIMBA (bottom). Line shade encodes redshift within each simulation's color, from $z=0$ (darkest) to $z=5$ (lightest).}
      \label{Fig: DMO_dist_z}
\end{figure}

Figures~\ref{Fig: DMO_dist_z0} and~\ref{Fig: DMO_dist_z} show the contribution of halos of different mass to the total matter budget in the dark-matter-only counterparts of the three simulations, i.e. the gravity-only reference for the baryonic distributions presented in Sect.~\ref{subsec: halo_contribution}.

Above $M_{\rm 200c} \simeq 2\times10^{11}\,\mathrm{M_\odot}$ the three models agree to within a few tenths of a percentage point in the cumulative share, and at $M_{\rm 200c} = 10^{11}\,\mathrm{M_\odot}$ to within two percentage points: at $z=0$ all three place $\simeq 31\%$ of the cosmic matter budget within $R_{\rm 200c}$ of halos above this mass. This is expected, since in the absence of baryonic physics the halo mass function depends only on cosmology, and it establishes that the three runs share a common gravity-only baseline. At lower masses SIMBA no longer resolves the halo population and its cumulative curve flattens, while TNG and EAGLE continue to rise, reaching $41\%$ and $42\%$ respectively at the low-mass limit of the sample.

The redshift dependence is that of the halo population itself. In TNG the share of the total matter budget within $R_{\rm 200c}$ of resolved halos falls from $41\%$ at $z=0$ to $24\%$ at $z=1$ and $7\%$ at $z=5$, and the differential peak migrates from $\log_{10}(M_{\rm 200c}/\mathrm{M_\odot}) \simeq 13.7$ to below $10$ over the same interval. This is the gravity-only counterpart of the trend described in Sect.~\ref{subsec: halo_contribution}.

These totals exceed the corresponding hydrodynamical budgets of Sect.~\ref{subsec: halo_contribution}, consistent with the feedback-driven suppression of halo masses quantified in Sect.~\ref{subsec: halo_mass_ratio}.

\section{The baryon fraction of the intergalactic medium}
\label{app: IGM_fb}

\begin{figure}
      \centering
      \includegraphics[width=\columnwidth]{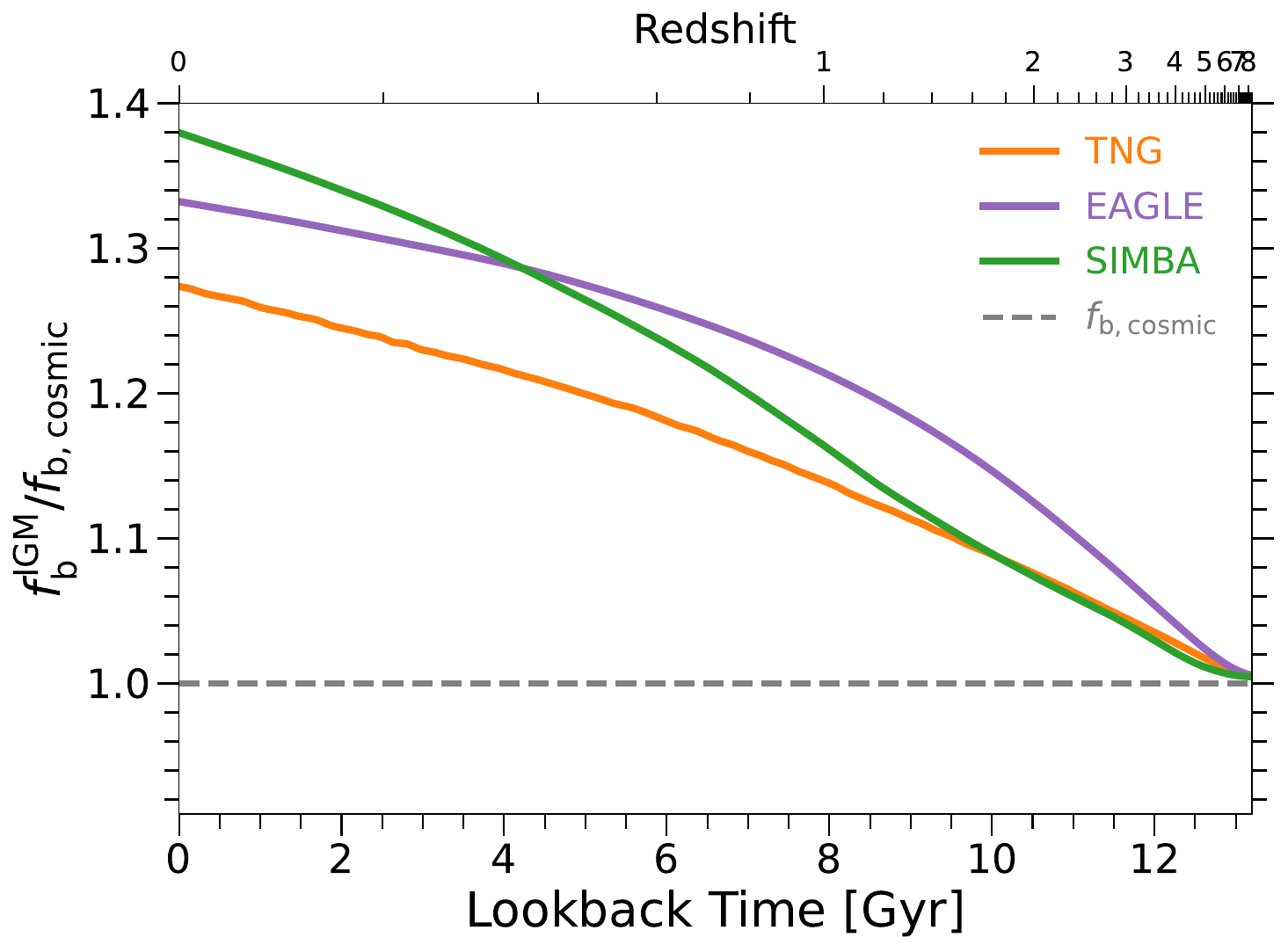}
      \caption{The baryon fraction of the material outside halos, $f_{\rm b}^{\rm IGM}$, normalized to the cosmic value, as a function of lookback time (lower x-axis) and redshift (upper x-axis) for the TNG (orange), EAGLE (purple), and SIMBA (green) simulations. The dashed gray line marks $f_{\rm b}^{\rm IGM} = f_{\rm b,cosmic}$. All three models converge to unity at high redshift, as they must once only a negligible fraction of the matter has collapsed into halos.}
      \label{Fig: IGM_fb}
\end{figure}

Sect.~\ref{subsec: baryon_budget} quantifies the \emph{shares} of the cosmic baryon and dark matter budgets that reside outside halos, $F_{\rm b}^{\rm IGM}$ and $F_{\rm DM}^{\rm IGM}$. A related but distinct quantity is the baryon fraction of the intergalactic material itself,
\begin{equation}
    f_{\rm b}^{\rm IGM} \equiv
    \frac{M_{\rm b,IGM}}{M_{\rm b,IGM} + M_{\rm DM,IGM}} ,
    \label{eq: fb_IGM}
\end{equation}
which measures how baryon-rich that material is per unit of its own mass, rather than what portion of the cosmic baryon budget it holds. The two are related by
\begin{equation}
    \frac{f_{\rm b}^{\rm IGM}}{f_{\rm b,cosmic}}
    = \frac{F_{\rm b}^{\rm IGM}}{F_{\rm m}^{\rm IGM}} ,
    \label{eq: fb_IGM_from_shares}
\end{equation}
where $F_{\rm m}^{\rm IGM} \equiv f_{\rm b,cosmic} F_{\rm b}^{\rm IGM} + \left(1 - f_{\rm b,cosmic}\right) F_{\rm DM}^{\rm IGM}$ is the share of the total matter budget lying outside halos. Equation~(\ref{eq: fb_IGM_from_shares}) shows that $f_{\rm b}^{\rm IGM}$ carries no information beyond the two shares already presented in Fig.~\ref{Fig: matter_fraction}; we show it separately here because it is the quantity referred to in Sect.~\ref{subsec: baryon_budget}, and because it makes the degree of baryon enrichment of the IGM directly visible.

Fig.~\ref{Fig: IGM_fb} shows this ratio as a function of cosmic time. All three models lie within $2\%$ of the cosmic value at $z\gtrsim6$, when only a negligible fraction of the matter has collapsed into halos, and rise monotonically thereafter, reaching $\simeq 1.3$ (TNG and EAGLE) and $\simeq 1.4$ (SIMBA) at $z=0$. The ordering of the models is the same as for $F_{\rm b}^{\rm IGM}$ itself, but the spread between the extreme models is roughly halved, from $15\%$ to $8\%$, because normalizing by $F_{\rm m}^{\rm IGM}$ divides out the (smaller) model-to-model differences in how much dark matter lies outside halos. For this reason we retain $F_{\rm b}^{\rm IGM}$, rather than $f_{\rm b}^{\rm IGM}$, as the quantity that displays the model differences more directly.

We further note that EAGLE maintains the most baryon-rich IGM at $z \gtrsim 0.4$, whereas SIMBA overtakes it toward the present day. This is consistent with the picture developed elsewhere in this paper, in which EAGLE redistributes baryons early through comparatively strong stellar feedback (Sect.~\ref{subsec: Pk}), while SIMBA's more ejective AGN feedback operates predominantly at late times (Sect.~\ref{subsec: baryon_budget}).

\section{Tracing individual halos through cosmic time}
\label{app: App1}
\begin{figure}
      \centering
      \includegraphics[width=0.24\textwidth]{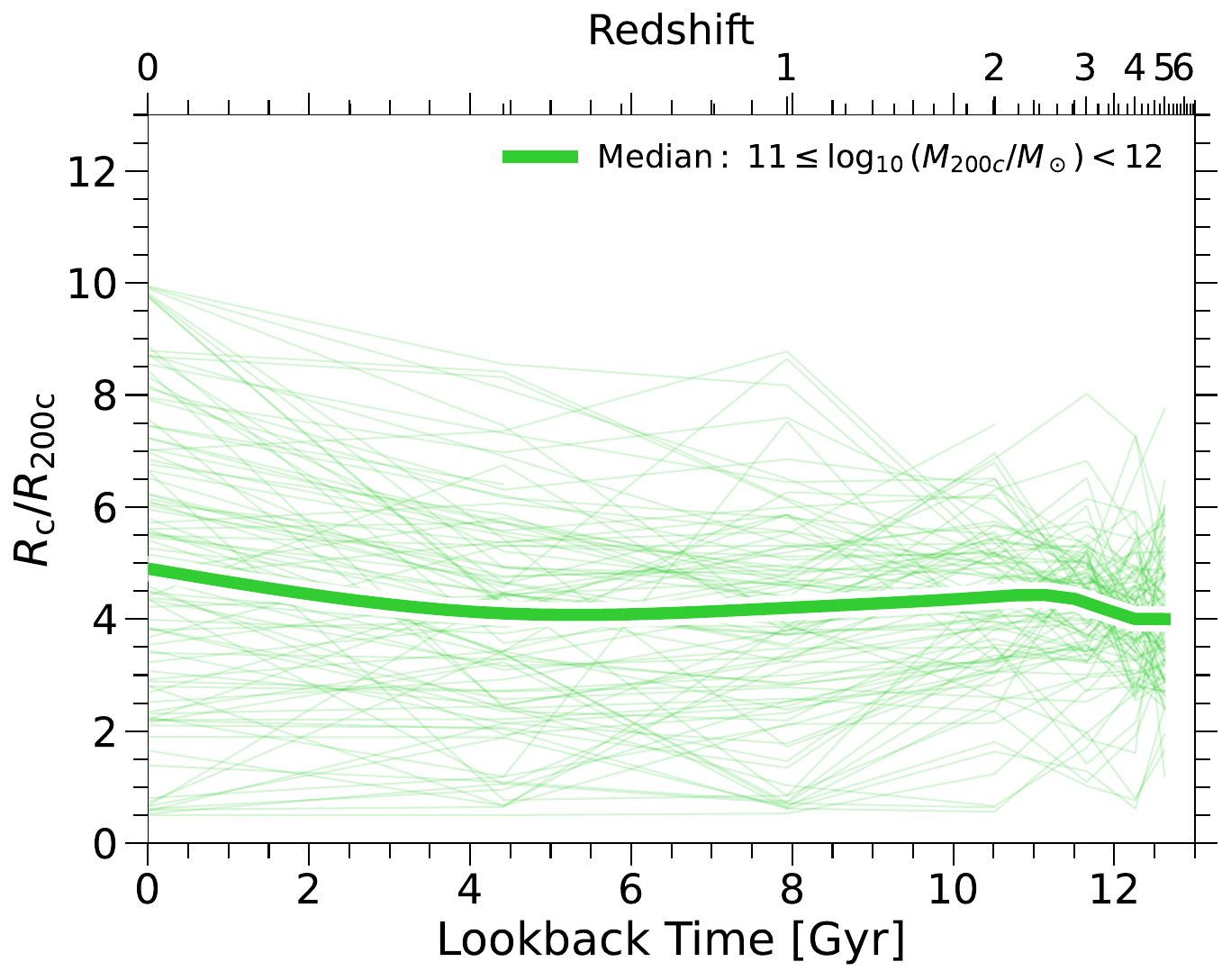}
      \includegraphics[width=0.24\textwidth]{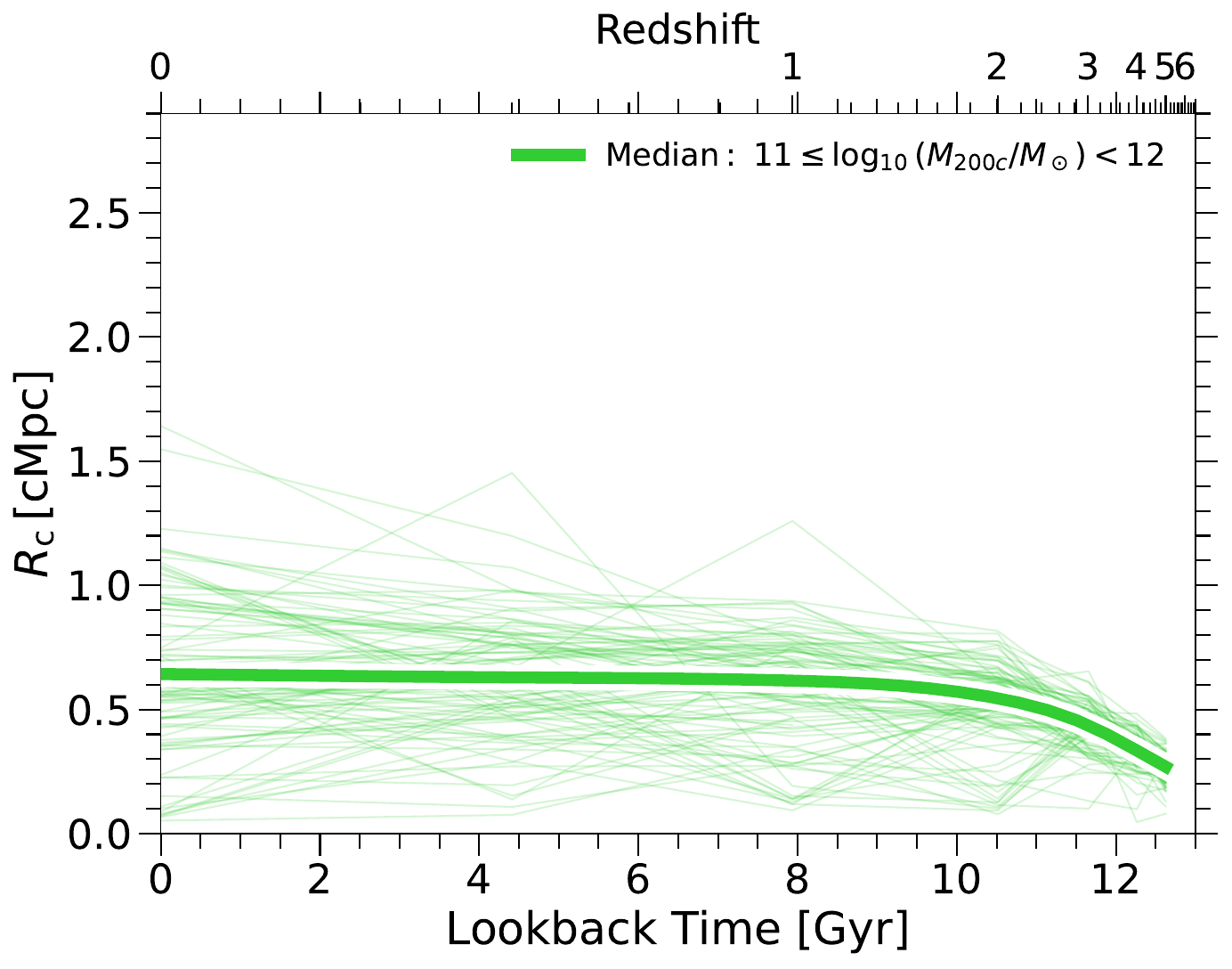}

      \includegraphics[width=0.24\textwidth]{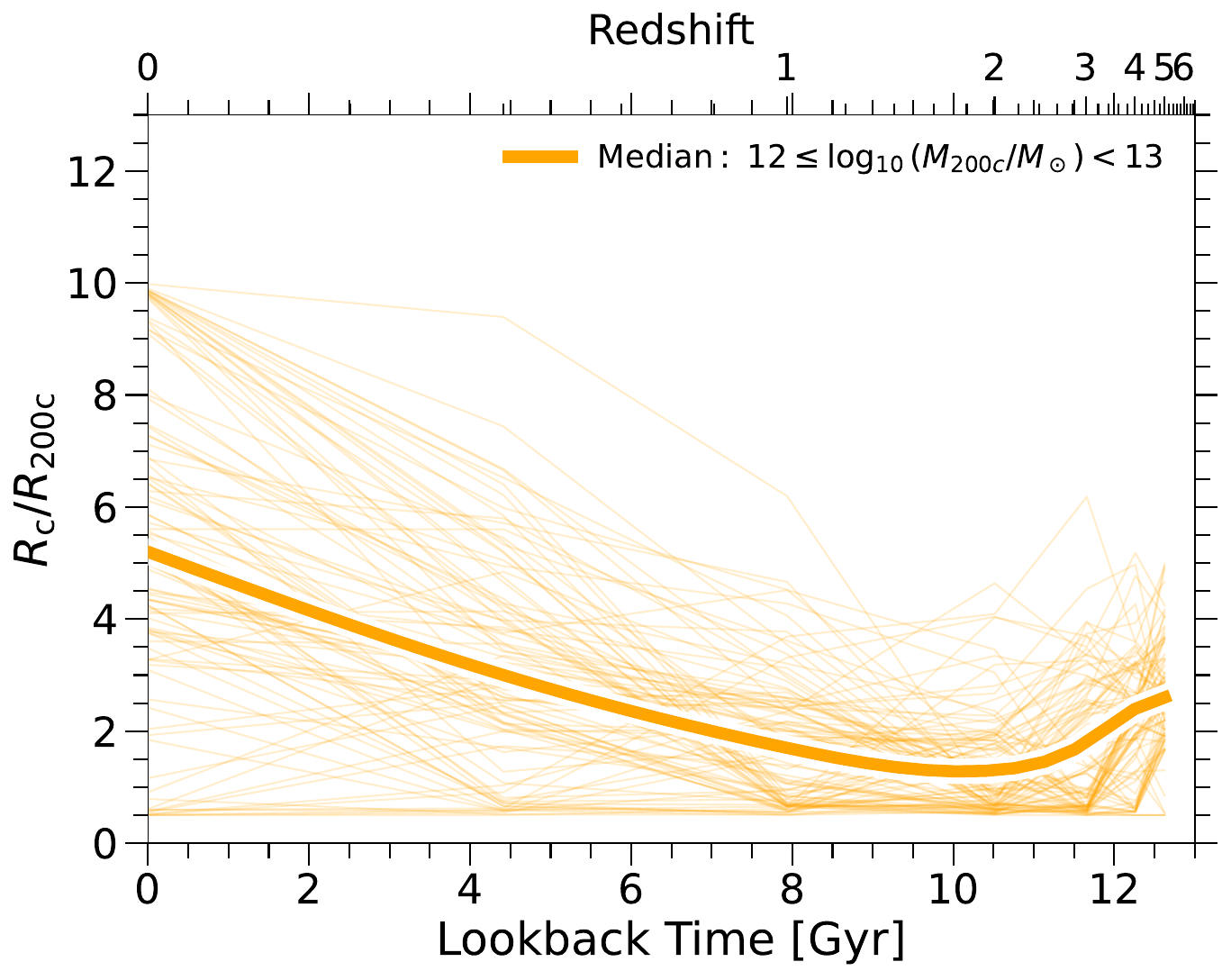}
      \includegraphics[width=0.24\textwidth]{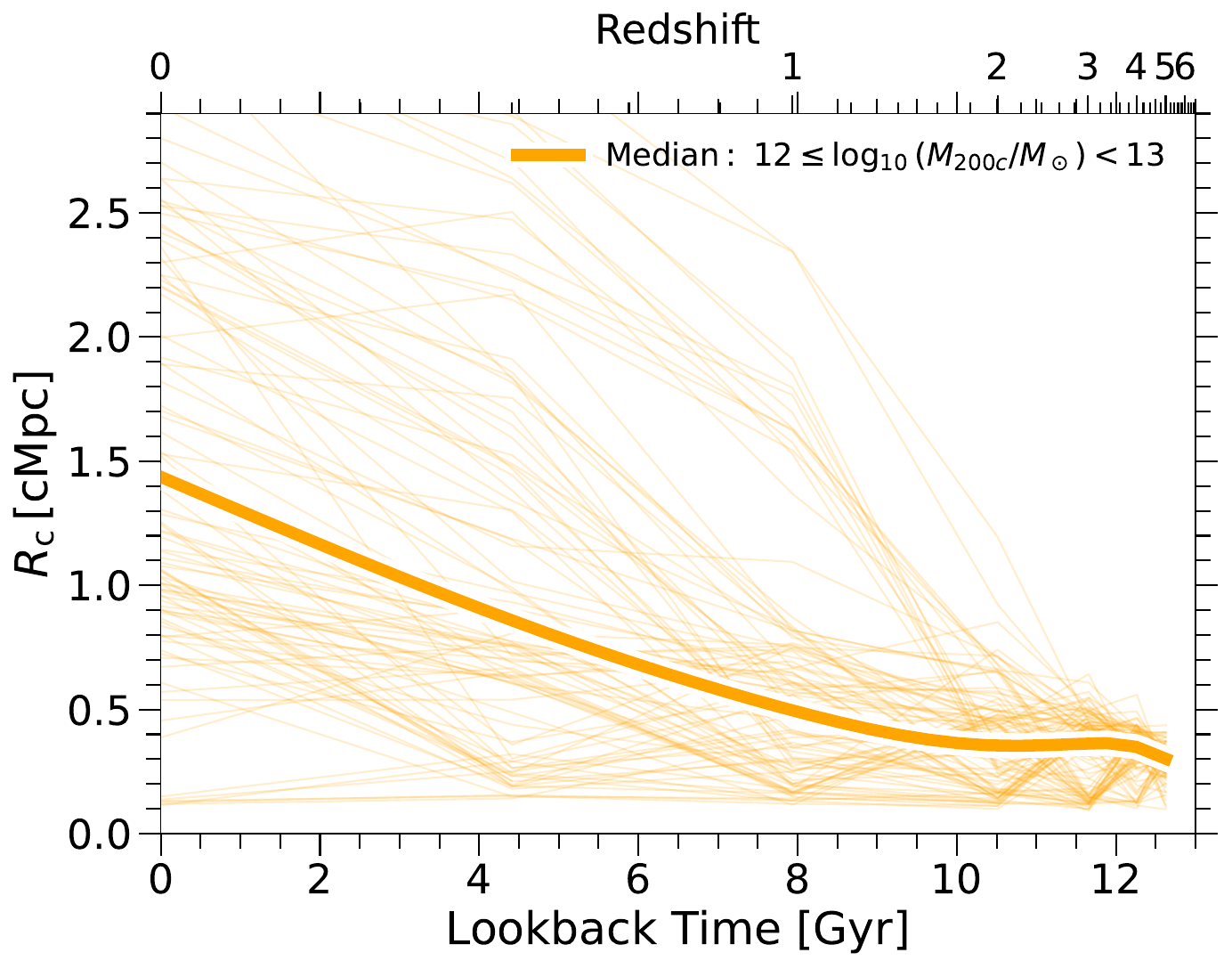}

      \includegraphics[width=0.24\textwidth]{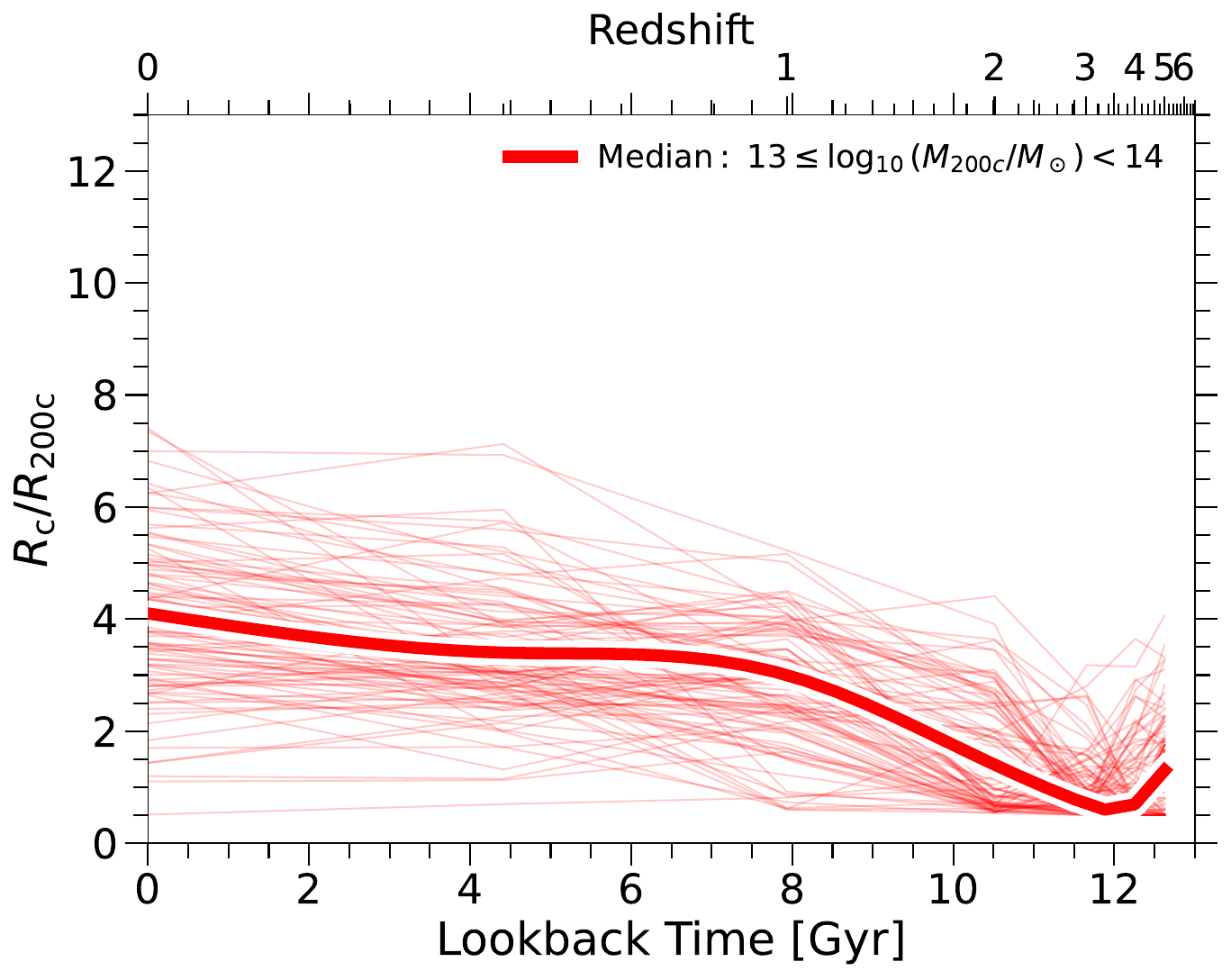}
      \includegraphics[width=0.24\textwidth]{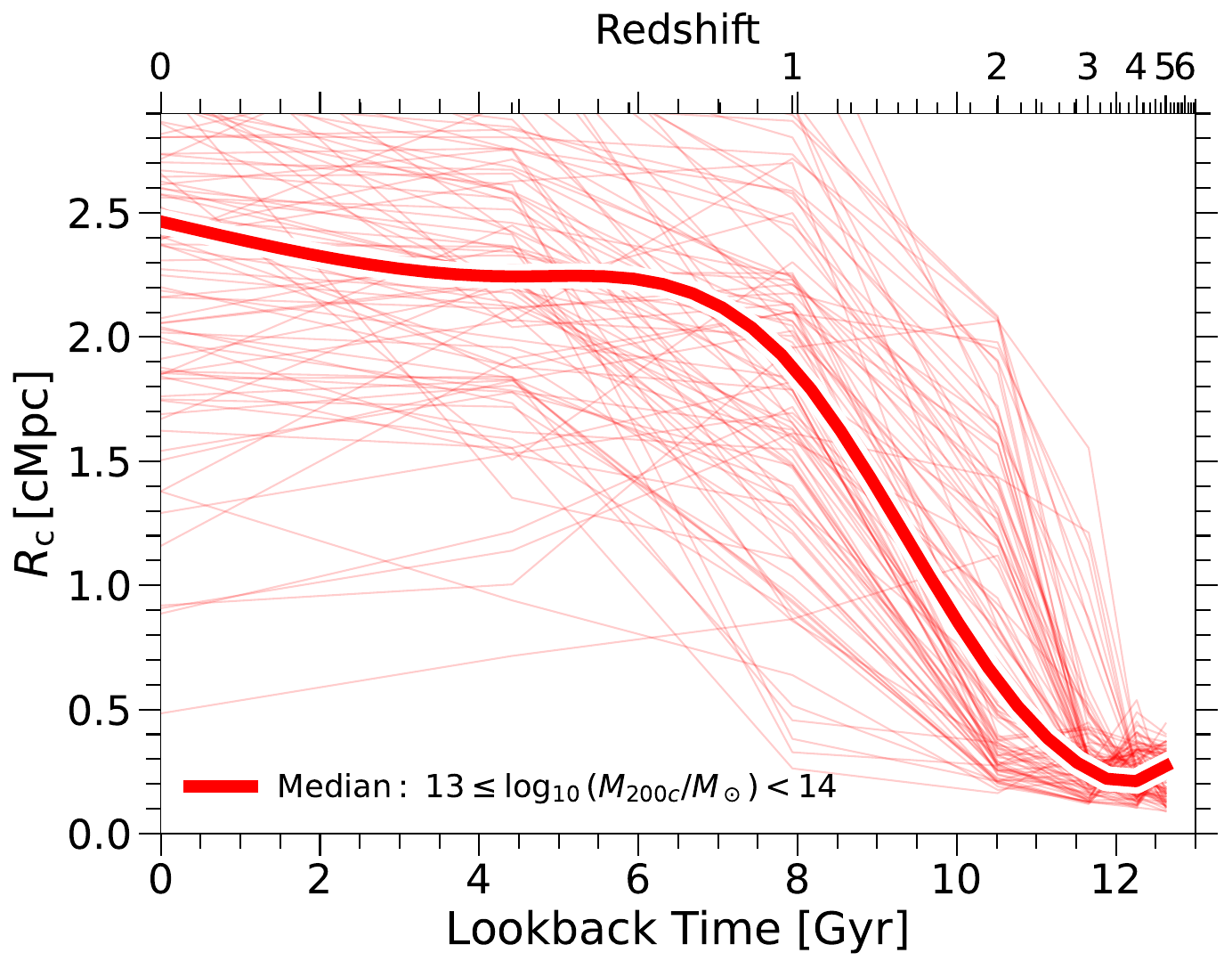}

      \caption{Closure radius along individual halo assembly histories in TNG, as a function of lookback time (lower x-axis) and redshift (upper x-axis). The left panels show the normalized closure radius, $R_{\rm c}/R_{\rm 200c}$, and the right panels the closure radius in comoving units, $R_{\rm c}\,[\rm cMpc]$. Within each column, panels run over three bins in present-day halo mass, increasing from $10^{11}$--$10^{12}\,\mathrm{M_\odot}$ to $10^{13}$--$10^{14}\,\mathrm{M_\odot}$. Thin lines follow individual main progenitors and thick lines their median.}
\label{Fig: Rc_vs_time_app}
\end{figure}

\begin{figure}
      \centering
      \includegraphics[width=0.24\textwidth]{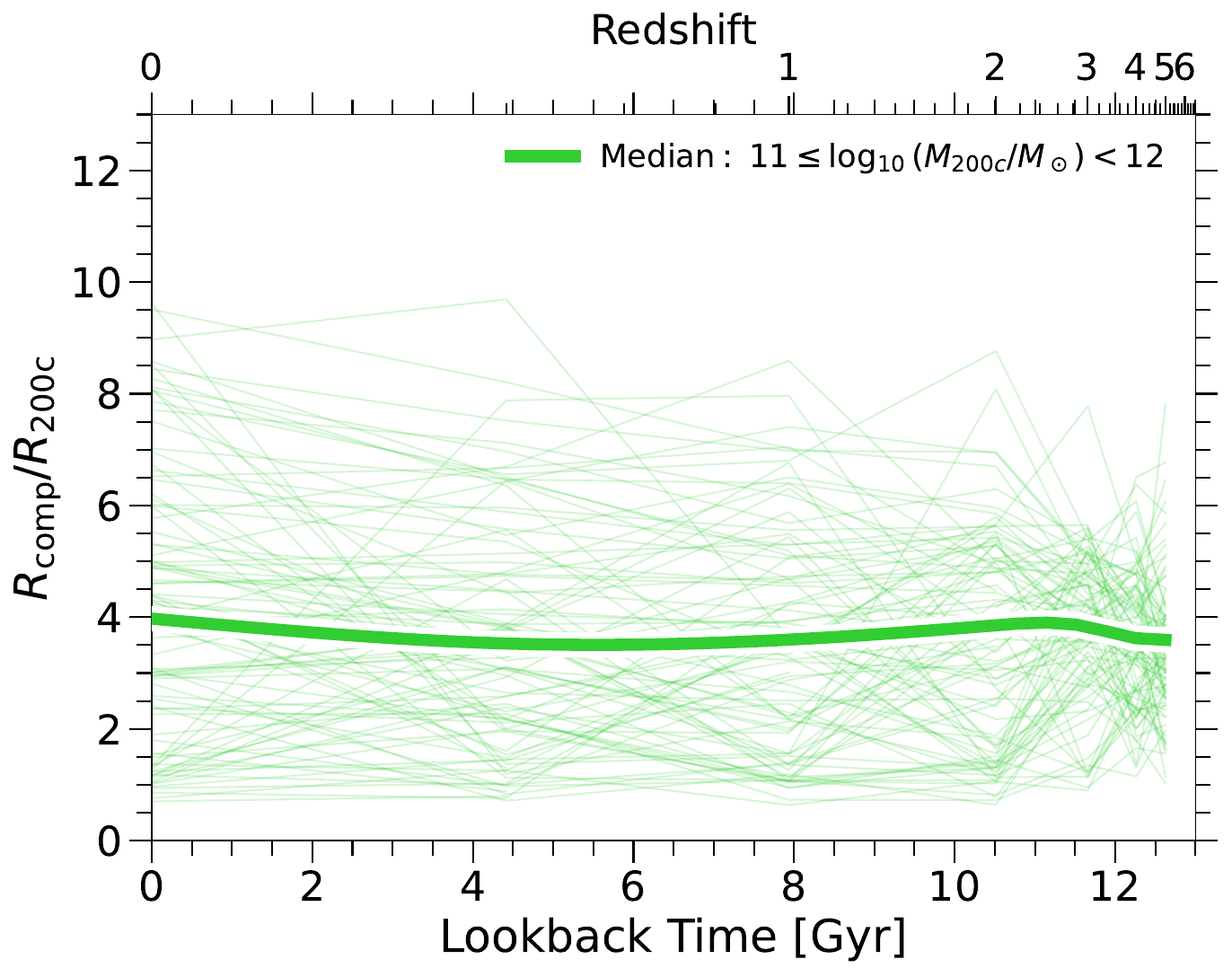}
      \includegraphics[width=0.24\textwidth]{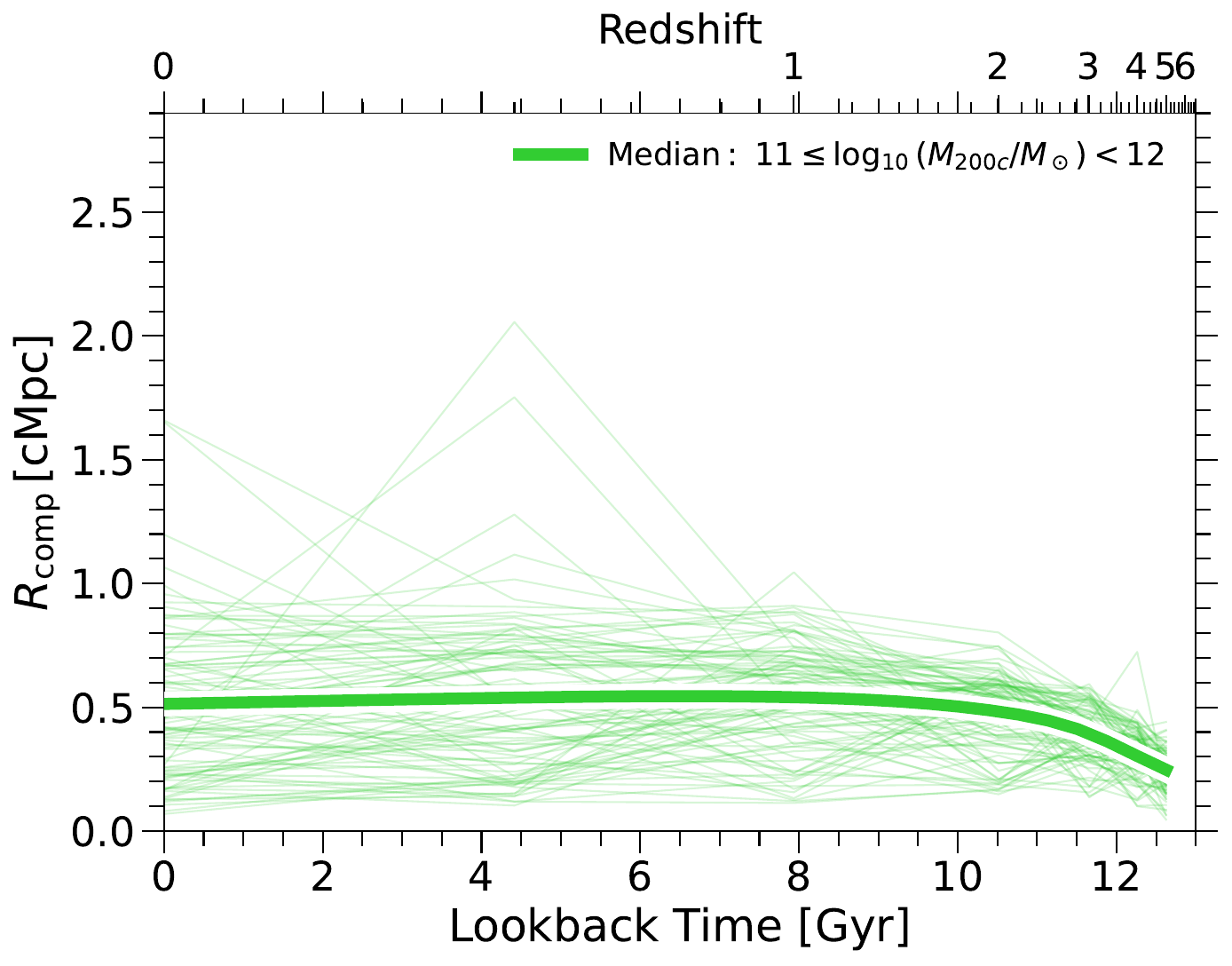}

      \includegraphics[width=0.24\textwidth]{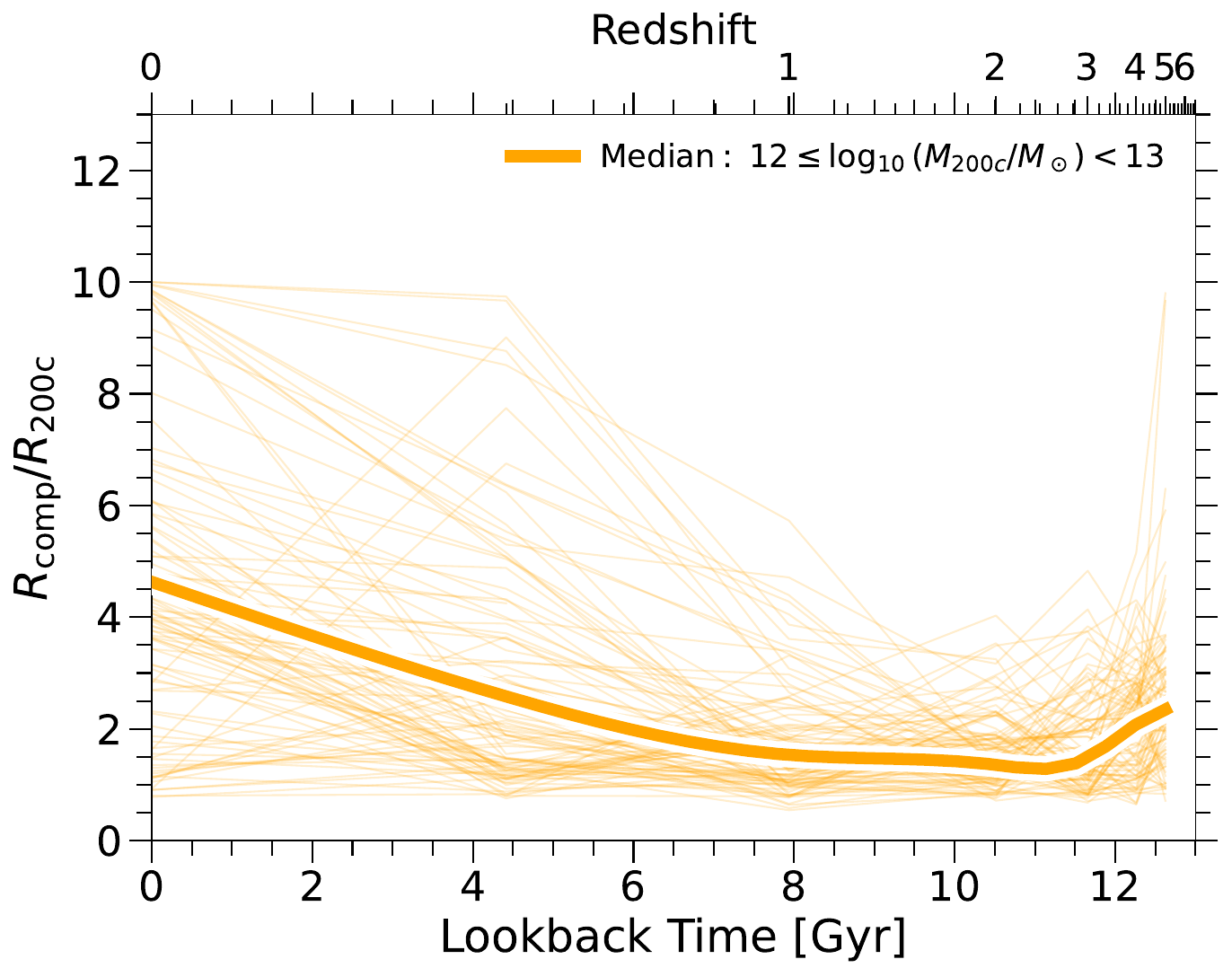}
      \includegraphics[width=0.24\textwidth]{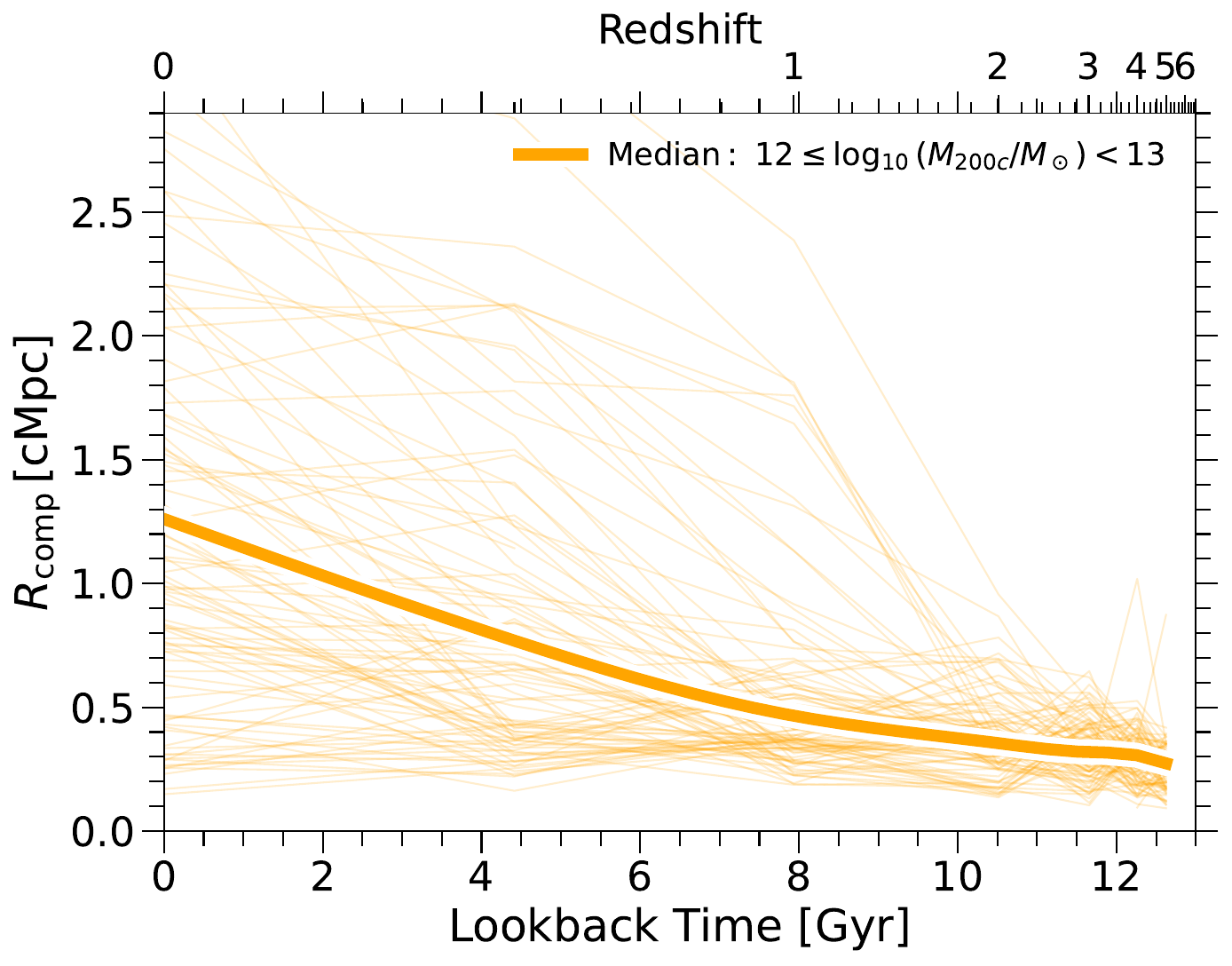}

      \includegraphics[width=0.24\textwidth]{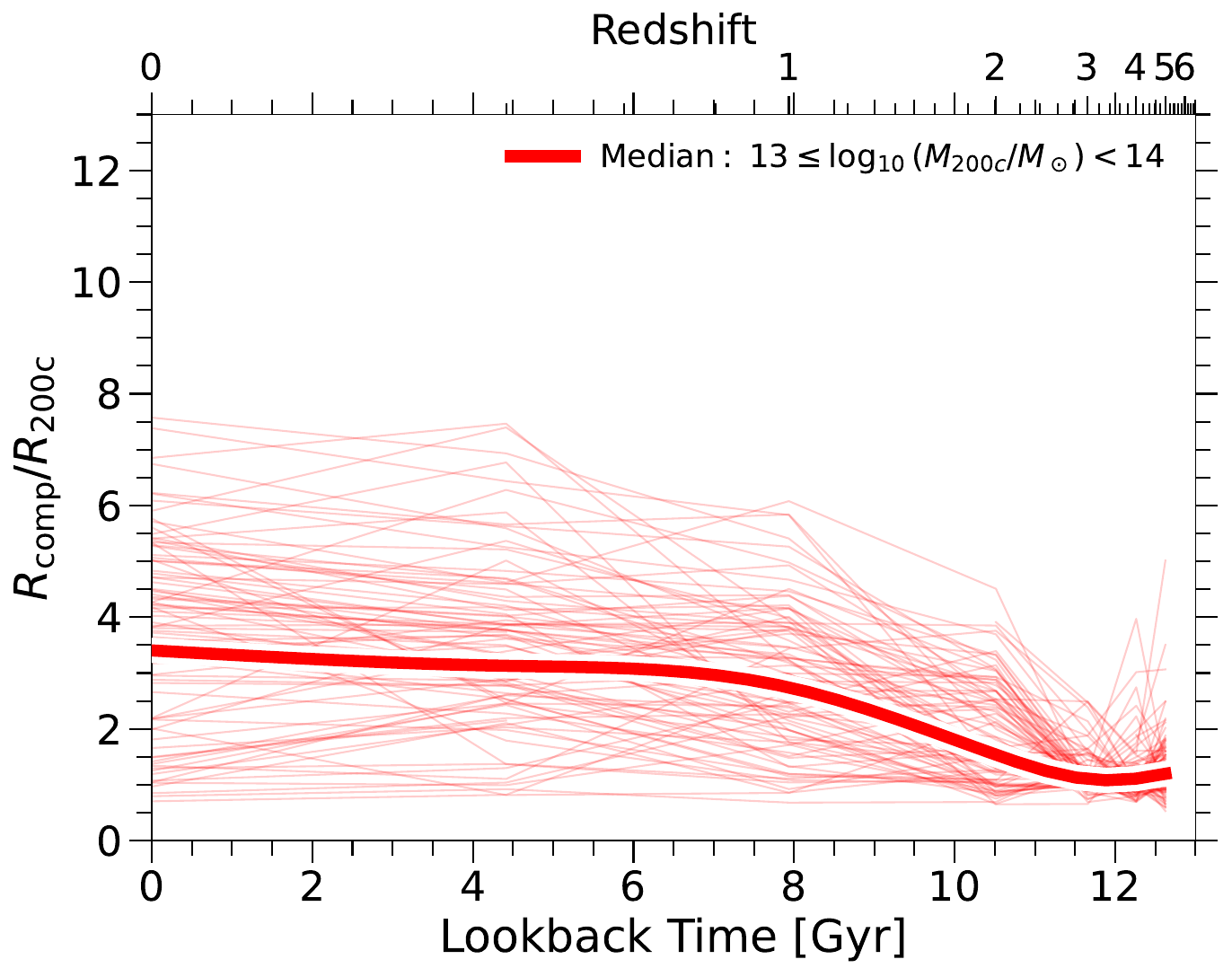}
      \includegraphics[width=0.24\textwidth]{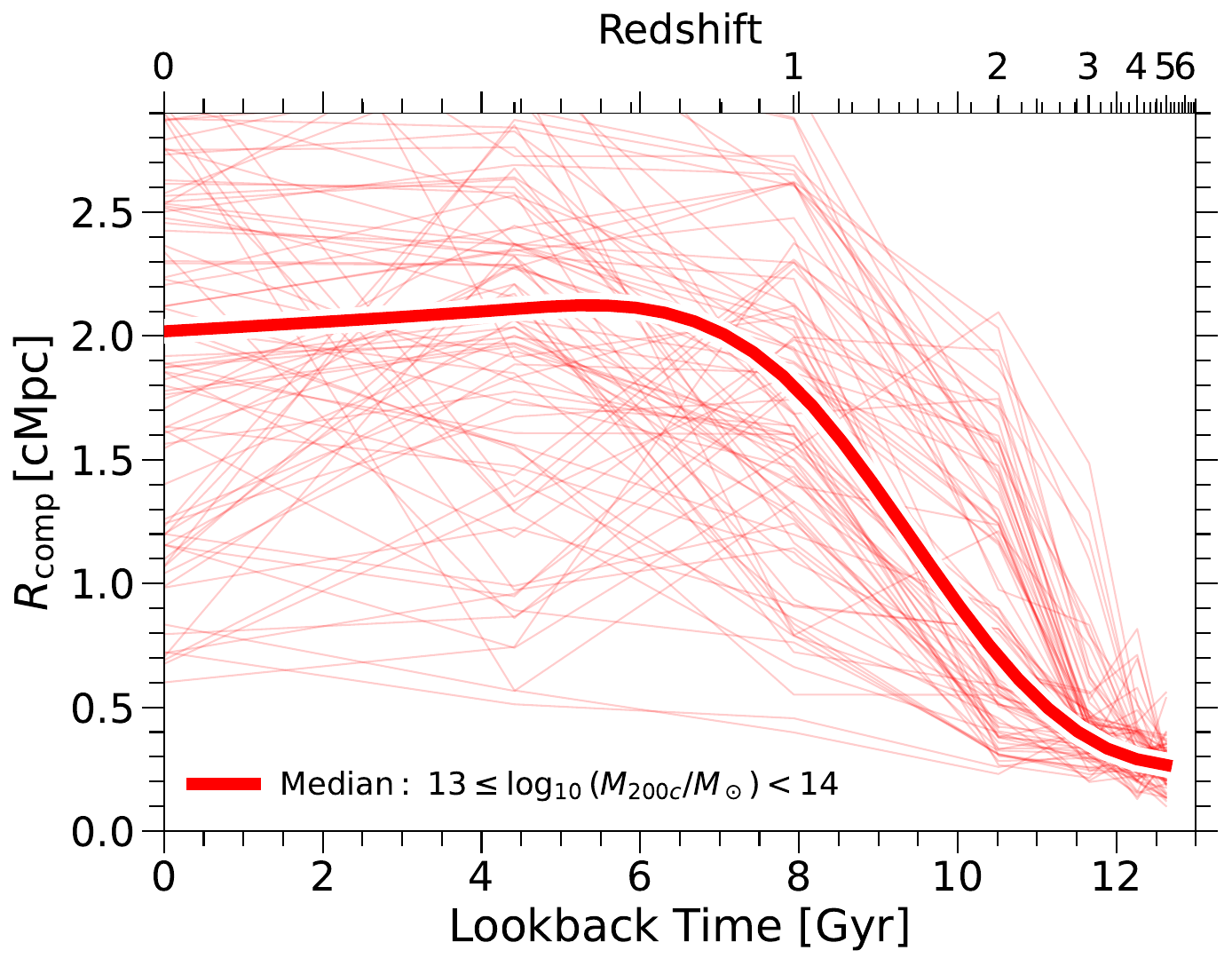}

      \caption{Compensation radius along individual halo assembly histories in TNG, as a function of lookback time (lower x-axis) and redshift (upper x-axis). The colors, layout, and line styles match those of Fig.~\ref{Fig: Rc_vs_time_app}.}
\label{Fig: Rcomp_vs_time_app}
\end{figure}

Figs.~\ref{Fig: Rc_vs_time_app} and \ref{Fig: Rcomp_vs_time_app} show the closure and compensation radii along individual main-progenitor branches, respectively, and complement the medians presented in Fig.~\ref{Fig: Rc_tree}. These figures aim to illustrate the diversity of individual halo histories and are plotted for the TNG simulation, as an example. We have verified that the same qualitative behavior is present in EAGLE too.

Individual histories are considerably more variable than those medians suggest. Within a single mass bin, individual halos show large variability, and many move across a large part of the plotted range between consecutive snapshots. This reflects the sensitivity of the closure radius to the instantaneous distribution of baryons around the halo: mergers, neighboring systems and individual feedback episodes can all move material across the radius at which the enclosed baryon fraction reaches the cosmic value. The medians are correspondingly much smoother, and it is these that we use throughout the main text.

\end{appendix}
\end{document}